\documentclass[12pt, b5paper,twoside]{book}
\usepackage[utf8x]{inputenc}
\usepackage[english]{babel}
\usepackage[b5paper]{geometry}
\usepackage{graphicx}

\usepackage{amsmath}
\usepackage{amsfonts}
\usepackage{amssymb}
\usepackage{xcolor}  %
\usepackage{bbding}
\usepackage{tikz}
\usetikzlibrary{arrows,positioning,shapes.geometric}
\usepackage{mwe}  %
\usepackage[export]{adjustbox}  %
\usepackage[frozencache]{minted}  %

\newcommand{\splitatcommas}[1]{%
  \begingroup
  \ifnum\mathcode`,="8000
  \else
    \begingroup\lccode`~=`, \lowercase{\endgroup
      \edef~{\mathchar\the\mathcode`, \penalty0 \noexpand\hspace{0pt plus 1em}}%
    }\mathcode`,="8000
  \fi
  #1%
  \endgroup
}

\newcommand{\set}[1]{\{\splitatcommas{#1}\}}

\usepackage{microtype}
\microtypesetup{protrusion=false}
\usepackage{hyperref}
\usepackage{breakcites}  %

\usepackage[autostyle, threshold=1]{csquotes}  %

\usepackage{accents}
\newcommand\tbar[1]{\accentset{\rule{.4em}{.8pt}}{#1}}

\usepackage{url}
\usepackage{pdflscape} %
\usepackage{afterpage}
\usepackage{rotating}
\usepackage{caption}
\usepackage{subcaption}
\usepackage{dcolumn}
\usepackage{booktabs} %
\newcommand{\vect}{\textbf}

\newcommand{\dint}{\mathrm{d}}  %
\newcommand{\cost}{C}

\newcolumntype{d}[1]{D{.}{.}{#1}}{}
\newcommand\mc[1]{\multicolumn{1}{c}{#1}}

\usepackage{makeidx}
\makeindex

\title{Wavelet-based spatial audio framework}
\author{Author: Davide Scaini}

\begin{document}
    \setlength{\lineskiplimit}{2pt}\setlength{\lineskip}{3pt} %

    \frontmatter
    \maketitle
    \cleardoublepage

    \noindent
    \begin{center}
        In theory there is no difference\\ between theory and practice;\\in practice there is.
    \end{center}
    \cleardoublepage
    \noindent
    \section*{\Large \sffamily  Acknowledgments}

    Manu for being part of this journey, with her personality and the desire to experiment every day how to live our thing. We have plans.

    Dani, for his guidance and persistence.
    Someone once said that only two things are infinite, the Universe and Dani's patience.
    I have extensive experimental data for the second.
    I owe you a lot.
    More than I can express.

    Barcelona Media for giving me the opportunity to start a new life in a different field in a wonderful city.

    Thanks to Ricardo for the thorough review of this manuscript.

    Dolby, in the persons of Toni and Claus, for believing in me and gently pushing me towards this goal.
    A sincere and happy ``thank you!'' is in order.
    I am very grateful.

    \vspace{2cm}
    \begin{center}
    \includegraphics{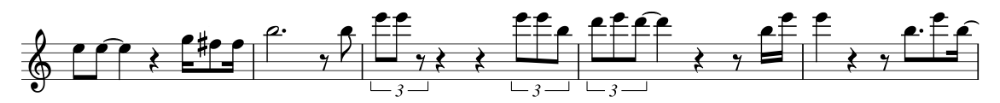}
    \end{center}

    \cleardoublepage

    \section*{\Large \sffamily Abstract}
    Ambisonics is a complete theory for spatial audio whose building blocks are the spherical harmonics.
    Some of the drawbacks of low order Ambisonics, like poor source directivity and small sweet-spot,
    are directly related to the properties of spherical harmonics.
    In this thesis we illustrate a novel spatial audio framework similar in spirit to Ambisonics
    that replaces the spherical harmonics by an alternative set of functions with compact support: the spherical wavelets.
    We develop a complete audio chain from encoding to decoding,
    using discrete spherical wavelets built on a multiresolution mesh.
    We show how the wavelet family and the decoding matrices to loudspeakers can be generated via numerical optimization.
    In particular, we present a decoding algorithm optimizing acoustic and psychoacoustic parameters
    that can generate decoding matrices to irregular layouts for both Ambisonics and the new wavelet format.
    This audio workflow is directly compared with Ambisonics.

    {\chapter*{Preface}
    It was 2012 when I started my Ph.D. in the Department of Information and Communication Technologies at the University Pompeu Fabra in Barcelona.
    At that time Dolby and DTS were launching their solutions for object-based audio, with Dolby Atmos and DTS X.
    At the same time Auro 3D took the opposite direction, making its bet on a new channel based format.
    The Academia was (and still is) more focused on sound-field reconstruction methods, like Ambisonics is.
    In 2012 Virtual Reality (VR) was not yet a trend, the Google Cardboard was launched two years later, in June 2014.
    Facebook and Google lead the dissemination of 3D video
    and ultimately 3D audio, making use of Ambisonics for binaural rendering.
    This was finally the point when the (mass-scale) industry met academia.

    Object-based formats are generic representations of sound scenes, and are a powerful tool for 3D soundscape creation.
    Given their rendering-agnostic construction, they can be plugged to almost any spatial audio rendering technology,
    from amplitude panners to sound-field reconstruction methods.
    Given this scenario, we felt that there was still room for improvement in existing technologies,
    making them more robust and easy to use, from an acoustic point of view.

    We started with Ambisonics and identified that the stage of decoding to speakers was still an open problem for experimenting with it.
    We decided to make a new decoder, that built on well known psychoacoustic principles, but at the same time bent slightly
    the dogmas of Ambisonics, like constant $r_E$ modulus and null $r_E$ transverse component, to get a better sounding rendition.
    And we released it as open-source software (2013).

    \sloppy Approximately at the same time Aaron Heller released his Ambisonics Toolbox (2014), that is a collection of tools to easily generate Ambisonics decoders.

    After almost 40 years from the creation of Ambisonics by Michael Gerzon, the researchers were (and are) still working some of its edges.

    Nevertheless, this research was not addressing the core drawbacks of Ambisonics,
    and we started looking for some alternatives to Spherical Harmonics.
    It was Pau Gargallo that suggested to look into spherical wavelets, since he was familiar with Schr{\"o}der and Sweldens' work.
    As a particle physicist I was not familiar with wavelets at all, and I started wandering into this new (for me) world.
    We started from the very early works of Haar, Gabor, Morlet, Meyer, Mallat all the way to Daubechies, then moved to
    the spherical manifold with the works of Wiaux and McEwans, looking at different sampling theorems on the sphere,
    and then Christian Lessig with his SOHO wavelets... and finally we landed where everything started (at least for us) with Schr{\"o}der and Sweldens.

    The final concept we developed allows to encode sound sources to a cloud of points and to reduce (or recover) the dimensionality of the cloud at will.
    The spatial downsampling is implemented as a linear transformation that can be fully reverted.
    This construction allows for different coexisting spatial representations, that can scale based on various requirements,
    for example transmission bandwidth or the complexity of the destination playback system.
    We call this construction a ``framework'', that is used to generate actual audio formats.
    In this thesis we illustrate the general framework and one special format designed to be compared with and evaluated against Ambisonics.

    Interestingly, for the decoding to speakers of this new format, the same principles used for Ambisonics decoding apply.

    We then tried to push the idea further, by generating our own spatial-audio-oriented wavelets.
    The idea was to numerically optimize the wavelets for some observables, e.g.\ pressure preservation across the down/upsampling process.
    This literally took years...

    It has been a long journey with many dead-ends, but we think we found something interesting and new in the spatial-audio field,
    that may inspire old and young researchers.
    }

    \cleardoublepage
    \tableofcontents

    \listoffigures
    \addcontentsline{toc}{chapter}{List of Figures}

    \listoftables
    \addcontentsline{toc}{chapter}{List of Tables}

    \mainmatter

\chapter{Introduction} \label{ch:intro}
This thesis is a first point of connection between two worlds,
the world of \emph{wavelets}, which is related to time-frequency analysis and signal processing,
and the world of \emph{spatial audio}, which embraces sound perception, sound recording, encoding and reproduction.

Section~\ref{sec:spatial-audio} briefly describes spatial audio,
Section~\ref{sec:spat-tech-prosandcons} gives some context about the available technologies.
Section~\ref{sec:why-this-thesis} outlines the motivations for this thesis.
Section~\ref{sec:contributions} lists the original contributions resulting from this thesis.
The final Section~\ref{sec:outline} describes the outline of the thesis.

\section{What is Spatial Audio?} \label{sec:spatial-audio}
Among the five senses that humans can experience,
\emph{hearing} or audition is the sense of sound perception.
Humans are able to identify the location of a sound in direction, distance and size.
\emph{Spatial audio} refers to the set of tools, technologies and theories for
creation or recreation of a subjective sound scene,
that has to produce all the spatial characteristics of a sound located in a 2D/3D space:
\emph{direction}, \emph{distance} and \emph{size}.
It is possible to classify the techniques to (re)create an auditory scene (2D or 3D) in three categories:
\begin{enumerate}
    \sloppy
    \item \emph{Discrete panning techniques}
    (e.g.~VBAP, ABAP, VBIP, ABIP):
    the known apparent direction of the source is used to feed a limited number of loudspeakers.

    \item \emph{Sound field reconstruction methods}
        (e.g.~Ambisonics, Wave Field Synthesis):
        the intent is to control the acoustical variables of the sound field (pressure, velocity) in the listening space.

    \item \emph{Head-related stereophony} (binaural, transaural):
        the aim is to measure (binaural recording) or (re)produce (binaural synthesis)
        the acoustic pressure at the ears of the listener.
\end{enumerate}

Besides the underlying theory of each technology,
the spatial audio techniques can be also classified
by analyzing how the whole encoding/decoding pipeline is structured:
\begin{itemize}
    \item \emph{Channel-based}:
        the whole encoding/decoding and recording/repro\-duc\-tion is based on a specific channel layout,
        e.g.~2.0, 5.1, 7.1, ..., Auro3D, Hamasaki 22.2.

    \item \emph{Layout-independent} (channel-agnostic):
        the recording and encoding format is independent from the reproduction layout
        (includes sound field reconstruction methods and object-based formats).
\end{itemize}

Malham~\cite{malham1999homogeneous} gives an interesting perspective on the existing (at that time)
surround sound systems, but also gives two criteria that can be applied to any existing or future technology:
the ideas of homogeneous and coherent sound reproduction systems.
Quoting from~\cite{malham1999homogeneous}:

\blockquote{
``An \emph{homogeneous} sound reproduction system is defined as one in which no direction is preferentially treated.
A \emph{coherent} system as one in which the image remains stable if the listener changes position within it,
though the image may change as a natural soundfield does.''
}

With this set of properties we can start categorizing the existing technologies.
For example: VBAP is a channel-based discrete panning technique and in general is not homogeneous.
Ambisonics is a theory that aims at reconstructing the sound field, is layout-independent, is coherent and homogeneous.

Why there is this variety of techniques?
Why not one method that works in all possible conditions?
Each technology has its strengths and weaknesses and a specific area of application.

\section{Benefits and Limitations} \label{sec:spat-tech-prosandcons}
In the following, we will give a non exhaustive list
of benefits and limitations of the mentioned techniques.
We are aware that every point in these lists of \emph{pros} and \emph{cons} is a simplification, that is open for discussion
and could generate many distinctions.
The detailed description and dissection of each existing technique is out of the scope of this introduction.

 \paragraph{Discrete panning techniques}
Pan-pot or stereo amplitude panning is a technique that, by changing the amplitude of a signal in a pair of loudspeakers,
is able to generate a virtual sound source along the arc connecting the two speakers.
The position of the virtual sound source depends on the difference in amplitude between the two speakers.
Vector-Base Amplitude Panning (VBAP) is an extension of the classical stereo pan-pot to multi-speakers layouts (2D and 3D)
and was introduced by Ville Pulkki~\cite{pulkki1997vbap}.
Many variants of this technique are available, for example: Vector-Base Intensity Panning (VBIP), Distance-Based Amplitude Panning (DBAP) \cite{Lossius2009DBAPD}.
Some of the advantages and disadvantages of these techniques are:
\begin{itemize}
    \item[+] Simple to implement.
    \item[+] The most common.
    \item[+] Easily to handle non-uniform layouts.
    \item[+] The experience holds decently also outside the sweet spot (true for layouts with many loudspeakers).
\end{itemize}
\begin{itemize}
    \item[--] Not a complete theory, e.g.\ recording in VBAP it is not possible.
    \item[--] Not homogeneous and not coherent.
    \item[--] Jump of the virtual source from speaker to speaker
            (apparent source size changes as a function of the virtual source position).
\end{itemize}

 \paragraph{Ambisonics}
Ambisonics was invented by Michael Gerzon of the Oxford Mathematical Institute who developed the theoretical
and practical aspects of the system in the early 1970s.
Ambisonics comprises both encoding, recording and reproduction (decoding) techniques that can be used live or in studio
to present a 2-dimensional (\emph{planar}, or horizontal-only) or
3-dimensional (\emph{periphonic}, or full-sphere) sound field~\cite{ambiboook}.
There are professional quality commercial microphones which can directly record
in first order Ambisonics.
In recent years higher order microphones have appeared,
e.g.~Eigenmike~\cite{eigenmike} and Zylia~\cite{zylia},
that enable the encoding of an Higher Order Ambisonics soundscape.
Some of the advantages and disadvantages of Ambisonics are:
\begin{itemize}
    \item[+] Nice physical formulation.
    \item[+] Aims to reproduce the sound field.
    \item[+] Is homogeneous and coherent.
    \item[+] Getting a lot of traction in Augmented/Virtual Reality (AR/VR) (for binaural reproduction over headphones),
            but has a minor role in the spatial audio industry as a whole.
\end{itemize}
\begin{itemize}
    \item[--] Poor localization at low orders.
    \item[--] Small sweet spot at low orders.
    \item[--] Difficult to handle irregular layouts. %
    \item[--] Not so common: reproduction over speakers mostly limited to universities and research centers.
    \item[--] Sound source distance is not included in basic the theory,
            nevertheless there are implementations which enable distance encoding~\cite{daniel2003spatial}.
\end{itemize}

 \paragraph{Wavefield Synthesis}
Wavefield Synthesis (WFS) was introduced in the 1980s by Dr. A. J. Berkhout~\cite{berkhout1988a},
a professor of seismics and acoustics at the Delft University of Technology.
WFS is a theory that allows to reconstruct a wave field within a volume given
the acoustic variables at its boundary~\cite{Berkhout1993, Ahrens2012}.
WFS exhibition systems typically focus on reproduction only, and the reproduced signals come from audio objects.
The typical WFS layouts are 2-dimensional, horizontal, and can be linear of circular.
One advantage over Ambisonics is that the distance of the source is embedded in the theory.
Some of the advantages and disadvantages of WFS are:
\begin{itemize}
    \item[+] Nice physical formulation.
    \item[+] Aims to reproduce the sound field.
    \item[+] Is homogeneous and coherent.
    \item[+] Sound source distance is embedded in the theory.
\end{itemize}
\begin{itemize}
    \item[--] Very limited diffusion, less common than Ambisonics (limited mostly to Universities and research centres);
                currently almost zero presence in the spatial audio industry.
    \item[--] Expensive and difficult to set up (speaker layouts have to be specifically tailored to WFS).
    \item[--] Computationally intense.
\end{itemize}

 \paragraph{Head-related stereophony}
The techniques inside the realm of the head-related stereophony go from binaural~\cite{rumsey2017binaural}
(recording with in-ear microphones and reproduction over headphones),
to transaural~\cite{bauck1996generalized} (reproduction over speakers), to generic `virtualization' techniques (e.g.~virtual surround).
All these techniques exploit the information about head and outer-ear (pinna) shape to create the right pressure at the eardrums
to simulate a sound in space.
Some of the advantages and disadvantages of these techniques are:
\begin{itemize}
    \item[+] It can be better than plain stereo over headphones.
    \item[+] Homogeneous (depending on the specific implementation).
    \item[+] Coherent, for synthesized sound fields with head-tracking.
    \item[+] Inexpensive, in terms of production, transmission and exhibition.
\end{itemize}
\begin{itemize}
    \item[--] Extremely complex to get an experience that works for everyone without tuning and training.
    \item[--] Once produced, it is very difficult to modify and manipulate without destroying the experience.
\end{itemize}

 \paragraph{Object-based formats}
In object based audio formats at each sound source is associated an audio track and some characteristics (metadata),
that can be position, distance, size, and possibly more~\cite{parmentier2015sound}.
Examples of commercial object-based formats are DTS X and Dolby Atmos.
This kind of technologies are by construction layout-independent, since the rendering stage is (or can be) completely disconnected from the format itself.
This means that one could choose different rendering techniques for the same object-based format.
Some of the advantages and disadvantages of object-based audio are:
\begin{itemize}
    \item[+] Naturally handles any type of speaker layout (number and position of speakers).
    \item[+] Rendering is separate from the encoding and transmission stage, so in principle any type of rendering can be used.
\end{itemize}
\begin{itemize}
    \item[--] The number of channels to be transmitted/stored depends on the number of objects (audio sources).
    \item[--] Since the rendering is done in real-time, the computational load scales with the number of objects.
\end{itemize}

\subsection{Differences in audio workflow}
The typical audio workflow can be simplified in three stages: encoding, transmission and decoding (or playback).
Each technology can attach a different meaning to each of these three stages.
Here we intend the encoding (or decoding) as `spatial encoding', i.e.\ the tools to encode a spatial signal into some format,
and not the encoding to (or decoding from) a bitstream.
In broad terms, and looking only at encoding and decoding, we will give some examples to clarify the differences in the audio workflow for different techniques.

In channel based formats the encoding involves the use of some panning law, that translates the spatial position
of a source into gains of that signal for each channel of the format.
The panning law defines the format.
The decoding stage is typically just the direct playback to speakers, since in a channel based format the channels correspond to actual speaker positions.

In channel agnostic (but not object based) formats the encoding format translates the three-dimensional position of the source to some other space,
and the gains are the weights of the functions or filters that map the 3D space to the new one.
In the decoding the process is reversed, and these `abstract' channels regain their meaning in the space of positions.

In object based formats the role of encoding and decoding is essentially reversed with respect to the channel based formats.
There is no spatial encoding, and the task of generating the gains for each speaker is given to the decoding stage,
where a panning law is used at playback time to convert the object metadata into gains.
Here the panning law does not define the format, and it is possible to use different panning laws for the same object based format,
providing that the panning law `understands' the object metadata.

\section{Motivation} \label{sec:why-this-thesis}
We have seen in Section~\ref{sec:spat-tech-prosandcons} that there are several techniques for spatial audio and each of them has its purpose.
The choice of the best technique is application dependent.
An interesting evaluation of stereophony, Ambisonics and WFS in the context of spatial music can be found in \cite{endabates}.
There is no `absolute best' solution that works for every condition and context.
This thesis rises from this realization and the consequent question:
\blockquote{
Is it possible to build a theory for spatial audio that is
channel agnostic, homogeneous and coherent, but also has good localization with few channels, easily handles irregular layouts
and holds well when moving out of the sweet spot?
In other words, \emph{is it possible to build a theory that combines the best of channel-based and channel-agnostic worlds?}
}

The question already sets some requirements and a focus on the problem.
A channel-agnostic format is preferred, since there are already many channel based formats that span a wide portion of the localization spectrum,
from the low channel count and limited localization like stereo 2.0 and 5.1, to the very high channel count of Hamasaki 22.2.
Moreover, the channel based formats are not flexible nor future proof.
In the realm of channel agnostic formats there are technologies like WFS and Ambisonics.
WFS is not suited to irregular layouts and typically uses a large number of speakers.
Ambisonics has a reasonable number of channels but it is not trivial to decode to irregular layouts and the sweet spot can be quite limited at low orders.
The Ambisonics channels can be understood as a series expansion of
the distribution of sources in terms of spherical harmonics (SH) \cite{jd-phd}:
the higher the order in the expansion, the more spatial detail and the bigger the sweet spot.
Each Ambisonics coefficient corresponds to a SH function.
The properties of SH are a key element in how Ambisonics works and sounds:
all SH have significant non-vanishing support in all points of the sphere
(except for a finite set of points),
and moreover the SH are completely delocalized,
meaning that it is not possible to assign
an individual spatial location to any SH by itself.
This implies that at low orders almost all speakers
contribute significantly to create a virtual source.
These properties of SH translate directly into the subjective characteristics of
Ambisonics as a spatial audio format, which is often reported to be smooth, diffuse and immersive,
but also confuse, imprecise, and delocalized.

In this thesis we develop a new spatial audio codification, similar in spirit
to Ambisonics but replacing the SH by a different and more localized, set of functions: the \emph{spherical wavelets} \cite{Schroeder95}.
The goal is to get better localization and a larger sweet spot with few channels,
that can be easily decoded to irregular layouts.

In this context, we encounter a first gap to fill: successfully decode Ambisonics and Higher Order Ambisonics to irregular layouts.
Part~\ref{part:ambisonics} focuses on this task.

Part~\ref{part:wavelets} is dedicated to the world of wavelets and comprises an introduction (Chapter~\ref{ch:wavelet-theory}),
the description of a wavelet based spatial audio format (Chapter~\ref{ch:wavelet-audio-chain}),
and finally our special wavelet transform (Chapter~\ref{ch:our-wavelets}).

Part~\ref{part:evaluation} is dedicated to the evaluation of the new format.

This thesis sets a new approach to spatial audio encoding, that bridges the channel-based approaches with the channel-agnostic ones,
widening the spectrum of existing spatial audio methods, possibly generalizing and incorporating already existing theories.

\begin{table}[t]
    \centering
    \begin{tabular}{cccc}
        \toprule
        Technology      & Encoding & Transmission & Decoding \\ \midrule
        Ambisonics      & \XSolidBrush   & \XSolidBrush   & \CheckmarkBold   \\
        Wavelet Framework  & \CheckmarkBold  & \CheckmarkBold  & \CheckmarkBold  \\
        \bottomrule
    \end{tabular}
    \caption{Table of contributions.} \label{tab:contributions}
\end{table}

\section{Outline}  \label{sec:outline}
After a first introductory Chapter~\ref{ch:intro}, the thesis est omnis divisa in partes tres:  %
Part~\ref{part:ambisonics}, on Ambisonics; Part~\ref{part:wavelets}, on wavelets and the wavelet spatial audio framework we designed,
and Part~\ref{part:evaluation}, describes an incarnation of this wavelet framework into a spherical audio format
which is evaluated against Ambisonics.

Part~\ref{part:ambisonics} is composed of three Chapters,
the first summarizes the background information and the following two describe the original contributions.
Chapter~\ref{ch:ambi-theory} briefly describes Ambisonics giving the basis for encoding and decoding Higher Order Ambisonics.
Chapter~\ref{ch:idhoa} is an original contribution to Ambisonics' decoding to irregular speakers' layouts.
Initially, we describe the physical and psychoacoustical variables to design the Ambisonics decoder.
We formulate the problem as an optimization problem, so later we define the optimization's cost function.
Finally we show the resulting performance of the decoder for a specific layout of speakers.
In Chapter~\ref{ch:idhoa-evaluation} we evaluate the performance of our decoder against some publicly available ones,
both objectively and subjectively.

Part~\ref{part:wavelets} is composed of three chapters,
the first summarizes the background information and the following two describe the original contributions.
Chapter~\ref{ch:wavelet-theory} is an introduction to Wavelet Theory, starting from a comparison with Fourier Transform
and then fast-forwarding into multiresolution, the Lifting Scheme and a construction of Spherical Wavelets.
Chapter~\ref{ch:wavelet-audio-chain} describes a method to generate spherical audio formats using wavelets built on a multiresolution mesh.
Chapter~\ref{ch:our-wavelets} illustrates a numerical method to obtain spherical wavelet filters optimized for spatial audio purposes.

The last Part~\ref{part:evaluation} is composed of three chapters.
Chapter~\ref{ch:comparison-swf-internals} evaluates different versions of Spherical Audio Formats defined by different wavelet families.
Chapter~\ref{ch:idhoa-evaluation-swf-ambi} inspects the properties of this new format against Ambisonics for a reference layout of speakers.
Chapter~\ref{ch:results-futurework} draws the conclusions and picture some directions for future work.

In Appendix~\ref{app:ipopt+pytorch} we present a method to implement optimization problems
in Python leveraging autodifferentiation.
In Appendix~\ref{app:dof-reduction} we describe the method used to reduce the dimensionality
of the spherical wavelet optimization problem.
Both Appendices are original contributions.

\subsection{Original Contributions} \label{sec:contributions}
Novel contributions produced in the context of this thesis are:
Chapter~\ref{ch:idhoa} description of a generic method for decoding of linear-encoding formats to irregular layouts,
the evaluation of this approach in Chapter~\ref{ch:idhoa-evaluation}, the new wavelet based spherical audio framework
described in Chapter~\ref{ch:wavelet-audio-chain},
the optimization of the wavelet filters for spatial audio purposes in Chapter~\ref{ch:our-wavelets},
and the whole evaluation, Part~\ref{part:evaluation}, which includes Chapters~\ref{ch:comparison-swf-internals},
\ref{ch:idhoa-evaluation-swf-ambi} and \ref{ch:results-futurework},
is an original contribution as well.

Table~\ref{tab:contributions} shows graphically the contributions to the Ambisonics and Wave\-let format audio chains,
that are the result of this work.

\subsection{Previously published material}
\sloppy The material presented in Part~\ref{part:ambisonics} is an updated version of two already published contributions:
a peer-reviewed  conference proceeding \cite{scaini2014decoding}
and a conference proceeding with a peer-reviewed abstract \cite{scaini2015an}.

Some of the material presented in Part~\ref{part:wavelets} is part of an article already submitted to
the \emph{Journal of the Audio Engineering Society},
with title ``Wavelet based spherical audio format'' focused on the audio workflow of the new wavelet framework \cite{ScainiArteagaWavelet}.
Another article devoted to the wavelet optimization method
is in preparation \cite{ScainiArteagaInPreparation}.

Other material published during the course of the Ph.D. which is not part of this thesis,
is: ``Layout Remapping Tool for Multichannel Audio Productions'' published in May 2013 at the
\emph{134$^{th}$ Audio Engineering Society Convention},
``Measurements of jet multiplicity and differential production cross sections
of Z+jets events in proton-proton collisions at sqrt(s)=7 TeV'' published in August 2014 in \emph{Physical Review D},
and ``Volumetric Security Alarm Based on a Spherical Ultrasonic Transducer Array'' published in May 2015 in \emph{Physics Procedia}.
    \part{Ambisonics} \label{part:ambisonics}

\chapter{Introduction to Ambisonics} \label{ch:ambi-theory}
Ambisonics is a theory for spatial audio recording and reproduction, developed by Michael Gerzon during the 1970s, that
aims at the encoding of the sound field and its accurate reconstruction in a point in space.
From a theoretical point of view, it is possible to define the Ambisonics channels as the coefficients of a
perturbative series expansion in terms of spherical harmonics (SH) of the sound field around the origin.

Zeroth order Ambisonics consists of one channel, the $W$ channel, is the omnidirectional component of the field,
and corresponds to the sound pressure.
First order Ambisonics (FOA) adds the $X$, $Y$ and $Z$ channels,
which are the directional components in three dimensions, and correspond to the three components of the pressure gradient,
which amount to the acoustic velocity at the origin (see Table~\ref{tab:1st-order}).
Together, these components approximate, at first order of the multipole expansion,
the sound field on a sphere around the listening point.
$L$-th order Ambisonics adds other coefficients to the multipole expansion which amount to quantities proportional
to the derivatives (up to $L$-th order) of the pressure field.

Higher Order Ambisonics (HOA) is made of $K=(L+1)^{2}$ channels, where $L$ is the Ambisonics order or spherical
harmonic degree (each $l$ order has $2l+1$ channels).
In the following we will refer to an arbitrary Ambisonics order, including HOA, simply as \emph{Ambisonics}.

Ambisonics has a series of remarkable properties, which make it a good format for spatial audio.
First, it is a complete theory of spatial audio, going from recording to reproduction.
Second, it is based on solid acoustic and mathematical grounds. %
Third, it is has a fixed number of channels, independently from the number of sound sources (in contrast to object-based methods).
Fourth, it is completely independent of the exhibition layout.
Fifth, it provides a smooth listening experience from all directions.
Finally, depending on the order, it requires only a moderate number of loudspeakers for exhibition,
if compared for example with Wave Field Synthesis.

\begin{table}[t]
\centering
\begin{tabular}{cccc}
\toprule
Ch. \# & (l,m) & Name & physical meaning \\
    \midrule
    0 & (0,0)  & W & pressure \\ \hline
    1 & (1,-1) & Y & particle velocity $y$ dir. \\
    2 & (1,0) & Z & particle velocity $z$ dir. \\
    3 & (1,1) & X & particle velocity $x$ dir. \\
\bottomrule
\end{tabular}
\caption{First Order Ambisonics Channels.}
\label{tab:1st-order}
\end{table}

\section{Encoding Higher Order Ambisonics} \label{sec:encoding-hoa}

The decomposition of a distribution of sources $S(\theta,\phi)$ over a sphere is expressed, in $L$-th order Ambisonics,
\begin{equation}
    \label{eq:ambiChannels}
    S(\theta,\phi;t) = \sum_{l=0}^{L} \sum_{m=-l}^{l} a_{l,m}(t) Y_{l,m} (\theta,\phi)
\end{equation}
where $Y_{l,m} (\theta,\phi)$ are the real-valued spherical harmonics
\footnote{Following the convention in \url{http://ambisonics.ch/standards/channels/glossary}.}
(which form a basis in $\mathcal S^2$) \cite{math-methods},
and $a_{l,m}$ are the coefficients or the projection of $S$ onto the basis $Y$:
\begin{align*}
 a_{l,m}(t) &= \int_{\Omega} S(\theta,\phi;t) Y_{l,m} (\theta,\phi) \dint\Omega \label{eq:ambi_coeffs} \\
		    &= \int_{0}^{2\pi} \int_{-\pi/2}^{\pi/2} S(\theta,\phi;t) Y_{l,m} (\theta,\phi) \cos(\theta) \dint\phi \dint\theta \notag
\end{align*}
where $\dint\Omega$ is the usual measure on the sphere, $\dint\Omega = \cos\phi \dint\phi \dint\theta $
in acoustic coordinates convention $\phi$ is azimuth and $\theta$ is elevation.
The distribution of sources can be linked to the perturbative decomposition of the pressure field around the origin \cite{DanielMoreau03}.
The coefficients $a_{l,m}(t)$ are the so-called \emph{Ambisonics channels}.

A plane wave $S_{\mathbf{\hat{u}}}(\theta,\phi)$, representing a virtual point source coming from direction\footnote{Hat denotes unit vectors in $\mathcal S^2$.}
 $\mathbf{\hat{u}}(\theta_{\mathbf{\hat{u}}},\phi_{\mathbf{\hat{u}}})$, can be approximated in terms of spherical harmonics up to $L$-th degree as:
\begin{equation}
    \label{eq:encoding}
    S_{\mathbf{\hat{u}}}(\theta,\phi) = \sum_{l=0}^{L} \sum_{m=-l}^{l} (a_{\mathbf{\hat{u}}})_{l,m} Y_{l,m}(\theta,\phi) ,
\end{equation}
where the Ambisonics coefficients $ (a_{\mathbf{\hat{u}}})_{l,m} $ are given in the normalization N3D \cite{jd-phd} by:
\begin{equation}
    \label{eq:Channels}
    (a_{\mathbf{\hat{u}}})_{l,m}= g_{\mathbf{\hat{u}}} Y_{l,m}(\theta_{\mathbf{\hat{u}}},\phi_{\mathbf{\hat{u}}}),
\end{equation}
where $g_{\mathbf{\hat{u}}}$ is the amplitude of the plane wave.
The importance of plane waves lies in the fact that any distribution of sources can be represented as a superposition of plane waves.

These expressions can be rewritten in matrix notation.
Equation~\eqref{eq:ambiChannels} can be reexpressed as:
\begin{equation}
S (\theta,\phi)= \mathbf{a} \cdot \mathbf{Y}(\theta,\phi)
\end{equation}
with
\begin{align*}
 \mathbf{a} &= (a_{0,0} \: \ldots a_{1,-1} \: \ldots \: a_{1,0} \: \ldots \: a_{1,1} \ldots \: \\
  & \underbrace{a_{l,-l} \: \ldots \: a_{l,0} \: \ldots \: a_{l,l}}_{2l+1} \: \ldots \: a_{L,-L} \: \ldots \: a_{L,0} \: \ldots \: a_{L,L} ) ^T
\end{align*}
and
\begin{align*}
 \mathbf{Y} &= (Y_{0,0} \: \ldots \: Y_{1,-1} \: \ldots \: Y_{1,0} \: \ldots \: Y_{1,1}  \: \ldots \: \\
& Y_{l,-l} \: \ldots \: Y_{l,0} \: \ldots \: Y_{l,l}  \: \ldots \: Y_{L,-L} \: \ldots \: Y_{L,0} \: \ldots \: Y_{L,L})^{T} \text{.}
\end{align*}
The encoding equation~\eqref{eq:Channels} for a point source can be reexpressed as:
 \begin{equation}
 	\mathbf a_{\mathbf{\hat{u}}} =  g_{\mathbf{\hat{u}}} \mathbf Y(\theta_{\mathbf{\hat{u}}},\phi_{\mathbf{\hat{u}}})
 \end{equation}
Explicitly, with the choice of the N3D convention mentioned above:
\begin{equation*}
    \left\{
        \begin{array}{ll}
            W_{\mathbf{\hat{u}}} = (a_{\mathbf{\hat{u}}})_{0,0} & = S\\
            Y_{\mathbf{\hat{u}}} = (a_{\mathbf{\hat{u}}})_{1,-1} & = S \sqrt{3} \cos(\theta_{\mathbf{\hat{u}}})\sin(\phi_{\mathbf{\hat{u}}}) \\
            Z_{\mathbf{\hat{u}}} = (a_{\mathbf{\hat{u}}})_{1,0} & = S \sqrt{3} \sin(\theta_{\mathbf{\hat{u}}}) \\
            X_{\mathbf{\hat{u}}} = (a_{\mathbf{\hat{u}}})_{1,1} & = S \sqrt{3} \cos(\theta_{\mathbf{\hat{u}}})\cos(\phi_{\mathbf{\hat{u}}}) \\
            \ldots &
        \end{array}
    \right.
\end{equation*}

\section{Decoding Higher Order Ambisonics} \label{sec:decoding-hoa}

\subsection{Basic Ambisonics Decoding}
The \emph{basic decoding} assumes phase coherence among signals emitted by the loudspeakers,
and the requirement is to accurately reconstruct the sound field at the origin up to order $L$ in the Ambisonics decoding,
from superposition of the plane waves emitted by the different loudspeakers in the layout\footnote{Note that it is assumed
that the loudspeakers emit plane waves and are placed at the same distance.
Compensation of level, delay and near-field effect \cite{daniel2003spatial} has to be addressed in another stage of the signal processing.}.
In this case the problem is linear and can be solved analytically with algebraic methods.

With more detail, let us define a set of directions $\Theta = \lbrace \mathbf{\hat{u}}_i \rbrace_{i=1,\ldots,N}$,
where $\mathbf{\hat{u}}_i (\theta_i, \phi_i) \in \mathcal S^2 $ correspond to the position of the loudspeakers, which sample the $K=(L+1)^2$
spherical harmonics up to $L$-th degree:
\begin{align*}
    \mathbf{Y}=(Y_{l_1,m_1} \: \ldots \: Y_{l_L,m_L})^T ,
\end{align*}
in $N$ directions:
\begin{align*}
   \mathbf{y}_{l,m}=(Y_{l,m}(\theta_1,\phi_1) \: \ldots \: Y_{l,m}(\theta_N,\phi_N))^T .
\end{align*}

The decoding equation requests that the original sound field represented by Eq.~\eqref{eq:ambiChannels} is accurately reproduced
up to order $L$ from the plane waves emitted from the different
$ \lbrace \mathbf{\mathbf{\hat{u}}}_i \rbrace_{i =1,\ldots,N}$ loudspeakers' directions, given by~\eqref{eq:encoding}:
\begin{equation}
	 \mathbf a = \sum_{j=1}^N g_j \mathbf Y(\theta_{j},\phi_{j}) ,
    \label{eq:reencoding}
\end{equation}
where $g_i$, the gain of each one of the loudspeakers, is the unknown in the above equation.
At first Ambisonics order this request amounts to reproducing correctly the first four
spherical harmonics,
corresponding to the sound pressure $p$
and normalized acoustic velocity at the origin $\vect v$.
Eq.~\eqref{eq:reencoding} can be reexpressed in matrix form more concisely as:
\begin{equation}
	 \mathbf a = \mathbf{C}\, \mathbf g,
\end{equation}
where the matrix $\mathbf{C} = \lbrace c_{k,j} \rbrace$ represents the sampled spherical harmonics basis
\begin{equation*} \label{eq:encodingAmb}
    \mathbf{C} =
    \begin{pmatrix}
        Y_{0,0}(\theta_1,\phi_1) & \cdots & Y_{0,0}(\theta_N,\phi_N) \\
        Y_{1,-1}(\theta_1,\phi_1) & \cdots & Y_{1,-1}(\theta_N,\phi_N) \\
        Y_{1,0}(\theta_1,\phi_1) & \cdots & Y_{1,0}(\theta_N,\phi_N) \\
        Y_{1,1}(\theta_1,\phi_1) & \cdots & Y_{1,1}(\theta_N,\phi_N) \\
        \vdots  & c_{k,j} & \vdots  \\
        Y_{L,-L}(\theta_1,\phi_1) & \cdots & Y_{L,-L}(\theta_N,\phi_N) \\
        \vdots  & \vdots & \vdots  \\
        Y_{L,0}(\theta_1,\phi_1) & \cdots & Y_{L,0}(\theta_N,\phi_N) \\
        \vdots  & \vdots & \vdots  \\
        Y_{L,L}(\theta_1,\phi_1) & \cdots & Y_{L,L}(\theta_N,\phi_N) \\
    \end{pmatrix}
    _{\substack{
    k=1,\ldots,K;\\
    i=1,\ldots,N}}
\end{equation*}
and $\mathbf g = (g_1 \ldots g_N)^T$ is the vector of gains.  %

The decoding matrix $\mathbf D$ is the matrix giving:
\begin{equation}
	 \mathbf g = \mathbf{D}\, \mathbf a.
\end{equation}
In most realistic cases the system is under-determined, given that the number of loudspeakers $N$
is generally greater than the number of Ambisonics channels $K$.
From all the possible solutions to the system, the pseudoinverse is the one that minimizes the energy emitted \cite{jd-phd}:
\begin{equation}
    \mathbf{D} =   \mathbf{D}_\text{pinv} = \mathbf{C}^T (\mathbf{C} \mathbf{C}^T)^{-1} .
    \label{eq:pinv}
\end{equation}

Eq.~\eqref{eq:pinv} represents the general solution for the basic decoding.
However, it is to be noted that in highly irregular layouts the inverse in Eq.~\eqref{eq:pinv}
is ill-conditioned and a regularization should be applied  \cite{ZotterEtAl12}.

The set of sampling directions $\Theta$ is said to be \emph{regular} for the basis $\mathbf{Y}$
if it preserves the orthonormality of the sampled basis \cite{jd-phd}.
This means that
$ \mathbf{C} \mathbf{C}^T /N = \mathbf{1}_K $, where  $ \mathbf{1}_K $ is the unity matrix of range $K$.
The set of sampling directions $\Theta$ is said to be \emph{semi-regular} for the basis $\mathbf{Y}$
if it preserves the orthogonality of the sampled basis. This means that $ \mathbf{C} \mathbf{C}^T$ is diagonal.
If the set of sampling directions $\Theta$ does not preserve any of the previous properties,
then it is said to be \emph{irregular}\footnote{An alternative definition, makes use of spherical $t$-designs, with $t \geq 2L+1$,
which identifies the optimal loudspeaker arrangement \cite{zotter2012all}.}.
As a further clarification, in Ambisonics terms a regular set of directions does not necessarily imply a `regularly spaced' set of directions.

If the set of sampling directions is regular, then then the decoding matrix becomes:
\begin{equation}
    \mathbf{D} =  \mathbf{D}_\text{proj} = \mathbf{C}^T/N  .
    \label{eq:proj}
\end{equation}
Namely, the decoding equations consist on a mere projection of the spherical harmonics on the corresponding loudspeaker direction.
The decoding matrix obtained via projection is often referred to as `na{\"i}ve decoding'.

\subsection{Modified Psychoacoustical Decodings}

Psychoacoustically, the basic decoding method is optimal at low frequencies, below 500~Hz approximately, and close to the sweet spot.
At higher frequencies or for a large listening area it is preferable to use modified psychoacoustic decodings \cite{jd-phd}.

Let $\{\mathbf{\hat{d}}_j\}_{j=1,\dots,n}$ be a set of $n$ directions sampling the sphere (useful later).
Given a decoding $\mathbf{D}$, the signal fed to the speaker $i$
while reproducing a virtual plane wave of unit amplitude coming from direction $j$ will
be labelled $s_{ij}$, and is actually given by $s_{ij} = (\mathbf{D} \, \mathbf{Y}(\theta_j,\phi_j))_i$.

The \emph{max-$r_E$ decoding} assumes incoherent sum of the speaker signals.
For regular layouts, the modified decodings can be computed by requiring that
the decoding reproduces the original energy and acoustic intensity at the origin.
Within the incoherent sum hypothesis, and assuming that each one of the incoming waves is a plane wave,
a statistical estimator of the signal energy $E_{j}$ at the origin is:
\begin{equation}
	E_{j} = \sum_{i=1}^n |s_{ij}|^2
\end{equation}
and a statistical estimator of the normalised acoustic intensity $\mathbf I_{j}$ is
\begin{equation}
	\mathbf I_{j} =  \frac{1}{E_{j}} \sum_{i=1}^n |s_{ij}|^2 \mathbf{\hat{u}}_i = r_E \mathbf{\hat{d}}_E.
\end{equation}

The max-$r_E$ decoding requests:
\begin{align*}
	E_{j} &= 1, \\
	\mathbf I_{j} &= \mathbf{\hat{d}}_j \implies
	\begin{cases}
		r_E = 1, \\
		\mathbf{\hat{d}}_E= \mathbf{\hat{d}}_j
	\end{cases}
\end{align*}
It is physically impossible to fulfill the condition $r_E = 1$ by summing incoherently the signal of several loudspeakers;
decodings will instead try to maximise this value (hence the name).
Psychoacoustically, this decoding reproduces the impression of the original sound at high frequencies, above 500~Hz approximately.

The \emph{in-phase} decoding imposes the additional restriction that there are no loudspeakers emitting in opposite phase.
This decoding gives a more robust localisation for listeners who are far from the sweet spot.

While there is experimental evidence that $\mathbf I_j$ gives a good indicator of the perceived source direction,
and that a frequency-weighted version of $r_E$ is a good indicator of the perceived source width for
broadband signals \cite{Frank13PhD}, a detailed analysis of the optimal localization criteria depending on
the frequency is out of the scope of this thesis.
However, let us stress that the key feature of the max-$r_E$ and in-phase decodings is the incoherent
summation hypothesis rather than the specific localization criteria.

For regular or semi-regular layouts, the intensity vector under the incoherent sum hypothesis is
parallel to the acoustic velocity vector in the coherent sum hypothesis and it is possible to obtain
the optimal max-$r_E$ or in-phase decodings by doing slight modifications of the regular decoding,
Eq.~\eqref{eq:proj} \cite{jd-phd}.
However for non-regular layouts the velocity and intensity vectors in the two hypothesis are not parallel,
the problem is fully nonlinear and algebraic methods are not helpful.
To solve this case a nonlinear method can be used.

\chapter{Ambisonics Decoding to Irregular Layouts} \label{ch:idhoa}
Ambisonics has some drawbacks which hinder its widespread adoption.
First, the directionality properties of sound sources encoded in low order Ambisonics is often regarded as poor.
Second, the sweet spot, the area where the reconstruction is optimal, is small.
These two drawbacks can be ameliorated by going to Higher Order Ambisonics (HOA).
Additionally, decoding Ambisonics to non-regular loudspeaker layouts is challenging.
The decoding equations for HOA have closed analytic expressions
only for regular loudspeaker arrays \cite{GerzonBarton92,jd-phd},
but most real-world layouts,
like the ubiquitous stereo, 5.1 and 7.1 surround configurations,
are non-regular from the Ambisonics point of view.
The generation of optimal and psychoacoustically correct decodings for irregular loudspeakers
layouts is a nonlinear problem which can be solved using numerical search algorithms.
In this Section we address this problem by presenting an algorithm for higher order Ambisonics decoding,
together with its open source implementation, IDHOA \cite{github-idhoa-new}.

There are several previous references in the literature calculating the decoding of Ambisonics for irregular loudspeaker arrays.
The papers \cite{wiggingsetal03,wiggingsthesis} and \cite{moorewakefield07,moorewakefield2011}
concentrate on decoding algorithm for the 5.1 ITU
layout based on a modified tabu search algorithm.
Tsang et al. similarly have a decoding strategy based on genetic algorithms and neural networks \cite{tsangcheung09}
while \cite{benjaminetal10,benjaminetal12} work with a preexisting nonlinear optimisation library.
The algorithm presented here follows the method of \cite{arteaga13}, but extends the technique up to 5th order Ambisonics.
The method differs from some of the above references in the decoding technique employed, the fitting functions employed and the
focus on 3D layouts (see Section~\ref{sec:idhoa-summary} for further details).
Besides the methodology, the implementation is completely different.
It is to be noted that there are also alternative decoding techniques which do not involve nonlinear search methods,
like decoding to an intermediate regular layout, and later on using VBAP for decoding \cite{batkekeiler2010,bohem2011}
(see however the comments in \cite{schmeleetal2013}).
Or, again, modifying the basic Ambisonics decoding to ensure preservation of the energy
with Energy Preserving Ambisonics Decoder (EPAD) \cite{ZotterEtAl12} and All-Round Ambisonic Decoding (AllRAD) \cite{zotter2012all, zotter2018allrad2}.
In \cite{ambiboook} the authors make a good comparison between different Ambisonics decoding techniques,
with particular focus on the projection method, mode matching decoder (MAD) \cite{poletti2005three}, the EPAD and All-RAD decoders.

The results presented in this Chaper are based on the paper \cite{scaini2014decoding}.

\section[Psychoacoustically Motivated Numerical Optimization\\ of an Ambisonics Decoder]
{Psychoacoustically Motivated Numerical Optimization of an Ambisonics Decoder} \label{sec:obj-function}
The decoding matrix is computed minimising a function by means of a multidimensional search algorithm.
In this Section the physical variables ($E$, $I^R$, $I^T$) used to calculate the function to be minimized
and the objective function itself are defined.

Assuming $n$ different directions sampling the sphere,
the energy and acoustic intensity generated at the origin are:
\begin{align}
	E_j &= \sum_{i=1}^n |s_{ij}|^2 \label{eq:E_definition} \\
	\mathbf{I}_j &=  \frac{1}{E_j} \sum_{i=1}^n |s_{ij}|^2 \mathbf{\hat{u}}_i , \label{eq:I_definition}
\end{align}
where $s_{ij}$ is the signal emitted by the loudspeaker $i$, when reproducing a sound source coming from direction $j$.

The vector $\mathbf{I}_j$ can be projected in the radial and transverse parts as follows:
\begin{subequations}\label{decomp}
\begin{align}
	I^R_j &= \mathbf{I} \cdot \mathbf{\mathbf{\hat{d}}}_j =  \frac{1}{E_j}  \sum_{i=1}^n |s_{ij}|^2 \, \mathbf{\hat{u}}_i \cdot \mathbf{\mathbf{\hat{d}}}_j,  \label{eq:IR_definition} \\
	I^T_j &= ||\mathbf{I} \times  \mathbf{\mathbf{\hat{d}}}_j|| = \frac{1}{E_j} \sum_{i=1}^n |s_{ij}|^2  \, ||\mathbf{\hat{u}}_i \times \mathbf{\mathbf{\hat{d}}}_j|| . \label{eq:IT_definition}
\end{align}
\end{subequations}

The radial part $ I^R_j$ represents the desired component of the intensity vector, and the tangential part $I^T_j$ represents the unwanted component.
In an ideal decoding, $E_j = 1$, $I^R_j = 1$ and $I^T_j = 0$, but for regular (ideal) layouts the values for $I^R_j$ are always $I^R_j < 1$ (see Table~\ref{tab:r_E}).

\begin{table}[t]
\centering
\begin{tabular}{ccc}
\toprule
Order & max-$r_E$ & in-phase \\ \midrule
1	& 0.577 & 0.500 \\
2	& 0.775 & 0.667 \\
3	& 0.861 & 0.750 \\
4	& 0.906	& 0.800 \\
5	& 0.932 & 0.833 \\ \bottomrule
\end{tabular}
\caption{Maximum theoretical values for $I^R_j$ (or $r_E$) for a regular layout for max-$r_E$ and in-phase decodings.}
\label{tab:r_E}
\end{table}

Note that the decomposition in Eq.~\eqref{decomp} is different
from the decomposition
used in several previous irregular decoding references \cite{wiggingsetal03,wiggingsthesis,moorewakefield07,moorewakefield2011}:
instead of maximising the norm of the intensity vector, and minimising the angle mismatch,
it proves more natural and effective to maximise the radial vector components and minimise the tangential components.

From these quantities different cost function terms can be defined:
\begin{align}
	\cost_E &= \frac{1}{n} \sum_{j=1}^n (1 - E_j)^{2} w_j, \label{eq:costE} \\
	\cost_\text{IR} &= \frac{1}{n} \sum_{j=1}^n (1- I^R_j)^2 w_j, \label{eq:costIR} \\
	\cost_\text{IT} &= \frac{1}{n} \sum_{j=1}^n (I^T_j)^2 w_j, \label{eq:costIT}
\end{align}
where $w_j$ is possible weighting function (see Section~\ref{subsec:config} for more details).
These contributions can be interpreted as follows: $\cost_E$ is  the mean quadratic deviation
from the correct level normalisation;  $\cost_\text{IR}$ is  the mean quadratic deviation from the optimal directionality;
$\cost_\text{IR} > 0$ means that the directionality of the sources is not optimal, and, finally,
$\cost_\text{IT}$ is  the mean quadratic value of the wrong component of the direction;

In the case of the in-phase decoding, there is an extra term in the cost function to take into account:
\begin{align}
    \begin{split}
	E^\text{ph}_j &= \sum_{i=1}^n |s_{ij}|^2 \theta( - s_{ij}),\\
	\cost_\text{ph}  &= \frac{1}{n} \sum_{j=1}^n (E^\text{ph}_j)^2 w_j,
    \end{split}
\end{align}
where $\theta(\cdot)$ is the Heaviside step function.

While physically it is possible to obtain $\cost_E = \cost_{\text{IT}} = 0$,
it is impossible to get the ideal decoding with $\cost_\text{IR} = 0$,
i.e. $I^R = 1$.

Similarly to the energy and intensity terms, it is possible to define some cost function terms for
the pressure and velocity (radial and transverse), which are the relevant quantities for the low frequencies decoders.
Being
\begin{align}
    P_j &= \sum_{i=1}^n s_{ij} \label{eq:P_definition} \\
	\mathbf{v}_j &=  \sum_{i=1}^n s_{ij} \mathbf{\hat{u}}_i , \label{eq:V_definition}
\end{align}
the pressure $P$ and the particle velocity $\mathbf{v}$,
we can separate the velocity in its radial and transverse components
\begin{subequations}\label{eq:v_decomp}
\begin{align}
	v^R_j &= \mathbf{v} \cdot \mathbf{\mathbf{\hat{d}}}_j =  \sum_{i=1}^n s_{ij} \, \mathbf{\hat{u}}_i \cdot \mathbf{\mathbf{\hat{d}}}_j,  \label{eq:VR_definition} \\
	v^T_j &= ||\mathbf{v} \times  \mathbf{\mathbf{\hat{d}}}_j|| = \sum_{i=1}^n s_{ij}  \, ||\mathbf{\hat{u}}_i \times \mathbf{\mathbf{\hat{d}}}_j|| . \label{eq:VT_definition}
\end{align}
\end{subequations}
and define the cost function terms as:
\begin{align}
	\cost_P &= \frac{1}{n} \sum_{j=1}^n (1 - P_j)^{2} w_j, \label{eq:costP} \\
	\cost_\text{VR} &= \frac{1}{n} \sum_{j=1}^n (1- v^R_j)^2 w_j, \label{eq:costVR} \\
	\cost_\text{VT} &= \frac{1}{n} \sum_{j=1}^n (v^T_j)^2 w_j, \label{eq:costVT}
\end{align}

Finally, the different cost function terms are combined to give the objective function to be minimised:
\begin{align}
    \begin{split}
        f &= \alpha_P \cost_P + \alpha_\text{VR} \cost_\text{VR} + \alpha_\text{VT} \cost_\text{VT} \\
        &+ \alpha_E \cost_E + \alpha_\text{IR} \cost_\text{IR} + \alpha_\text{IT} \cost_\text{IT} \\
        &+ \alpha_\text{ph} \cost_\text{ph}.
    \end{split}
    \label{eq:objective-func}
\end{align}
The values of the coefficients $\alpha_P$, $\alpha_\text{VR}$, $\alpha_\text{VT}$ (basic), $\alpha_E$, $\alpha_\text{IR}$, $\alpha_\text{IT}$ (max-$r_E$) and $\alpha_\text{ph}$ (in-phase) can be selected at will.
\section{The IDHOA Decoder}
\label{sec:IDHOA}

\subsection{The Decoder Strategy}
\label{subs:decoder-strategy}
The IDHOA decoder calculates decoding matrices for Ambisonics up to order 5.
The decoder is based on the minimization of the objective function in \eqref{eq:objective-func}.
Our implementation of IDHOA \cite{github-idhoa-new} makes use of Python \cite{CS-R9526}, IPOPT \cite{ipopt} and PyTorch \cite{paszke2017automatic}.
The flow of the algorithm can be summarized as follows:
\begin{enumerate}
    \item \emph{Initialization:} Operations that are performed only once when the algorithm is launched.
    \item Given the loudspeakers' layout, calculate $\mathbf{D}_{\text{pinv}}$ Eq.~\eqref{eq:pinv} and $\mathbf{D}_{\text{proj}}$ Eq.~\eqref{eq:proj}.

    \item Calculate the various physical variables that constitute the objective function, $p$, $E$, $\vect v$, $\vect I$,
          over the $n$ sampling directions.
          Calculate the objective function, which is $f=f \left( \mathbf{D}_{\text{pinv}} \right) $ and $f=f\left( \mathbf{D}_{\text{proj}} \right)$.

    \item Select the $\mathbf{D}_{\textbf{init}}$ matrix that minimizes $f$, $f= \min \lbrace f(\mathbf{D}_\text{proj}), f(\mathbf{D}_\text{pinv}) \rbrace$,
          to be used as the initial point for the minimization algorithm $\mathbf{D}_\text{init}$.
    \item \emph{Fix constraints (optional):} constrain some parameters to have a fixed value (e.g.~lock to zero).
    \item \emph{Minimization stage:} Call to the external minimization algorithm, passing $\mathbf{D}_\text{init}$ and $f$.
        When the minimization algorithm terminates, it returns a $\widetilde{\mathbf{D}}$.
\end{enumerate}

\subsection{Configuration of IDHOA}
\label{subsec:config}
In the IDHOA decoder it is possible to tune several parameters to obtain the desired Ambisonics decoder.
In the following we will detail the compulsory and optional ones.

\paragraph{Layout} First to be provided, are the coordinates of the target layout $(\theta,\phi)$.
It is also possible to provide cartesian $(x,y,z)$ coordinates.
Cartesian coordinates will be converted to $(\theta,\phi)$, stripping out the distance $R$ information.
The hypothesis is that the distance will be addressed in another stage of the decoding, with proper delays and near-field filters.

\paragraph{Basic settings}
\emph{Degree.} The decoder allows to generate the Ambisonics decoding coefficients up to fifth order,
setting the \texttt{DEG} variable to the desired order.
\emph{Decoding scheme.} In the first implementations of IDHOA described in published papers,
the \texttt{DEC} variable allowed to choose between the three different decodings:
basic, $ \text{max-r}_E$, in-phase.
In the recent years we decided to simplify this approach and have only a \emph{universal} decoding,
where the different requests for the optimization are directly balanced with the $\alpha_E$, $\alpha_P$, $\dots$, coefficients of~\eqref{eq:objective-func}.
This choice has two motivations: firstly, in HOA it is possible to reconstruct pressure and energy
at the same time\footnote{It is possible also at FOA, but the price to pay to have a linear (pressure) and quadratic (energy) quantities
(which are function of the speakers' gains) that sum to 1, is having negative gains.
This is typically a bad idea, because negative gains mean out-of-phase speakers that results in a limited (depending on frequency) area
where the signals sum properly, i.e.\ small sweet spot.},
so separating the decoder in two (or three) completely different decoders can be avoided,
or the transition between the different decors can be made smoother.
Secondly, the rigid separation in different cost functions depending on the decoding scheme was a limitation during the
experimentation with wavelet format decoding described in Part~\ref{part:wavelets}.

\paragraph{Optional parameters}
The weighting function $w_{j} $ in the definition of the various $\cost$ terms (Eqs.~\eqref{eq:costE} and following)
is an optional biasing factor which allows to improve the decoding performance in some regions of the sphere
(at the expense of other regions).
A non-biased decoding is given by $w_{j} = 1$.
Some examples of possible weighting functions are reported in~\cite{arteaga13}.
For example, it is possible to use a function that masks automatically the areas with no loudspeakers;
being $\mathbf{\hat{u}}$ the direction of the loudspeaker, and $\mathbf{\mathbf{\hat{d}}}$ the direction where the
function is being evaluated:
\begin{equation*}
    w_{j} =
    \begin{cases}
    1 &	\text{if } d(\mathbf{\hat{u}},\mathbf{\mathbf{\hat{d}}}) < \tilde{d}, \\
    \beta &	\text{if } d(\mathbf{\hat{u}},\mathbf{\mathbf{\hat{d}}}) \geq \tilde{d} ,
    \end{cases}
\end{equation*}
where $\beta$ is a parameter that can be tuned\footnote{With $\beta = 0$ the weighting behaves like a binary mask,
with $\beta > 0$ the process of masking is smoother, keeping partially into account the areas without speakers.}
with $0 \leq \beta \leq 1$, $d$ is the angular distance calculated with the Haversine formula~\cite{Haversine}, and $\tilde{d}$ is a
fixed threshold (a reasonable value can be 1.5 times the mean great-circle distance between the loudspeakers).
This approach is quite flexible and can be modified to fit other preferences.

Another interesting feature is the preservation of the natural left/right symmetry of the speakers layout in the generated decoding matrix.
One part of the algorithm searches for left-right speakers and pairs them, reducing the effective number of degrees of freedom,
and fixing this symmetry into the optimized decoding matrix $\mathbf D$.

In our Python implementation, these parameters are set in the \texttt{layout\_name.ini} file.

\section{Performance of the Decoder}
\label{sec:results}

The tests reported here were carried out on the layout of the 3D audio studio in Barcelona Media,
equipped with 23 loudspeakers, placed in an irregular hemispherical
configuration\footnote{The layout is explicitly given as an example in IDHOA code~\cite{github-idhoa}.}.

The results reported in this Section make reference to the version of IDHOA originally published here \cite{scaini2014decoding}.
The Table~\ref{tab:result} shows
the values of the objective function $f$ for max-$r_E$ and in-phase decodings
at first and third order, for the \emph{na{\"i}ve} and optimized \emph{decodings}.
The na{\"i}ve refers to the decoding equation~\eqref{eq:proj}, corrected for the desired
modified decoding (max-$r_E$ or in-phase).
From this table it is possible to extract a qualitative fact: the value of the
objective function decreases during the optimization process.
The quantity $n_{active}$ is the number of speakers left active in the decoding.
Originally the ``muting'' of speakers and/or Ambisonics channels was obtained by running the minimization several times,
and each time locking to zero the coefficients under a certain threshold.
This procedure was necessary since the used minimization algorithm, \texttt{Sbplx} (reimplementation of \texttt{Subplex} \cite{Rowan:1990sbplx})
from \texttt{NLopt} library \cite{nlopt}, is a local derivative-free algorithm.
In the new implementation, the minimization algorithm uses jacobian and hessian and gets to the similar results in one run.
The change in the algorithm impacted also the execution times reported in the Table, reducing them by a factor 10 approximately.

\begin{table}[t]
\centering
\begin{tabular}{ccccc}
\toprule
Decoding 		   & $f$ Na{\"i}ve & $f$ Opt. & time    & $n_{active}$ \\ \midrule
$1^{st}$ max-$r_E$ &    74.9	& 24.36   & 235~$s$ & 12~(/23) \\
$1^{st}$ in-phase  &    66.4	& 24.36   & 286~$s$ & 13~(/23) \\ \hline
$3^{rd}$ max-$r_E$ &    193.2	& 13.94   & 736~$s$ & 21~(/23) \\
$3^{rd}$  in-phase &    115.5	& 13.36   & 814~$s$ & 20~(/23) \\ \bottomrule
\end{tabular}
\caption{Objective function $f$ value for different decodings at first and third order.
Moreover it is reported the lasted time for the algorithm to reach the minimum,
and the number of active speakers at the end of the evaluation (out of 23).}
\label{tab:result}
\end{table}

\begin{figure*}
\centering
       \begin{subfigure}[b]{0.46\textwidth}
               \includegraphics[width=\textwidth]{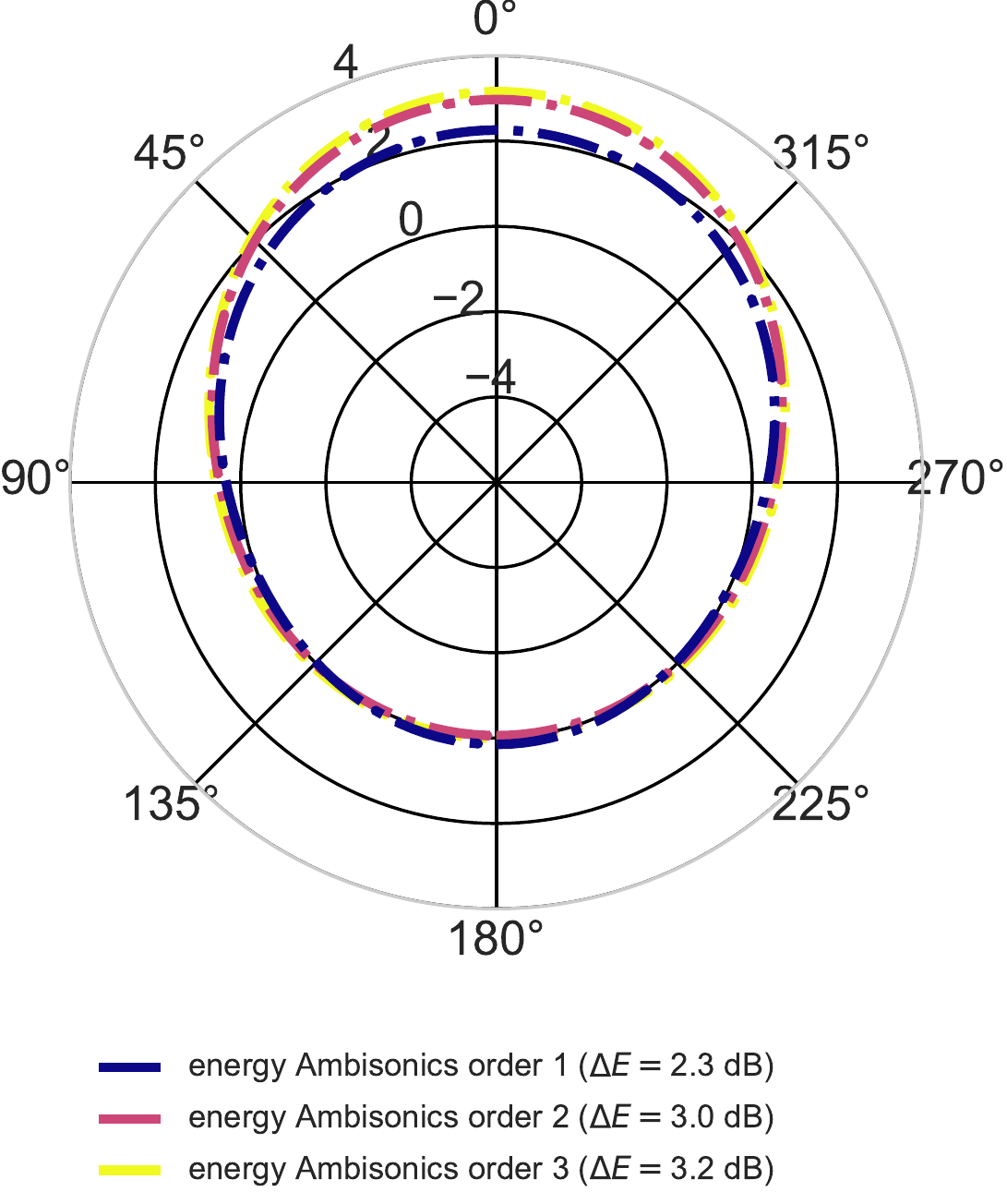}
               \caption{Reconstructed energy for na{\"i}ve decoding, horizontal plane. Values are expressed in dB.}
               \label{fig:naive-energy-h}
       \end{subfigure}
       \begin{subfigure}[b]{0.46\textwidth}
               \includegraphics[width=\textwidth]{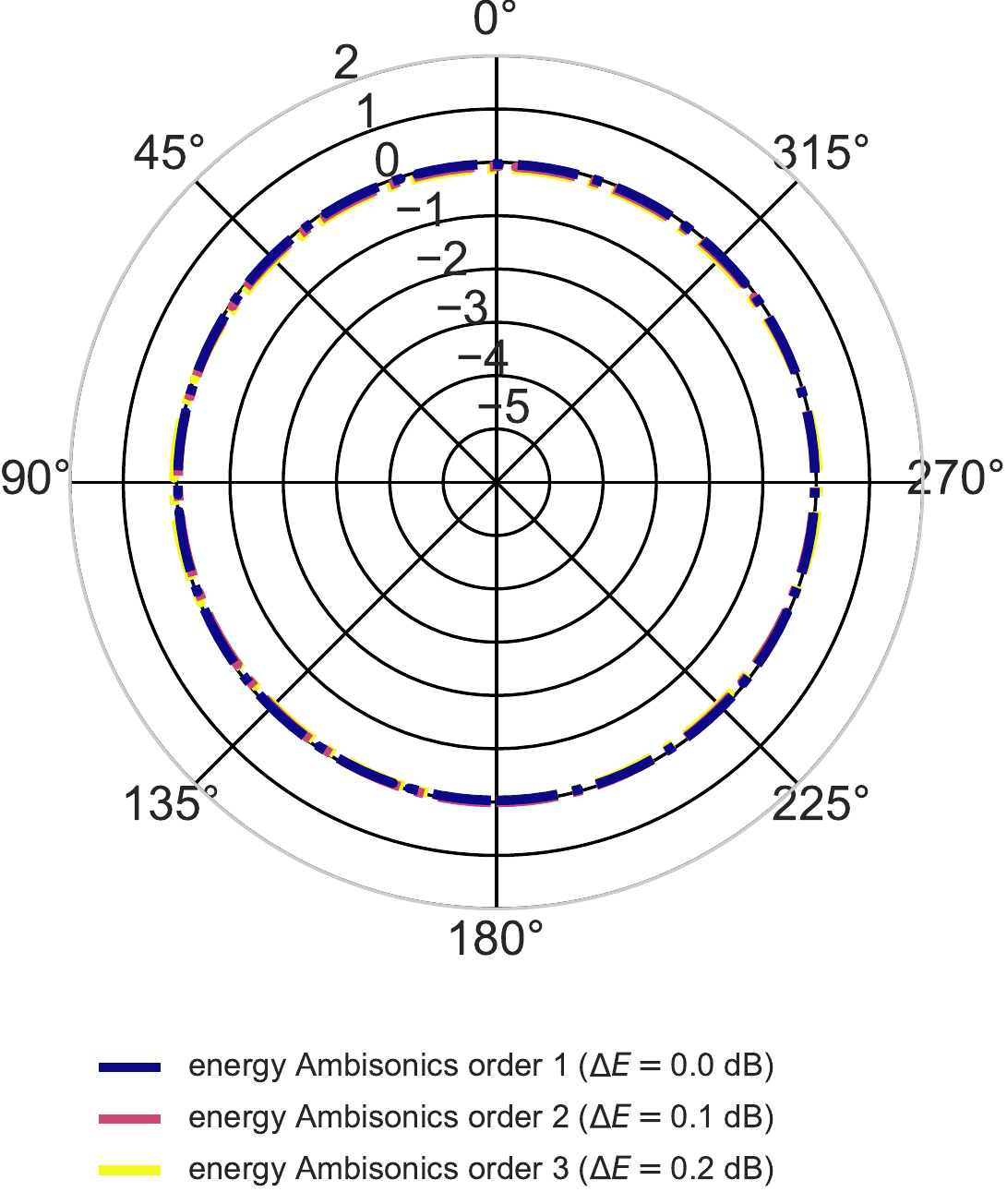}
               \caption{Reconstructed energy for optimized decoding, horizontal plane. Values are expressed in dB.}
               \label{fig:idhoa-energy-h}
       \end{subfigure}
       \begin{subfigure}[b]{0.46\textwidth}
               \includegraphics[width=\textwidth]{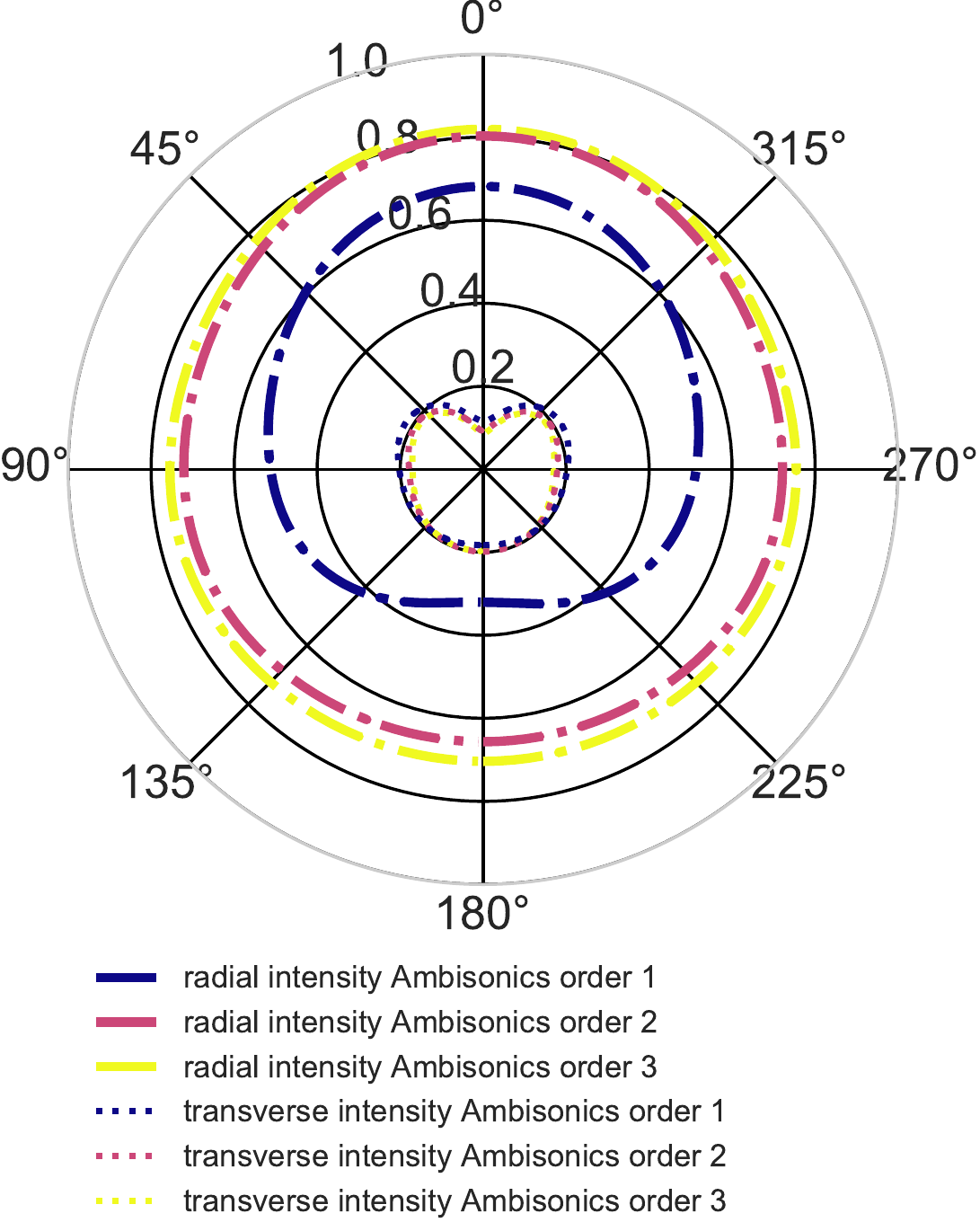}
               \caption{Reconstructed intensity for na{\"i}ve decoding, horizontal plane.}
               \label{fig:naive-intensity-h}
       \end{subfigure}
       \begin{subfigure}[b]{0.46\textwidth}
               \includegraphics[width=\textwidth]{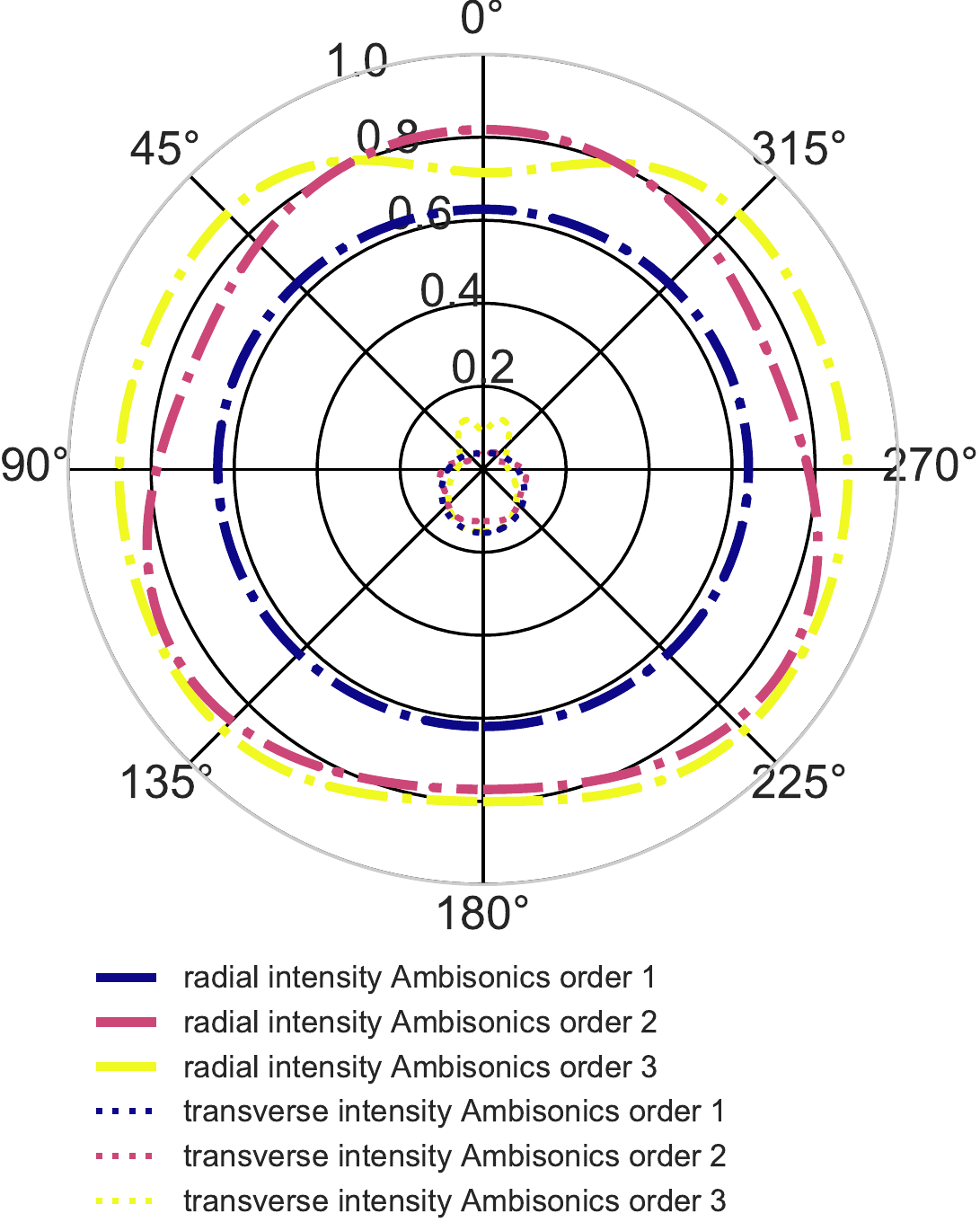}
               \caption{Reconstructed intensity for optimized decoding, horizontal plane.}
               \label{fig:idhoa-intensity-h}
       \end{subfigure}
   \caption{First, second and third order Ambisonics.
       Comparison between na{\"i}ve and optimized decodings for a panning around the horizontal plane, (front $0^\circ$ -- left $90^\circ$).
       The figures on the left side show the na{\"i}ve decoding,
       while those on the right side show the optimized decodings.
       }\label{fig:bm-ambi-h}
\end{figure*}

\begin{figure*}
\centering
       \begin{subfigure}[b]{0.46\textwidth}
               \includegraphics[width=\textwidth]{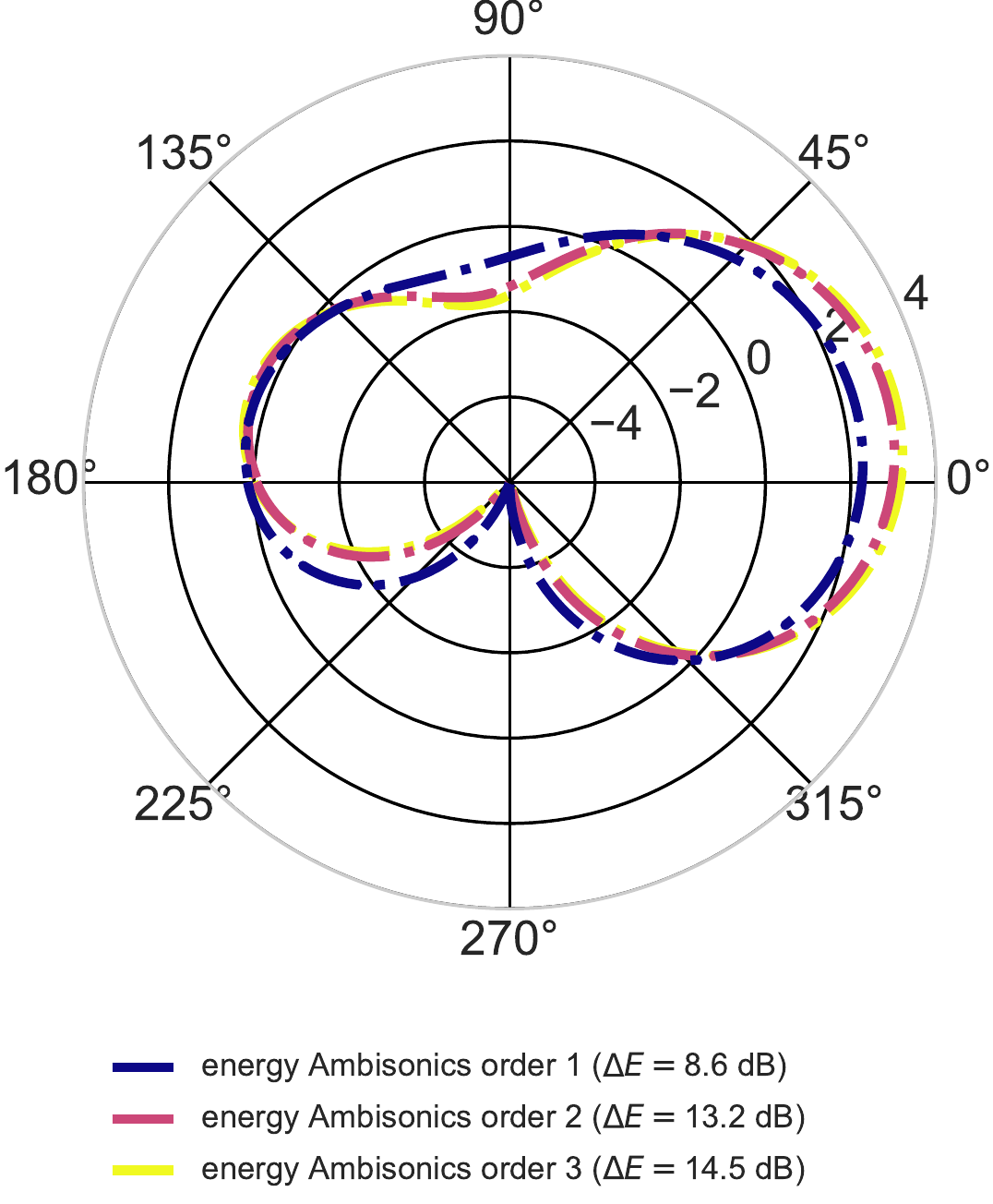}
               \caption{Reconstructed energy for na{\"i}ve decoding, vertical plane. Values are expressed in dB.}
               \label{fig:naive-energy-v}
       \end{subfigure}
       \begin{subfigure}[b]{0.46\textwidth}
               \includegraphics[width=\textwidth]{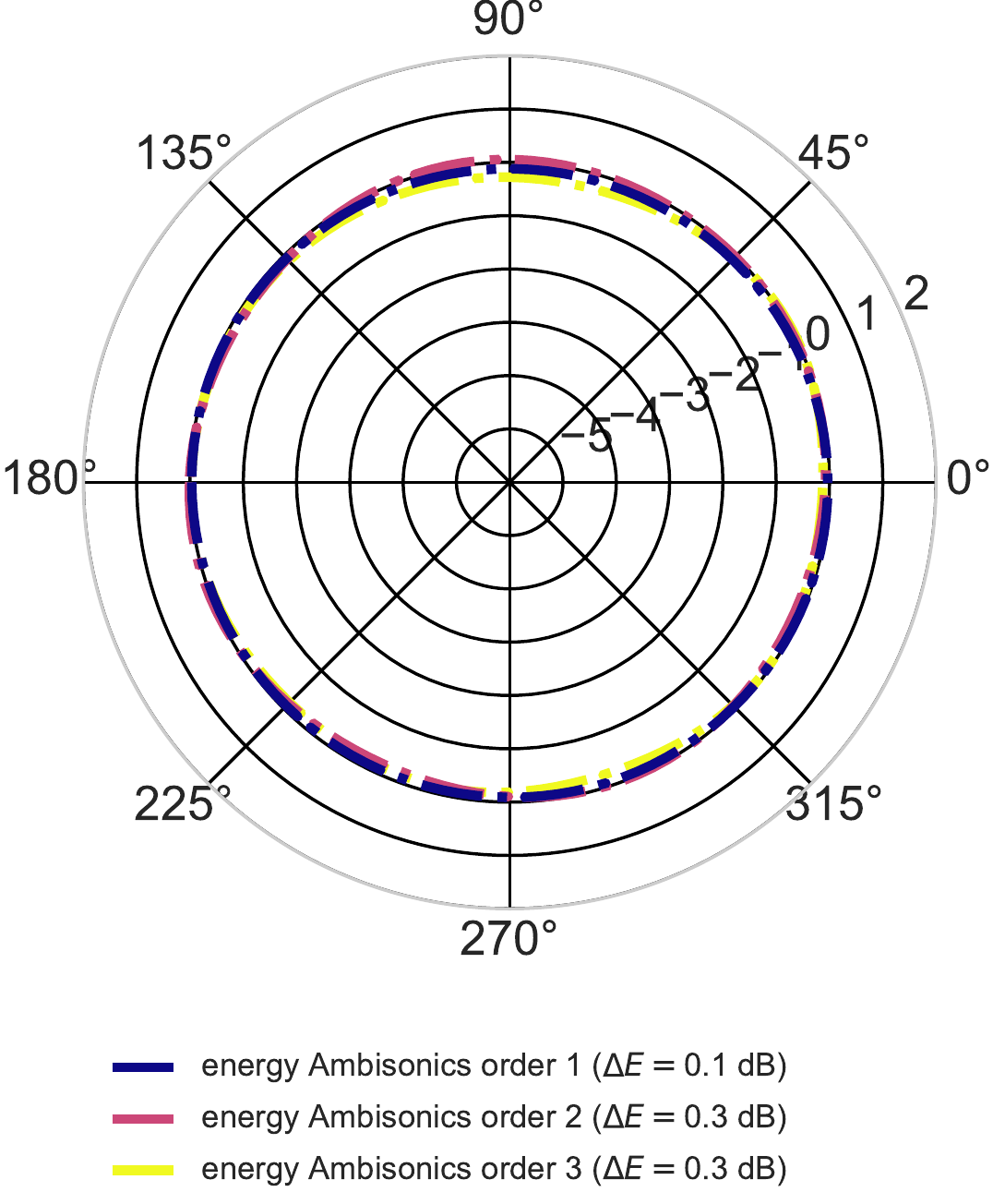}
               \caption{Reconstructed energy for optimized decoding, vertical plane. Values are expressed in dB.}
               \label{fig:idhoa-energy-v}
       \end{subfigure}
       \begin{subfigure}[b]{0.46\textwidth}
               \includegraphics[width=\textwidth]{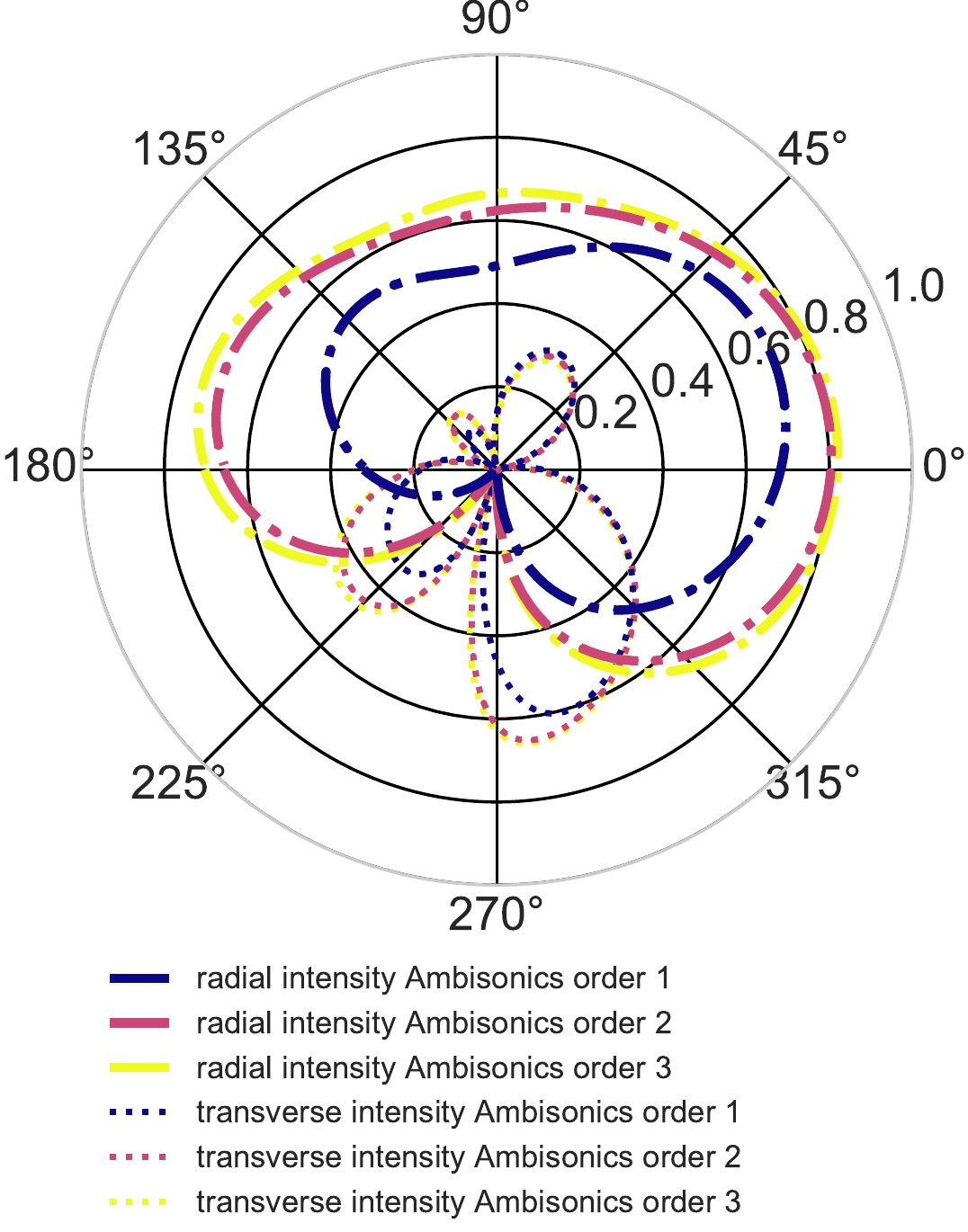}
               \caption{Reconstructed intensity for na{\"i}ve decoding, vertical plane.}
               \label{fig:naive-intensity-v}
       \end{subfigure}
       \begin{subfigure}[b]{0.46\textwidth}
               \includegraphics[width=\textwidth]{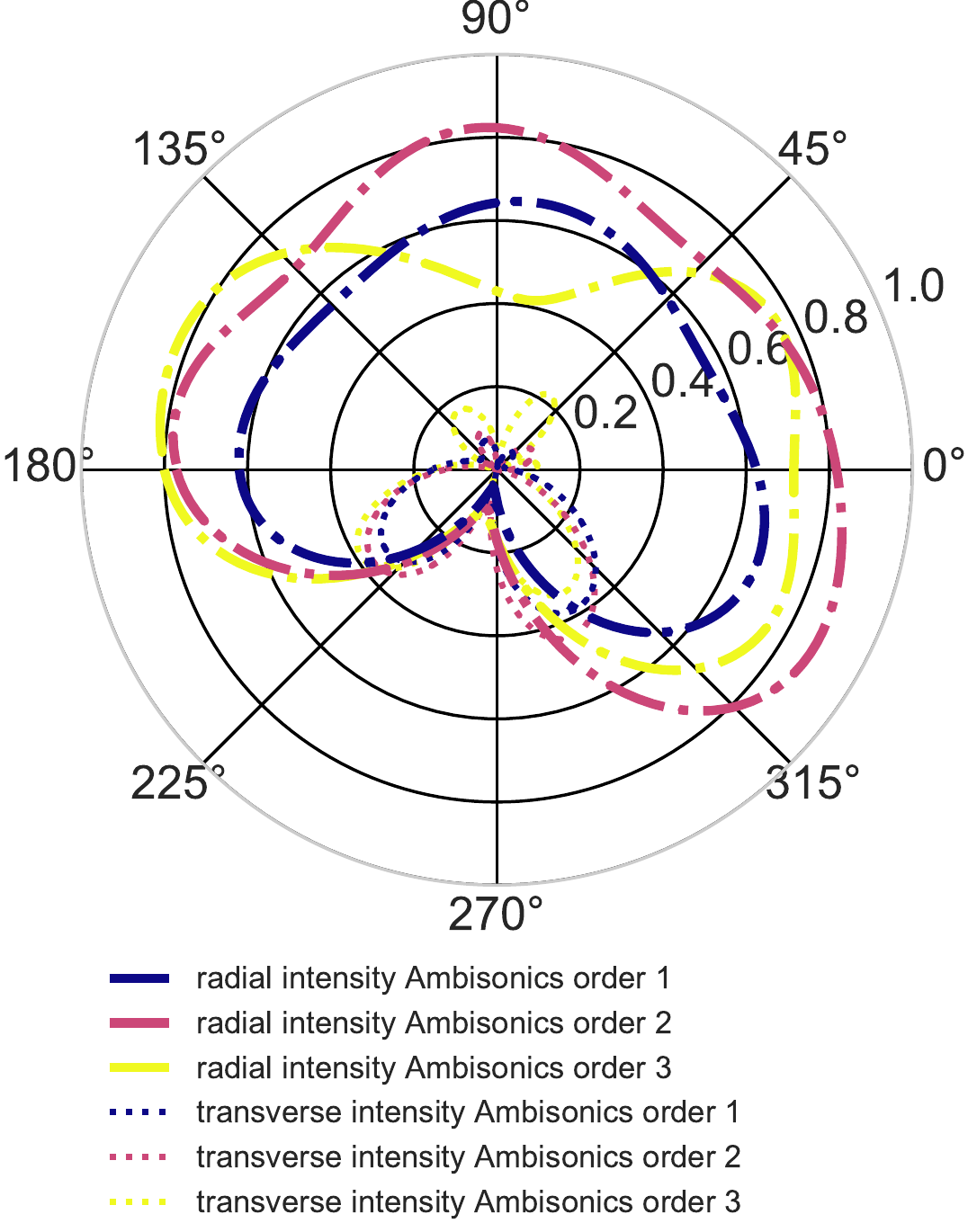}
               \caption{Reconstructed intensity for optimized decoding, vertical plane.}
               \label{fig:idhoa-intensity-v}
       \end{subfigure}
   \caption{First, second and third order Ambisonics.
       Comparison between na{\"i}ve and optimized decodings for a panning around the vertical plane, (front $0^\circ$ -- up $90^\circ$).
       The figures on the left side show the na{\"i}ve decoding,
       while those on the right side show the optimized decodings.
       }\label{fig:bm-ambi-v}
\end{figure*}

Comparing Figures~\ref{fig:naive-energy-h} and~\ref{fig:idhoa-energy-h} it is possible to note that
the energy is properly reconstructed by IDHOA at all three Ambisonics orders considered here.
Looking at Figures~\ref{fig:naive-intensity-h}, \ref{fig:idhoa-intensity-h}, \ref{fig:naive-intensity-v} and
\ref{fig:idhoa-intensity-v} it is possible to highlight the effect of
increasing the Ambisonics order: the radial intensity improves at higher orders, getting close to 0.8
in all directions covered by loudspeakers already for second order Ambisonics.
Figures~\ref{fig:bm-ambi-h} and \ref{fig:bm-ambi-v} are all obtained with the last version of IDHOA code,
with a combination of coefficients for the cost function that mixes all the three decoding schemes.

The decoder was tested carrying out some informal listening tests, where the subjects involved noted
an improved localization with HOA with respect to FOA.
Furthermore, the subjects reported Ambisonics to display smoother pannings than VBAP does.
A quantitave subjective test is reported in Chapter~\ref{ch:idhoa-evaluation} for 2D 5.0 arrays.

\section{Summary} \label{sec:idhoa-summary}
The described IDHOA decoder has been released as open-source code under the GPLv3 license, and can be downloaded
at \cite{github-idhoa-new}.
The code generates a set of decoding coefficients for each loudspeaker allowing to decode Ambisonics signals up to fifth order.
The strategy adopted for the search of the decoder aims to maximise the directionality of the decoded sounds on a number of
sampled directions over the sphere, minimise the directional mismatch and ensure the correct sound level.

Some remarkable properties of the decoder are:
\begin{itemize}
    \item IDHOA can generate basic, max-$r_E$, in-phase (and any almost continuous combinations of them)
        \emph{periphonic} decoders up to \emph{fifth order} of Ambisonics.
    \item Automatic disconnection of loudspeakers\footnote{When there are more loudspeakers that the minimal number
        for a given order in Ambisonics, often the best decoding strategy is to use a subset of all loudspeakers.},
        and/or Ambisonics order muting\footnote{It is possible for IDHOA to attempt to decode a given
        Ambisonics order with less speakers than channels. The orders that cannot be properly decoded are automatically muted.}.
    \item Automatic recognition of exact or approximate left/right symmetry in the layout.
    \item Optional weighting of some sectors of the space,
        to avoid trying to optimize large sectors with no speakers.
    \item Optional horizontal plane and frontal area weighting,
        to provide a better imaging in the frontal area and/or the horizontal plane.
\end{itemize}

The developed decoder successfully minimises the objective function, optimizing the intensity vector
and ensuring the correct energy reproduction.
Informal listening tests confirm the improvement from the na{\"i}ve to the optimized decoding,
and detect a clear improvement in the localisation with HOA compared to FOA.

\chapter{IDHOA Evaluation} \label{ch:idhoa-evaluation}
In Chapter~\ref{ch:idhoa} we presented an algorithm for decoding higher order Ambisonics
for irregular real-world 3D loudspeaker arrays,
implemented in the form of IDHOA,
an open source project.
IDHOA has many features tailored for the reproduction of Ambisonics in real audio venues.
In order to benchmark the performance of the decoder against other decoding solutions,
we restrict the decoder to 2D layouts,
and in particular to the well studied stereo, 5.1 and 7.1 surrounds.

We report on the results of the objective evaluation of the IDHOA decoder in these layouts,
and of the subjective evaluation in 5.1 by benchmarking IDHOA against different decoding solutions.

This Chapter is based on the paper \cite{scaini2015an}.

\section{Objective Evaluation} \label{sec:objective-eval}
We generated dual-band decodings, i.e.~basic decoding strategy for low frequencies, and max-$r_\text E$
for high frequencies, for the stereo, 5.1 and 7.1 surround layouts, at different Ambisonics orders.

For each decoding tested, a point source has been encoded in different directions around the circle, and the following has been considered:
\begin{enumerate}
\item The sound level generated from each direction (values of $E$ and $p$), see Eqs.~\eqref{eq:E_definition}, \eqref{eq:I_definition}.
\item The amount of directionality of the sound generated (values of $v_\text R$ and $I_\text R$), see Eqs.~\eqref{eq:VR_definition}, \eqref{eq:IR_definition}.
\item The correctness of position of the sound source (values of $v_\text T$ and $I_\text T$), see Eqs.~\eqref{eq:VT_definition}, \eqref{eq:IT_definition}.
\item The amount of crosstalk for sources panned exactly at the loudspeaker positions.
\end{enumerate}
For the 5.1 layout we have additionally compared the decoders generated with selected reference state-of-the-art decoders.
The criteria for decoder selection was first: public availability, and second: the decoders had to target the standard ITU angular position\footnote{For example,
the \emph{Vienna decoders}~\cite{GerzonBarton92} do not address the standard ITU layout.}.

\begin{table}[t]
 \centering
 \begin{tabular}{ lcccc }
  \toprule
   & \multicolumn{2}{c}{Mean LF} & \multicolumn{2}{c}{Mean HF} \\ \cmidrule(r){2-3} \cmidrule(l){4-5}
  Decoding & $v_\text R$ & $v_\text T$ & $I_\text R$ & $I_\text T$ \\
  \midrule
  5.0 \textit{BHL1} & 1.00 & 0.00 & 0.69 & 0.10  \\
  5.0 \textit{idhoa1} & 1.00 & 0.00 & 0.69 & 0.15  \\
  5.0 \textit{FAA2} & 1.00 & 0.00 & 0.65 & 0.01  \\
  5.0 \textit{idhoa2} & 1.00 & 0.02 & 0.78 & 0.13  \\
  5.0 \textit{idhoa3} & 1.00 & 0.02 & \textbf{0.80} & 0.14  \\ \hline
  7.0 \textit{idhoa3} & 1.01 & 0.01 & 0.87 & 0.06 \\
  \bottomrule
 \end{tabular}
 \caption{Mean values around the circle for the radial and transversal components of the velocity
 (basic component, low frequencies) and radial and transversal components of the intensity (max-rE component, high frequencies)
 for the different decodings.
 The best mean value for the radial intensity for the 5.0 layout is reached by the IDHOA decoding at third order, \textit{idhoa3}.}\label{tbl:mean}
\end{table}

\begin{table}[t]
 \centering
 \begin{tabular}{ l *4{d{3.1}} }
  \toprule
   & \multicolumn{4}{c}{Crosstalk HF (dB)}
   \\ \cmidrule(r){2-5}
   Decoding & \mc{C} & \mc{L/R} & \mc{Ls/Rs} & \mc{Lb/Rb} \\
  \midrule
   2.0 \textit{idhoa1} & \mc{--} & -7.2 & \mc{--} & \mc{--} \\ \hline
   5.0 \textit{BHL1} &  \multicolumn{1}{r}{\textbf{1.7}} & -0.8 & -6.7 & \mc{--} \\
   5.0 \textit{idhoa1} & 1.9 & 2.5 & -7.3 & \mc{--} \\
   5.0 \textit{FAA2} & 1.9 & -1.8 & -4.7 & \mc{--} \\
   5.0 \textit{idhoa2} & \multicolumn{1}{r}{\textbf{1.7}} & -4.5 & -11.3 & \mc{--}\\
   5.0 \textit{idhoa3} & 2.6 & \multicolumn{1}{r}{\textbf{--5.8}} & \multicolumn{1}{r}{\textbf{--13.0}} & \mc{--} \\ \hline
   7.0 \textit{idhoa3} & 2.5 & -8.1 & -3.8 & -6.8 \\
  \bottomrule
 \end{tabular}
 \caption{Total crosstalk for the different loudspeakers (crosstalk defined here as the portion of energy emitted
 by other loudspeakers for signals panned exactly at the loudspeaker positions, as compared to the energy emitted by that loudspeaker).
 The best values for the total crosstalk for the 5.0 layout are highlighted in bold.}
\label{tbl:crosstalk}
\end{table}

 \paragraph{Layout 5.0: first order decoding}
Regarding the first order decoder generated by IDHOA (from here on referred as \textit{idhoa1}),
the low frequency portion of the decoder reproduces correctly the pressure and velocity
and is identical to the analytic decoder generated by the analytic method (pseudoinverse).
The high frequency portion of the decoder reconstructs correctly the energy and attempts
to maximize the radial component of the intensity, at the expense of some localization mismatch.
We activated an option in IDHOA to privilege the frontal region
over the lateral and rear regions during the optimization of the decoding.

As reference decoder we have chosen the one published in~\cite{benjaminetal10},
later on referred as \textit{BHL1} decoding.
This decoder addresses the standard ITU 5.0 layout
described before %
and it was obtained with a similar minimization process as in IDHOA.
It is to be remarked that surprisingly the \textit{BHL1} decoder is normalized to pressure
(assuming coherent addition of the signals) in the high frequency band.

As shown in Tables~\ref{tbl:mean} and~\ref{tbl:crosstalk} the performance of both decoders
is similar with respect to directionality properties.
The main difference is the pressure normalization at high frequencies of the \textit{BHL1} decoding,
which leads to a lower output level at high frequencies.

 \paragraph{Layout 5.0: second order decoding}
The 5.0 layout has as many loudspeakers as channels in second order Ambisonics,
meaning that in principle the analytic inversion method could be used to reconstruct
the second order Ambisonics components from the five loudspeaker signals.
However, the analytic solution, while nominally correct, relies on extremely large phase
cancellations which are not desired in practice.
As an alternative we designed the low frequency portion of the \textit{idhoa2} decoding by requesting the correct pressure
and velocity (which can already be obtained at first order), and secondarily by optimizing
the intensity vector.
The high frequency portion of the decoding was created using similar criteria as
the first order counterpart.

The second order decoder chosen as reference is the decoder derived by Fons Adriaensen
that comes with the Ambdec decoding software \cite{ambdec}, later called simply \textit{FAA2} decoding.
This decoder is one of the few public second order decoders and probably the most widespread since
it is shipped with the Ambdec software.

\begin{figure*}
    \centering
    \begin{subfigure}[c]{0.46\textwidth}
        \includegraphics[width=\textwidth]{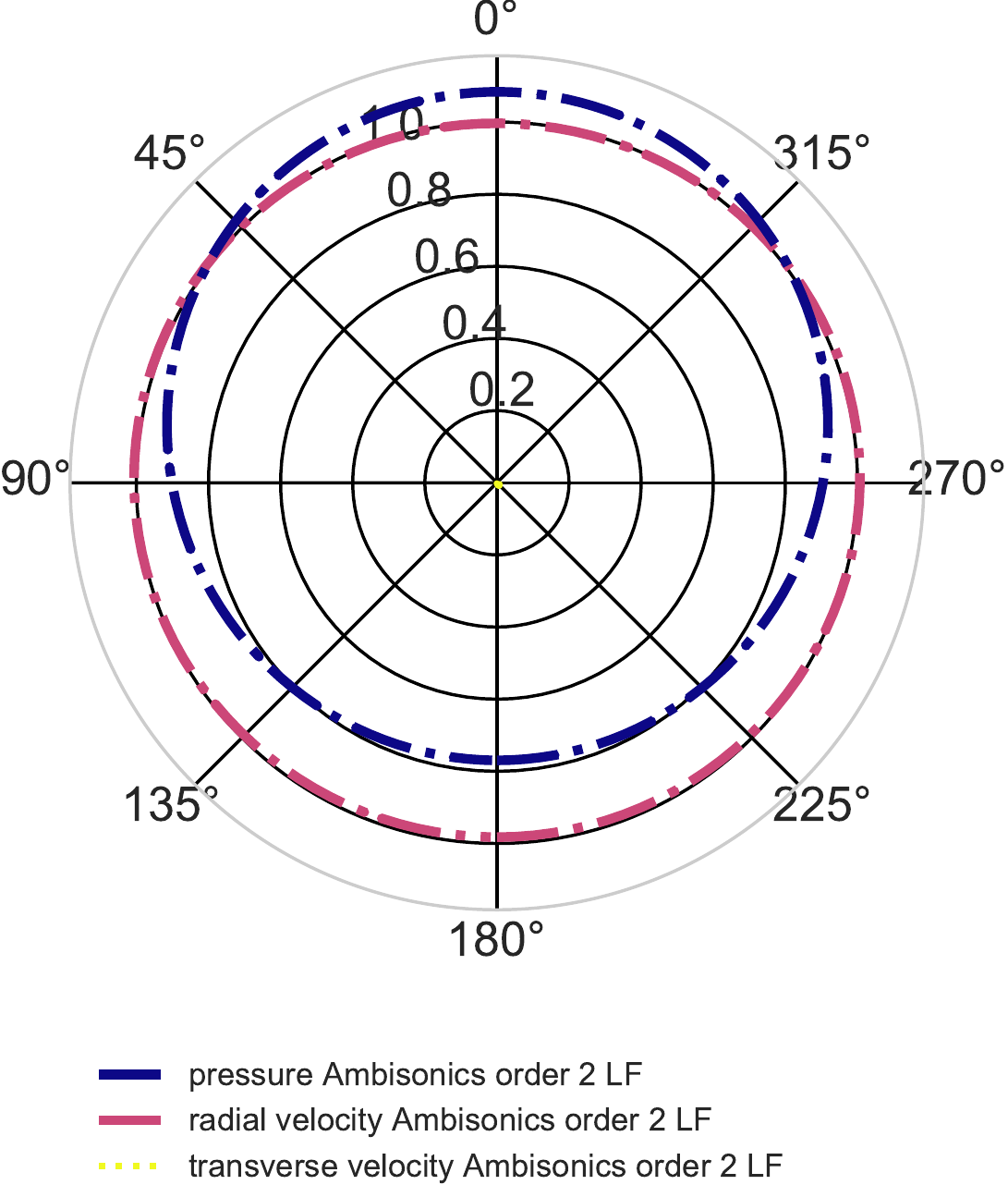}
        \caption{\textit{FAA2} decoding for LF.}
        \label{fig:2fons-lf}
    \end{subfigure}
    \begin{subfigure}[c]{0.46\textwidth}
        \includegraphics[width=\textwidth]{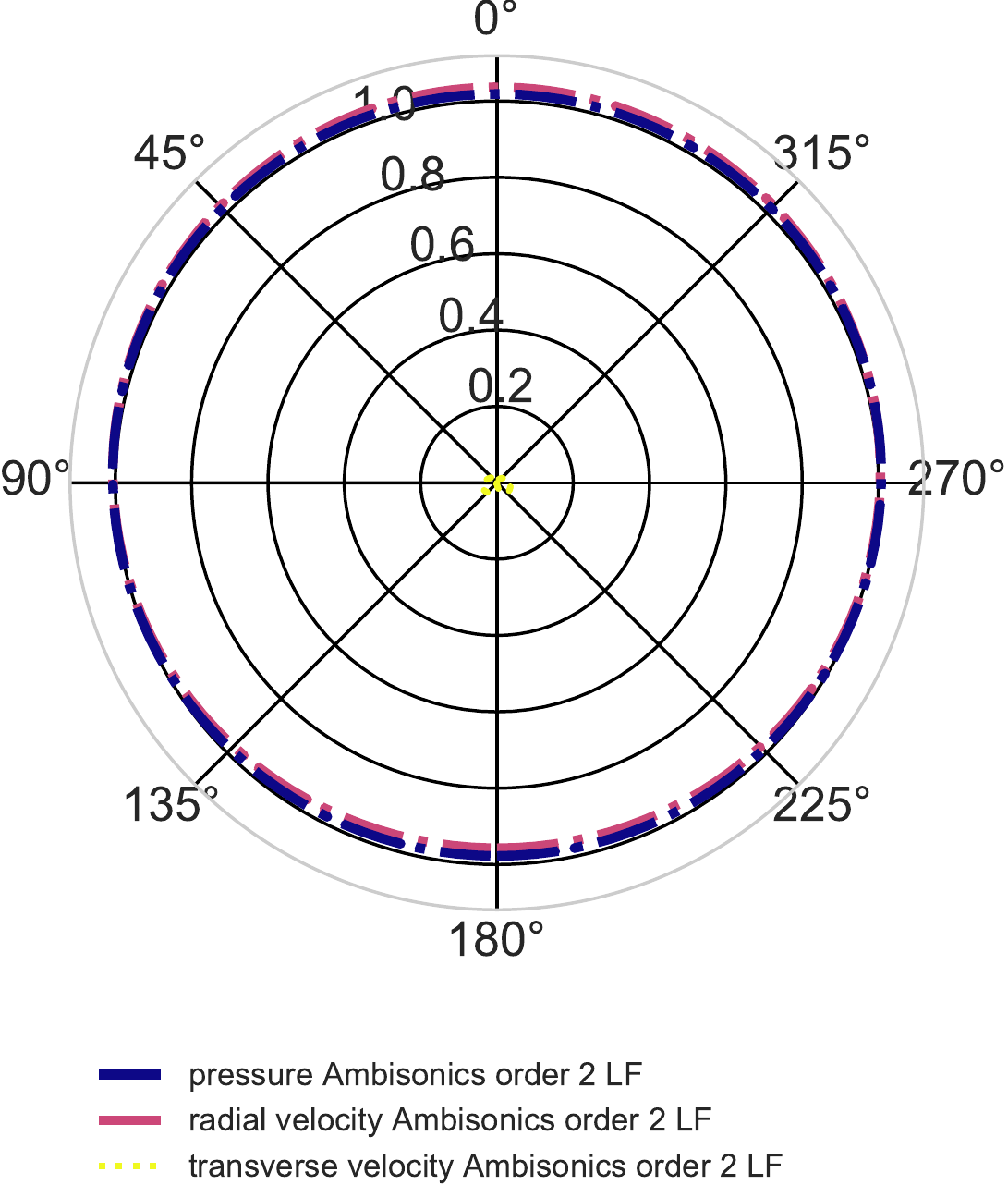}
        \caption{\textit{idhoa2} decoding for LF.}
        \label{fig:2idhoa-lf}
    \end{subfigure}
    \begin{subfigure}[c]{0.46\textwidth}
        \includegraphics[width=\textwidth]{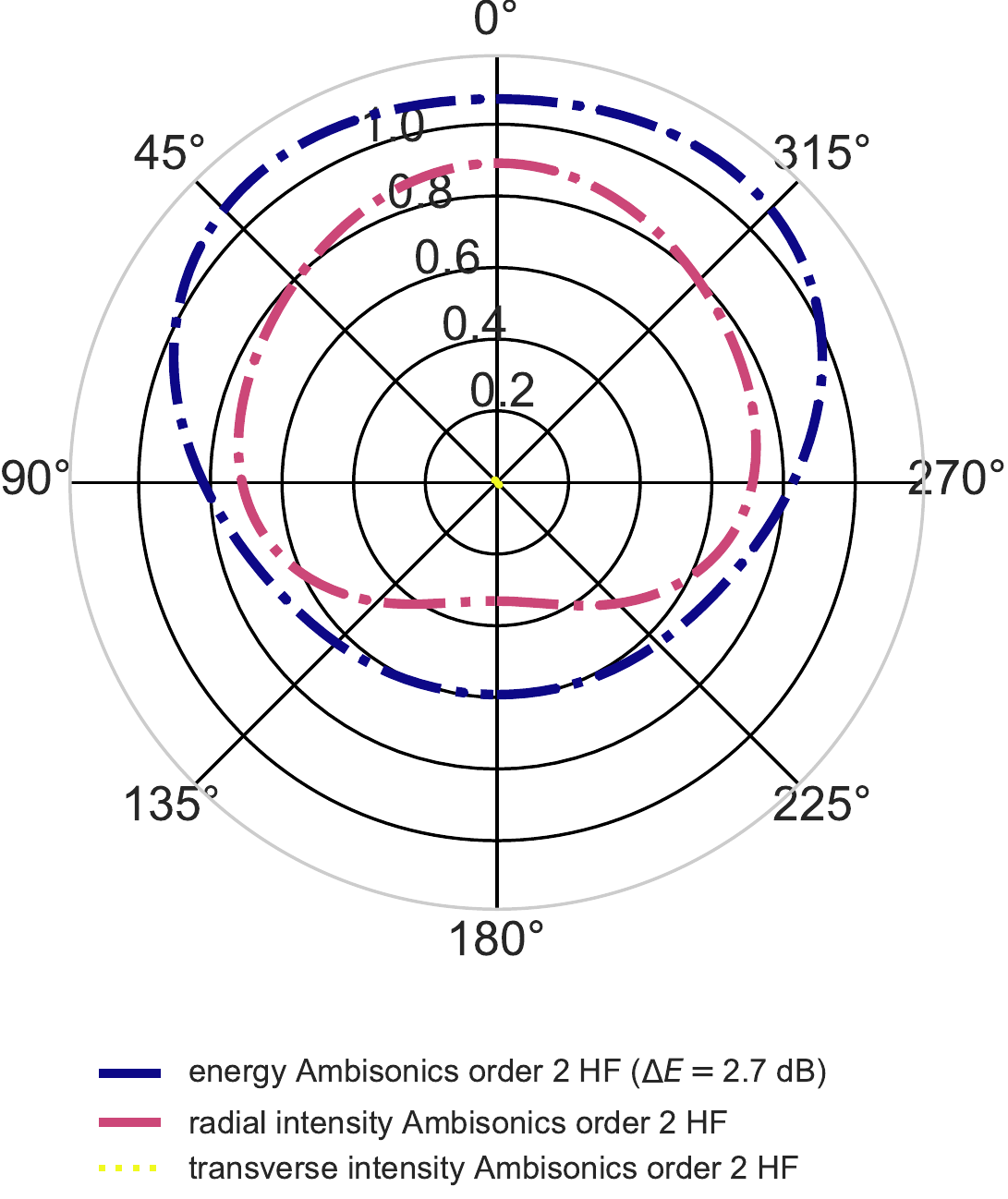}
        \caption{\textit{FAA2} decoding for HF.}
        \label{fig:2fons-hf}
    \end{subfigure}
    \begin{subfigure}[c]{0.46\textwidth}
        \includegraphics[width=\textwidth]{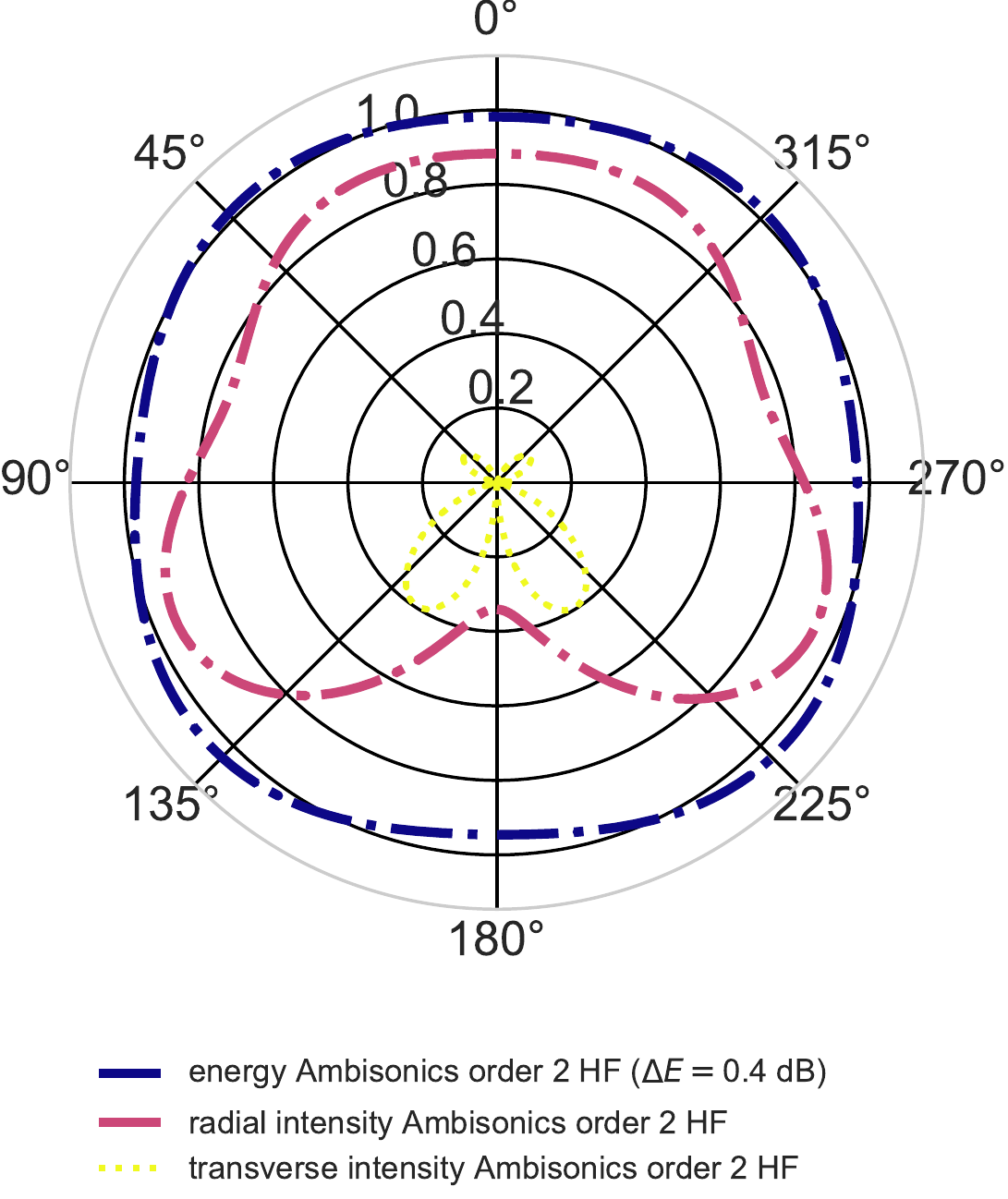}
        \caption{\textit{idhoa2} decoding for HF.}
        \label{fig:2idhoa-hf}
    \end{subfigure}

    \caption{Second order Ambisonics decoders.
        In the left column the \textit{FAA2} decoding
        shipped with ambdec decoder software by Fons Adriaensen, and in the
        right column the one generated with IDHOA. Plots~\subref{fig:2fons-lf} and
        \subref{fig:2idhoa-lf} show the magnitude of pressure (dotted black) and radial
        (dashed red) and transverse (continuous green) components of velocity as a
        function of the polar angle in the horizontal plane. Plot~\subref{fig:2fons-hf}
        and~\subref{fig:2idhoa-hf} show the magnitude of energy (dotted black) and
        radial (dashed red) and transverse (continuous green) components of intensity
        vector as a function of the polar angle in the horizontal plane.
        }\label{fig-ambdec-idhoa-comparison}
\end{figure*}

\begin{figure*}
    \centering
    \begin{subfigure}[c]{0.46\textwidth}
        \includegraphics[width=\textwidth]{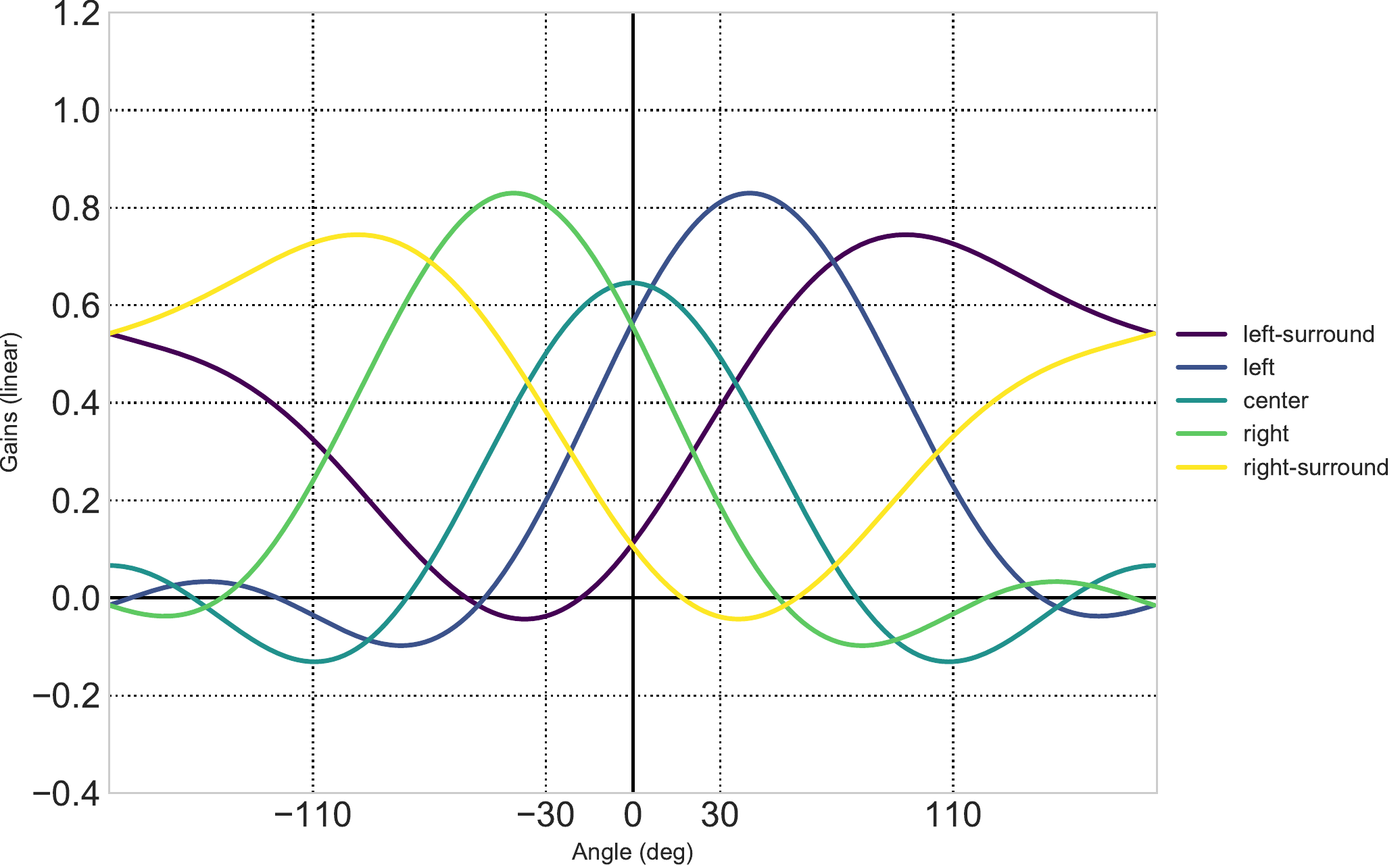}
        \caption{Gains as a function of the angle, for the HF \textit{FAA2} decoder.}
        \label{fig:2fons-gains-hf}
    \end{subfigure}
    \begin{subfigure}[c]{0.46\textwidth}
        \includegraphics[width=\textwidth]{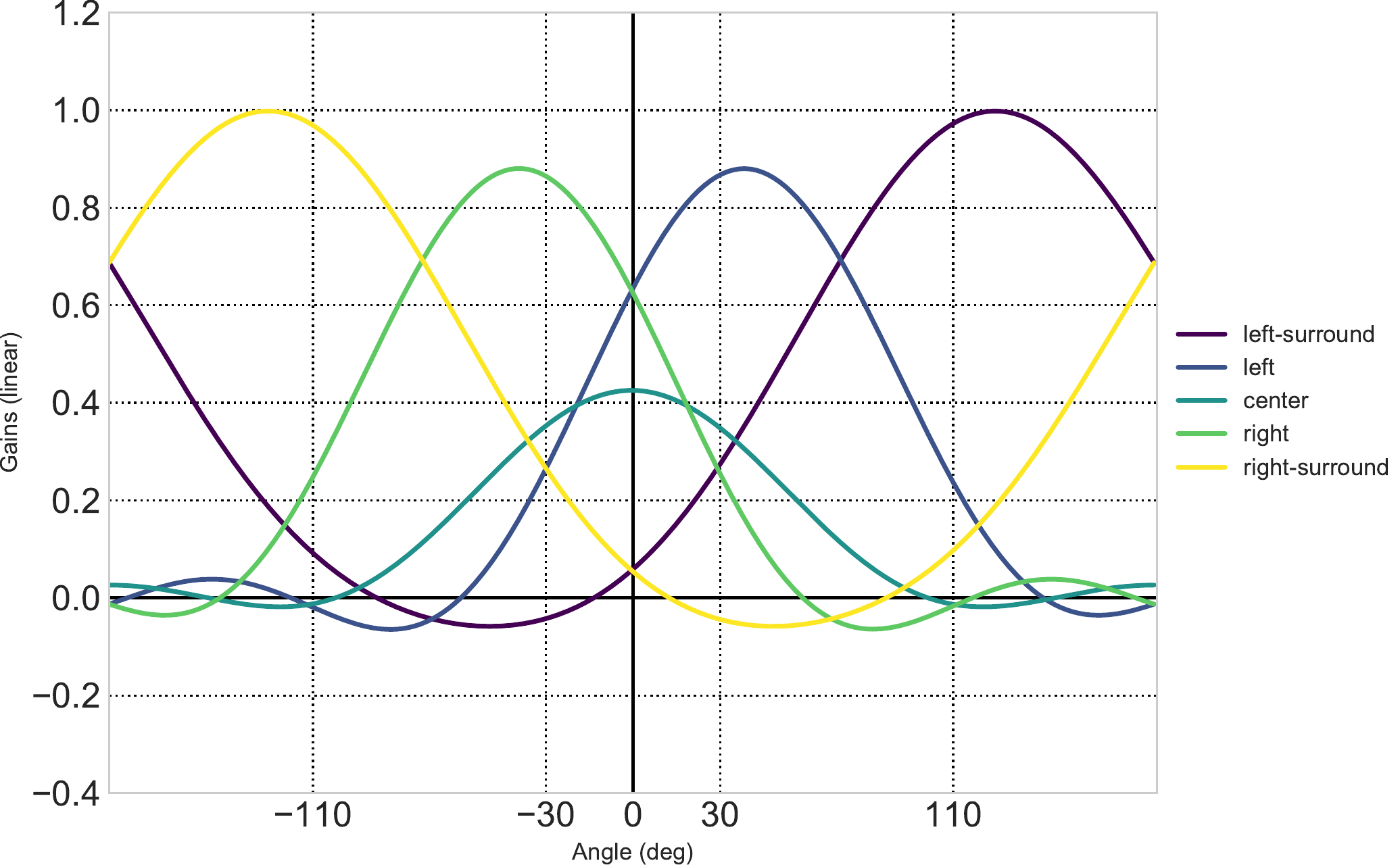}
        \caption{Gains as a function of the angle, for the HF \textit{idhoa2} decoder.}
        \label{fig:2idhoa-gains-hf}
    \end{subfigure}

        \caption{Second order Ambisonics decoders.
        In the left column the \textit{FAA2} decoding
        shipped with ambdec decoder software by Fons Adriaensen, and in the
        right column the one generated with IDHOA.
        Plots \subref{fig:2fons-gains-hf} and~\subref{fig:2idhoa-gains-hf} show the gains of the
        five different loudspeakers as a function of the source position.
        }\label{fig-ambdec-idhoa-comparison-gains}
\end{figure*}

\begin{figure*}
    \centering
    \begin{subfigure}[c]{0.46\textwidth}
        \includegraphics[width=\textwidth]{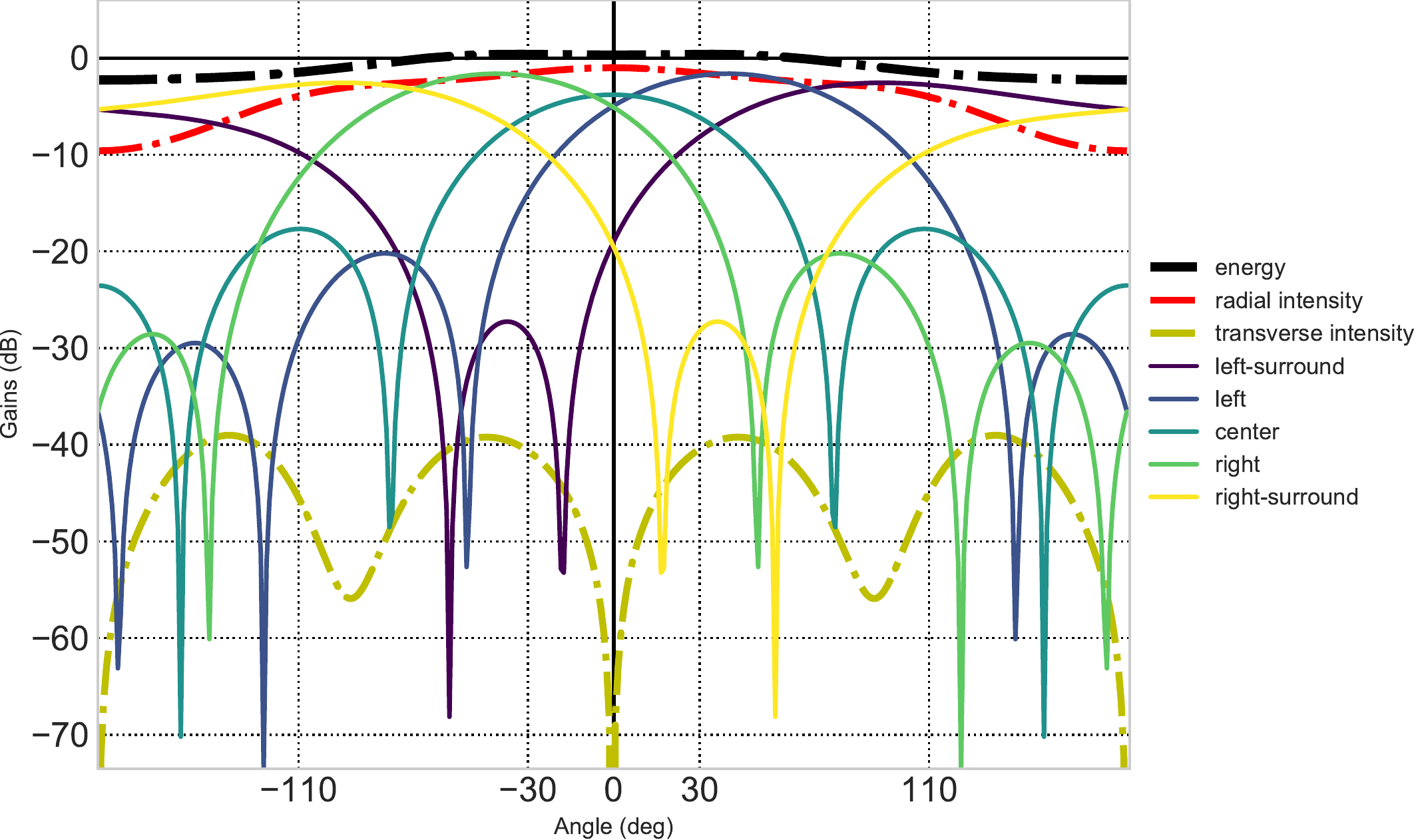}
        \caption{Gains in logarithmic scale as a function of the angle, for the HF \textit{FAA2} decoder.}
        \label{fig:2fons-gains-hf-dB}
    \end{subfigure}
    \begin{subfigure}[c]{0.46\textwidth}
        \includegraphics[width=\textwidth]{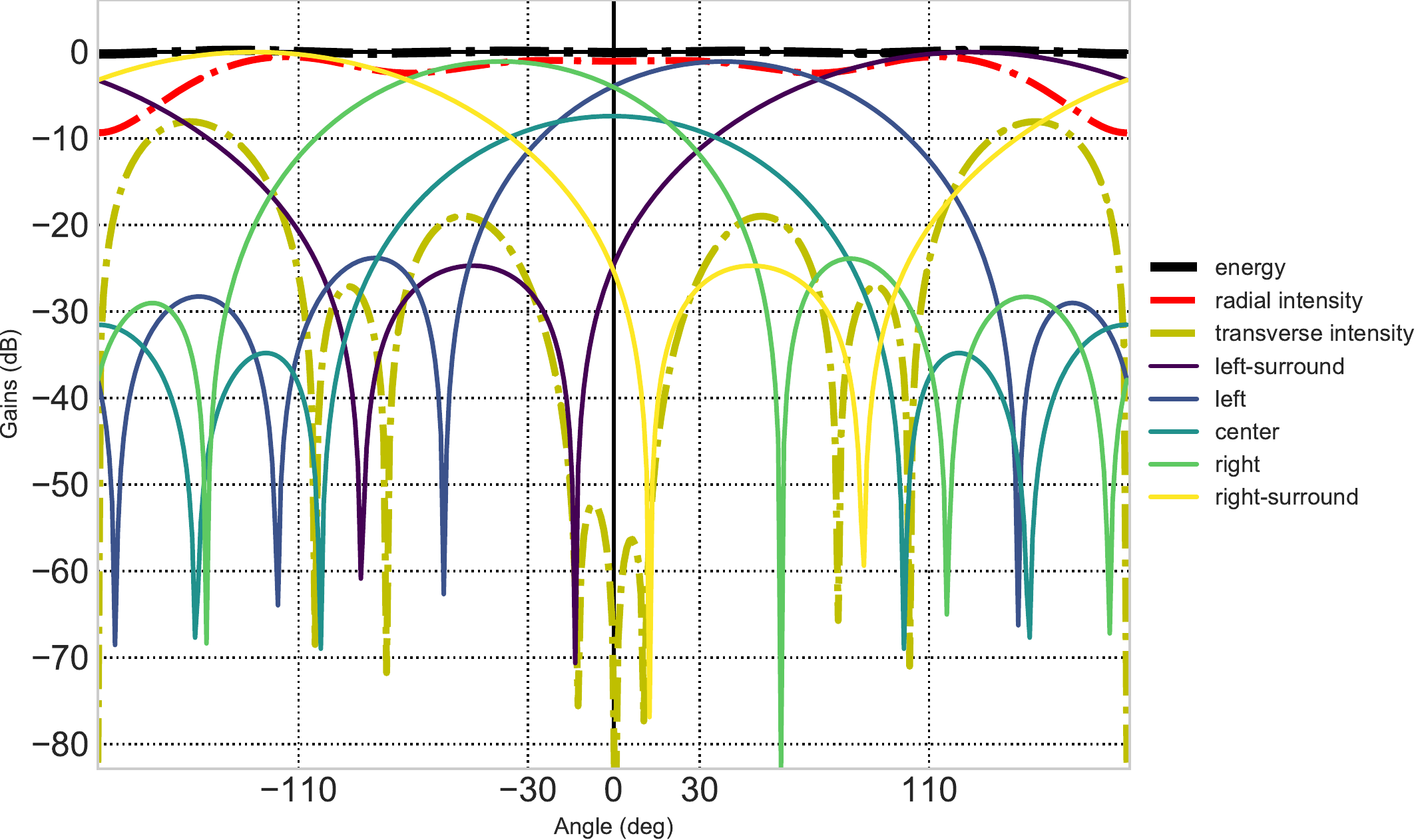}
        \caption{Gains in logarithmic scale as a function of the angle, for the HF \textit{idhoa2} decoder.}
        \label{fig:2idhoa-gains-hf-dB}
    \end{subfigure}

        \caption{Second order Ambisonics decoders.
        In the left column the \textit{FAA2} decoding
        shipped with ambdec decoder software by Fons Adriaensen, and in the
        right column the one generated with IDHOA.
        Plots \subref{fig:2fons-gains-hf-dB} and~\subref{fig:2idhoa-gains-hf-dB} show the gains (in logarithmic scale) of the
        five different loudspeakers as a function of the source position.
        On the same plot is reported the reconstructed energy and intensity.
        }\label{fig-ambdec-idhoa-comparison-gains_dB}
\end{figure*}

Figure~\ref{fig-ambdec-idhoa-comparison} shows the comparison between the
second order decoders \textit{FAA2} and \textit{idhoa2}.
The \textit{FAA2} decoder (Figures~\ref{fig:2fons-lf}, \ref{fig:2fons-hf} and~\ref{fig:2fons-gains-hf}) shows a
3~dB front dominance both at low and high frequencies.
At HF the radial part of the intensity vector is
maximum, 0.9, at $0^\circ$ and then decreases below 0.8 at $\pm 45^\circ$
while the localization error, represented by the tangential part of intensity,
is always very small.

While we could possibly have generated a very similar decoder by tuning IDHOA parameters,
we decided to generate a different decoder, assuring the energy to be
preserved in all directions, and asking for better localization (greater radial
intensity) at the expense of some directionality mismatch,
see Figures~\ref{fig:2idhoa-lf}, \ref{fig:2idhoa-hf} and~\ref{fig:2idhoa-gains-hf}.
The HF plot highlights clearly the differences between the two decoders:
in \textit{idhoa2} the choice is to have good source size at the expense
of some source localization mismatch where no loudspeakers are present.
The reasoning behind allowing for some error in sound position is that: if the apparent source size
is already ``big'' then a localization error is not going to be relevant.
Hence, we preferred to reduce the apparent size first and then,
if the apparent source size gets sufficiently ``small'',
optimize for its position\footnote{\emph{Optimize} an Ambisonics decoding implies always a trade off and it is the result
of a deliberate (arbitrary but informed) choice.}.
\textit{FAA2} decoder prefers to keep small the angular error along all the circle
but with an increased source size.
This difference is also evident from the mean values reported in Table~\ref{tbl:mean}.

Finally, let us note that the \textit{idhoa2} decoder has smaller values for the crosstalk of the lateral loudspeakers
than the \textit{FAA2} decoder, see Table~\ref{tbl:crosstalk}).

 \paragraph{Layout 5.0: third order decoding}
In principle the third order decoding goes beyond what it can be reproduced with a 5.1 layout,
since there are more channels than loudspeakers.
Trying to decode third order Ambisonics in a 5.1 layout can lead to some spatial aliasing,
which can manifest in the form of ``holes in the middle'' of the loudspeaker layout.

However, IDHOA can generate a meaningful third order decoder to a 5.0 layout
leading to a decoding that has better directionality properties near the loudspeaker positions,
at the expense of showing the individual character of each loudspeaker.
This is in contrast to traditional Ambisonics decodings, which tend to
provide an approximately constant radial intensity in all directions.
This way, the resulting behaviour comes closer to traditional pairwise panning.

Table~\ref{tbl:mean} and~\ref{tbl:crosstalk} show that the third order decoding provides
a marginally better mean directionality, with somewhat reduced crosstalk between the loudspeakers.

\begin{figure*}
 \centering
    \begin{subfigure}[c]{0.46\textwidth}
        \includegraphics[width=\textwidth]{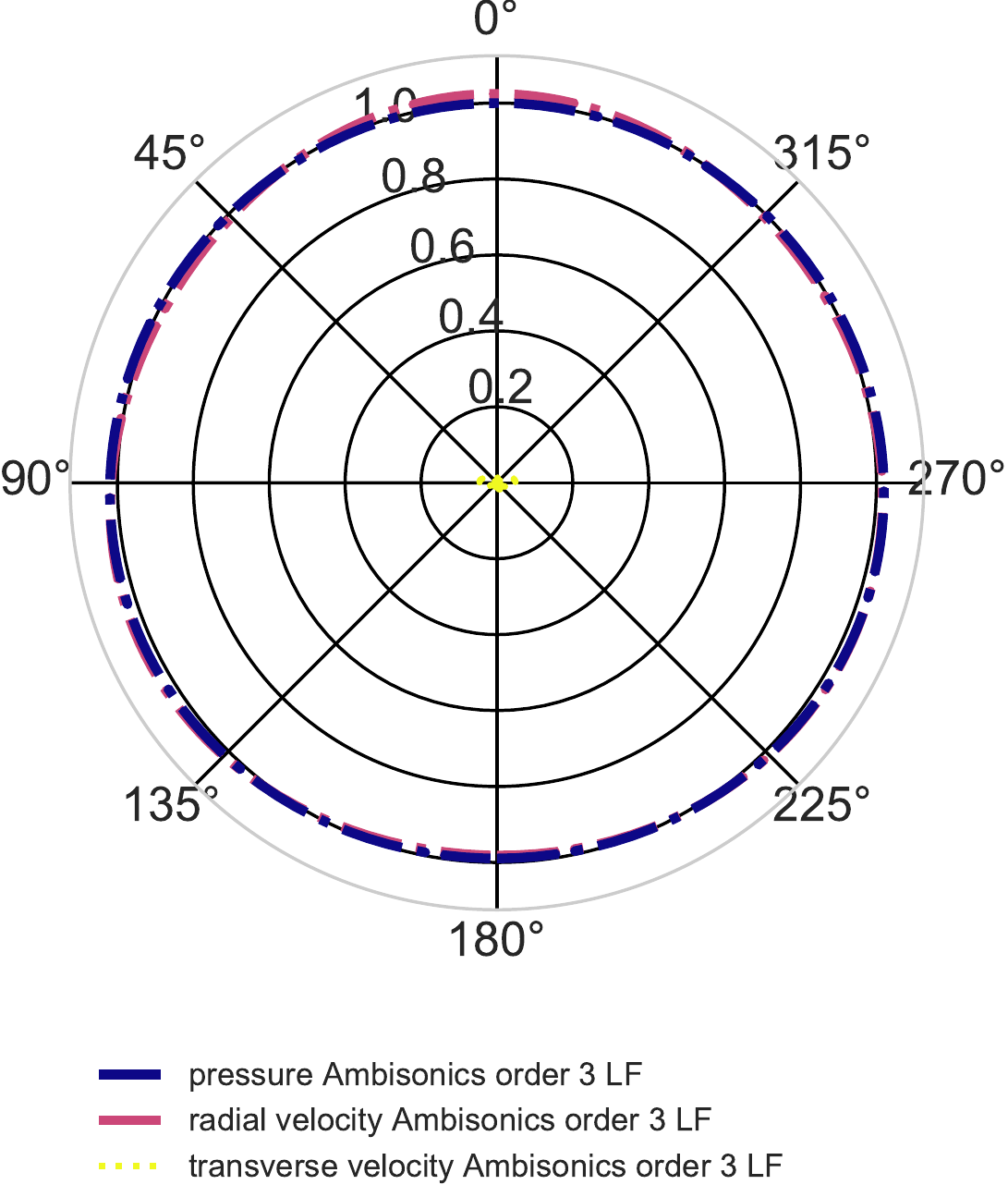}
        \caption{7.0 \textit{idhoa3} decoding for LF.}
        \label{fig:71-lf}
    \end{subfigure}
 		\begin{subfigure}[c]{0.46\textwidth}
        \includegraphics[width=\textwidth]{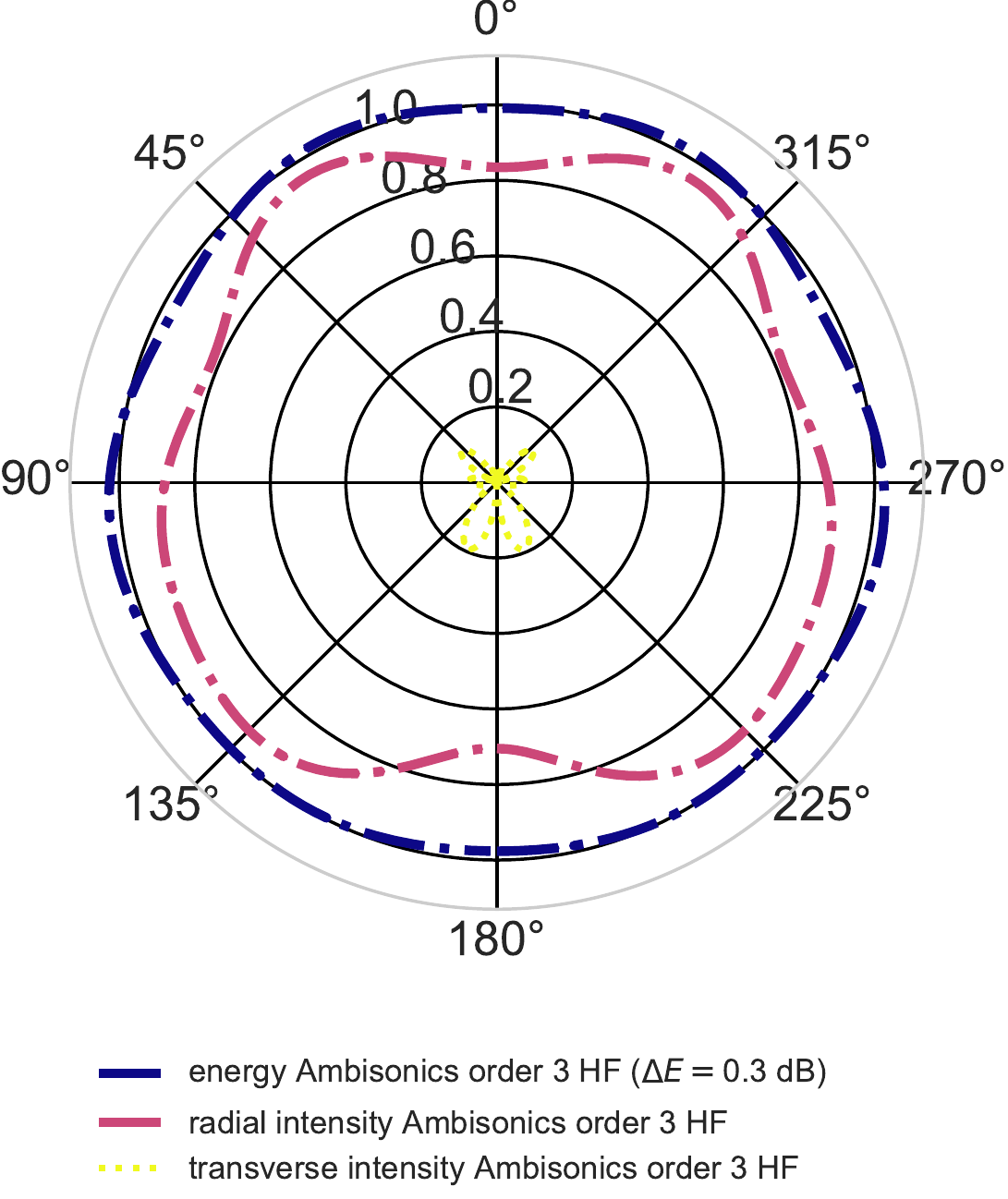}
        \caption{7.0 \textit{idhoa3} decoding for HF.}
        \label{fig:71-hf}
    \end{subfigure}
    \begin{subfigure}[c]{0.46\textwidth}
        \includegraphics[width=\textwidth]{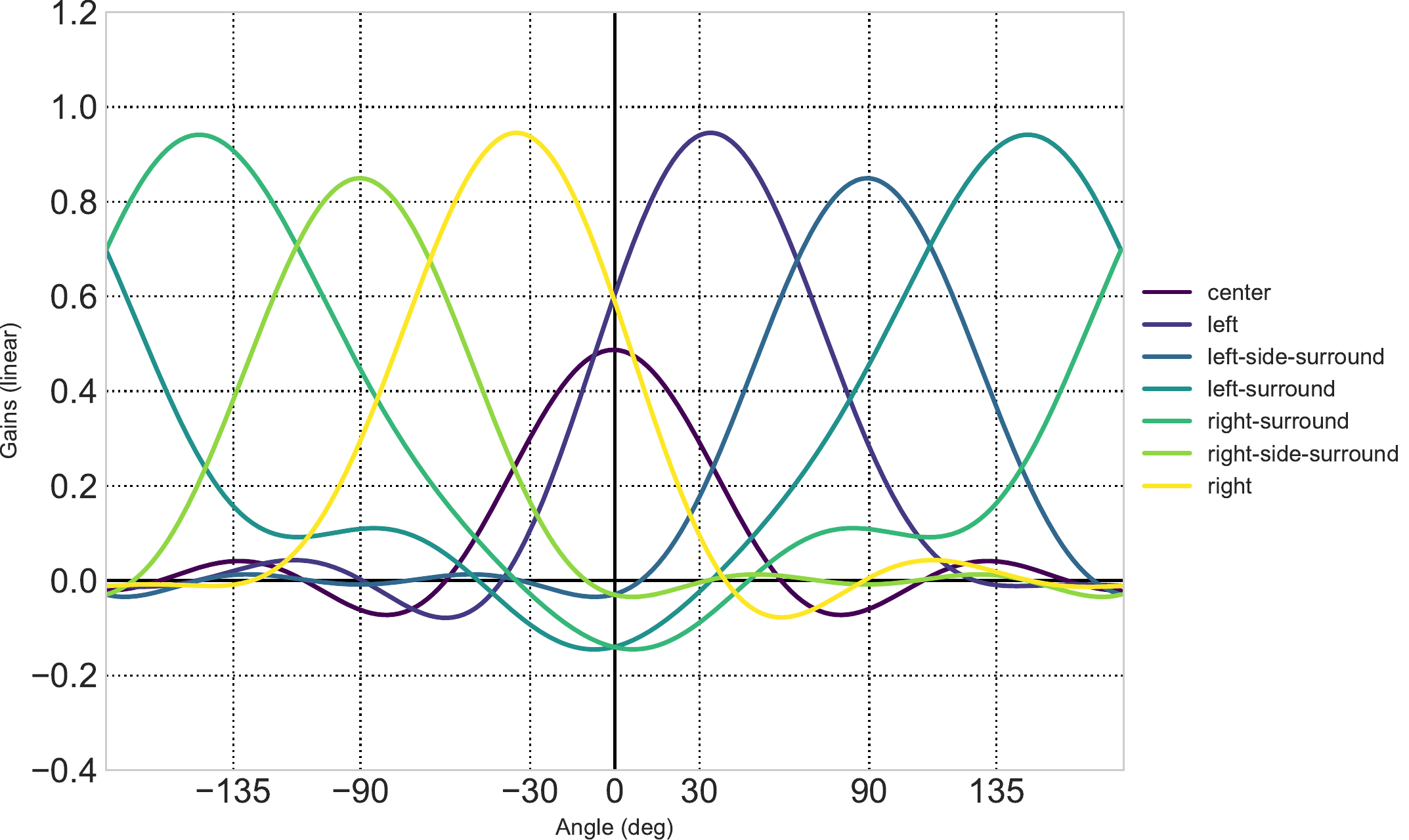}
        \caption{Gains as a function of the angle, for the HF IDHOA decoder.}
        \label{fig:71-gains-hf}
    \end{subfigure}
    \begin{subfigure}[c]{0.46\textwidth}
        \includegraphics[width=\textwidth]{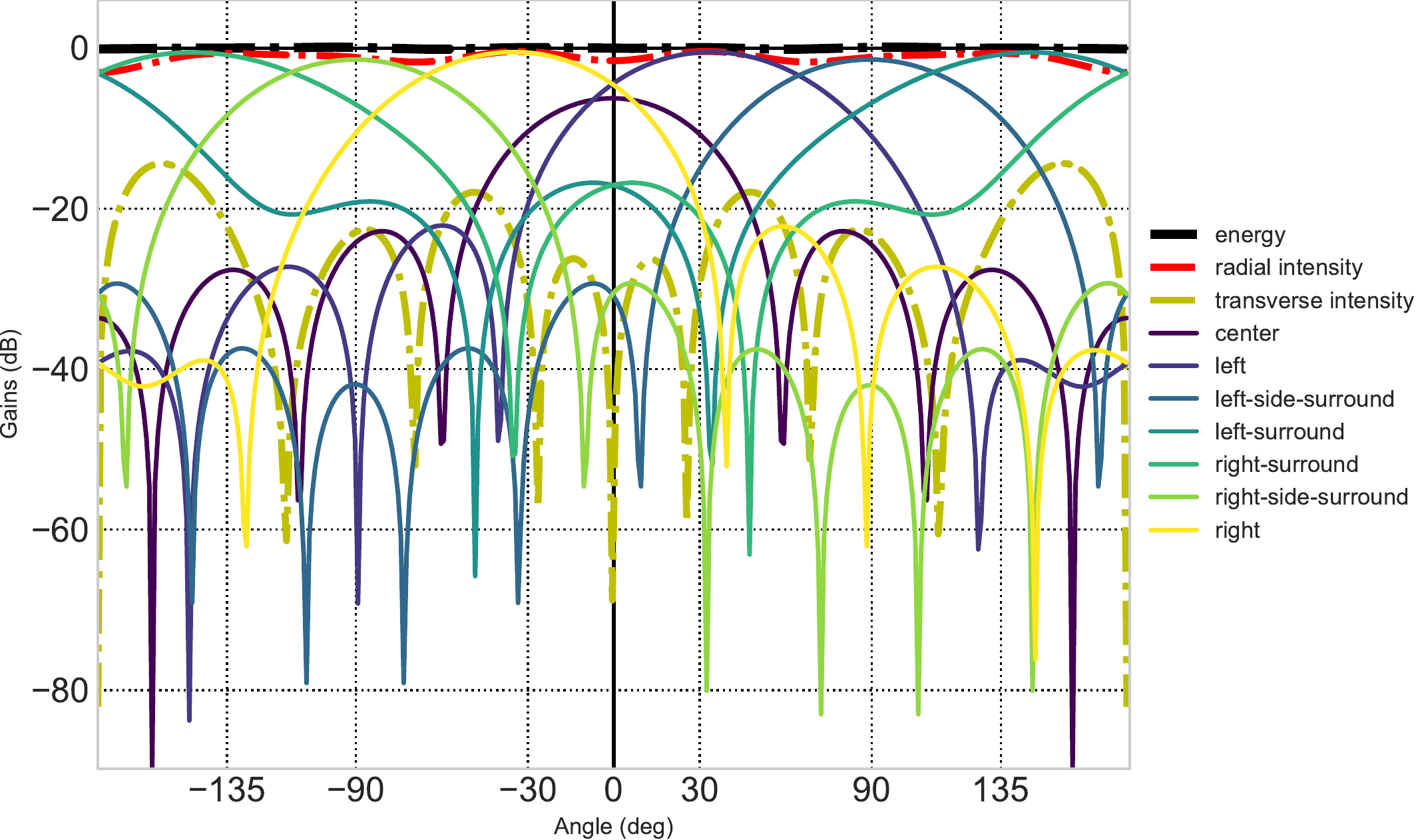}
        \caption{Gains in logarithmic scale as a function of the angle, for the HF IDHOA decoder.}
        \label{fig:71-gains-hf-dB}
    \end{subfigure}
\caption{Third order decoding to 7.0 layout, obtained with IDHOA software.
}\label{fig-71}
\end{figure*}

 \paragraph{Layout 7.0: third order decoding}
The 7.0 layout has enough loudspeakers to decode, in principle, Ambisonics
up to third order.
The distribution of loudspeakers is not regular, from an Ambisonics
point of view, but is indeed more homogeneous than 5.0 layout, leading to a better Ambisonics decoding.
For a comparison between 5.0 and 7.0 look Table~\ref{tbl:mean}: the
intensity components indicate better focused and localized sources in 7.0 than 5.0.
Figures\ref{fig:2idhoa-hf} and~\ref{fig:71-hf} show that in 7.0 the localization properties are more
uniform than in 5.0 and the minimum value for the radial intensity is larger than 0.7, which
is already considered to be good.

\begin{figure*}
 \centering
    \begin{subfigure}[c]{0.46\textwidth}
        \includegraphics[width=\textwidth]{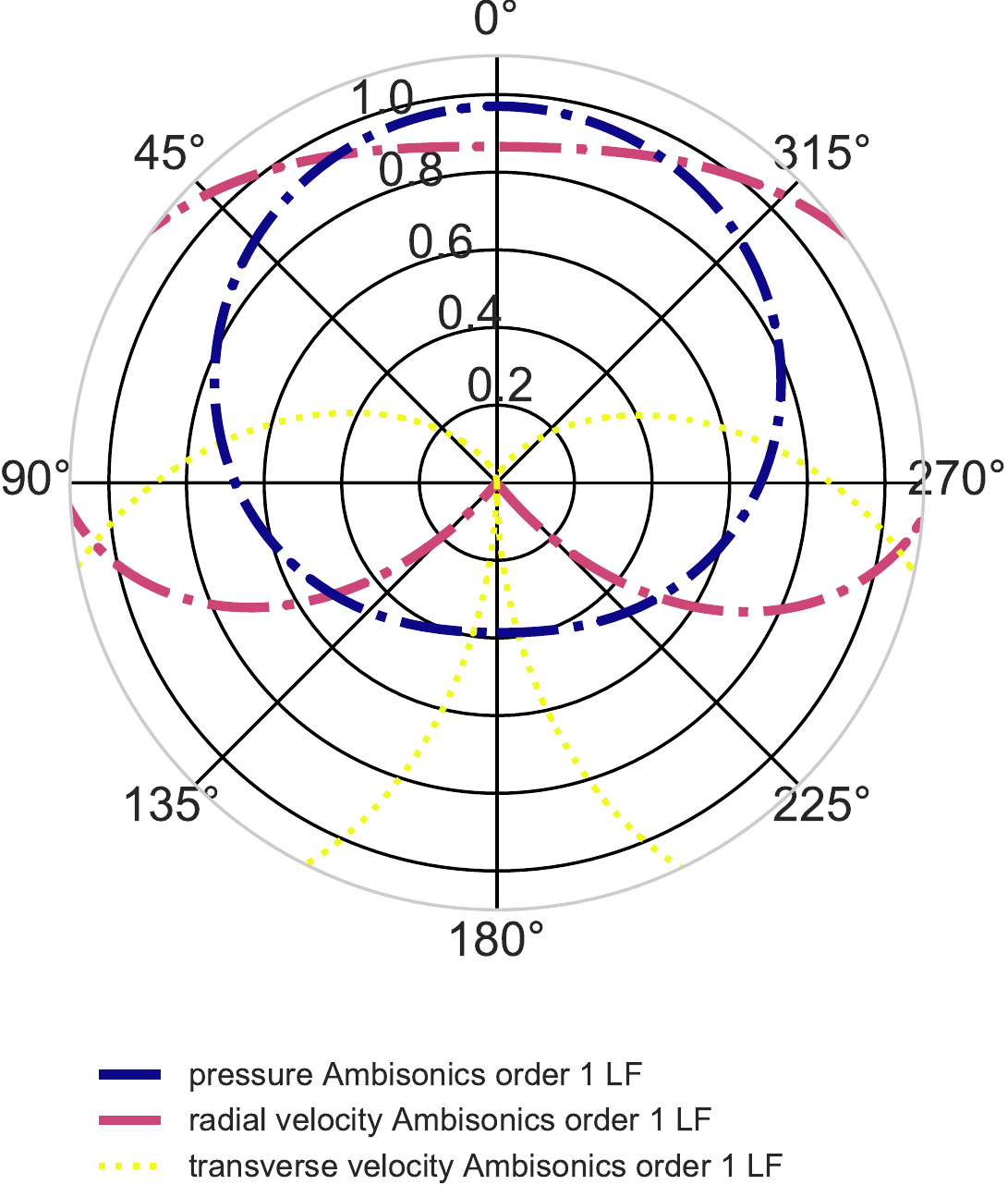}
        \caption{Stereo \textit{idhoa1} decoding for LF.}
        \label{fig:20-lf}
    \end{subfigure}
 		\begin{subfigure}[c]{0.46\textwidth}
        \includegraphics[width=\textwidth]{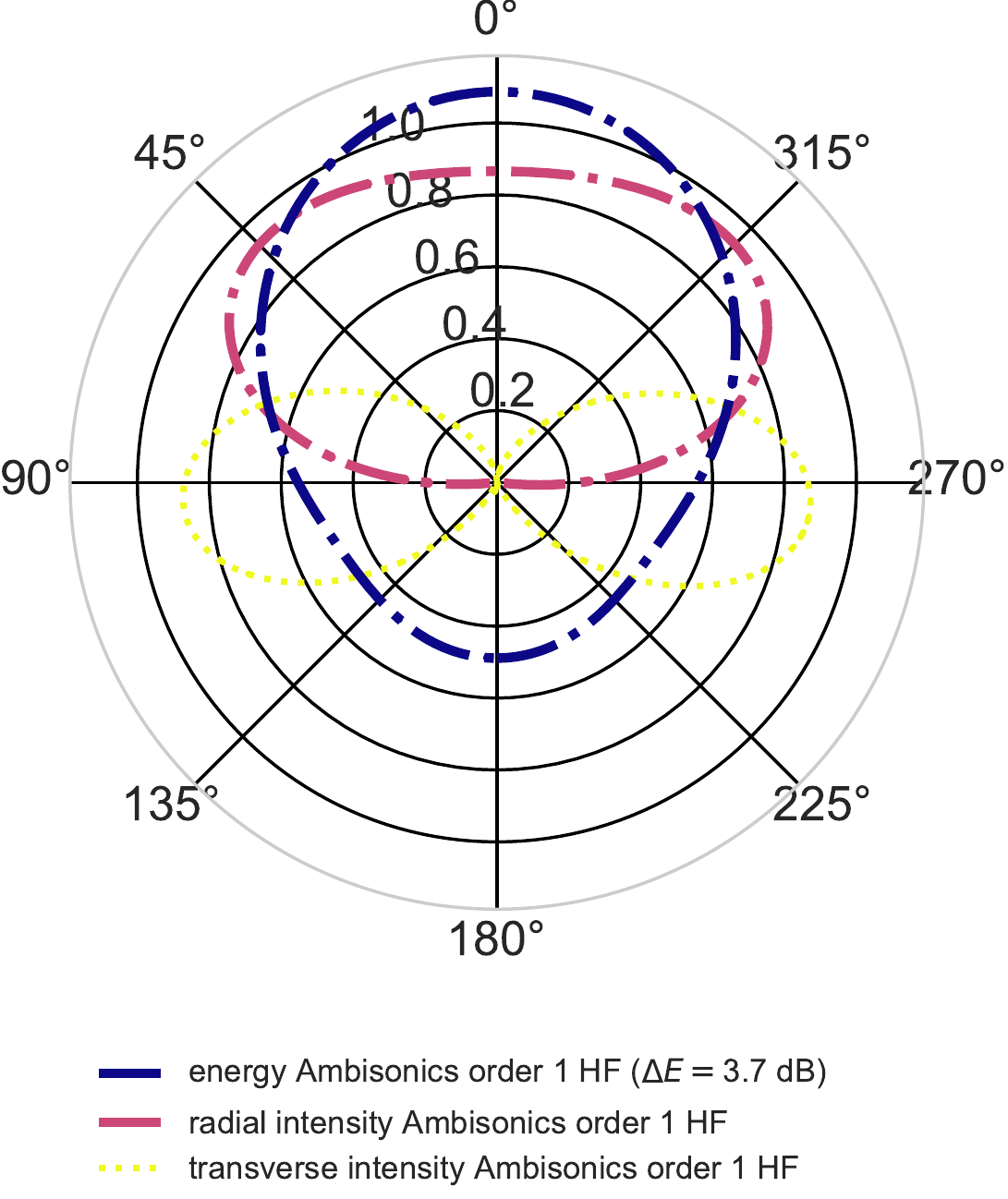}
        \caption{Stereo \textit{idhoa1} decoding for HF.}
        \label{fig:20-hf}
    \end{subfigure}
    \begin{subfigure}[c]{0.65\textwidth}
        \includegraphics[width=\textwidth]{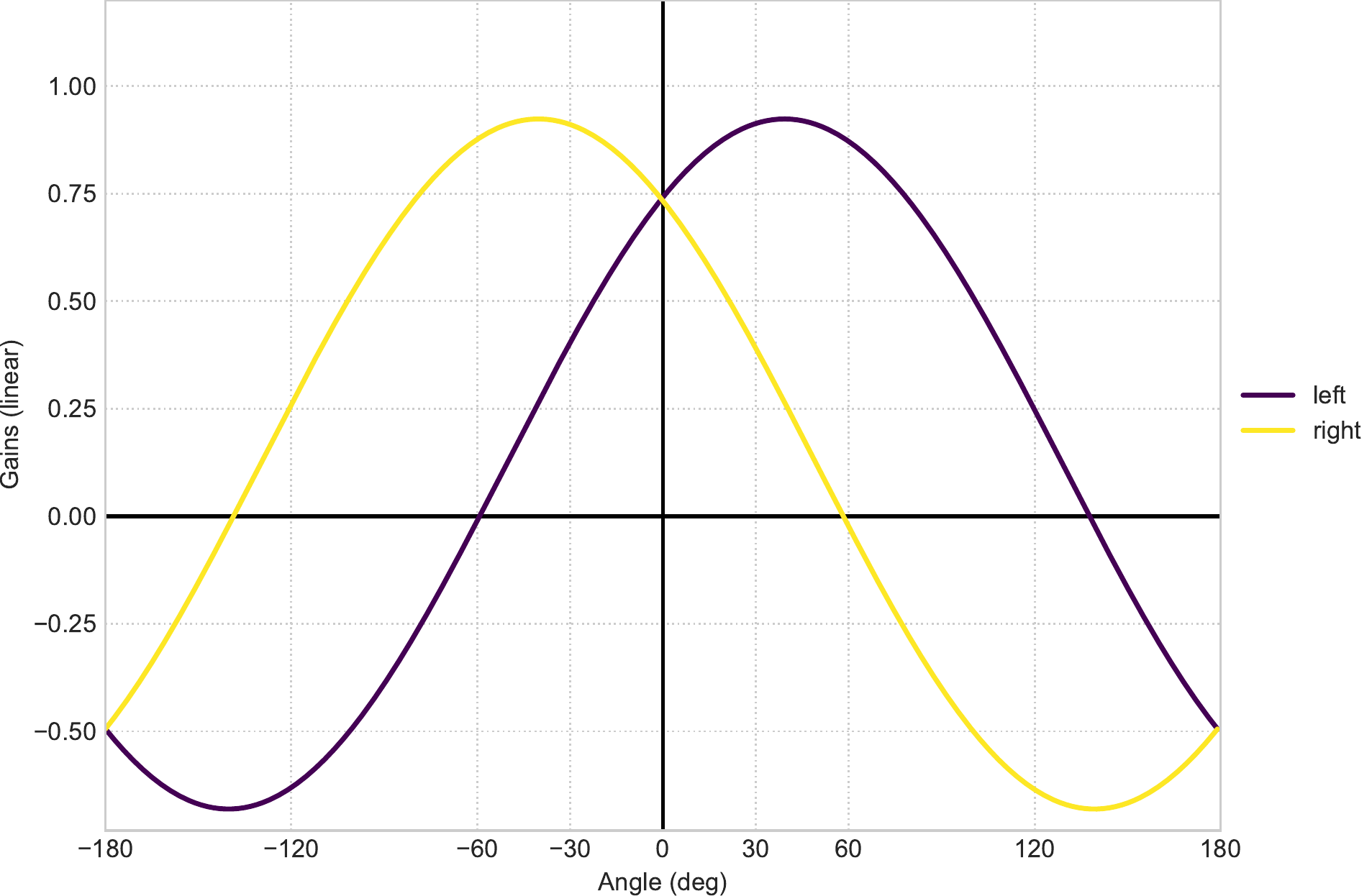}
        \caption{Gains as a function of the angle, for the HF IDHOA decoder.}
        \label{fig:20-gains-hf}
    \end{subfigure}

\caption{First order decoding to 2.0 layout, obtained with IDHOA software.
}\label{fig-20}
\end{figure*}

 \paragraph{Layout 2.0: first order decoding}
We also produced a decoder for a stereo  layout using IDHOA software, requesting an
average of $-3$~dB trim in the rear part.
This choice is common but arbitrary,
other choices are possible and motivated by the amount of information to be mapped from
the back to the front.
In Figure~\ref{fig-20} it is possible to see how the trim in the rear region is
realized, while a good localization is achieved between $\pm 30^\circ$.

Anyway, in our opinion, this has to be considered as an exercise in style,
since for such low number of degrees of freedom manual methods are to be preferred,
given that manual fine tuning might be more predictable and adjust better to the individual preferences.

\section{Subjective Evaluation} \label{sec:subjective-eval}

\subsection{Methodology} \label{subs:methodology}
By using a cohort of 14
subjects with age comprised between 20 and 40 years (13 males, 1 female, all of them with at least some degree of listening experience),
we compared, by means of a ``MUlti Stimulus test with
Hidden Reference and Anchor'' (MUSHRA) test \cite{mushra}, five different Ambisonics
decoders, two state-of-the-art (first and second order), and three generated
with IDHOA (first, second and third order).

The tests were performed in a treated listening room equipped with Genelec 8040B loudspeakers.
The speakers' feed signals are compensated for distance, near-field effect and equalized for room coloration.

The different 5.0 decodings are assessed with respect to
the following criteria:
\begin{enumerate}
    \item The amount of directionality of the sound generated and the correctness of position of the sound source.
    \item Smoothness of panning.
    \item Global spatial perception.
\end{enumerate}

We run three different tests:
\begin{enumerate}
    \item \emph{Localization test.} Source positioned at $0^\circ$, $\pm 30^\circ$, $\pm 110^\circ$.
            Reference is the loudspeaker itself\footnote{We run some preliminary tests
            with a source at $90^\circ$ but all the decoders performed quite badly.
            For this reason we concentrated on positions where a loudspeaker is present in the Ambisonics setup.}.

    \item \emph{Panning test.} Circular panning, one round and two rounds in ten seconds.
            Reference is a standard amplitude panning with size
            (maximum cross-talk with adjacent channels is approximately $-12~dB$).
            The speakers layout of the reference, in the absence of a rotating
            loudspeaker, has been designed to be a custom 8.0 setup (5.0 with three more channels at $\pm 90^\circ$ and $180^\circ$).

    \item \emph{Global perception test.} Custom object-based mix of a pop song rendered through 5.0 Ambisonics
            and 8.0 reference layout with amplitude panning.
\end{enumerate}

For the first two tests the types of sources used are broadband noise (pink noise),
voice (male English voice speaking, recorded in anechoic room) and music (fragment of flamenco with voice and instruments).

Each subject had to evaluate 7 different dual-band decoders, with crossover frequency set to 400~Hz, (5 Ambisonics, 1 reference,
1 anchor) with respect to the reference.
The anchor has been chosen to be a
basic single-band ``na\"\i ve'' or ``projection'' decoding.

For test 1 (localization test), each subject evaluated the 7 decoders in 9 trials
(3 positions times 3 signals).
For test 2 (panning test), each subject evaluated the 7 decoders in 4 trials (2
pannings times 2 signals). Test 3 (assessing the global spatial
perception), had only one trial per subject.

The tests where in one trial the reference is evaluated less than 90
over 100 are discarded from the analysis,
leaving 11 subjects in the worst case.

The three tests were done in succession and lasted between 30 minutes and 2 hours, depending on the listener.

For each one of the three tests we carried out the following comparisons:
\begin{enumerate}
 \item \textit{BHL1} vs.\ \textit{idhoa1}.
 \item \textit{FAA2} vs.\ \textit{idhoa2}.
 \item \textit{idhoa2} vs.\ \textit{idhoa1}.
 \item \textit{idhoa3} vs.\ \textit{idhoa2}.
\end{enumerate}
After checking the normality of the data with the Kolmogorov-Smirnov test,
each pairwise comparison is done using the two-tailed paired t-test method
on the averages of the trials of each subject.
The statistical influence of multiple comparisons is considered and corrected with
the Holm-Bonferroni method \cite{HolmBonferroni}.

\subsection{Tests Results}

\begin{table*}
 \centering
\begin{small}
   \begin{tabular}{l @{} d{3.6} @{} d{2.5} @{} c @{}  d{2.0} }
    \toprule
    & \multicolumn{4}{c}{Test 1 (localization)} \\
    \cmidrule(r){2-5}
    Comparison & \mc{\shortstack{Orig.\\$p$-value}} & \mc{\shortstack{Corr.\\$p$-value}} & S & \mc{Diff.} \\
    \midrule
    \textit{idhoa1} vs.\ \textit{BHL1} &
    0.004 & 0.008 & ** & 14 \\
    \textit{idhoa2} vs \textit{FAA2} &
    0.000014 & 0.00006 & *** & 14 \\
    \textit{idhoa2} vs \textit{idhoa1} &
    0.002 & 0.007 & ** &8 \\
    \textit{idhoa3} vs \textit{idhoa2} &
    0.09 & 0.09 & \mc{--} & 2 \\
    \bottomrule
  \end{tabular}
\end{small}
\vspace{0.5cm}

\begin{small}
   \begin{tabular}{l @{} d{3.6} @{} d{2.5} @{} c @{}  d{2.0} }
    \toprule
    & \multicolumn{4}{c}{Test 2 (panning)} \\
    \cmidrule(r){2-5}
    Comparison & \mc{\shortstack{Orig.\\$p$-value}} & \mc{\shortstack{Corr.\\$p$-value}} & S & \mc{Diff.} \\
    \midrule
    \textit{idhoa1} vs.\ \textit{BHL1} &
    0.04 & \mc{--} & \mc{--} &16 \\
     \textit{idhoa2} vs \textit{FAA2} &
    0.1 & \mc{--} & \mc{--} &9 \\
    \textit{idhoa2} vs \textit{idhoa1} &
    0.04 & 0.08 & \mc{--} & 6  \\
    \textit{idhoa3} vs \textit{idhoa2} &
    0.11 & \mc{--} & \mc{--} & 3 \\
    \bottomrule
  \end{tabular}
\end{small}
\vspace{0.5cm}

\begin{small}
   \begin{tabular}{l @{} d{3.6} @{} d{2.5} @{} c @{}  d{2.0} }
    \toprule
    & \multicolumn{4}{c}{Test 3 (global perception)} \\
    \cmidrule(r){2-5}
    Comparison & \mc{\shortstack{Orig.\\$p$-value}} & \mc{\shortstack{Corr.\\$p$-value}} & S & \mc{Diff.} \\
    \midrule
    \textit{idhoa1} vs.\ \textit{BHL1} &
    0.0008 & 0.003 & ** & 23 \\
     \textit{idhoa2} vs \textit{FAA2} &
    0.4 & \mc{--} & \mc{--} &-2 \\
    \textit{idhoa2} vs \textit{idhoa1} &
    0.3 & 0.96 & \mc{--} & 4 \\
    \textit{idhoa3} vs \textit{idhoa2} &
    0.98 & \mc{--} & \mc{--} & 0 \\
    \bottomrule
  \end{tabular}
\end{small}

 \caption{Significance analysis of the four comparisons in the three tests. The original $p$-value lists the result of the two-tail paired t-test. The corrected value corresponds to the result of the Holm-Bonferroni correction. The column ``S'' indicates the significance. The ``Diff.'' column indicates the average difference in MUSHRA points. }
 \label{tab:subjective}
\end{table*}

\begin{figure*}
 \centering
    \begin{subfigure}[b]{0.8\textwidth}
        \includegraphics[width=\textwidth]{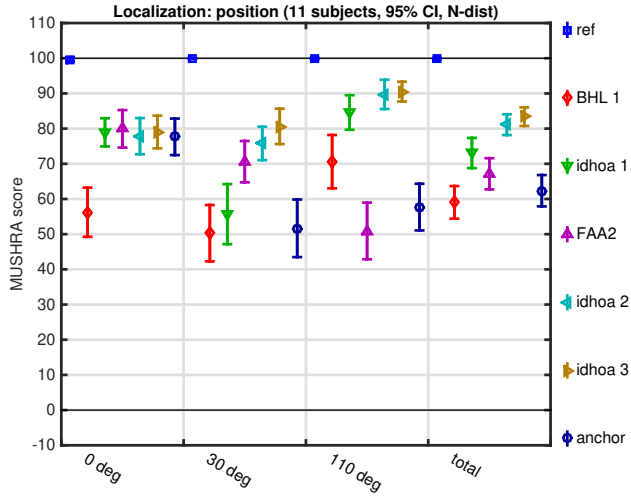}
        \caption{Test 1 (localization). Trials grouped for source direction.}
        \label{fig:local-direction}
    \end{subfigure}
    \begin{subfigure}[b]{0.8\textwidth}
        \includegraphics[width=\textwidth]{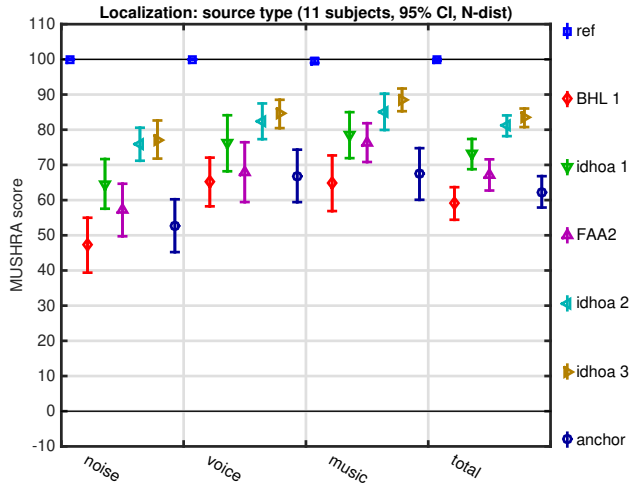}
        \caption{Test 1 (localization). Trials grouped for source type.}
        \label{fig:local-type}
    \end{subfigure}

    \caption{Listening tests results.
        Figure~\subref{fig:local-direction} shows the
        listening test scores for source position and size evaluation grouped for
        source position, while in~\subref{fig:local-type} the scores are grouped for
        source type.
        Error bars correspond to two times the standard deviation of the mean.}\label{fig:list-test-result-1}
\end{figure*}

\begin{figure*}
 \centering
    \begin{subfigure}[b]{0.8\textwidth}
        \includegraphics[width=\textwidth]{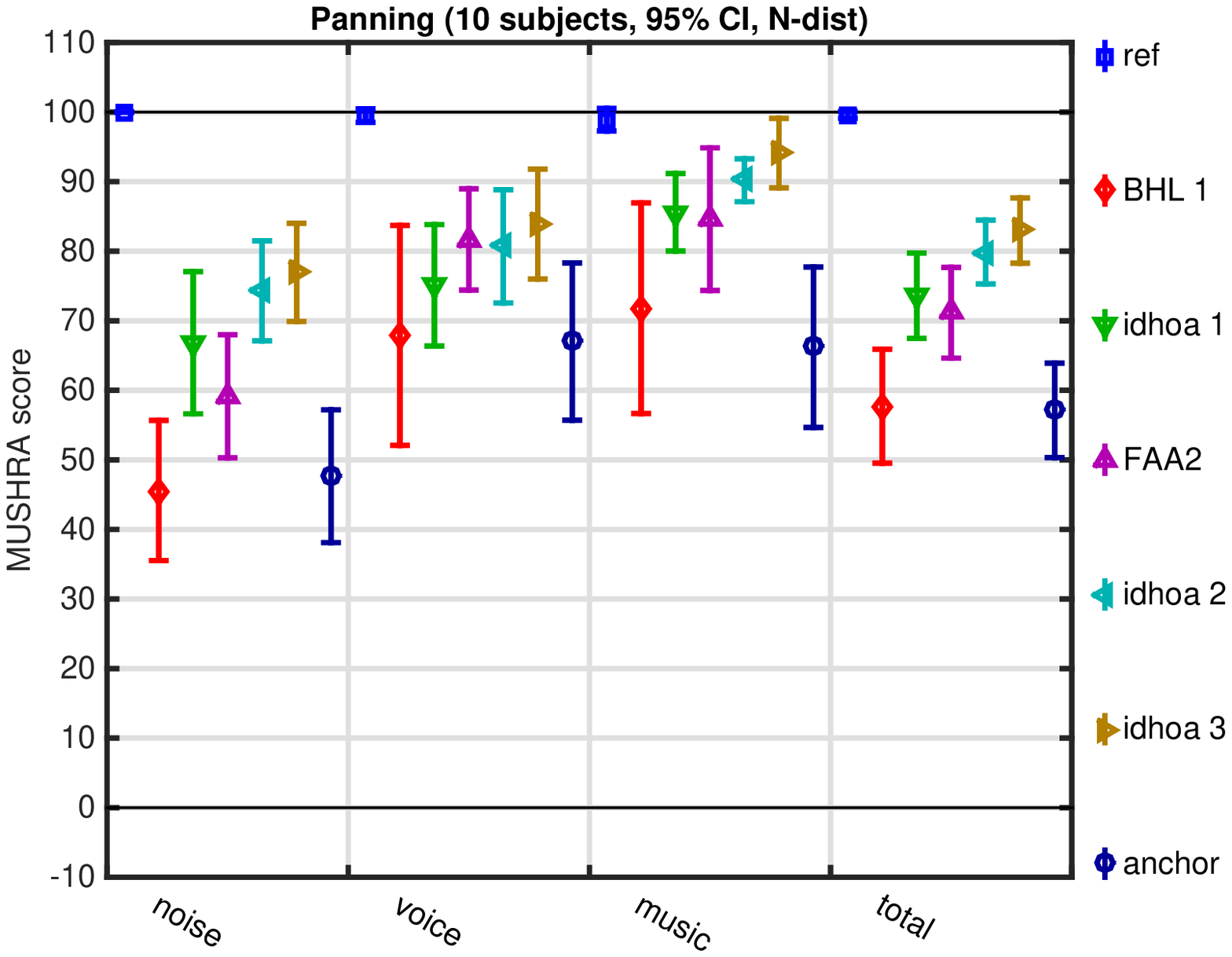}
        \caption{Test 2 (panning). Trials grouped for source type.}
        \label{fig:panning}
    \end{subfigure}
    \begin{subfigure}[b]{0.8\textwidth}
        \includegraphics[width=\textwidth]{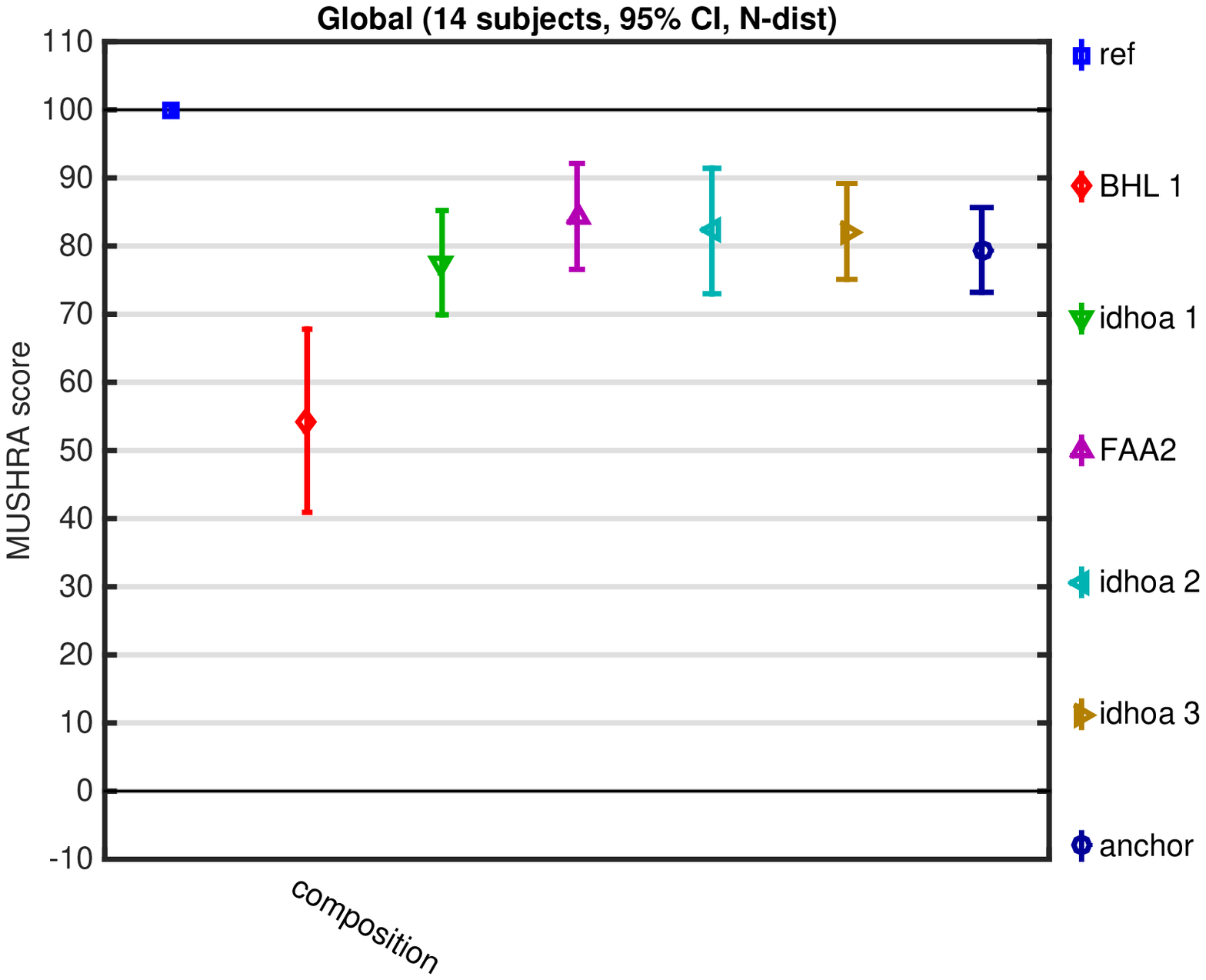}
        \caption{Test 3 (global perception).}
        \label{fig:global}
    \end{subfigure}

    \caption{Listening tests results.
        Figure~\subref{fig:panning} shows the results for panning quality
        grouped for source type.
        Figure~\subref{fig:global} reports the scores obtained by
        the different decoders evaluated with a ``pop song'' spatial composition.
        Error bars correspond to two times the standard deviation of the mean.}\label{fig:list-test-result-2}
\end{figure*}

Figures~\ref{fig:list-test-result-1} and \ref{fig:list-test-result-2} show the results of the three different tests.
Figures~\ref{fig:local-direction}
and~\ref{fig:local-type} show the results of the localization test,
grouped for source direction and source type, respectively.
Figure~\ref{fig:list-test-result-2} shows the results for the panning and the global perception tests.

In general the BLH1 decoder exhibits good
properties at localizing the sources in correct place [e.g.\ see the
$110^\circ$ set of trials in Figure~\ref{fig:local-direction}] but the lower loudness, due
to the pressure normalization at high frequencies, was negatively evaluated by all listeners.
Globally, the BLH1 decoder was not better evaluated than the anchor, probably due to the loudness issue.

The \textit{FAA2} decoder suffers especially when the source is at $110^\circ$.
This suggests that the request for zero angular error in every direction at the expense of spatial sharpness and crosstalk,
as \textit{FAA2} does, is detrimental to localization performances.

Some decoders, particularly \textit{BHL1} and \textit{idhoa1}, perform especially bad at $30^\circ$.
Particularly problematic for Ambisonics is the frontal region at $0^\circ$,
where all the three speakers are active at the same time.

In the global analysis, averaging all the measurements, it is possible to highlight
a trend for the decoders where \textit{idhoa1} and
\textit{FAA2} are almost equivalent, and \textit{idhoa2} and 3 are better evaluated than the
former. %

All the listeners reported that the differences between the decoders where
evident when listening to the broadband noise, while much more subtle when
using ``natural'' signals, especially music.
Figures~\ref{fig:local-type} and~\ref{fig:panning} show that the MUSHRA scores are
higher for voice and music than for noise, both in localization and panning.

Since the Kolmogorov-Smirnov test showed no significant deviations from normality,
data are analyzed performing a pairwise comparison between four combinations of decoders using the two-tailed
paired t-test method, as explained in Section~\ref{subs:methodology}, and
results are summarized in Table~\ref{tab:subjective}.

In the localization test, the pairwise comparison reveals that there is significant difference between \textit{idhoa1} and \textit{BHL1},
\textit{idhoa2} and \textit{FAA2}, where the IDHOA decoders are significantly better rated than the alternatives.
When comparing \textit{idhoa2} and \textit{idhoa1}, the former gets significantly better evaluation than the latter.
While checking for \textit{idhoa3} against \textit{idhoa2}, no significant difference is found.
Not surprisingly, this trend follows closely the values for $I_\text{R}$ reported in Table~\ref{tbl:mean}.

In the panning test none of the four comparisons gives significant difference among the decoders.
Nevertheless, the tendency is completely analogous to the localization test, hinting that similar significant results
could perhaps be obtained with increased statistics. %

For the global evaluation test only the comparison \textit{idhoa1} versus \textit{BHL1} results significant,
and the former is significantly better rated than the latter.
Again, this could be due to the level difference.

\section{Summary} \label{sec:evaluation-summary}
IDHOA can produce a wide variety of decoders both in 2D and 3D,
allowing for a fine control over the loudness and localization properties
by tuning a small set of parameters.

The energy and intensity plots show that the decodings generated with IDHOA have
--to some extent-- better directionality properties than the state-of-the-art decoders at the expense of some error in localization.

Subjective testing has shown that localization properties of the decodings generated by IDHOA
 are better evaluated than the state-of-the-art decoders, %
with a similar trend for the panning properties (although results are not significant in this case).
On the other hand, no significant differences have been found in the global evaluation test (except for the state-of-the-art %
first order decoder, which can probably be attributed to a level mismatch).
This might indicate that the chosen fragment is not representative enough to show differences between the decoders.

The code used to generate the Ambisonics decodings and the decodings themselves
are publicly available in \cite{github-idhoa-new}.

    \part{Wavelets} \label{part:wavelets}

\chapter{Introduction to Wavelet Theory} \label{ch:wavelet-theory}
In this Chapter we want to give sufficient background information to later understand the idea behind our wavelet optimization,
even if our approach aims at capturing the main concepts of wavelets, e.g.\ locality, more than a formal wavelet construction.

The contributions to Wavelet Theory come from very different areas of Science and Engineering.
Because the wavelets come from very different areas of expertise,
there are many ways to motivate their construction and understand their properties.
This fragmented and diverse development also lead to many wavelet transforms and wavelet-generation schemes.
Moreover, one of the main concepts behind the wavelet transforms, especially for compression,
is to adapt the basis of the analysis functions to the signal to be analyzed.
For this reason there are almost as many wavelet families as applications or problems, increasing the wild diversity of wavelet approaches.

\section{Introduction to Wavelet Transforms}  \label{sec:intro-to-wavelets}

In this Section we will gradually introduce the concept of Wavelet transform by similarity and difference with the Fourier transforms.

\paragraph{Fourier Transform}
The Fourier transform (FT) of a one dimensional square integrable signal $s(t)$ is given by
\begin{equation}
    S \left( f \right) = \int_{-\infty}^{+\infty} s(t) e^{-2i\pi ft} dt
    \label{eq:fourier-transform}
\end{equation}
the inverse transform is given by
\begin{equation}
     s \left( t \right) = \int_{-\infty}^{+\infty} S \left( f \right) e^{2i\pi ft} df .
    \label{eq:inverse-fourier-transform}
\end{equation}
\sloppy Equation~\eqref{eq:fourier-transform} gives a representation of the frequency content of the signal $s(t)$
but gives no information about its localization in time, vice versa for Eq.~\eqref{eq:inverse-fourier-transform}.
The bases of the Fourier transform are the sine and cosine.
The FT, then, maps time (or space) to frequency (and vice versa) but, because of the infinite support of the FT
basis functions, it is impossible to have information on time and frequency at the same time.
Note that the same happens with the Spherical Harmonics (and so with Ambisonics), just the two domains connected by the SH
are space, $(\theta, \phi)$, and `angular frequency', $(l, m)$.

\paragraph{Windowed Fourier Transforms}
One way to have information on both domains at the same time
while preserving the linearity of the operator is to introduce a window, giving birth to the
windowed Fourier transform (WFT), also known as short-time Fourier transform (STFT).
Being $w(t)$ a window function (real, for simplicity) with a finite integral and compact support
(i.e.\ non-zero over a finite interval\footnote{This
condition is often relaxed by asking some fast decay, for example exponential.}),
the WFT of the signal $s(t)$ is defined as:

\begin{equation}
    S_{W} \left( \tau, f \right) = \int_{-\infty}^{+\infty} s(t) w(t-\tau) e^{-2i\pi ft} dt
\end{equation}

The application of a window has several consequences.
The transform is function of two variables, the frequency $f$ and the position at which the window is applied $\tau$.
The filter function $w(t)$ is a window in time but also a window in the frequency spectrum $f$ around $\tau$.
The shape of the filter in frequency domain is $W(f)$, which is the FT of $w(t)$.
One thing that is often forgotten (and it is good to keep in mind also when we will talk about wavelets) is that
the shape of $W(f)$ in general is very different from $w(t)$,
and only for a limited set of functions it is possible to get $w(t) \propto W(f)$.
The choice of the window in the time domain affects the shape of the window in frequency domain:
typically there will be a main lobe and some ``spill'' at low and high frequencies.
If we define the spread in frequency, \emph{bandwidth}, of the window $w(t)$ as:
\begin{equation*}
    \Delta f^2 = \frac{ \int f^2 |W(f)|^2 df }{ \int |W(f)|^2 df},
\end{equation*}
while the spread in time can be defined as:
\begin{equation*}
    \Delta t^2 = \frac{ \int t^2 |w(t)|^2 dt}{ \int |w(t)|^2 dt}
\end{equation*}
(by Parseval's theorem both denominators are equal, and are the energy of $w(t)$)
then the Heisenberg inequality bounds their product\footnote{The lower bound $\Delta f \: \Delta t = \frac{1}{4\pi}$
is reached by the Gabor transform that uses a Gaussian as $w(t)$ window function.
Note that the FT of a Gaussian is a Gaussian, so the $W(f)$ is a Gaussian too.
The Gabor transform has a set of nice properties that come from the choice of the Gaussian as $w(t)$.}
\begin{equation*}
    \Delta f \: \Delta t \geq \frac{1}{4\pi} .
\end{equation*}
The Heisenberg inequality has two consequences.
First, it is not possible to have infinite precision in both time and frequency.
This means that it is not possible to separate two impulses that are closer than $\Delta t$
or separate two tones that are closer than $\Delta f$.
The second effect is that, once the window is chosen, the resolution limit is the same over all times and frequencies.
For example, let us imagine that we choose a window in time to have good relative resolution at mid frequencies,
then we will get poor relative resolution in frequency for the low frequencies,
but for high frequencies we will get very good relative resolution in frequency and a very bad one in time.
(In typical audio applications more than one STFT is run in parallel with different window sizes).

\paragraph{Wavelet Transform} As already said, the paths that lead to the Wavelet Transform come from very different directions,
but the ideas and motivations behind it are the same and can be reduced to two:
\begin{enumerate}
    \item Constant resolution along frequency: $ \frac{\Delta f }{f} = c $, with $c$ a constant
        (in signal processing is known as \emph{constant-Q} analysis).
    \item Modeling a signal using a basis that is similar to the signal, resulting in less coefficients and better compression.
\end{enumerate}

\begin{figure}[t!]
    \centering
    \begin{subfigure}[c]{0.45\textwidth}
        \includegraphics[width=\textwidth]{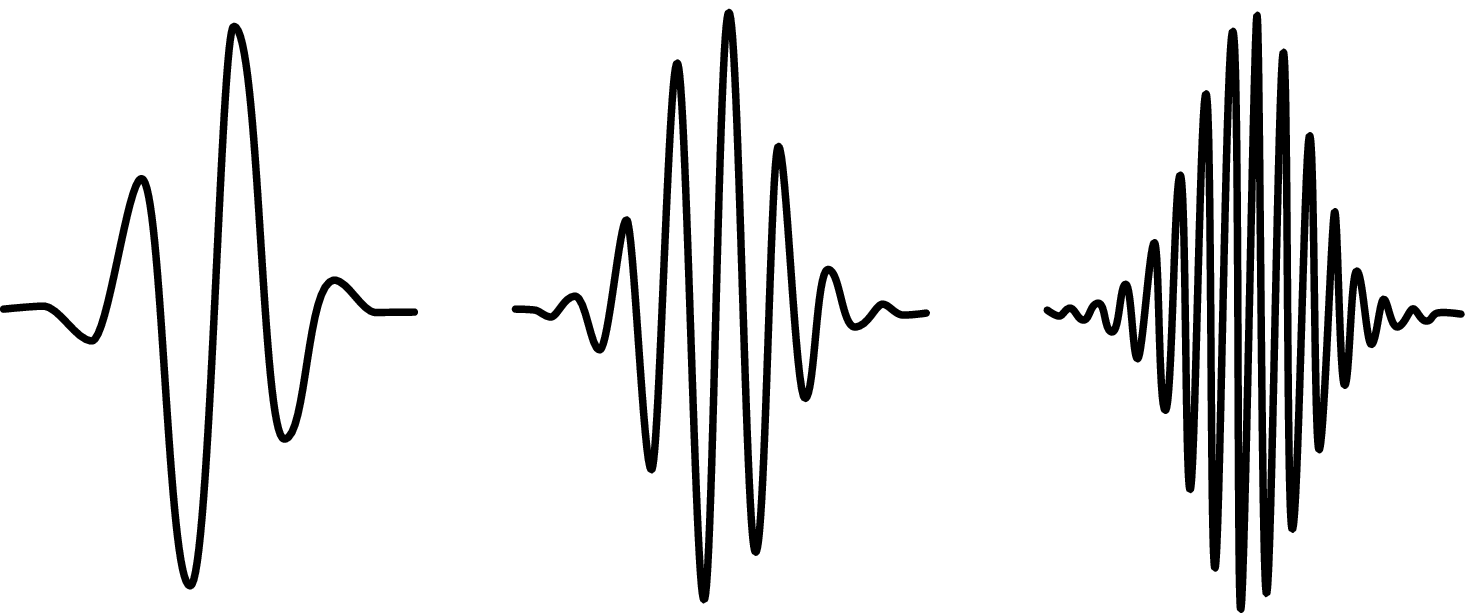}
    \end{subfigure}
    \hspace{0.4cm}
    \begin{subfigure}[c]{0.45\textwidth}
        \includegraphics[width=\textwidth]{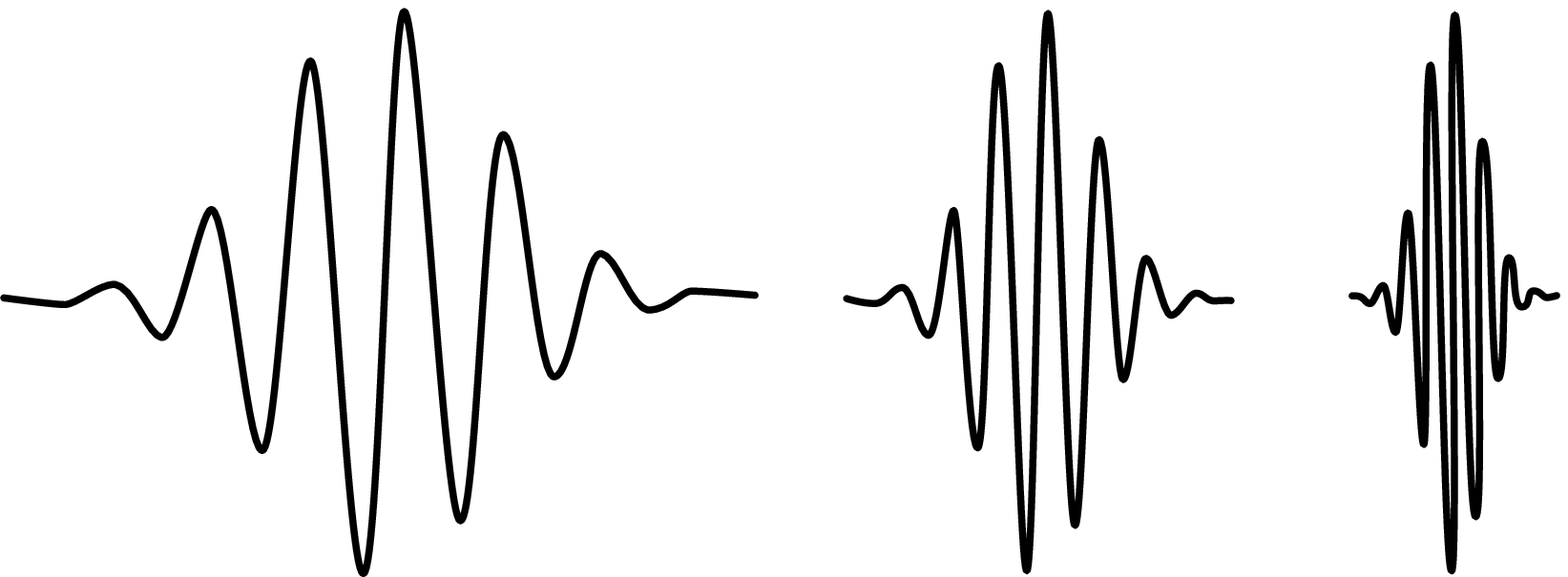}
    \end{subfigure}
    \\
    \begin{subfigure}[c]{0.46\textwidth}
        \includegraphics[width=\textwidth]{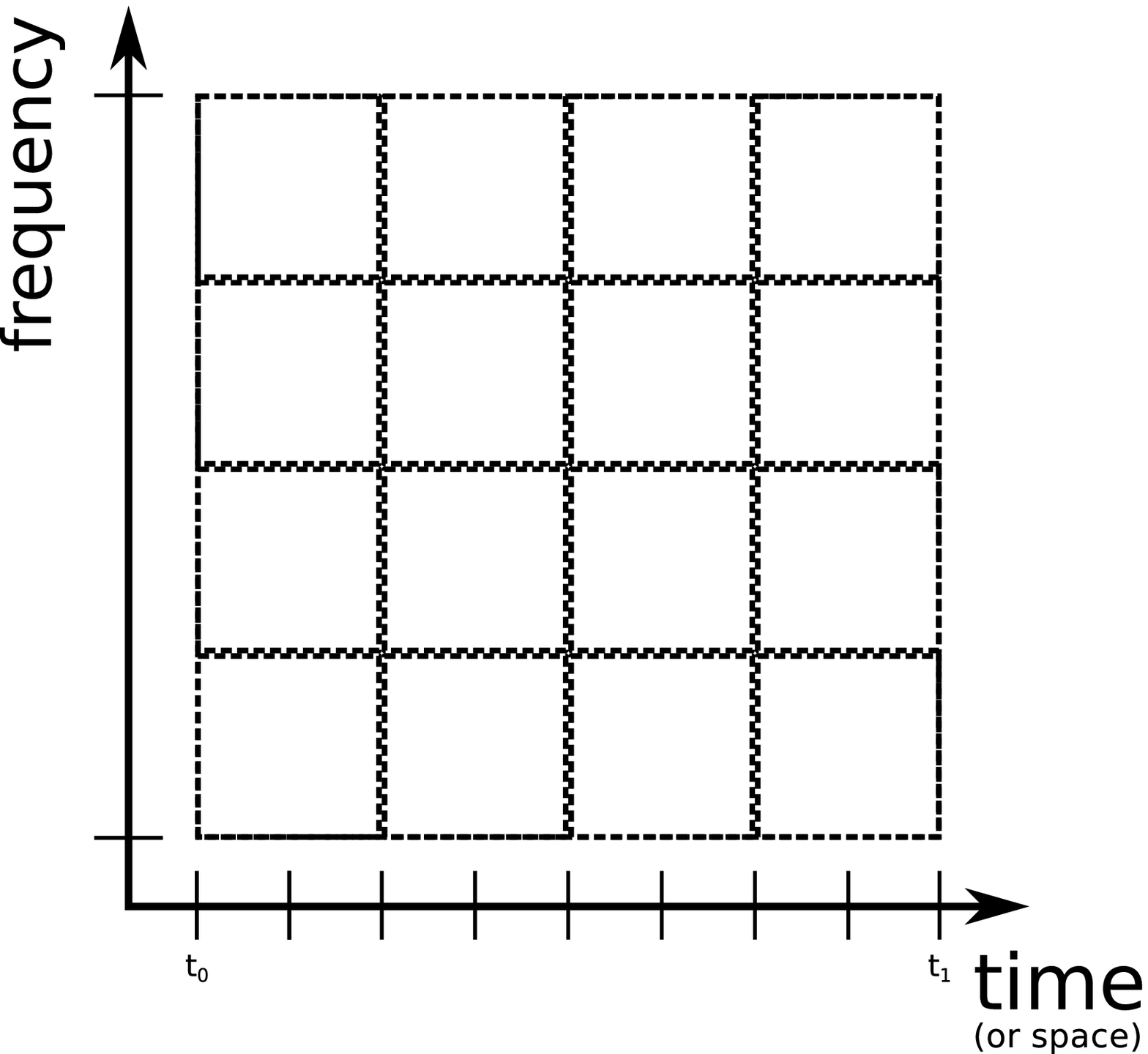}
        \caption{STFT - The oscillations increase in frequency inside a window of fixed length.}
    \end{subfigure}
    \hspace{0.2cm}
    \begin{subfigure}[c]{0.46\textwidth}
        \includegraphics[width=\textwidth]{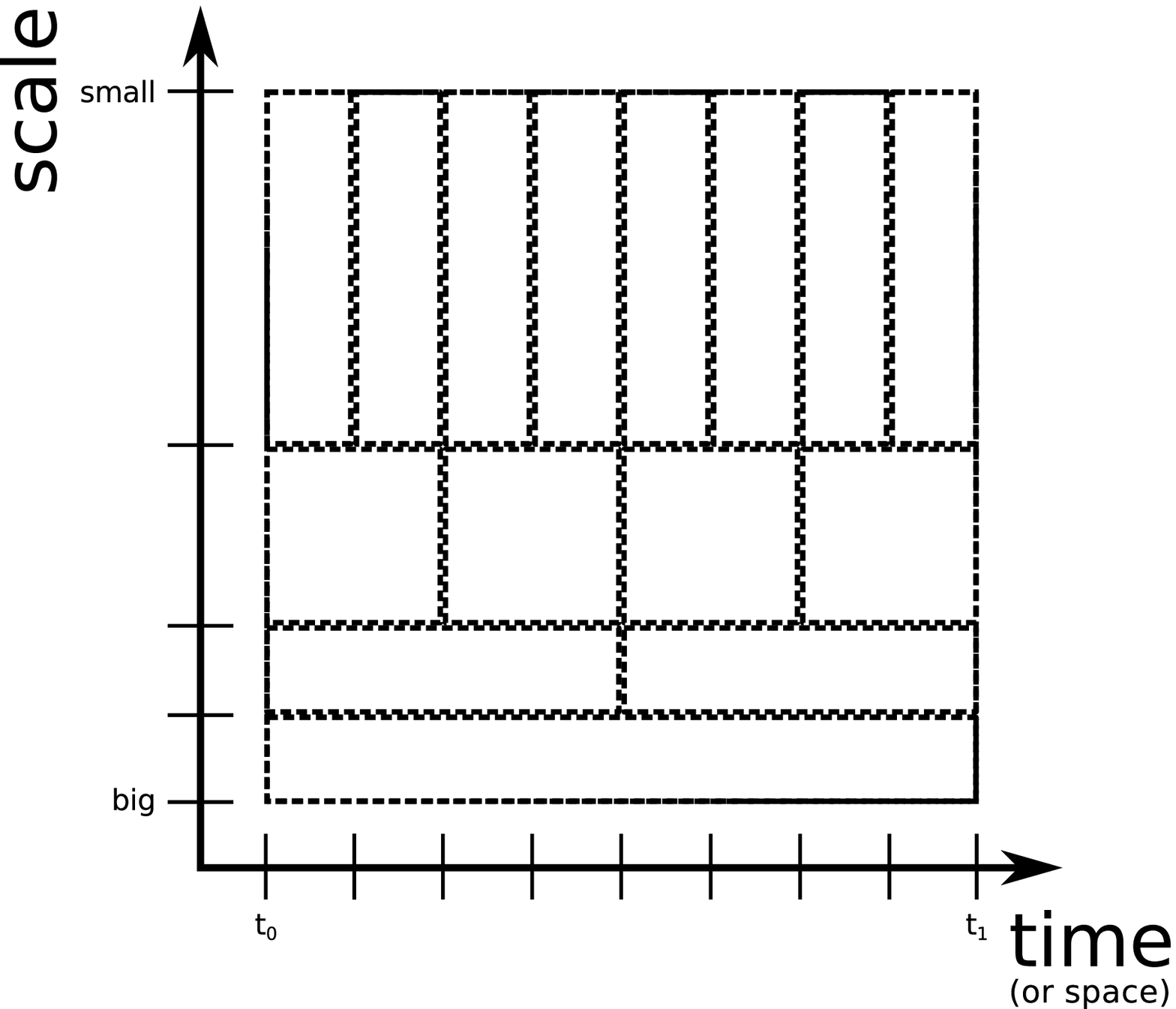}
        \caption{WT - The oscillating function is compressed/stretched (scaled or dilated).}
    \end{subfigure}

    \caption{Simplified illustration of the STFT fixed window paradigm,
    versus the idea of dilation and scale in WT.
    The idea for the illustrations in the upper row is taken from \cite{hpintrowv92}.} \label{fig:comp-stft-dwt}
\end{figure}

The first concept is visually rendered in Figure~\ref{fig:comp-stft-dwt}: instead of changing the frequency of the basis function
inside a window of fixed length, the idea is to compress or stretch (scale) a time-limited oscillating function effectively changing its
support and frequency at the same time.

If we call $\psi$ this ``time-limited oscillating function'' then we can write this idea as:
\begin{equation}
    CWT_s (a, b) = \frac{1}{\sqrt{|a|}} \int_{-\infty}^{+\infty} s(t) \psi^* \left( \frac{t-b}{a} \right) dt .
    \label{eq:cwt-explicit}
\end{equation}
$CWT_s (a, b)$ is the Continuous Wavelet Transform (CWT) of the signal $s(t)$ and is function of two variables:
$a$, called \emph{dilation} (scale) and $b$, the \emph{translation}.
Dilating a wavelet means stretching it (if  $\vert a \vert < 1 $) or compressing it (if $\vert a \vert > 1 $).
We can restrict to $a>0$ without loss of generality.

Typically Eq.~\eqref{eq:cwt-explicit} is written in a more compact form as:
\begin{equation*}
    CWT_s (a, b) = \int_{-\infty}^{+\infty} s(t) \psi^*_{a,b}(t)  dt
    \label{eq:cwt-compact}
\end{equation*}
with
\begin{equation*}
    \psi_{a,b}(t) = \frac{1}{\sqrt{\vert a \vert}} \psi\left( \frac{t-b}{a} \right) \text{, with } a,b \in \mathbb{R}
    \label{eq:mother-wavelet}
\end{equation*}
being the \emph{mother wavelet}, which has to satisfy a couple of properties.
The first is called the \emph{admissibility condition}
\begin{equation*}
    \int_{-\infty}^{+\infty} \psi(t) dt = 0,
\end{equation*}
and the second property asks for $\psi$ to be square integrable (i.e.\ have finite energy)
\begin{equation*}
    \int_{-\infty}^{+\infty} \psi^2(t) dt < \infty.
\end{equation*}
The admissibility condition is often reported for the Fourier transform of $\psi(t)$, $\Psi(\omega)$, and translates into
\begin{equation*}
    \int \frac{|\Psi(\omega)|}{|\omega|} d\omega < +\infty .
\end{equation*}
with $\omega = 2 \pi f$.
The admissibility condition implies that $\Psi(\omega)$ vanishes at the zero frequency, (otherwise the integral would blow up at $\omega = 0$)
\begin{equation*}
    |\Psi(\omega)^2|_{\omega=0} = 0.
\end{equation*}
This means that the wavelets must have a band-pass like spectrum.

The inverse formula, the Inverse Continuous Wavelet Transform (ICWT), is given by:
\begin{equation}
s(t) = \frac{1}{c_\psi} \int_{-\infty}^{+\infty} \int_{-\infty}^{+\infty} (CWT_s)_{a,b} \psi_{a,b}(t) \frac{da db}{a^2}
\label{eq:cwt-reco}
\end{equation}
where $c_\psi = 2\pi \int_{-\infty}^{+\infty} \frac{\vert \Psi(\omega) \vert^2}{\vert \omega \vert} d\omega $, and $\Psi(\omega)$ is the Fourier transform of $\psi(t)$.
Equation \eqref{eq:cwt-reco} can be interpreted in two ways:
\begin{itemize}
    \item As a way of reconstructing $s(t)$ once its wavelet transform $(CWT_s)_{a,b}$ is known;
            this formula is known as the \emph{reconstruction formula} or \emph{scheme} (or \emph{resolution of the identity}).
    \item As a way to write $s$ as a superposition of wavelets $\psi_{a,b}$.
            The coefficients in this superposition are exactly given by the wavelet transform of $s$.
\end{itemize}
Note that the $\psi_{a,b}$ are defined over every point in the $(a,b)$ space, and so they are highly redundant.
Is it possible to discretize the $(a,b)$ space so that the $\psi_{a,b}$ form a true orthonormal basis?

\paragraph{Discretized Wavelet Transform}
Let's start from discretizing the dilation parameter $a$ as a power of a fixed dilation step $a_0 > 1$: $a = a_0^m$, with $m \in \mathbb{Z}$.
The parameter $m$ will control the dilation, while the translation will be $b = n b_0 a_0^m$ with $n \in \mathbb{Z}$,
so that is adapted to the width of the wavelet.
This gives
\begin{equation}
    \psi_{m,n}(t) = \psi_{(a_0^m, nb_0 a_0^m)}(t) = a_0^{-m/2} \psi(a_0^{-m} t - nb_0)
    \label{eq:dyadic}
\end{equation}
and the discretized wavelet coefficients are
\begin{equation*}
    d_{m,n} = \int s(t) \psi_{m,n}^* (t) dt .
\end{equation*}
The choice of the $a_0$ and $b_0$ parameters defines the type of the wavelet family.
A common choice that goes under the name of \emph{dyadic sampling} is $a_0 = 2$ and $b_0 = 1$,
giving dyadic sampling along frequency and time respectively.
The question translates now in if the $\psi_{m,n}$ and the inverse wavelet transform
\begin{equation}
    \tilde s(t) = c \sum_{m,n} d_{m,n} \psi_{m,n}(t)
    \label{eq:dwt-approx-reco}
\end{equation}
form a sort of discrete approximation\footnote{In math terms if a family of wavelets $\psi_{m,n}$ constitutes a \emph{frame}.}
of Eq.\eqref{eq:cwt-reco}, so that $\tilde s(t) \approx s(t)$.
The short answer is \emph{yes}: it is possible to design some $\psi_{m,n}$ so that the Eq.~\eqref{eq:dwt-approx-reco}
is actually an equation that defines the Discrete Wavelet Transform (DWT).
The important thing to notice is that the wavelet transform can be redundant (when $\psi_{m,n}$ is a frame\footnote{
A set of non-zero vectors $\left\{ \phi_i  \right\}_{i \in J}$ constitutes a frame in the Hilbert space $\mathcal{H}$,
if exist an $A > 0$ and a $B < \infty$ such that, for all $f \in \mathcal{H}$:
$A \left\| f \right\| ^2 \leq \sum_{i \in J} \lvert \langle \phi_i | f \rangle \rvert^2 \leq B \left\| f \right\|^2  $.
When $A=B$ the frame is called a tight frame.
})  %
or not (when $\psi_{m,n}$ is a basis),
and this redundacy can be (somewhat) tuned and can be actually an interesting feature for signal analysis.
It is also interesting to exploit this redundancy in (numerical) reconstruction, because
for a given reconstruction precision, the redundacy allows to calculate the wavelet coefficients with less precision than the one needed
with zero redundacy (othonormal bases), at the cost of having more coefficients \cite{daub90wt}.

\paragraph{Scaling Function}
Before reaching a more modern way to build wavelets (via the \emph{multiresolution analysis}),
we have to introduce a couple of notions, at least intuitively.
The first is the concept of \emph{scaling function} (or \emph{smoothing function}), $\phi(t)$, introduced by Mallat \cite{mallat89}.
If we say that $m=0$ is the lowest value for the dilation parameter or, in other words, the lowest \emph{level} at which
we are decomposing the signal $s(t)$, we need something that takes what remains of $s(t)$ at the point we stopped the decomposition.
Since the wavelet functions are band-pass like filters, we need a low-pass kind of function.
The filter that fulfills this role is the scaling function $\phi(t)$ and has the property that $\int \phi (t) dt = 1$.
Similarly to what happens with wavelets, \eqref{eq:dyadic}, the scaling functions families are also
dilated and translated copies of a an original scaling function
\begin{equation*}
    \phi_{m,n}(t) = a_0^{-m/2} \phi(a_0^{-m} t - nb_0) \text{, with } n \in \mathbb{Z} .
\end{equation*}
The Eq.~\eqref{eq:dwt-approx-reco}, in this context, could become something like this
\begin{equation*}
    \tilde s(t) = c_{0,0} \phi_{0,0}(t) + \sum_{m=0}^{J-1} \sum_{n\in \mathbb{Z}} d_{m,n} \psi_{m,n}(t)
\end{equation*}
with $J$ the maximum level of decomposition.

\paragraph{Second Generation Wavelets}
The second concept is the distinction between the ``First Generation wavelets'' and the ``Second Generation wavelets''.
The main differences are two, one is about the relation between scaled/translated wavelets and
the other concerns the framework to actually build the wavelet filters.

Regarding the relation between scaled/translated wavelets,
Schr{\"o}der and Sweldens explain perfectly in~\cite{Schroeder95} the shift in paradigm from the wavelet scheme
described in this Section, called First Generation wavelets, to the Second Generation:
\blockquote{
    ``In the classic wavelet setting, i.e., on the real line, wavelets are defined as the dyadic translates and dilates of one particular, fixed function.
    They are typically built with the aid of a scaling function.
    Scaling functions and wavelets both satisfy refinement relations (or two scale relations).
    This means that a scaling function or wavelet at a certain level of resolution ($j$) can be written as a linear combination
    of scaling basis functions of the same shape but scaled at one level finer (level $j+1$)\footnote{To make the dissertation
    more agile, we give a universal definition of the refinement relations (that works for first and second
    generation wavelets) in Section~\ref{sec:multiresolution}, instead of following the historical development
    of the theory.}
    [...] The basic philosophy behind second generation wavelets is to build wavelets with all desirable properties
    (localization, fast transform) adapted to much more general settings than the real line.
    [...] Adaptive constructions rely on the realization that translation and dilation are not fundamental to obtain the wavelets with the desired properties.
    The notion that a basis function can be written as a finite linear combination of basis functions at a finer,
    more subdivided level, is maintained and forms the key behind the fast transform.
    The main difference with the classical wavelets is that the filter coefficients of second generation wavelets are not the same throughout,
    but can change locally to reflect the changing (non translation invariant) nature of the surface and its measure.''
}

Concerning the actual method to operatively build the wavelets, they say:
\blockquote{
    ``The tool that we use to build wavelets transforms is called the \emph{lifting scheme}.
    The main feature of the lifting scheme is that all constructions are derived in the \emph{spatial domain}.
    This is in contrast to the traditional approach wich relies heavily on the frequency domain\footnote{See for example \cite{Daubechies1992}.}.
    Staying in the spatial domain leads to two major advantages.
    First, it does not require the machinery of Fourier analysis as a prerequisite.
    This leads to a more intuitively appealing treatment better suited to those interested in applications,
    rather than mathematical foundations.
    Secondly, lifting leads to algorithms that can easily be generalized
    to complex geometric situations which typically occur in computer graphics.
    This will lead to so called \emph{Second Generation Wavelet}.
    [...]
    Even though the wavelets which result from using the lifting scheme in the more general settings will not
    be translates and dilates of one function anymore they still have all the powerful properties
    of first generation wavelets: fast transforms, localization and good approximation.''
}

The operative construction of second generation wavelets via the lifting scheme is beautifully described in the same manuscript \cite{Sweldens_buildingyour}.
For an exhaustive mathematical definition of the lifting scheme, the interested reader should refer to~\cite{Sweldens1998}.
We will give a concise definition in Section~\ref{sec:lifting-scheme}.

With this brief, and possibly agile, introduction we can move to a proper and more modern definition of wavelet transforms.

\section{Multiresolution Analysis}  \label{sec:multiresolution}
The starting point to define wavelets is a mathematical framework called \emph{multiresolution analysis}.
To define the multiresolution analysis, we have to define first a nested set of closed vector subspaces
$ V^0 \subset \cdots \subset V^j \subset \cdots \subset V^n$.
The higher the space index, the finer is the space.
For each $j$, the basis functions of $V^j$ are called \emph{scaling functions},
and are denoted like this: $\phi^j_k$ with $k \in \mathbb{K}\left( j \right)$,
where $\mathbb{K}$ is and index set with $\mathbb{K}\left( j \right) \subset \mathbb{K} \left( j+1 \right)$.
Since the vector spaces are nested, it is possible to write each $\phi^j_k$
as a function of the next level $\phi^{j+1}$, and obtain these \emph{refinement relations}:
\begin{equation}
    \phi^j_k = \sum_l p^{j+1}_{l,k} \phi^{j+1}_l
\end{equation}
where $n > j \geq 0$, $k \in \mathbb{K}\left( j \right)$ and $l \in \mathbb{K}\left( j+1 \right)$.
Note: in this refinement relation there is no explicit mention to how dilation and translation are implemented, e.g.\ \eqref{eq:dyadic}.
This definition is valid also in the second generation wavelets,
where the dilation and translation relations are not maintained between different levels.
Additionally, the vector spaces used to build the multiresolution are very generic.
The construction of wavelets in the multiresolution framework follows the same procedure for 1D, 2D or $n$-dimensional spaces.

Adopting a more compact and convenient matrix notation, putting together the different scaling functions
$\phi^j_k$ for the level $j$ as one row vector:
\begin{equation*}
    \Phi^j = \begin{pmatrix} \phi^j_1  & \cdots & \phi^j_{m^j} \end{pmatrix}
\end{equation*}
where $m^j$ is the dimension of $V^j$ (here we assume that $V^j$ has a finite basis).

The \emph{wavelet spaces}, $W^j$, are defined to be the {orthogonal complement} of $V^j$ in $V^{j+1}$, so that $V^j \oplus W^j = V^{j+1}$.
Meaning that $W^j$ includes all the functions in $V^{j+1}$ that are orthogonal to
all those in $V^j$ under some inner product (typically $\mathcal{L}^2$).
The functions that form a basis of $W^j$ are called \emph{wavelets}, and are denoted with $\psi^j_p$.
The corresponding refinement equations for the wavelets are:
\begin{equation}
    \psi^j_k = \sum_l q^{j+1}_{l,k} \phi^{j+1}_l
\end{equation}
Similarly to the scaling functions, we can group them in a row vector:
\begin{equation*}
    \Psi^j = \begin{pmatrix} \psi^j_1 & \cdots & \psi^j_{n^j} \end{pmatrix}
\end{equation*}
where $n^j$ is the dimension of $W^j$, with $m^{j+1} = m^j + n^j$.

With this matrix notation it is possible to rewrite the \emph{refinement relations}:
\begin{equation}
    \Phi^j = \Phi^{j+1} \mathbf{P}^{j+1},
    \label{eq:scaling-phi-matrix-eq}
\end{equation}
and, in similar fashion, a matrix $\mathbf{Q}$ must exist to satisfy:
\begin{equation}
    \Psi^j =  \Phi^{j+1} \mathbf{Q}^{j+1}.
    \label{eq:wavelet-psi-matrix-eq}
\end{equation}
The biorthogonality conditions then become
\begin{equation}
    \langle \Phi^j | \Psi^j \rangle = \mathbf{0}
    \label{eq:othog-matrix}
\end{equation}
where
\begin{equation*}
    \langle \Phi^j | \Psi^j \rangle_{kl} = \langle \phi_{k}^j | \psi_l^j \rangle,
\end{equation*}
and $\langle \phi | \psi \rangle$ denotes the inner product.
Substituting~\eqref{eq:wavelet-psi-matrix-eq} into this last equation~\eqref{eq:othog-matrix}, it gives
\begin{equation*}
    \langle \Phi^j | \Phi^{j+1} \rangle \mathbf{Q}^{j+1} = \mathbf{0}
\end{equation*}
this is an homogeneous system of linear equations
and there is no unique solution to it.
The matrix $\mathbf{Q}$ is a basis of the space of all possible solutions. %
So there is no unique $\mathbf{Q}$, meaning that there are many different wavelet bases for a given wavelet space $W^j$.
To determine uniquely the $\mathbf{Q}$ matrices we have to impose further constraints to the orthogonality alone.
The discussion to the different constraints options and resulting wavelets available in literature
is out of the scope of this introduction to wavelets, more information can be found in~\cite{Stollnitz95pt2}.

Each multiresolution analysis is accompanied by a dual multiresolution analysis consisting of nested spaces $\tilde{V}^j$  %
with bases given by dual scaling functions $\widetilde{\Phi}^j$,
which are biorthogonal to the scaling functions:  %
\begin{equation}
    \langle \widetilde{\Phi}^j | \Phi^j \rangle = \mathbf{1}.
\end{equation}
The dual scaling functions satisfy similar \emph{refinement relations} as of Eq.~\eqref{eq:scaling-phi-matrix-eq}:
\begin{equation}
    \widetilde{\Phi}^j =   \widetilde{\Phi}^{j+1} [\mathbf{A}^{j+1}]^T.
    \label{eq:dual-scaling-phi-matrix-eq}
\end{equation}
Similarly, for any given wavelet basis there is a dual basis $\widetilde{\Psi}^j$ and the two are
biorthogonal with respect to each other: $ \langle \widetilde{\Psi}^j | \Psi^j \rangle = \mathbf{1} $.
This also implies $ \langle \widetilde{\Psi}^j | \Phi^j \rangle = \langle \widetilde{\Phi}^j | \Psi^j \rangle = \mathbf{0} $.
And similarly to Eq.~\eqref{eq:wavelet-psi-matrix-eq}, a matrix $B$ will exists, so that:
\begin{equation}
    \widetilde{\Psi}^j = \widetilde{\Phi}^{j+1} [\mathbf{B}^{j+1}]^T.
    \label{eq:dual-wavelet-psi-matrix-eq}
\end{equation}
Combining the fact that $\phi_k^j \in V_j \oplus W_j$ with the biorthonormality relations
leads to the inverse refinement equations for the original scaling function:
\begin{equation}
    \Phi^{j+1} = \Phi^j \mathbf{A}^{j+1} +  \Psi^j \mathbf{B}^{j+1}.
    \label{eq:inverse_refine}
\end{equation}

The scaling function coefficients $\mathbf c^j$ and wavelet coefficients $\mathbf d^j$ of any function $f$
can be obtained by inner product
with the dual scaling function and dual wavelets respectively:
\begin{equation}
\begin{split}
    \mathbf c^j &= \langle \widetilde\Phi^j | f \rangle, \\
    \mathbf d^j &= \langle \widetilde\Psi^j | f \rangle.
\end{split}
\label{eq:f_proj_cd}
\end{equation}

These operators $\mathbf{A}^j$, $\mathbf{B}^j$ and $\mathbf{P}^j$, $\mathbf{Q}^j$ are the
\emph{decomposition} and \emph{reconstruction filters}, respectively.
The biorthogonality properties imply that the operators $\mathbf{A}^j$, $\mathbf{B}^j$, $\mathbf{P}^j$ and $\mathbf{Q}^j$
need to verify the following properties:
\begin{subequations} \label{eq:inv1}
\begin{align}
    \langle \widetilde{\Psi}^j | \Phi^j \rangle = \mathbf{0} &\implies \mathbf{B}^j \mathbf{P}^j = \mathbf{0} \label{eq:bp}, \\
    \langle \widetilde{\Phi}^j | \Psi^j \rangle = \mathbf{0} &\implies \mathbf{A}^j \mathbf{Q}^j = \mathbf{0} \label{eq:aq},\\
    \langle \widetilde{\Phi}^j | \Phi^j \rangle = \mathbf{1} &\implies \mathbf{A}^j \mathbf{P}^j = \mathbf{1} \label{eq:ap}, \\
    \langle \widetilde{\Psi}^j | \Psi^j \rangle = \mathbf{1} &\implies \mathbf{B}^j \mathbf{Q}^j = \mathbf{1} \label{eq:qb}
\end{align}
\end{subequations}
that can be rewritten in matrix notation like:
\begin{equation}
    \begin{bmatrix}
        \mathbf A^j \\
        \mathbf B^j
    \end{bmatrix}
    \begin{bmatrix}
        \mathbf P^j & \mathbf Q^j
    \end{bmatrix}
    =
    \begin{bmatrix}
        \mathbf{AP} & \mathbf{AQ} \\
        \mathbf{BP} & \mathbf{BQ}
    \end{bmatrix}
    =
    \begin{bmatrix}
        \mathbf 1 & \mathbf 0 \\
        \mathbf 0 & \mathbf 1
    \end{bmatrix}
    \: \text{and} \:
    \begin{bmatrix}
        \mathbf P^j & \mathbf Q^j
    \end{bmatrix}
    \begin{bmatrix}
        \mathbf A^j \\
        \mathbf B^j
    \end{bmatrix}
    = \mathbf 1
    \label{eq:biorthogonality-relations}
\end{equation}
Additionally, applying the biorthogonality properties to \eqref{eq:inverse_refine} it means that:
\begin{equation}
    \mathbf{P}^j \mathbf{A}^j +  \mathbf{Q}^j\mathbf{B}^j = \mathbf{1}.
    \label{eq:inv2}
\end{equation}
Equations \eqref{eq:inv1} and \eqref{eq:inv2} can be combined by stating that the decomposition and reconstruction filters
are globally inverse one to the other:
\begin{equation*}
    \begin{pmatrix}
        \mathbf A^j\\
        \mathbf B^j
    \end{pmatrix}
    = \begin{pmatrix}\mathbf{P}^{j} & \mathbf{Q}^{j} \end{pmatrix}^{-1}.
\end{equation*}

\section{Subdivision Mesh} \label{sec:meshinterpolation}
Now we are going to introduce a concept that is crucial for the practical construction of wavelets throughout the rest of the thesis.
A \emph{subdivision mesh} is a method of representing a smooth surface (in this case, the sphere)
as the limit of a series of increasingly finer polygonal meshes.
The mesh is built recursively starting from a primitive polygonal mesh (e.g.\ an octahedron),
and subdividing (i.e.\ adding new vertices) this original mesh according to some rule, called subdivision scheme.
Examples of subdivision schemes are Loop \cite{loopphd}, Catmull-Clark \cite{Catmull1978},
Doo-Sabin \cite{Doo1978ASA}.
At each iteration a finer (more dense) mesh is obtained.
In this work we will call each iteration a \emph{level}, and the primitive mesh is referred to as \emph{level 0}.
The wavelet framework is built out of the subdivision mesh.
A given wavelet level will be associated to a certain mesh level.
We will only consider subdivision meshes up to some given level $n$,
corresponding to the finest mesh.

To close the circle,
when the function space is a finite vector space defined over the finest space $V^n$,
and the dimensionality of the function space coincides with the dimensionality of the finest mesh,
the scaling functions at the finest level $n$ can be taken to delta functions:
$\phi_k^n(p) = \delta(p-k)$.
In this case the wavelets and dual wavelets can be computed from the decomposition and reconstruction filters at each level as follows:
\begin{align}
    \begin{split}
        \phi^j_k(p) &= \left(\mathbf{P}^n \, \cdots \, \mathbf{P}^{j+2} \, \mathbf{P}^{j+1}\right)_{pk} \\
        \psi^j_k(p) &= \left(\mathbf{P}^n \, \cdots \, \mathbf{P}^{j+2} \, \mathbf{Q}^{j+1}\right)_{pk} \\
        \tilde\phi^j_k(p) &= \left(\mathbf{A}^{j+1} \, \mathbf{A}^{j+2} \, \cdots \, \mathbf{A}^{n}\right)_{kp} \\
        \tilde\psi^j_k(p) &= \left(\mathbf{B}^{j+1} \, \mathbf{A}^{j+2} \, \cdots \, \mathbf{A}^{n}\right)_{kp}
    \end{split}
    \label{eq:wavs-from-matrices}
\end{align}
In fact, by using the procedural approach that we will outline in Section~\ref{sec:encoding},
it is perfectly possible to ignore the scaling functions, the wavelets and their duals
and work only with the scaling and wavelet coefficients and the decomposition and reconstruction filters.

\section{Second Generation Wavelets via the Lifting Scheme} \label{sec:lifting-scheme}
From the construction of wavelets in the scale and dilate paradigm we introduced in Section~\ref{sec:intro-to-wavelets}
to the method we implemented for our specific application, that will be discussed in Chapter~\ref{ch:our-wavelets},
there is quite a leap forward in methods and meanings.
To cover this distance we introduce a procedural method to build wavelets, the Lifting Scheme.

\subsection{Lifting Scheme}
Given an initial set of biorthogonal filter operators  $\left\{ \tbar{\mathbf A}^j, \tbar{\mathbf P}^j,
\tbar{\mathbf B}^j, \tbar{\mathbf Q}^j \right\}$, then a new set of biorthogonal filters $\left\{ \mathbf A, \mathbf P,
\mathbf B, \mathbf Q \right\}$ can be found as:
\begin{align}
    \begin{split}
        \mathbf P^j &= \tbar{\mathbf P}^j  \\
        \mathbf A^j &= \tbar{\mathbf A}^j + \mathbf S^j \tbar{\mathbf B}^j  \\
        \mathbf Q^j &= \tbar{\mathbf Q}^j - \tbar{\mathbf P}^j \mathbf S^j  \\
        \mathbf B^j &= \tbar{\mathbf B}^j
    \end{split}
    \label{eq:lifting-relations}
\end{align}
where $\mathbf S^j$ is the lifting operator.
Indicating explicitly the dimensions of the operators:
\begin{align*}
    \begin{split}
        \mathbf P^j_{M \times K} &= \tbar{\mathbf P}^{j}_{M \times K}  \\
        \mathbf A^j_{K \times M} &= \tbar{\mathbf A}_{K \times M}^{j} + \mathbf S^j_{K \times (M-K)} \tbar{\mathbf B}^{j}_{(M-K) \times M}  \\
        \mathbf Q^j_{M \times (M-K)} &= \tbar{\mathbf Q}_{M \times (M-K)}^{j} - \tbar{\mathbf P}^{j}_{M \times K} \mathbf S_{K \times (M-K)}  \\
        \mathbf B^j_{(M-K) \times M} &= \tbar{\mathbf B}^{j}_{(M-K) \times M}
    \end{split}
\end{align*}
It is quite straightforward to prove this by writing the lifting scheme in matrix notation (we assume for simplicity that all matrices are real valued):
\begin{equation*}
    \begin{bmatrix}
        \mathbf A^j \\
        \mathbf B^j
    \end{bmatrix}
    =
    \begin{bmatrix}
        \mathbf 1 & \mathbf S^j \\
        \mathbf 0 & \mathbf 1
    \end{bmatrix}
    \begin{bmatrix}
        \tbar{\mathbf A}^j \\
        \tbar{\mathbf B}^j
    \end{bmatrix}
    \label{eq:lifting-matrix-notation-1}
\end{equation*}
and
\begin{equation*}
    \begin{bmatrix}
        \mathbf P^j \\
        \mathbf Q^j
    \end{bmatrix}
    =
    \begin{bmatrix}
        \mathbf 1 & \mathbf 0 \\
        -\mathbf S^j & \mathbf 1
    \end{bmatrix}
    \begin{bmatrix}
        \tbar{\mathbf P}^j \\
        \tbar{\mathbf Q}^j
    \end{bmatrix}
    \label{eq:lifting-matrix-notation-2}
\end{equation*}
and if we recall the bihorthogonality relations ~\eqref{eq:biorthogonality-relations}, we get
\begin{align*}
    \begin{bmatrix}
        \mathbf A^j \\
        \mathbf B^j
    \end{bmatrix}
    \begin{bmatrix}
        \mathbf P^j & \mathbf Q^j
    \end{bmatrix}
    &=
    \begin{bmatrix}
        \mathbf 1 & \mathbf S^j \\
        \mathbf 0 & \mathbf 1
    \end{bmatrix}
    \begin{bmatrix}
        \tbar{\mathbf A}^j \\
        \tbar{\mathbf B}^j
    \end{bmatrix}
    \begin{bmatrix}
        \tbar{\mathbf P}^j & \tbar{\mathbf Q}^j
    \end{bmatrix}
    \begin{bmatrix}
        \mathbf 1 & -\mathbf S^{j \intercal} \\
        \mathbf 0 & \mathbf 1
    \end{bmatrix}
    \\
    &=
    \begin{bmatrix}
        \mathbf 1 & \mathbf S^j \\
        \mathbf 0 & \mathbf 1
    \end{bmatrix}
    \begin{bmatrix}
        \mathbf 1 & -\mathbf S^{j \intercal} \\
        \mathbf 0 & \mathbf 1
    \end{bmatrix}
    \\
    &=
    \begin{bmatrix}
        \mathbf 1 & \mathbf 0 \\
        \mathbf 0 & \mathbf 1
    \end{bmatrix}.
    \hspace{4cm} \blacksquare
\end{align*}

From the relations
\eqref{eq:scaling-phi-matrix-eq}
\eqref{eq:wavelet-psi-matrix-eq}
\eqref{eq:dual-scaling-phi-matrix-eq}
\eqref{eq:dual-wavelet-psi-matrix-eq}
and the definition of lifting scheme~\eqref{eq:lifting-relations}, it is possible to see how the lifting scheme
impacts the scaling functions, the wavelets, and their duals:
\begin{align*}
        \Phi^j &= \tbar{\Phi}^{j}  \\
        \widetilde\Phi^j &= \tbar{\mathbf A}^j \widetilde\Phi^{j+1} + \mathbf S^j \tbar{\mathbf B}^j \widetilde\Phi^{j+1} = \tbar{\mathbf A}^j \widetilde\Phi^{j+1} + \mathbf S^j \widetilde\Psi^j  \\
        \Psi^j &= \tbar{\mathbf Q}^j \Phi^{j+1} - \tbar{\mathbf P}^j \mathbf S^j \Psi^{j+1} = \tbar{\Psi}^j - \mathbf S^j \tbar{\Phi}^j \\
        \widetilde\Psi^j &= \tbar{\mathbf B}^j \widetilde\Phi^j
\end{align*}

A note of caution: the notation we are using, while being convenient, %
it partially hides the inner workings of $\mathbf S$.
For an explicit element-by-element definition of the lifting scheme (with \emph{index notation}), we refer to \cite{Sweldens1998}.
The great benefit of the lifting scheme is that, starting from some simple or even trivial filters $\left\{ \tbar{\mathbf A}^j, \tbar{\mathbf P}^j,
\tbar{\mathbf B}^j, \tbar{\mathbf Q}^j \right\}$ it is possible to build more complex ones just by tuning the operator $\mathbf S^j$.
By controlling the operator $\mathbf S$, it is possible to control the properties of the wavelets and dual scaling functions that
are built from the original scaling function $\tbar{\Phi}^{j}$.
Essentially, the new functions and operators are improved, \emph{lifted}, versions of the old ones, and can have custom properties.
Once the operator $\mathbf S$ is set, the lifting scheme keeps the biorthogonality of the old filters to the new ones.
Before getting to one practical example on how to build non-trivial wavelets
from trivial ones, we will define the fast lifted wavelet transform and the dual lifting scheme.

\subsection{Fast Lifted Wavelet Transform} \label{subsec:lifted-wavelet-transform}
Another benefit of the lifting scheme, is that it allows to write the wavelet transform just using the old filters
and the $\mathbf S$ filter, without the need to explicitly derive the new ones.
The forward lifted transform is
\begin{align}
    \begin{split}
        \mathbf c^j = \mathbf A^j \mathbf c^{j+1} &= \tbar{\mathbf A}^j \mathbf c^{j+1} + \mathbf S^j \mathbf d^j  \\
        &= \tbar{\mathbf A}^j \mathbf c^{j+1} + \mathbf S^j \tbar{\mathbf B}^j \mathbf c^{j+1}
    \end{split}
    \label{eq:forward-lifted-transform}
\end{align}
The signal $\mathbf c^j$ is called \emph{coarse signal}, and is calculated via the $\tbar{\mathbf A}^j \mathbf c^{j+1}$ and then lifted
with the $\mathbf d^j$, called \emph{details}.
Often in the literature this operation of lifting is called an \emph{update}, and is denoted with the operator $U$.
In the Eq.~\eqref{eq:forward-lifted-transform} the new filter $\mathbf A^j$ is never calculated explicitly.

The inverse lifted transform then becomes
\begin{align*}
    \mathbf c^{j+1} = \mathbf P^j \mathbf c^j + \mathbf Q^j \mathbf d^j
    = \tbar{\mathbf P}^j \left( \mathbf c^j - \mathbf S^j \mathbf d^j \right)  + \tbar{\mathbf Q}^j \mathbf d^j
\end{align*}

It is possible to see that the operations in these transforms can be done in-place,
meaning that the only required storage is for the original signal $\mathbf c$ at the finest level.
There is no need to calculate and store the full matrices, resulting in an easy implementation and a fast algorithm.
We are not going in the details of the in-place implementation since, for our specific study and application, the efficiency is
not a primary requirement, and we actually compute the full matrices.

\subsection{Dual Lifting Scheme} \label{subsec:dual-lifting}
We have seen that via the lifting scheme it is possible to build an improved version of the starting matrices
$\tbar{\mathbf A}$ and $\tbar{\mathbf Q}$, while the $\tbar{\mathbf P}$ and $\tbar{\mathbf B}$ remain unchanged.
It is possible to lift the dual step, and build a dual lifting scheme:
\begin{align}
    \begin{split}
        \mathbf P^j &= \tbar{\mathbf P}^j + \tbar{\mathbf Q}^j \widetilde{\mathbf S}^j  \\
        \mathbf A^j &= \tbar{\mathbf A}^j  \\
        \mathbf Q^j &= \tbar{\mathbf Q}^j  \\
        \mathbf B^j &= \tbar{\mathbf B}^j - \widetilde{\mathbf S}^j \tbar{\mathbf A}^j
    \end{split}
    \label{eq:dual-lifting-scheme}
\end{align}
where the operators that are improved are the $\tbar{\mathbf P}$ and $\tbar{\mathbf B}$, while the other two remain unchanged.
The dual operator $\widetilde{\mathbf S}$ is often called \emph{prediction} operator.
Indicating explicitly the dimensions of the operators:
\begin{align*}
    \begin{split}
        \mathbf P^j_{M \times K} &= \tbar{\mathbf P}^j_{M \times K} + \tbar{\mathbf Q}_{M \times (M-K)}^j \widetilde{\mathbf S}^j_{(M-K) \times K} \\
        \mathbf A^j_{K \times M} &= \tbar{\mathbf A}_{K \times M}^j  \\
        \mathbf Q^j_{M \times (M-K)} &= \tbar{\mathbf Q}_{M \times (M-K)}^j  \\
        \mathbf B^j_{(M-K) \times M} &= \tbar{\mathbf B}^j_{(M-K) \times M} - \widetilde{\mathbf S}^j_{(M-K) \times K} \tbar{\mathbf A}_{K \times M}^j
    \end{split}
\end{align*}

In the following we will give an example of a common trivial wavelet, and build a lifted set of wavelets, the interpolating wavelet.
This interpolating wavelet is going to be useful later in the dissertation.

\subsection{The Lazy Wavelet} \label{subs:lazy-wavelet}
As a trivial set of filters to start the lifting process, it is possible to define two operators $\mathbf{E}$, $\mathbf{D}$
that essentially split the signal $\mathbf c^j$ into \emph{even} and \emph{odd} samples.
These two operators are obviously orthogonal (as before we will assume that we are dealing with real operators)
\begin{equation*}
    \begin{bmatrix}
        \mathbf E \\ \mathbf D
    \end{bmatrix}
    \begin{bmatrix}
        \mathbf E^\intercal & \mathbf D^\intercal
    \end{bmatrix}
    =
    \begin{bmatrix}
        \mathbf 1 & \mathbf 0 \\
        \mathbf 0 & \mathbf 1
    \end{bmatrix}
    \hspace{1cm} \text{and} \hspace{1cm}
    \begin{bmatrix}
        \mathbf E^\intercal & \mathbf D^\intercal
    \end{bmatrix}
    \begin{bmatrix}
        \mathbf E \\ \mathbf D
    \end{bmatrix}
    = 1.
    \label{eq:even-odd-othogonality}
\end{equation*}
This means that the Lazy wavelet operators are exactly $\mathbf E$ and $\mathbf D$ only:
\begin{equation*}
    \mathbf A^j_{Lazy} = \mathbf P^{j \intercal}_{Lazy} = \mathbf E
    \hspace{1cm} \text{and} \hspace{1cm}
    \mathbf B^j_{Lazy} = \mathbf Q^{j \intercal}_{Lazy} = \mathbf D
\end{equation*}
The Lazy wavelet transform is a transform that splits and merges back the signal, without actually doing anything.
This trivial set of filters is sufficient to define more interesting transforms, like the the interpolating wavelet transform.

\subsection{Interpolating Wavelet Transform via the Lifting Scheme} \label{subs:dual-lifting}
First we have to note that any operator $\mathbf W$ can be split into two operators, one that acts on the even samples
and one on the odd ones, like this:
\begin{equation*}
    \mathbf W = \mathbf W_e \mathbf E + \mathbf W_d \mathbf D,
\end{equation*}
with
\begin{equation}
    \mathbf W_e = \mathbf W \mathbf E^\intercal
    \hspace{1cm} \text{and} \hspace{1cm}
    \mathbf W_d = \mathbf W \mathbf D^\intercal .
    \label{eq:operator-splitting}
\end{equation}

\begin{figure}
    \centering
    \includegraphics[width=0.7\textwidth]{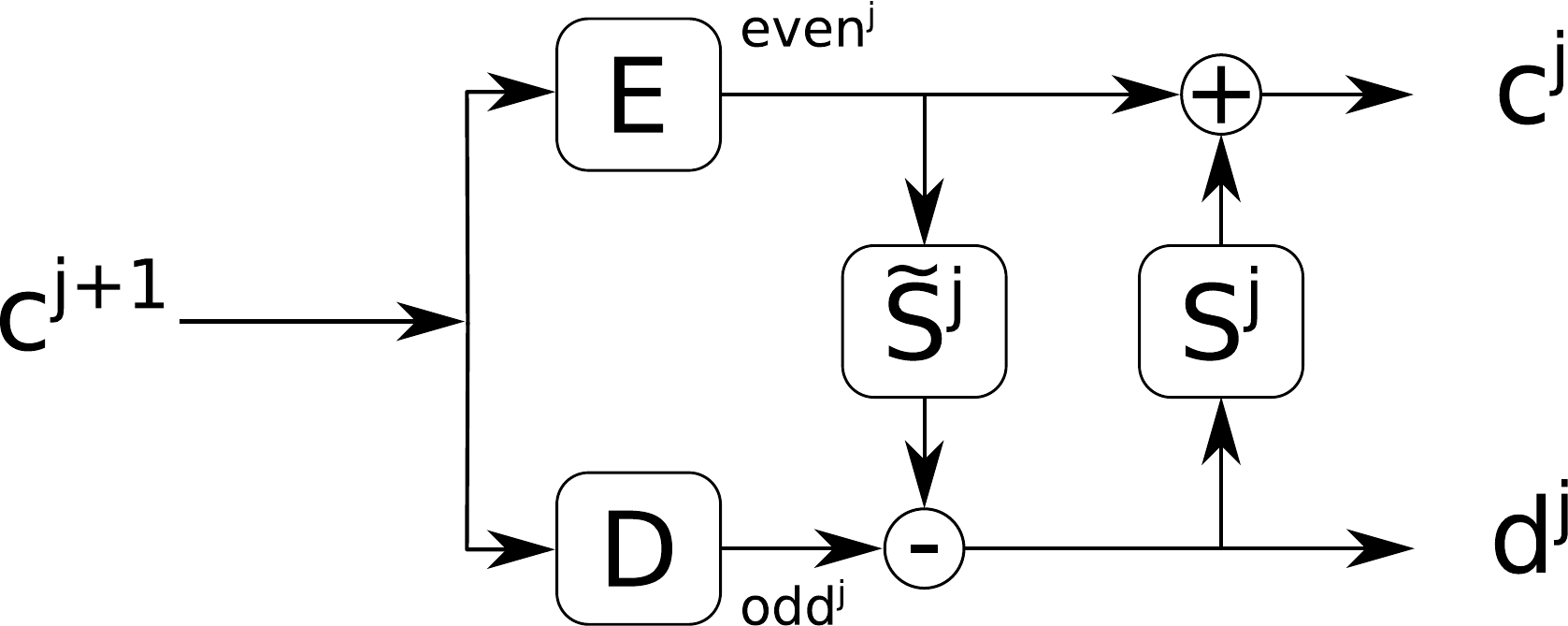}
    \caption{The interpolating wavelet transform is composed by three stages: the Lazy wavelet transform, the dual lifting and the normal lifting.}
    \label{fig:interpolating-lifting-decomposition}
\end{figure}

By definition, an \emph{interpolating filter} $P^j_{\text{INT}}$ is a filter that satisfies this equation
\begin{equation*}
    \mathbf P^j_{\text{INT}} \mathbf E^\intercal = 1.
\end{equation*}
The filter corresponding to the dual scaling function is then $ \mathbf A^j_{\text{INT}} = \mathbf E $.
If we define $\widetilde{\mathbf S}^j = \mathbf P^j_{\text{INT}} \mathbf D^\intercal $,
then from \eqref{eq:operator-splitting} any interpolating filter can be written as
$ \mathbf P^j_{\text{INT}} = \mathbf E + \widetilde{\mathbf S}^j \mathbf D $.
This expression is equivalent to applying the dual lifting scheme to the Lazy wavelet,
and so we can write the set of interpolating biorthogonal filters as
\begin{align}
    \begin{split}
        \mathbf P^j_{\text{INT}} &= \mathbf E^\intercal + \mathbf D^\intercal \widetilde{\mathbf S}^j  \\
        \mathbf A^j_{\text{INT}} &= \mathbf E  \\
        \mathbf Q^j_{\text{INT}} &= \mathbf D^\intercal  \\
        \mathbf B^j_{\text{INT}} &= \mathbf D - \widetilde{\mathbf S}^j \mathbf E
    \end{split}
    \label{eq:dual-lifting-from-lazy}
\end{align}
Note that the $\mathbf A^j_{\text{INT}}$ and $\mathbf Q^j_{\text{INT}}$ are essentially Dirac deltas.
These filters do not form a multiresolution analysis in $\mathcal{L}^2$, since the duals do not belong to $\mathcal{L}^2$.
We can apply the lifting scheme to the interpolating filters, so to improve the $\mathbf A^j_{\text{INT}}$ and $\mathbf Q^j_{\text{INT}}$, obtaining
\begin{equation}
    \begin{split}
        \mathbf P^j &= \mathbf P^j_{\text{INT}} = \mathbf E^\intercal + \mathbf D^\intercal \widetilde{\mathbf S}^j  \\
        \mathbf A^j &= \mathbf A^j_{\text{INT}} + \mathbf S^j \mathbf B^j_{\text{INT}} =
                        (\mathbf 1 - \mathbf S^j \widetilde{\mathbf S}^j) \mathbf E + \mathbf S^j \mathbf D^j  \\
        \mathbf Q^j &= \mathbf Q^j_{\text{INT}} - \mathbf P^j_{\text{INT}} \mathbf S^j =
                       - \mathbf E^\intercal \mathbf S^j + \mathbf D^\intercal (\mathbf 1 - \widetilde{\mathbf S}^j \mathbf S^j) \\
        \mathbf B^j &= \mathbf B^j_{\text{INT}} = \mathbf D - \widetilde{\mathbf S}^j \mathbf E
    \end{split}
    \label{eq:lifting-from-lazy}
\end{equation}
The new filters we obtain are then result of applying first the Lazy wavelet transform (that performs the splitting),
then the dual lifting step and finally the regular lifting.
Quite often in literature the union of these three operations it is identified with the \emph{lifting scheme},
even if the lifting is actually only one stage.
In Figure~\ref{fig:interpolating-lifting-decomposition} we illustrate the flow of these operations.
From the scheme in Figure~\ref{fig:interpolating-lifting-decomposition} we can also manually derive the two filters $\mathbf A^j$ and $\mathbf B^j$ just following the arrows.
The filter $\mathbf A^j$, for example, connects the $\mathbf c^{j+1}$ with the $\mathbf c^j$,
and there are three paths that connect the two: from $\mathbf c^{j+1}$ via $\mathbf E$ directly to $\mathbf c^j$,
from $\mathbf c^{j+1}$ via $\mathbf D$ and $\mathbf S^j$ and the last one goes through $\mathbf E$, $\widetilde{\mathbf S}^j$,
back via $\mathbf S^j$ and to $\mathbf c^j$.
And we get $\mathbf A^j = \mathbf E - \widetilde{\mathbf S}^j \mathbf E + \mathbf S^j \mathbf D $.
Similarly, to build the filter $\mathbf B^j$ that connects the $\mathbf c^{j+1}$ with the $\mathbf d^j$ there are two paths:
one that goes directly through $\mathbf D$, and one that gets to $\mathbf d^j$ via $\mathbf E$ and $\widetilde{\mathbf S}^j$.
Giving exactly $\mathbf B^j = \mathbf D - \widetilde{\mathbf S}^j \mathbf E$.

\begin{figure}
    \centering
    \includegraphics[width=\textwidth]{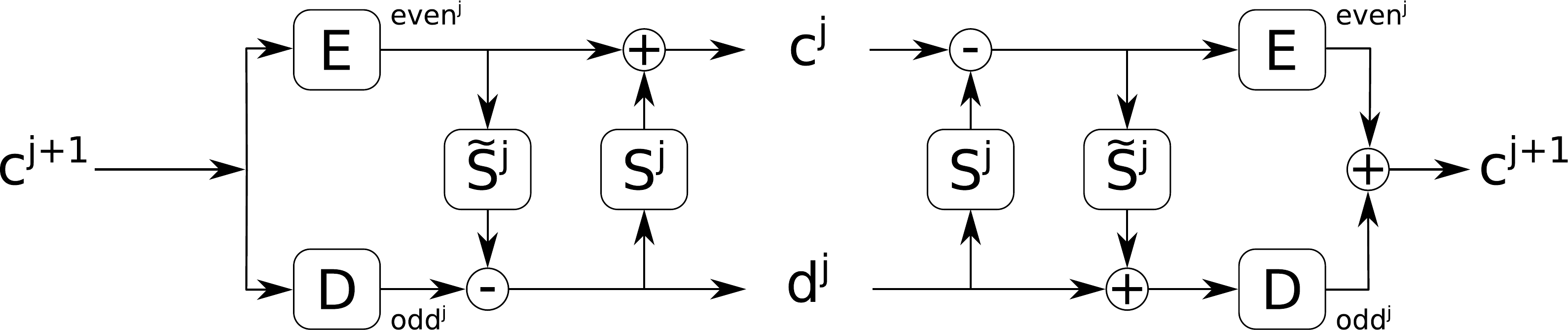}
    \caption{The complete interpolating wavelet transform with decomposition and reconstruction.
            The reconstruction is made by the same steps as decomposition, but performed backwards.}
    \label{fig:lifting-scheme-complete}
\end{figure}

Figure~\ref{fig:lifting-scheme-complete} shows
the scheme for the complete interpolating wavelet transform with decomposition and reconstruction.
It is possible to notice that the reconstruction stage is made by exactly the same steps as the decomposition, but performed backwards.

\begin{figure}
    \centering
    \includegraphics[width=\textwidth]{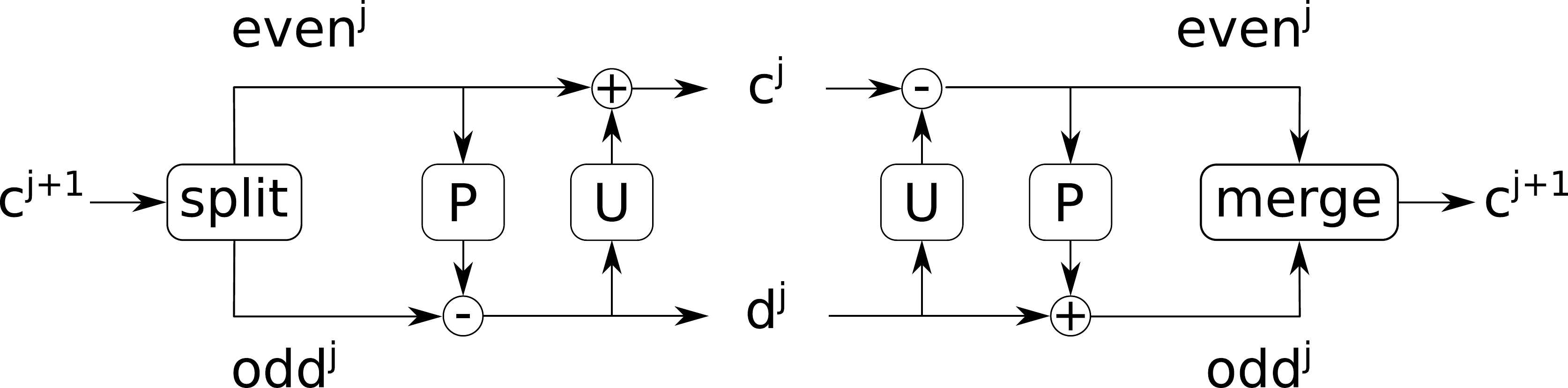}
    \caption{Alternative writing of the complete interpolating wavelet transform scheme.
            This form of writing is probably the most common in literature.}
    \label{fig:lifting-scheme-alternative}
\end{figure}

In Figure~\ref{fig:lifting-scheme-alternative} we show an alternative way of depicting the lifting scheme that is quite common in literature.
The operator $\widetilde{\mathbf S}^j$ is called \emph{prediction} operator $P$, while the $\mathbf S^j$ is called \emph{update} $U$.
Here we preferred to keep the operator $\mathbf P$ as the upsampling operator from $\mathbf c^j$ to $\mathbf c^{j+1}$ and
avoid confusion with different typesets of $P$.
In this same Figure it is possible to notice that the operators $\mathbf E$ and $\mathbf D$ are replaced by two boxes called
\emph{split} and \emph{merge}.
Often they are depicted with boxes containing a $\left[ \downarrow 2  \right]$ for the split stage (meaning downsample by a factor of 2)
and a $\left[ \uparrow 2 \right]$ for the merge (meaning upsample by a factor of 2).
To give a broader picture, since the wavelet literature its very diverse in origin and scope,
we wanted to collect here some different notations used in other frameworks.
Nevertheless the meaning is exactly the same.

\begin{figure}
    \centering
    \includegraphics[width=0.4\textwidth]{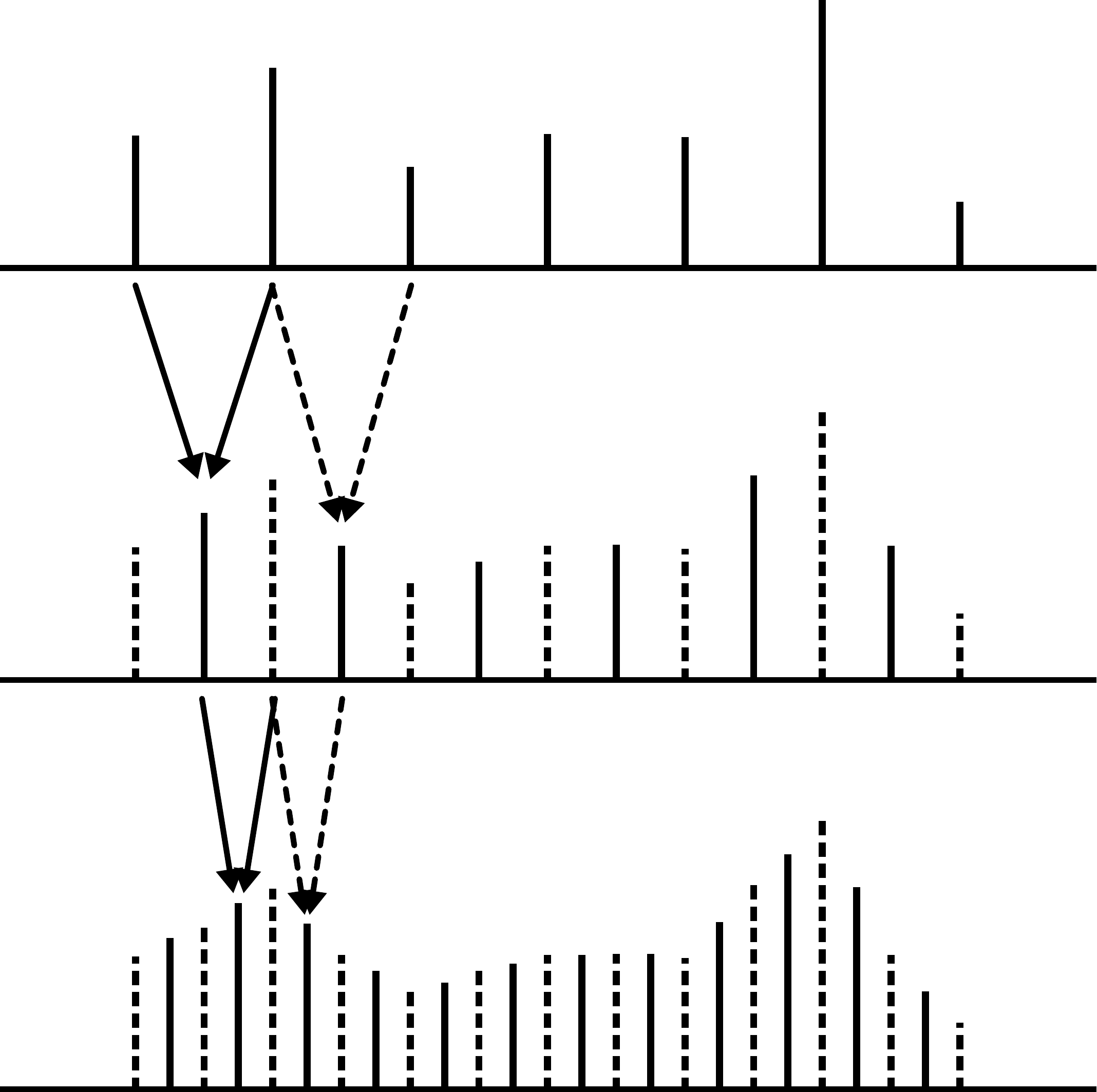}
    \hspace{1cm}
    \includegraphics[width=0.4\textwidth]{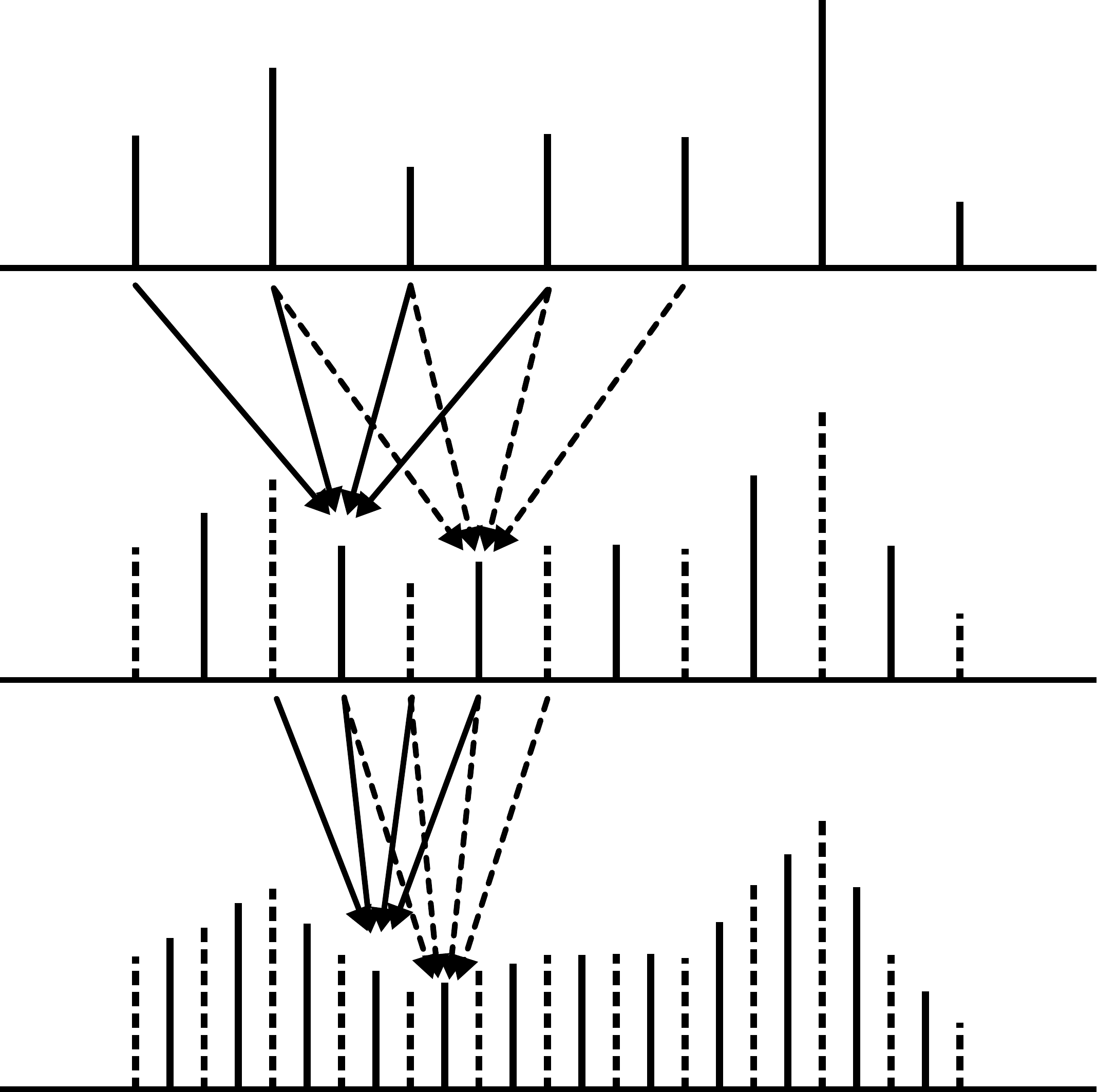}
    \caption{Graphic representation of linear (left) and cubic (right) prediction operators in the reconstruction stage.
            The idea for the graphical illustration is taken from \cite{Sweldens_buildingyour}.}
    \label{fig:graphic-prediction}
\end{figure}

We have described the lifting scheme as a way to build improved operators starting from trivial ones.
Another way to understand the lifting scheme is by interpreting the $\widetilde{\mathbf S}^j$ and $\mathbf S^j$
as prediction and update operators.
Assuming that the original signal has some sort of local correlation, once the signal is split
into odd and even subsets, these two signals are highly correlated.
This means that given one, it should be possible to predict the other with a certain accuracy.
And this is what the prediction operator does, e.g.\ getting an odd sample using its even neighbour(s).
Intuitive examples of prediction operator are the polinomial prediction operators, e.g.\ linear, quadratic or cubic ones,
where 2, 3 or four even neighbours respectively are used to predict one odd sample.
A graphic illustration is reported in Figure~\ref{fig:graphic-prediction}.
The update operator is designed so to preserve the overall average of the signal.
The idea is that the coarser signals $\mathbf c^j$ have the same average value of the original signal,
and going down to last possible level $\mathbf c^0$ this will capture its constant offset (or average).
This is equivalent to ask for zero average details $\mathbf d^j$.

Note: recall the intuitive definition given in Section~\ref{subs:lazy-wavelet} for odd and even operators,
that essentially split a signal in two signals based on the sample index.
When moving to spaces that are more complex than the line or the plane, e.g.\ generic meshes, this simple definition
can be generalized and the lifting scheme can still be applied in the very same way.
As an example, we will see now how the lifting scheme generalizes on a spheric mesh
and how we can build interpolating wavelets on this sphere.

\section{Spherical Wavelets via the Lifting Scheme} \label{sec:spherical-lifting-scheme}
To build wavelets on spaces other than the line, we need a data structure with hierarchical subdivision
and a (re)definition of odd and even samples (or vertices).
In the following we will use a subdivision mesh based on the interative Loop subdivision of an octahedron.
The same construction will be later used during the evaluation in Part~\ref{part:evaluation}.
\begin{figure}
    \hfill
    \includegraphics[width=0.8\textwidth]{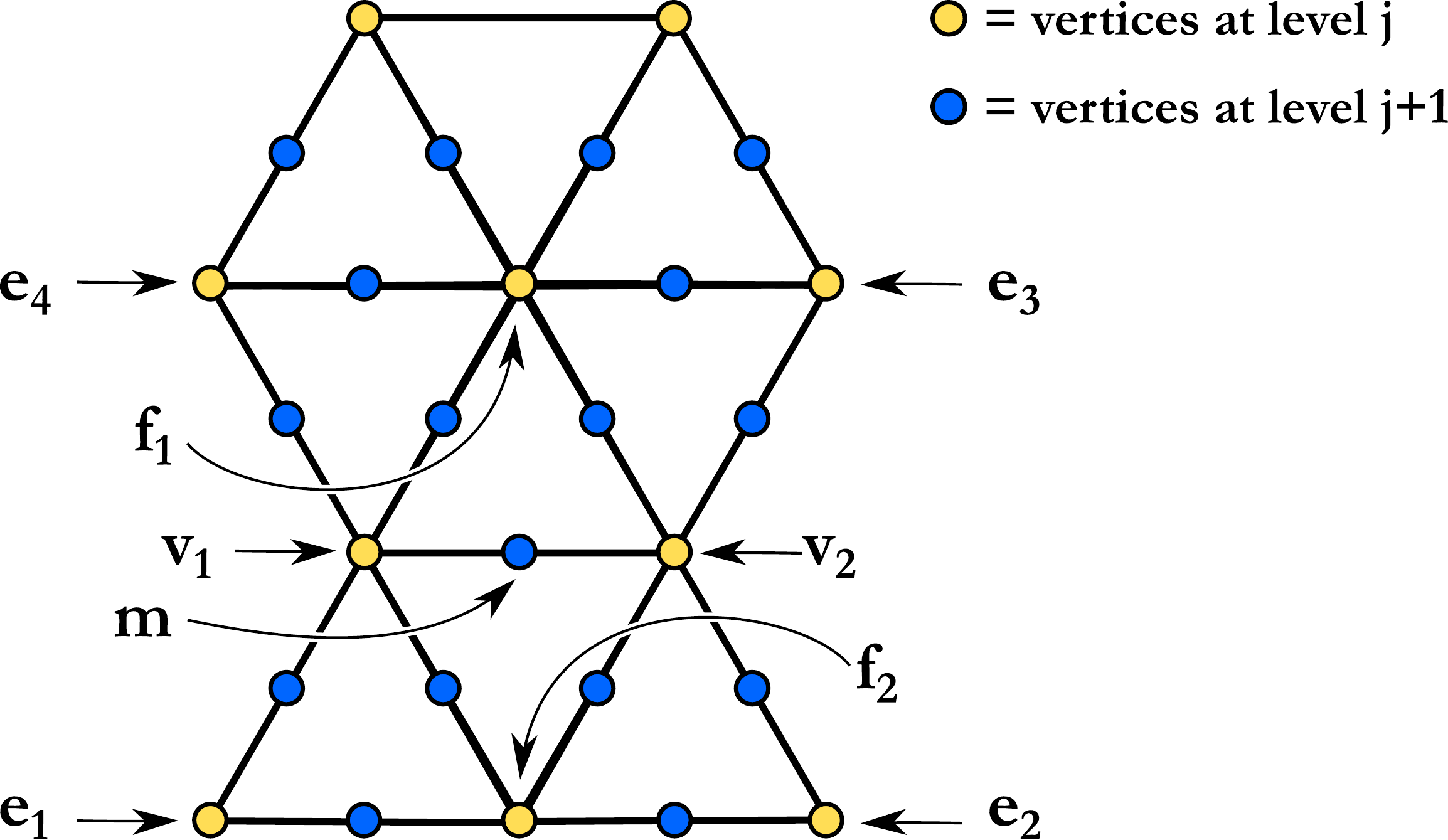}
    \caption{Mesh neighbours.
             Given $m$ a point on the $j+1$ mesh, we define three levels of neighbourhood:
             the closest ones $\left\{ v_1, v_2 \right\}$, then $\left\{ f_1, f_2 \right\}$ and $\left\{ e_1, e_2, e_3, e_4 \right\}$.}
    \label{fig:mesh-neighbours}
\end{figure}
In Figure~\ref{fig:mesh-neighbours} we show an example of a mesh obtained via Loop subdivision.
The yellow dots represent the vertices of the mesh at level $j$, while the blue dots (plus the yellow ones) represent the vertices of the mesh at level $j+1$.
If we take into consideration the point $m$ of the $j+1$ level, that would be the odd vertex in the previous dissertation,
then its neighbours are defined at different distances.
The vertices $\left\{ v_1, v_2 \right\}$ represent the even vertices, and are at equal distance from $m$.
With two points we can already build the linear interpolating wavelet transform.
If we want to build interpolating wavelets on this mesh with higher predictivity, we have to define other (further) neighbours
searching for close vertices at level $j$.
The next closest points are the two $\left\{ f_1, f_2 \right\}$, and the successive ones are the four $\left\{ e_1, e_2, e_3, e_4 \right\}$.

Defined these three sets of neighbours,
to operatively build the wavelets we have to define the lifting operators $\widetilde{\mathbf S}$ and/or the $\mathbf S$.
In the paper from \cite{Schroeder95} the dual lifting operators to generate linear, quadratic and butterfly wavelet filters are defined.
Here we are interested in the linear interpolating wavelet transform, which is not much different to the real line case,
since the prediction operator uses only the immediate neighbours.
Recalling the expressions in \eqref{eq:dual-lifting-from-lazy}, the dual lifting step (or the prediction step)
for the interpolating wavelets is for the analysis (lifting of $\mathbf{B}_{\text{INT}}$ and $\mathbf{P}_{\text{INT}}$)
\begin{equation}
\begin{aligned}
    \mathbf c_{j,k}^{\text{dual}} &= \mathbf c_{j+1,k}                                                     \\
    \mathbf d_{j,m}^{\text{dual}} &= \mathbf c_{j+1,m} - 1/2 (\mathbf c_{j+1,v_1} + \mathbf c_{j+1,v_2} )  \\
\end{aligned}
\hspace{1cm}
\begin{aligned}
    \text{even} \hspace{0.5cm} &\mathbf{A}_{\text{INT}}  \\
    \text{odd}  \hspace{0.5cm} &\mathbf{B}_{\text{INT}}  \\
\end{aligned}
    \label{eq:sph-dual-lift-analysis}
\end{equation}
and for the synthesis
\begin{equation}
\begin{aligned}
    \mathbf c_{j+1,k} &= \mathbf c_{j,k}^{\text{dual}}                                               \\
    \mathbf c_{j+1,m} &= \mathbf d_{j,m}^{\text{dual}} + 1/2 (\mathbf c_{j,v_1}^{\text{dual}} + \mathbf c_{j,v_2}^{\text{dual}} )  \\
\end{aligned}
\hspace{1cm}
\begin{aligned}
    \text{even} \hspace{0.5cm} &\mathbf{Q}_{\text{INT}} \\
    \text{odd}  \hspace{0.5cm} &\mathbf{P}_{\text{INT}} \\
\end{aligned}
    \label{eq:sph-dual-lift-synthesis}
\end{equation}
The dual lifting weights are then $\tilde{s}_{j,v_1,m} = \tilde{s}_{j,v_2,m} = 1/2$.

This construction can be lifted and the weights for the lifting step or update step are chosen so that the wavelet has zero integral
\begin{equation*}
    \mathbf s_{j,k,m} = I_{j+1, m} / 2 I_{j,k} \text{, with } I_{j,k} = \int_{\mathbb{S}^2} \psi_{j,k} d\omega
\end{equation*}
and the integrals $I_{j,k}$ can be approximated at the finest level, and calulated resursively at coarser levels using the refinement relations.
The update step updates the coarse coefficients via the details, as seen in \eqref{eq:lifting-relations}.
Explicitly, for the analysis (lifting of $\mathbf{A}_{\text{INT}}$)
\begin{align}
    \begin{split}
        \mathbf c_{j,v_1}^{\text{lift}} &= \mathbf c_{j,v_1}^{\text{dual}} + \mathbf s_{j,v_1,m} \mathbf d_{j,m}^{\text{dual}}  \\
        \mathbf c_{j,v_2}^{\text{lift}} &= \mathbf c_{j,v_2}^{\text{dual}} + \mathbf s_{j,v_2,m} \mathbf d_{j,m}^{\text{dual}}
    \end{split}
    \label{eq:sph-lift-analysis}
\end{align}
and for during synthesis the analysis step is essentially undone (lifting of $\mathbf{Q}_{\text{INT}}$)
\begin{align}
    \begin{split}
        \mathbf c_{j,v_1}^{\text{dual}} &= \mathbf c_{j,v_1}^{\text{lift}} - \mathbf s_{j,v_1,m} \mathbf d_{j,m}^{\text{dual}}  \\
        \mathbf c_{j,v_2}^{\text{dual}} &= \mathbf c_{j,v_2}^{\text{lift}} - \mathbf s_{j,v_2,m} \mathbf d_{j,m}^{\text{dual}}   .
    \end{split}
    \label{eq:sph-lift-synthesis}
\end{align}
The signal flow is then: apply the prediction step \eqref{eq:sph-dual-lift-analysis}, then the update \eqref{eq:sph-lift-analysis} for the analysis stage,
and then undo the update \eqref{eq:sph-lift-synthesis} and undo the prediction \eqref{eq:sph-dual-lift-synthesis} for the synthesis.

\begin{figure}
    \centering
    \begin{minipage}[t]{0.46\textwidth}
        \includegraphics[width=\textwidth,valign=t]{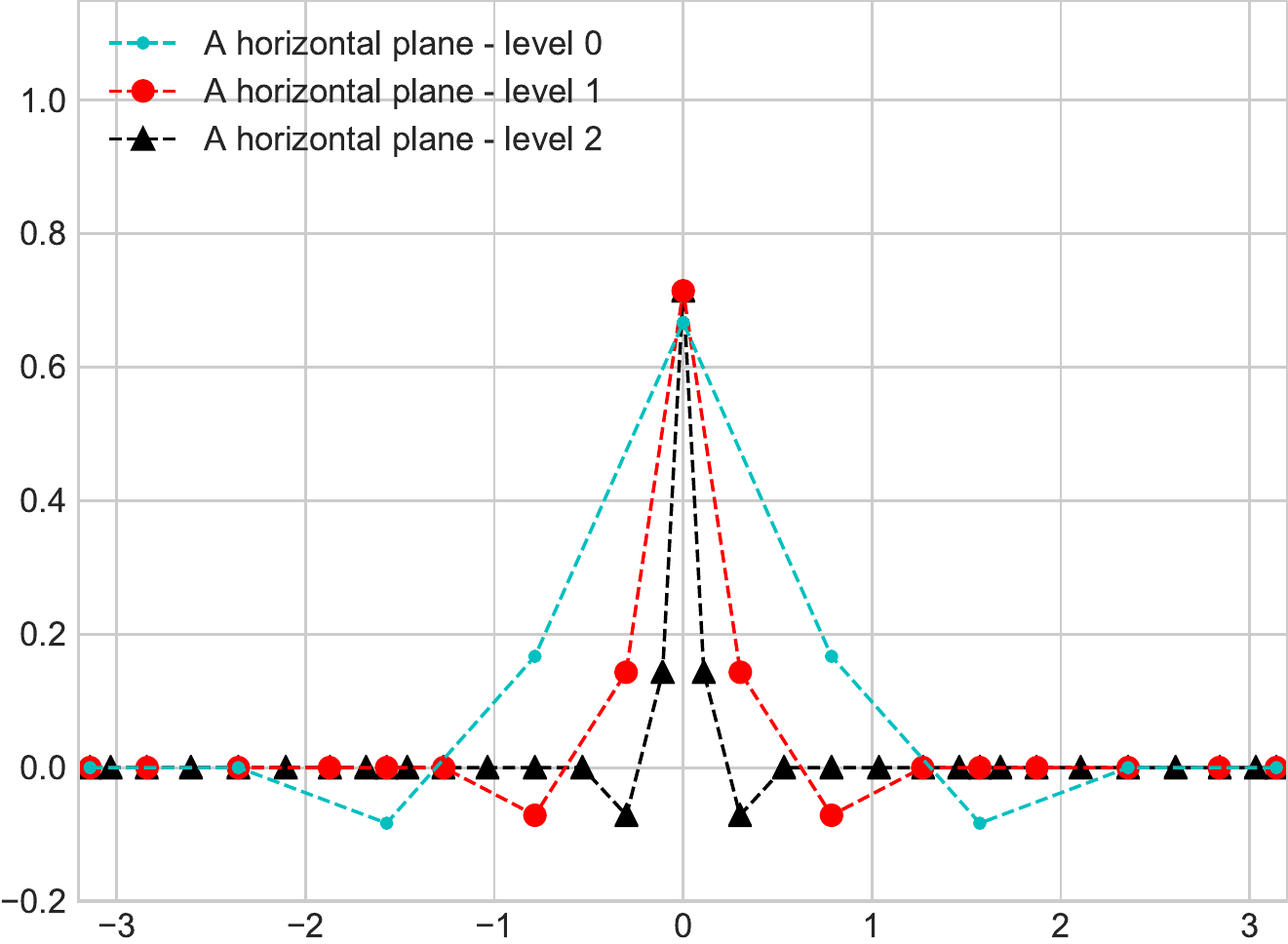}
        \caption{Interpolating dual scaling filter at levels 0, 1, 2 (given by one row of $\mathbf A^{1, 2, 3}$).}
        \label{fig:A-interpolating}
    \end{minipage}
    \hfill
    \begin{minipage}[t]{0.46\textwidth}
        \includegraphics[width=\textwidth,valign=t]{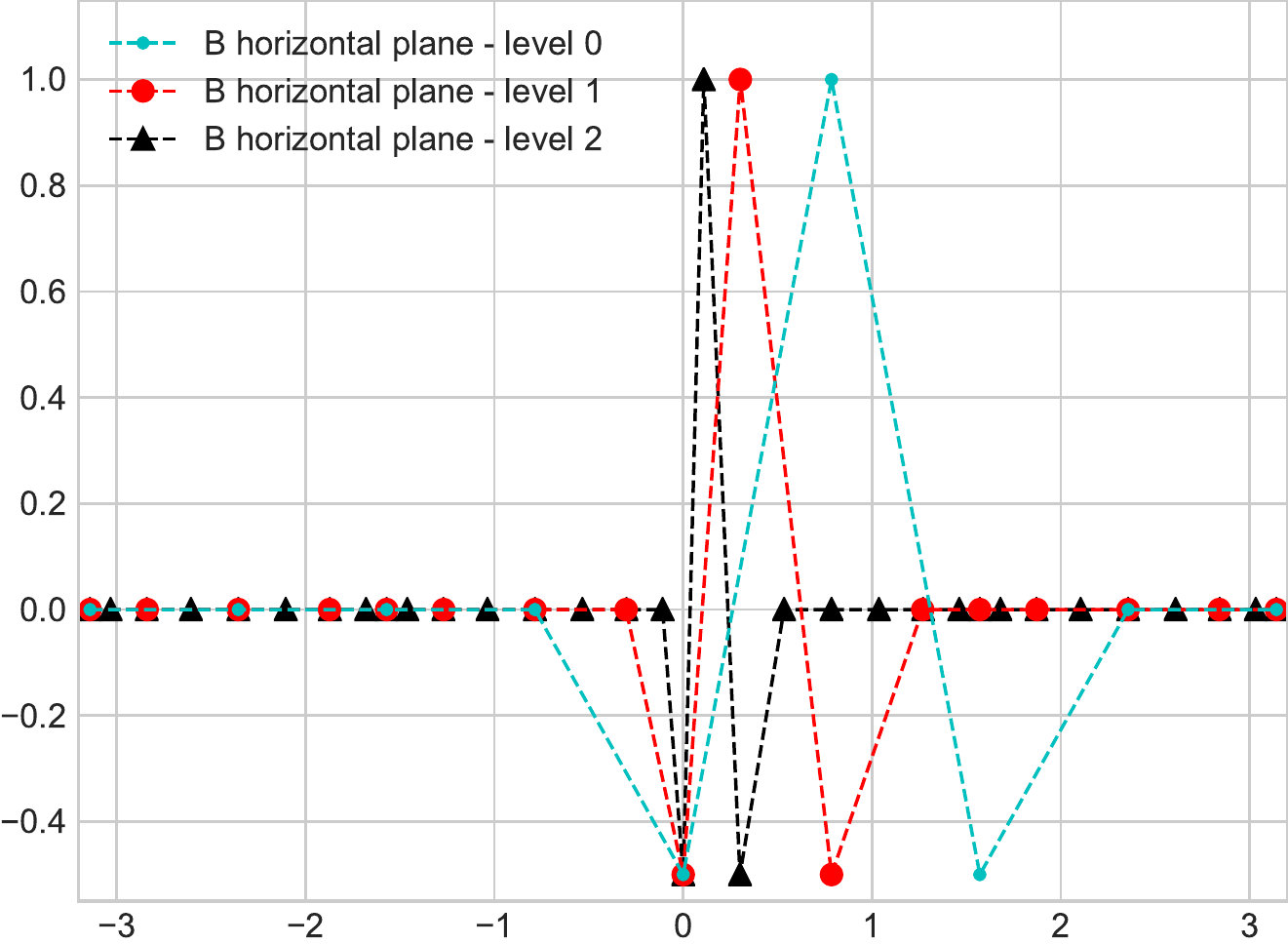}
        \caption{Interpolating dual wavelet filter at levels 0, 1, 2 (given by one row of $\mathbf B^{1, 2, 3}$).}
        \label{fig:B-interpolating}
    \end{minipage}

    \begin{minipage}[t]{0.46\textwidth}
        \includegraphics[width=\textwidth,valign=t]{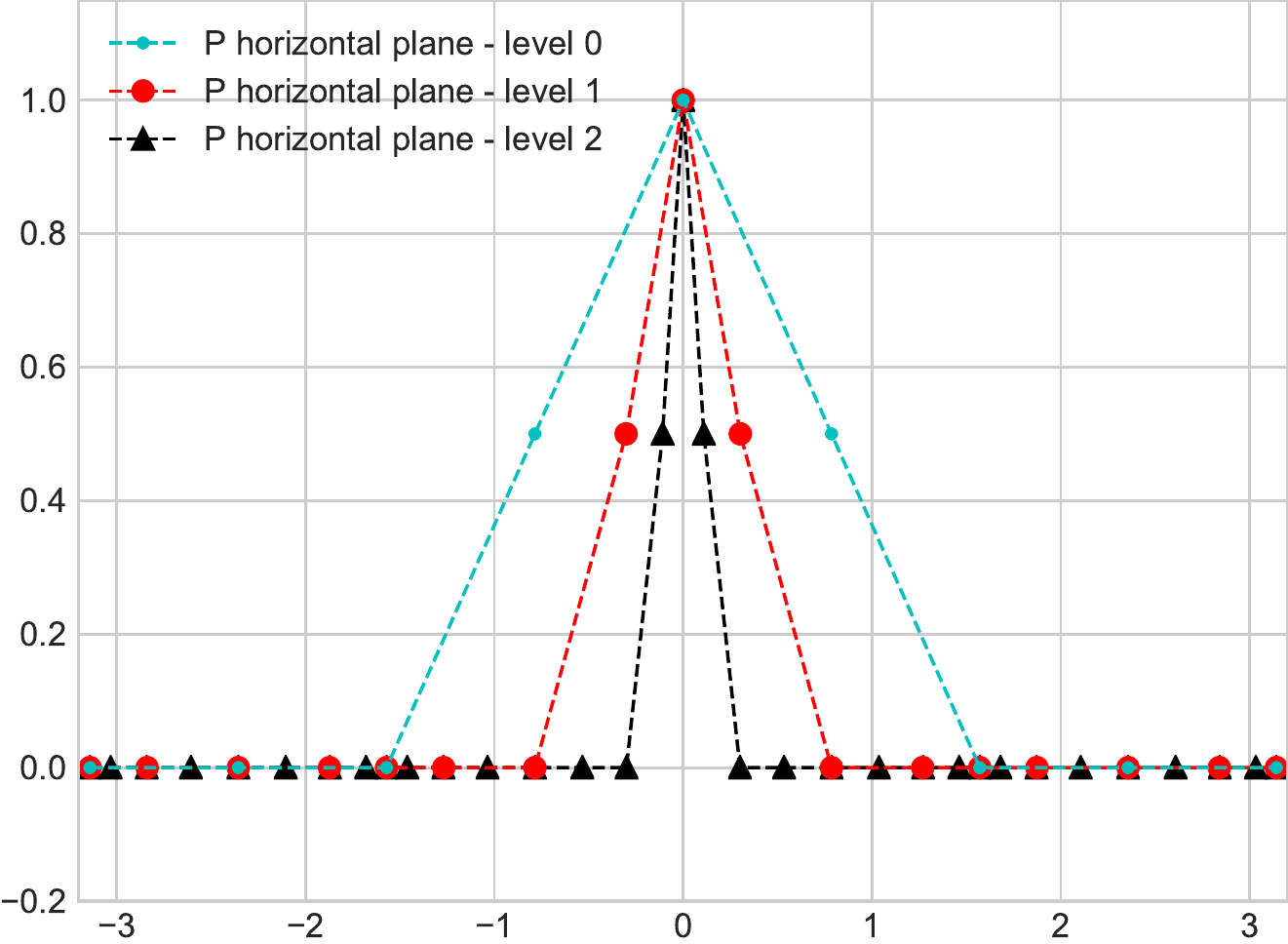}
        \caption{Interpolating scaling filter at levels 0, 1, 2 (given by one column of $\mathbf P^{1, 2, 3}$).}
        \label{fig:P-interpolating}
    \end{minipage}
    \hfill
    \begin{minipage}[t]{0.46\textwidth}
        \includegraphics[width=\textwidth,valign=t]{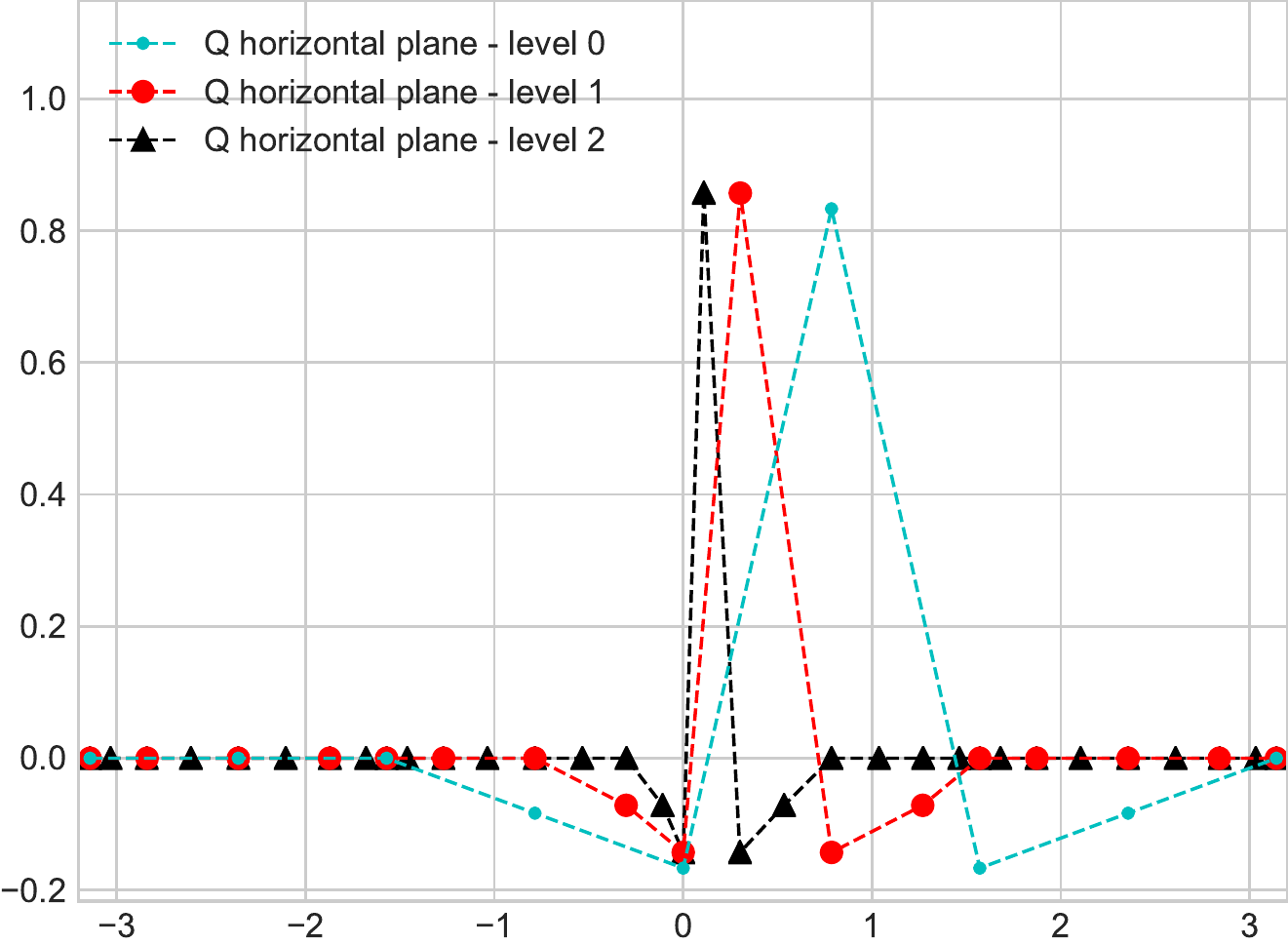}
        \caption{Interpolating wavelet filter at levels 0, 1, 2 (given by one column of $\mathbf Q^{1, 2, 3}$).}
        \label{fig:Q-interpolating}
    \end{minipage}
\end{figure}

The resulting interpolating wavelets and scaling functions, for a subdivision mesh starting from an octahedron
and refined via the Loop subdivision method, are depicted in Figures~\ref{fig:A-interpolating},
\ref{fig:B-interpolating}, \ref{fig:P-interpolating} and \ref{fig:Q-interpolating}.
Note that in this particular space, the sphere, because there are no boundaries the wavelets are all identical at the same level.
For this reason we show only one filter per level.
In these Figures it is possible to realize that higher level filters are actually shrunk versions of lower levels' ones.

Note: in the lifting scheme we never build explicitly the scaling functions or wavelets,
but we can get them by running a delta into the graph and running it ad infinitum, as described in~\eqref{eq:wavs-from-matrices}.

\chapter{Wavelet-based Spherical Audio Framework}  \label{ch:wavelet-audio-chain}
In this Chapter we will describe the full audio chain for a wavelet based audio format.
Section~\ref{sec:wavelet-format} describes the multiresolution scheme over a subdivision mesh,
defining our multiresolution framework for spatial audio.
Section~\ref{sec:encoding} depicts the strategy for the encoding of an audio source over the subdivision mesh.
Up to this point, the setup is completely general, and can be used to generate a wide diversity of formats.
For this reason we call this audio encoding scheme a \emph{framework} for wavelet audio formats.
In Section~\ref{sec:swf} we particularize this wavelet format to the spherical domain.
Section~\ref{sec:swf-decoding} regards the decoding of the new spherical wavelet format.

This Chapter is based on the submitted paper ``Wavelet based spherical audio format'' \cite{ScainiArteagaWavelet}.

\section{Multiresolution Framework for Audio} \label{sec:wavelet-format}
In this Section we will describe the basics of wavelet multiresolution analysis,
with emphasis on the practical aspects that are relevant for an audio encoding/decoding chain.

Many parts of this Section will consist in a particularization of the material already covered in Section~\ref{sec:multiresolution}

\begin{figure}
    \centering
\begin{tikzpicture}[>=latex']
    \tikzset{block/.style= {draw, rectangle, align=center,minimum width=1cm,minimum height=0.6cm},
    }
    \node [block]  (start) {$\mathbf{f}$};

    \node [coordinate, right = 0.5cm of start] (C1R){};
    \node [coordinate, above = 0.5cm of C1R] (C1RU){};
    \node [block, right = 1cm of C1RU] (C1){$\mathbf{c}^{n-1}$};

    \node [coordinate, right = 0.5cm of start] (D1R){};
    \node [coordinate, below = 0.5cm of D1R] (D1RD){};
    \node [block, right = 1cm of D1RD] (D1){$\mathbf{d}^{n-1}$};

    \node [coordinate, right = 0.5cm of C1] (C2R){};
    \node [coordinate, above = 0.5cm of C2R] (C2RU){};
    \node [block, right = 1cm of C2RU] (C2){$\mathbf{c}^{n-2}$};

    \node [coordinate, right = 0.5cm of C1] (D2R){};
    \node [coordinate, below = 0.5cm of D2R] (D2RU){};
    \node [block, right = 1cm of D2RU] (D2){$\mathbf{d}^{n-2}$};

    \node [draw=none,fill=none, right = 3cm of D1] (dots) {\ldots};
    \path[draw]
        (start) -- (C1R)
        (C1R) -- (C1RU)
        (D1R) -- (D1RD);

    \path[->]
        (C1RU) edge node [midway,above]{$\mathbf{A}^{n}$} (C1)
        (D1RD) edge node [midway,below]{$\mathbf{B}^{n}$} (D1);

    \path[draw]
        (C1) -- (C2R)
        (C2R) -- (C2RU)
        (C2R) -- (D2RU);

    \path[->]
        (D2RU) edge node [midway,below]{$\mathbf{B}^{n-1}$} (D2)
        (C2RU) edge node [midway,above]{$\mathbf{A}^{n-1}$} (C2);
\end{tikzpicture}
    \caption{Scheme of signal decomposition and encoding to wavelet space.}
    \label{scheme:decomposition}
\end{figure}
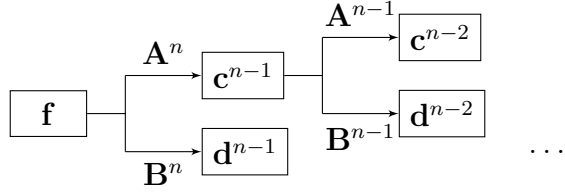
\subsection{Decomposition}
\label{sec:wvencoding}
Being $\mathbb{M}$ some mesh in $\mathbb{R}^3$, obtained with some subdivision scheme,
then the set of data $\mathbf{f} = \left(f_1 \ \cdots \ f_N \right)^T $ defined over the finest level of this mesh,
is called \emph{fine data}.
The process of subdivision separates the fine data $\mathbf{f}$
into two signals (sets of data), a \emph{coarse} approximation $\mathbf{c}$
and an additional information called \emph{details} $\mathbf{d}$.
The decomposition is defined then as:\footnote{For the definition of scaling and wavelet filters,
we use an adaptation of the the notation in~\cite{Olsen2007liftingOpt}.}
\begin{align}
\begin{split}
    \mathbf{c} &= \mathbf{A} \mathbf{f}, \\
    \mathbf{d} &= \mathbf{B} \mathbf{f},
\end{split}
\end{align}
where $\mathbf{A}$ and $\mathbf{B}$ are the \emph{decomposition} or \emph{analysis filters} introduced in Section~\ref{sec:multiresolution}.

The filters $\mathbf{A}$ and $\mathbf{B}$ connect levels: from the finest level $n$ to a coarser level $n-1$.
There are as many decomposition filters, or encoding matrices, as mesh levels minus one.
The signal $\mathbf{c}$ represents a spatially low-passed and downsampled version of $\mathbf{f}$.
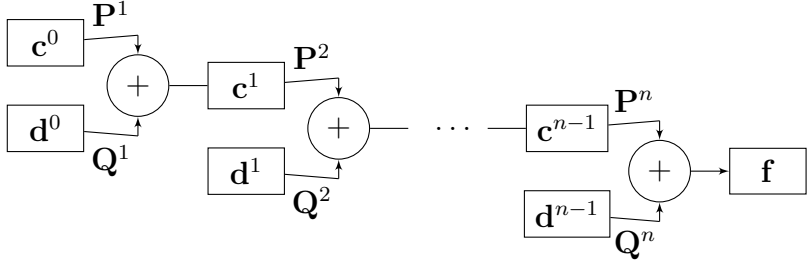
\begin{figure*}[bth] \label{fig:reconstruction}
    \centering
\begin{tikzpicture}[>=latex']
    \tikzset{block/.style= {draw, rectangle, align=center,minimum width=1cm,minimum height=0.6cm},
    }
    \node [block]  (c0) {$\mathbf{c}^0$};
    \node [block, below=0.5cm of c0]  (d0) {$\mathbf{d}^0$};
    \node [coordinate, below=0.25cm of c0] (below-c0){};
    \node [circle, right=0.8cm of below-c0, fill=none, draw, minimum height=0.4cm] (circle-plus-0){$+$};
    \node [coordinate, above=0.25cm of circle-plus-0.north] (above-plus0){};
    \node [coordinate, below=0.25cm of circle-plus-0.south] (below-plus0){};

    \path[draw]
        (c0) -- node [midway,above]{$\mathbf{P}^1$} (above-plus0)
        (d0) -- node [midway,below]{$\mathbf{Q}^1$} (below-plus0);

    \path[->]
    (above-plus0) edge (circle-plus-0)
    (below-plus0) edge (circle-plus-0);

    \node [block, right=0.5 of circle-plus-0.east] (c1) {$\mathbf{c}^1$};
    \node [block, below=0.5cm of c1]               (d1) {$\mathbf{d}^1$};
    \node [coordinate, below=0.25cm of c1] (below-c1){};
    \node [circle, right=0.8cm of below-c1, fill=none, draw, minimum height=0.4cm] (circle-plus-1){$+$};
    \node [coordinate, above=0.25cm of circle-plus-1.north] (above-plus1){};
    \node [coordinate, below=0.25cm of circle-plus-1.south] (below-plus1){};

    \path[draw]
        (c1) -- node [midway,above]{$\mathbf{P}^2$} (above-plus1)
        (d1) -- node [midway,below]{$\mathbf{Q}^2$} (below-plus1);

    \path[->]
    (above-plus1) edge (circle-plus-1)
    (below-plus1) edge (circle-plus-1);

    \path[draw] (circle-plus-0) -- (c1);

    \node [coordinate, right=0.5 of circle-plus-1.east] (coorddots) {};
    \path[draw] (circle-plus-1) -- (coorddots);
    \node [draw=none,fill=none, right=0.2cm of coorddots] (dots) {\ldots};

    \node [block, right=0.5 of dots.east] (cn) {$\mathbf{c}^{n-1}$};
    \node [block, below=0.5cm of cn]      (dn) {$\mathbf{d}^{n-1}$};
    \node [coordinate, below=0.25cm of cn] (below-cn){};
    \node [circle, right=0.8cm of below-cn, fill=none, draw, minimum height=0.4cm] (circle-plus-n){$+$};
    \node [coordinate, above=0.25cm of circle-plus-n.north] (above-plusn){};
    \node [coordinate, below=0.25cm of circle-plus-n.south] (below-plusn){};

    \path[draw]
        (cn) -- node [midway,above]{$\mathbf{P}^n$} (above-plusn)
        (dn) -- node [midway,below]{$\mathbf{Q}^n$} (below-plusn);

    \path[->]
    (above-plusn) edge (circle-plus-n)
    (below-plusn) edge (circle-plus-n);

    \path[draw] (dots) -- (cn);

    \node [block, right=0.5cm of circle-plus-n] (fn) {$\mathbf{f}$};
    \path[->] (circle-plus-n) edge (fn);
\end{tikzpicture}
    \caption{Scheme of reconstruction of the original signal from the wavelet space.}
    \label{scheme:reconstruction}
\end{figure*}

\subsection{Reconstruction}
\label{sec:wvdecoding}
The upsampling process increases the spatial resolution of the coarse data $\mathbf{c}$
to the fine data $\mathbf{f}$, and if the details $\mathbf{d}$ are available, then
the reconstruction process will give back the original fine data:
\begin{equation}
    \mathbf{f} = \mathbf{P} \mathbf{c} + \mathbf{Q} \mathbf{d}.
\end{equation}
where $\mathbf{P}$ and $\mathbf{Q}$ are the \emph{reconstruction} or \emph{synthesis filters}
introduced in Section~\ref{sec:multiresolution}.
The filters $\mathbf{P}$ and $\mathbf{Q}$ connect levels: from the coarser level $n-1$ to the finest level $n$.
There are as many reconstruction filters, or decoding matrices, as mesh levels minus one.

\subsection{Wavelet Transform}  \label{subsec:wavelet-transform}
Consider the finest subdivision level $n$, consisting of $N$ points,
and the next coarser level $n-1$, consisting of $M$ points,
with $N>M$.
Consider a data vector
$\mathbf{f} = [f_1, \ldots, f_N]^T$, defined at the finest level $n$.
The subdivision process would proceed as follows:
\begin{align*}
    \mathbf{c}_{M \times 1}^{n-1} &= \mathbf{A}_{M \times N}^{n} \mathbf{f}_{N \times 1},  \\
    \mathbf{d}_{\left( N - M \right) \times 1}^{n-1} &= \mathbf{B}_{\left( N - M \right) \times N}^{n} \mathbf{f}_{N \times 1},
\end{align*}
where the subindices indicate the dimension of each matrix.
The sum of elements in $\mathbf{c}$ and $\mathbf{d}$ is $M + \left( N - M \right) = N$.
Now if the even coarser subdivision level $n-2$ has $K$ points, the decomposition can continue:
\begin{align*}
    \mathbf{c}_{K \times 1}^{n-2} &= \mathbf{A}_{K \times M}^{n-1} \mathbf{c}_{M \times 1}^{n-1}, \\
    \mathbf{d}_{\left( M - K \right) \times 1}^{n-2} &= \mathbf{B}_{\left( M - K \right) \times M}^{n-1} \mathbf{c}_{M \times 1}^{n-1},
\end{align*}
and so on.
See Figure~\ref{scheme:decomposition} for a representation.

If the decomposition is followed up to the coarsest level available (\emph{level~0}),
there will be a list of $n-1$ detail signals or wavelet coefficients, $ \mathbf{d}^0, \ldots, \mathbf{d}^{n-1} $,
and one last coarse signal or scaling function coefficients $\mathbf{c}^0$;
the representation $\{\mathbf{c}^0, \mathbf{d}^0, \ldots, \mathbf{d}^{n-1}\}$ constitutes the \emph{wavelet transform}.
In compact form, the wavelet transform can be computed recursively as:
\begin{align}
\begin{split}
    \mathbf c^{j-1} &= \mathbf A^{j} \mathbf c^{j}, \\
    \mathbf d^{j-1} &= \mathbf B^{j} \mathbf c^{j},
\label{eq:wt}
\end{split}
\end{align}
with $\mathbf c^n = \mathbf f$ and $j=n,\ldots,1$.

To perform the inverse wavelet transform, namely, to reconstruct the original signal,
the procedure is recursive but this time starting from the coarsest \emph{level~0} and going to the finest \emph{level~$n$}:
\begin{align*}
    \mathbf{c}^1_{J\times1} &=
    \mathbf{Q}^1_{J\times\left( J-L \right)} \mathbf{d}^0_{\left( J-L \right)\times1} + \mathbf{P}^1_{J\times L} \mathbf{c}^0_{L\times 1} \\
    \mathbf{c}^2_{K\times1} &=
    \mathbf{Q}^2_{K\times\left( K-J \right)} \mathbf{d}^1_{\left( K-J \right)\times1} + \mathbf{P}^2_{K\times J} \mathbf{c}^1_{J\times 1},
\end{align*}
see Figure~\ref{scheme:reconstruction} for a diagram.
In compact form, the inverse wavelet transform can be represented recursively as:
\begin{align}
    \mathbf c^{k} = \mathbf P^{k} \mathbf c^{k-1} +  \mathbf Q^{k} \mathbf d^{k-1},
    \label{eq:iwt}
\end{align}
with $k=1,\ldots, n$.

The filters $\mathbf{A}^j$, $\mathbf{B}^j$, $\mathbf{P}^j$ and $\mathbf{Q}^j$
are the building blocks of the proposed spatial audio format.
They are not arbitrary: to define a wavelet framework they need to follow relations \eqref{eq:inv1} and \eqref{eq:inv2}.
Several methods to build these filters are available in literature.
In Section~\ref{sec:lifting-scheme} we describe a method based on the Lifting Scheme,
and in Chapter~\ref{ch:our-wavelets} we illustrate our optimization scheme.
In the following we will assume that some filters with appropriate characteristics are available.

A note of caution: the matrices $\mathbf{A}^j$ and $\mathbf{B}^j$ connect the level $j$ with the level $j-1$,
while the matrices $\mathbf{P}^j$ and $\mathbf{Q}^j$ connect the level $j-1$ with the $j$.
The index of the matrices is the same, but the dimension of their output signal is different.

\section{Audio Source Encoding} \label{sec:encoding}

\subsection{First Step: Source Interpolation}
In spatial audio, any source can have an arbitrary position in spherical space,
with continuous azimuth and elevation coordinates $\left(\theta, \phi \right)$.
This point source needs to be represented on the finest mesh,
which is defined only over a discrete set of points.
The first step of the encoding process is the interpolation
of the point source to the finest mesh.
The most natural interpolation for triangular meshes
is the tri-linear interpolation over the three vertices of the triangle.
In spatial audio this tri-linear interpolation is
the basis of Vector-Base Amplitude Panning~\cite{pulkki1997vbap} (VBAP).
In this thesis we use a tri-linear (or VBAP-like) interpolation to represent
the point source in the subdivision mesh (other choices are also possible).

\subsection{Second Step: Wavelet Encoding}
As a result of the interpolation, a set of coefficients
$\mathbf{f} = (f_1 \ \cdots \ f_N)^T$
will be available at the finest mesh (order $n$).
If the source is a point source,
at most three of these coefficients will be non-zero.
The second step is to apply the wavelet transform,
recursively downsampling the subdivision mesh,
by repeated application of the decomposition filters $\mathbf A$ and $\mathbf B$,
as explained in Section~\ref{sec:wavelet-format} and described in Figure~\ref{scheme:decomposition}.
The result of the wavelet transform will be the set of signals
$ \{ \mathbf c^0, \mathbf{d}^0, \dots, \mathbf{d}^{n-1} \}$,
having the same  total dimensionality $N$ as the original.

\subsection{Third Step: Wavelet Truncation}
At this stage, one of the most common techniques in different fields is zeroing all the details coefficients
smaller than some fixed threshold in order to achieve compression.
In spatial audio, the goal is also to transmit a low-dimensional field,
but to ensure a smooth and consistent playback experience, and in analogy to Ambisonics,
we will follow a different path by limiting the decomposition up to a given order.

Therefore, the third step in the encoding is truncating the decomposition to order $\ell$,
with $0 \leq \ell < n$,
which amounts to zeroing the detail coefficients with order equal or greater than $\ell$.
Namely, the truncated decomposition at order $l$ be
$\{ \mathbf c^{0}, \mathbf{d}^0, \dots, \mathbf{d}^{\ell-1}, \mathbf{0}^{\ell}, \dots, \mathbf{0}^{n-1} \}$, or,
more simply,
$\{ \mathbf c^{0}, \mathbf{d}^0, \dots, \mathbf{d}^{\ell-1} \}$.
(In the case $\ell=0$ all wavelet coefficients will be zero, with only the coarsest scaling coefficient remaining.)

\section{Spherical Wavelet Format} \label{sec:swf}

A \emph{Spherical Wavelet Format} (SWF) is defined to be each one of the spherical audio encodings determined by:
\begin{enumerate}
    \item A recursive subdivision mesh over the sphere, ranging from the coarsest level 0 to the finest level $n$.
    \item A set of filters $\{\mathbf A^j, \mathbf B^j, \mathbf P^j, \mathbf Q^j| j \in [1, n]\}, $ defining a wavelet space,
          and verifying the set of equations \eqref{eq:inv1} and \eqref{eq:inv2}.
    \item A truncation level $\ell \in [0, n]$, defining the order of the wavelet decomposition.
\end{enumerate}
The signals in SWF can be represented in two alternative equivalent ways: either as the 
scaling function coefficients at the coarser level plus a set of successively finer wavelet details,
$ \{ \mathbf c^0, \mathbf{d}^0, \dots, \mathbf{d}^{\ell-1} \}$,
or as the scaling functions coefficients at the truncation level $\ell$: $\{ \mathbf c^\ell \}$.
In this second representation only the downsampling $\mathbf{A}^j$ and upsampling $\mathbf{P}^j$ filters are strictly needed.

There are many different ways to generate the filters defining the wavelet space.
One method is the Lifting Scheme, described in Section~\ref{sec:lifting-scheme}.
Other methods build on the lifting scheme, generating optimized filter for specific applications e.g.\ \cite{Kammoun2012}.
In Chapter~\ref{ch:our-wavelets} we describe our audio-tailored method.

Another possibility is given by the application of VBAP as the interpolator between mesh levels.
VBAP implicitly defines a set of downsampling, $\mathbf{A}^j$ and upsampling $\mathbf{P}^j$ filters
for any subdivision mesh. This filters can be computed by considering a given mesh at level $j$ as a set of sound sources,
and rendering those meshes to the finer level ($\mathbf{P}^j$ is generated) or coarser level ($\mathbf{A}^j$ is generated).
In this case the filters created via VBAP have a maximum length of 3 points for $\mathbf{A}^j$, because VBAP activates at most three neighbouring points of the mesh.
In the case of $\mathbf{P}^j$ VBAP generates a trivial filter, which is a block matrix with an identity on the first block and zeros elsewhere
(since the points of level $j$ are contained in the level $j+1$ mesh), so there is no effective upsampling.
These VBAP-inspired wavelets are close in spirit to the interpolating wavelets \cite{Schroeder95},
with the difference of having a different set of neighbours and with the dual and direct spaces swapped.
We will call this version of SWF based on VBAP: VBAP-SWF.

\begin{figure}
    \centering
    \includegraphics[width=0.9\textwidth]{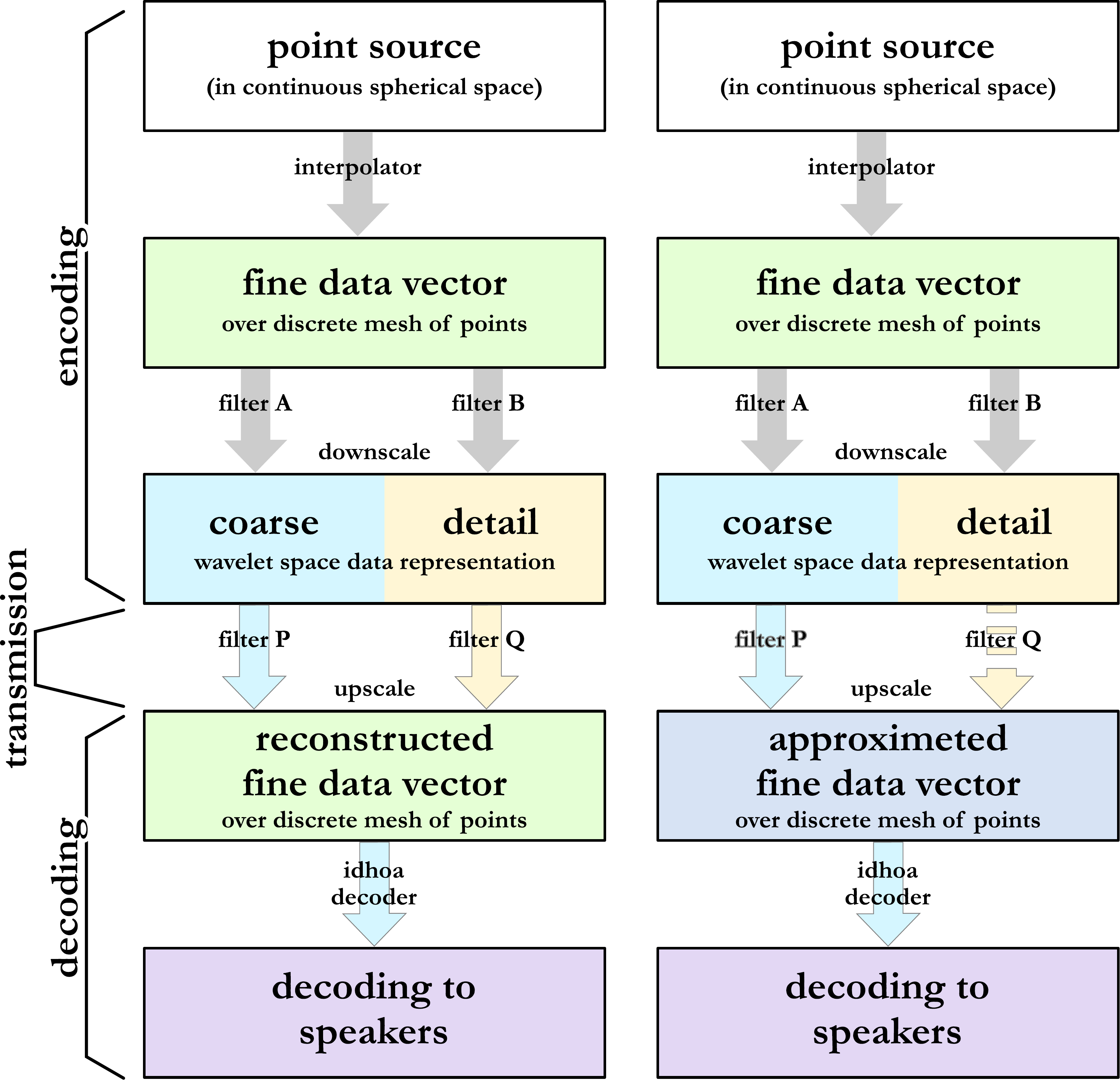}
    \caption{Encoding, transmission and decoding of a SWF,
    without wavelet truncation (on the left),
    and with wavelet truncation (on the right).}
    \label{fig:encoding-decoding-workflow}
\end{figure}

\section{Spherical Wavelet Format Decoding} \label{sec:swf-decoding}
\label{sec:decoding}

Let us consider a SFW encoding of an arbitrary sound source distribution.
If this wavelet encoding has not been truncated,
a trivial decoding
can be generated by inverting the wavelet transform,
making repeated use of Eq.~\eqref{eq:iwt}.
This will lead to the original source $\mathbf c^n = \mathbf f$.
By associating a loudspeaker to each point of the finer mesh,
the values of $\mathbf f$ can be interpreted as the loudspeaker signals.
This is represented on the first four blocks on the left column of Figure~\ref{fig:encoding-decoding-workflow}.
Essentially, this trivial decoding reproduces the VBAP decoding of the original sound source
to the finer mesh.

A more interesting case happens when the wavelet encoding has been truncated to order $\ell < n$:
the encoding can be partially inverted to order $\ell$, again by repeated use of Eq.~\eqref{eq:iwt}.
The truncated signal can be represented by the scaling coefficients at order $\ell$,
$\mathbf c^{\ell}$.
If the nodes of the $\ell\text{-th}$ mesh are interpreted as speakers,
then the values of $\mathbf{c}^\ell$ can be considered as the speaker feeds
corresponding to a decoding to a regular layout with one loudspeaker in each one of the vertices of the mesh at $\ell\text{-th}$ level.

Actually, the signal at level $\ell$ can be upsampled by applying repeatedly the reconstruction filter $\mathbf P$,
ignoring the details beyond level $\ell$:
\begin{align*}
    \mathbf {\tilde{c}}^{\ell+1} = \mathbf P^{\ell+1} \mathbf{c}^{\ell},
\end{align*}
where the tilde indicates that the reconstruction comes from a truncated representation.
Ultimately,
the signal can be reconstructed to the finest level. The resulting signal $\mathbf{\tilde f} = \mathbf{\tilde c}^n$,
corresponds to a spatial low-pass version
of the original signal $\mathbf{f}$.

This way, by interpreting the coefficients $\mathbf{\tilde c}^k$ as loudspeaker signals
associated to the vertices of the meshes at level $k$,
decodings to the different layouts corresponding to the mesh geometry at different levels are generated,
from the coarse level $\ell$ to the finest level $n$.
This is represented on the first four blocks on the right column of Figure~\ref{fig:encoding-decoding-workflow}.
This decoding procedure is the wavelet equivalent to the basic decoding of Ambisonics to a regular layout.
It is to be noted however that in wavelets, differently to Ambisonics,
there is no guarantee that the pressure, the acoustic velocity or any other relevant
acoustical or psychoacoustical parameter
are correctly reconstructed at the origin,
unless the wavelet family has been designed in some special way.

Here we propose to optimize the matrix responsible for the decoding to speakers leveraging some
acoustical and psychoacoustical observables, as we did for Ambisonics with IDHOA
(see Chapter~\ref{ch:idhoa}).

\subsection{Decoding of SWF to an Irregular Loudspeaker Layout using IDHOA} \label{sec:wv-idhoa}
The goal is, given a loudspeaker layout and a given SWF signal,
to generate the speaker signals $\mathbf s$
as a linear combination of the scaling function coefficients at the truncation order $\ell$:
\begin{align}
    \mathbf s = \mathbf{D}^{\ell+1} {\mathbf c}^{\ell}
\end{align}
This is represented on the bottom blocks of Figure~\ref{fig:encoding-decoding-workflow}.
The decoding matrix has as many rows as speakers and as many columns as channels in the SWF.
The scaling function channels at level $\ell$, $\mathbf c^{\ell}$,
can be computed out of the channels in the wavelet transform representation, $\{ \mathbf c^{0}, \mathbf{d}^0, \dots, \mathbf{d}^{\ell-1} \}$,
by applying repeatedly Eq.~\eqref{eq:iwt}.
Actually, the decoding matrix can be also computed from an upsampled version of the coarse channels:
\begin{align*}
    \mathbf s = \mathbf{D}^{j +1} \tilde {\mathbf c}^{j}, \quad \ell < j \leq n.
\end{align*}
The decoded $\mathbf s$ produced via $\mathbf D^{\ell +1}$ and $\mathbf{D}^{j +1}$ should be equivalent,
but in practice, since the different decoding matrices are obtained a via separate non-linear optimization processes,
there might be differences.

We have developed a numerical optimization method to find the optimal decoding of a given SWF, based on IDHOA.
With respect to the implementation described in Chapter~\ref{ch:idhoa}, IDHOA has been adapted to accept
any input in any format, as long as a sound source can be encoded in a sufficient number of points on the surface of the sphere.
With `sufficient' we mean that the number of points where the sound source can be encoded has to be greater than the number of speakers
of the destination layout.
The same perceptual criteria used for Ambisonics are still valid for Wavelets decoding:
optimize pressure, energy, radial velocity, radial intensity,
while the transverse velocity and transverse intensity are minimized.
A quadratic term is responsible for penalizing negative gains.
This way IDHOA can now produce decoding matrices for both Ambisonics and SWF.

Schematically:
\begin{enumerate}
    \item \emph{Initialization:} Operations that are performed only once when the algorithm is launched.
    \item Given the loudspeakers' layout, calculate $\mathbf{D}_\text{init}$.
          In this case we might not have a hint on the decoding matrix, like the projection matrix in Ambisonics,
          so we can initialize $\mathbf{D}_\text{init}$ completely random.

    \item Calculate the various physical variables that constitute the cost function: $p$, $E$, $\vect v$, $\vect I$,
            given by Eqs.~\eqref{eq:P_definition}, \eqref{eq:E_definition}, \eqref{eq:V_definition}, \eqref{eq:I_definition}.
          In the case of SWF, the set of $n$ points where is actually possible to evaluate these variables
          coincides with the set of points of the finest mesh at which the wavelets are available.
          The set of sampling points is not arbitrary anymore like for Ambisonics, where the functions defining the format are continuous
          and it is possible to calculate the encoding matrix at any point in space.
          Calculate the objective function, which is $f=f(\mathbf{D}_\text{init})$.

    \item \emph{Fix constraints (optional):} constrain some parameters to have a fixed value (e.g.~lock to zero).
    \item \emph{Minimization stage:} Call to the external minimization algorithm, passing $\mathbf{D}_\text{init}$ and $f$.
          When the minimization algorithm terminates, it returns a $\widetilde{\mathbf{D}}$.
\end{enumerate}

The IDHOA code for SWF decoding will be published in the same location of \cite{github-idhoa-new}.
\section{Summary}
In this Chapter we have described a method for spatial audio encoding and decoding leveraging the multiresolution paradigm.
The encoding and decoding filters are built directly on the multiresolution mesh
(in Section~\ref{sec:spherical-lifting-scheme} and Chapter~\ref{ch:our-wavelets} we show two different methods to actually build them).
The final step of decoding to speakers is left to IDHOA, leveraging the same observables used for Ambisonics decoding.

\chapter{Building Spherical Wavelets via Numerical Optimization}  \label{ch:our-wavelets}

\section{Motivation}
The application of wavelets in spatial audio is very different from the other fields
such as computer graphics or image compression. On the one hand, the number of coefficients is very limited, e.g.~between 4 and 20,
          while typically for analysis or compression the number of coefficients can be of the order of thousands or millions;
          this fact impacts the maximum length of the filters and their shape.
On the other hand, the tuning of the analysis and synthesis filters  is not targeted to the minimization of some reconstruction error,
        but to specific characteristics:
        pressure preservation, smooth filtering, or
        limiting the negative components of the filters (which correspond to out-of-phase contributions).

In the conclusions of \cite{Dremin:2001kv}, they say:
\blockquote{
``[...] the wavelet applications in various fields are numerous and give nowadays very fruitful
outcome. [...] The potentialities of wavelets are still not used at their full strength.
However one should not cherish vain hopes that this machinery works automatically in all
situations by using its internal logic and does not require any intuition.
According to~\cite{meyer94}, ``no `universal algorithm' is appropriate for the extreme diversity of the situations encountered''.
Actually it needs a lot of experience in choosing the proper wavelets, in suitable formulation of the
problem under investigation, in considering most important scales and characteristics describing
the analyzed signal, in the proper choice of the algorithms (i.e., the methodology) used, in studying
the intervening singularities, in avoiding possible instabilities etc.
By this remark we would not like to prevent newcomers from entering the field but, quite to the contrary, to attract those who
are not afraid of hard but exciting research and experience.''
}

We found that the standard lifting scheme,
while it makes it very easy to tune the characteristics of the synthesis operator,
makes the tuning of the analysis operator much more challenging,
often leading to very non-smooth constructions.
Looking at Eq.~\eqref{eq:lifting-from-lazy} it is apparent that in the construction of $\mathbf A^j$ and $\mathbf Q^j$,
both lifting and dual lifting steps get mixed together.
There is no way of constructing, for example, a $\mathbf P^j$ without affecting the $\mathbf A^j$ (and vice versa).

\section{Numerical Optimization}
We instead designed a brute force approach based on numerical optimization which aims
at optimizing simultaneously both $\mathbf A^j$ and $\mathbf P^j$ operators,
retaining the idea of locality of the wavelets and the symmetries of the multiresolution.

The unknowns of the problem are all the four operators: $\mathbf{A}^j$, $\mathbf{B}^j$, $\mathbf{P}^j$ and $\mathbf{Q}^j$.
Since our main interest is on the (scaling function) operators $\mathbf A$ and $\mathbf P$,
the optimization problem has been split into two stages: \emph{stage 1} with $\mathbf{A}$, $\mathbf{P}$ as unknowns,
and \emph{stage 2} with $\mathbf{B}$, $\mathbf{Q}$ unknowns.
All the unknowns grow quadratically with the number of points in the mesh, and so their constraints,
i.e.~biorthogonality relations, Eq.~\eqref{eq:biorthogonality-relations}.
For this reason we exploit the regularity of the mesh, that reflects in symmetries in the target matrices,
to reduce the dimensionality of the problem (see Appendix~\ref{app:dof-reduction}).
The structure of the mesh also suggests some relations of neighbourhood, as we discussed in Section~\ref{sec:spherical-lifting-scheme}
with the help of Figure~\ref{fig:mesh-neighbours}.
The optimization is then performed level by level, from level 0 to level $\ell$,
leaving all previous coarser levels frozen in each subsequent step.

For the first stage, the optimization of scaling function operators proceeds level by level from $j=1$ to $j=\ell$;
at any given level $j$, the cost function is made of 4 terms:
\begin{align}
    C = \alpha_\Lambda C_\Lambda + \alpha_{p1} C_{p1} + \alpha_{p2} C_{p2} + \alpha_\text{neg}C_\text{neg},
\end{align}
which are optimized with respect to operators $\mathbf{A}^j$ and $\mathbf{P}^j$.
The first term is related to the \emph{shape} of the low-pass filtering induced by $\mathbf{A}$ and $\mathbf{P}$.
Being $\Lambda$ the desired target for the subsequent application of $\mathbf{A}$ and $\mathbf{P}$,
the associated cost term is:
\begin{align}
    C_\Lambda = \frac{1}{N_k N_m} \sum_{k,m }\left[ \sum_l {p}_{kl}^j {a}_{lm}^j - \Lambda_{km} \right]^2,
\end{align}
where $a^j_{lm} \in \mathbf{A}^j$ and $p^j_{kl} \in \mathbf{P}^j$
and $N_k$ and $N_m$ indicate the number of elements in each one of the terms in the sum.
The second term asks for \emph{pressure preservation} during decomposition,
and is:
\begin{align}
    C_{p1} = \frac{1}{N_m}\sum_m \left[ \sum_l a^j_{lm} - 1 \right]^2,
\end{align}
The third term asks for \emph{pressure preservation} across decomposition and reconstruction, also among different levels,
and is:
\begin{align}
C_{p2} = \frac{1}{N_m} \sum_{j'=1}^j \sum_m \left[ \sum_k \left( \mathbf{P}^j \cdots \mathbf{P}^{j'} \mathbf{A}^{j'} \cdots \mathbf{A}^j \right)_{km} - 1 \right]^2 .
\label{eq:press-pres-among-levels}
\end{align}
where the terms $\mathbf{P}^{j'} \mathbf{A}^{j'}$ with $j' < j$ are optimized in the previous level optimization
and are left fixed in the current one.
For example, at level 2, the term in  Eq.~\eqref{eq:press-pres-among-levels} would read
\begin{align*}
C_{p2}^{j=2} = \frac{1}{N_m} \sum_m \left[ \sum_k \left( \mathbf{P}^2 \mathbf{A}^2 \right)_{km} - 1 \right]^2 + \\
    \frac{1}{N_m} \sum_m \left[ \sum_k \left( \mathbf{P}^2 \mathbf{P}^1 \mathbf{A}^1 \mathbf{A}^2 \right)_{km} - 1 \right]^2
\end{align*}
with only $\mathbf{A}^2$ and $\mathbf{P}^2$ being optimized.
The fourth and last term asks for positive panning laws, penalizing negative coefficients,
and is a condition on the matrices alone:
\begin{align}
    C_\text{neg} &=  \frac{1}{N_l N_m} \sum_{jk} ({a}^j_{lm})^2 \theta(- {a}_{lm}) \\
    &\quad + \frac{1}{N_k N_l} \sum_{kl} ({p}^j_{kl})^2 \theta(- {p}^j_{kl}).
\end{align}
where $\theta(\cdot)$ is the Heaviside function.
Besides, there is an orthonormality constraint
\begin{align}
    \mathbf{A}^j \mathbf{P}^j = \mathbf{1} .
\end{align}

For the second stage, once $\mathbf A$ and $\mathbf P$ are set, the wavelet operators $\mathbf B$ and $\mathbf Q$
can be obtained almost algebraically, from the requirement
$\mathbf{Q}^j\mathbf{B}^j + \mathbf{P}^j \mathbf{A}^j = \mathbf{1}$, Eq.~\eqref{eq:inv2},
and the constraints given by Eqs.~\eqref{eq:bp}, \eqref{eq:aq} and \eqref{eq:qb}.
\section{Example of Optimized Filters}
In this Section we will specify some of the free parameters in the optimization of $\mathbf A$ and $\mathbf P$,
and illustrate the resulting optimized filters.

\paragraph{Define $\Lambda$} The only explicit free parameter to set is the target $\Lambda$.
Since we desire the operators to act locally, we have to limit the distance of the non-zero neighbours.
In other words, the $\Lambda$ matrix will be mostly zero valued, with the largest value on the diagonal, say $\gamma$.
In this work the $\Lambda$ is designed so that the only non zero neighbours are the $\left\{ v_1, v_2, f_1, f_2 \right\}$
defined in Figure~\ref{fig:mesh-neighbours}.
These vertices are set to have the same value, $1/2 \gamma$, so that the $\Lambda$ matrix results normalized.

\paragraph{Independent parameters and free parameters}
In the optimization of $\mathbf A$ and $\mathbf P$ all the entries of the matrices can be left completely free
or they can be bound by the symmetries of the mesh, as already mentioned
(a more detailed discussion can be found in Appendix~\ref{app:dof-reduction}).
Moreover, we can decide how many of the independent parameters remaining from the symmetries reduction are effectively let free.
By design, we would like the filter to go to zero for the points of the mesh far from the point under consideration.
One option is to constrain them to be zero.
With these choices we can considerably reduce the number of degrees of freedom of the problem.

\begin{figure}
    \centering
    \begin{minipage}[t]{0.46\textwidth}
        \includegraphics[width=\textwidth,valign=t]{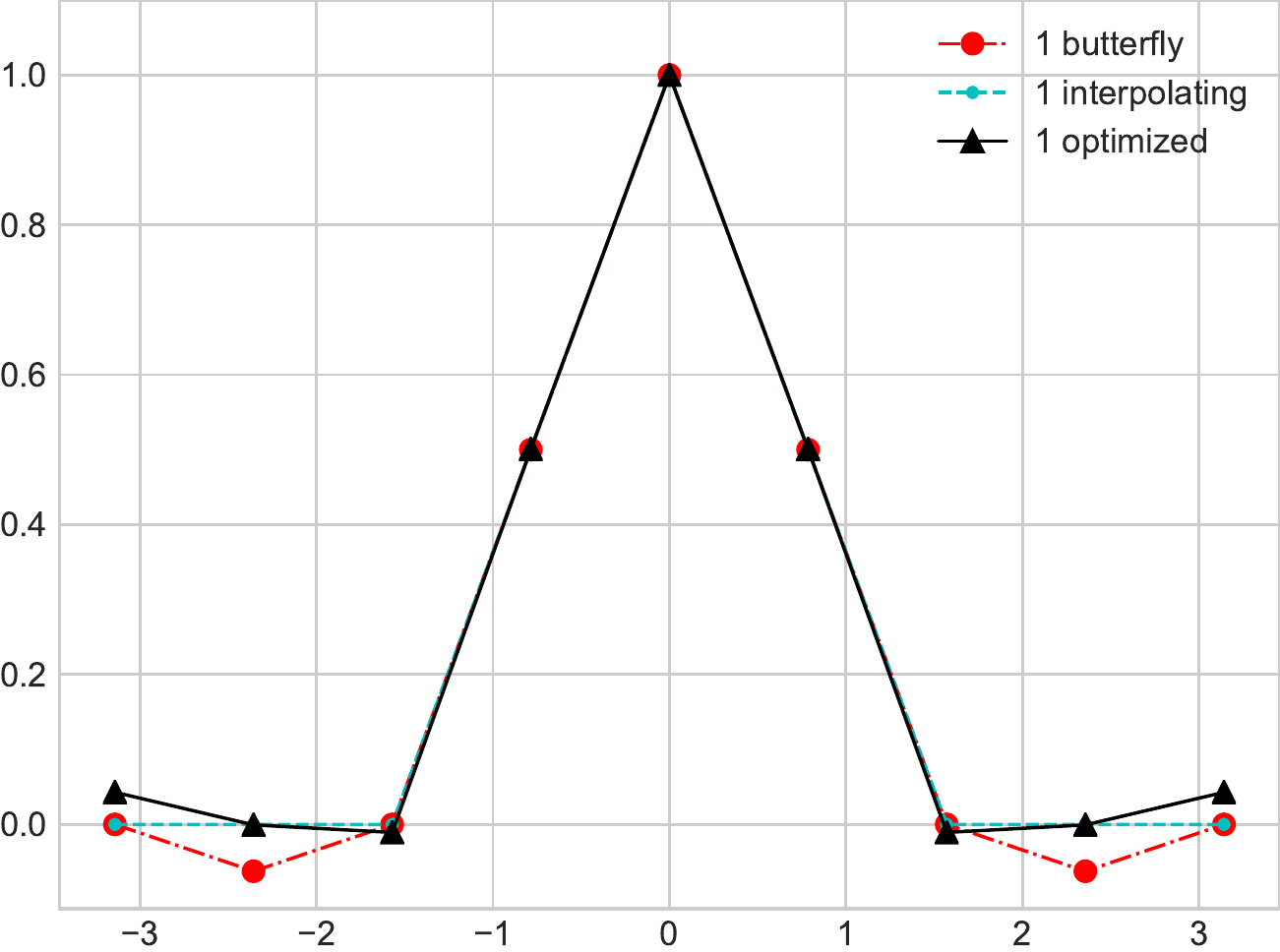}
        \caption{Horizontal representation of one filter of $\mathbf A^1$, for the butterfly, interpolating and optimized filters.}
        \label{fig:A1-opt}
    \end{minipage}
    \hfill
    \begin{minipage}[t]{0.46\textwidth}
        \includegraphics[width=\textwidth,valign=t]{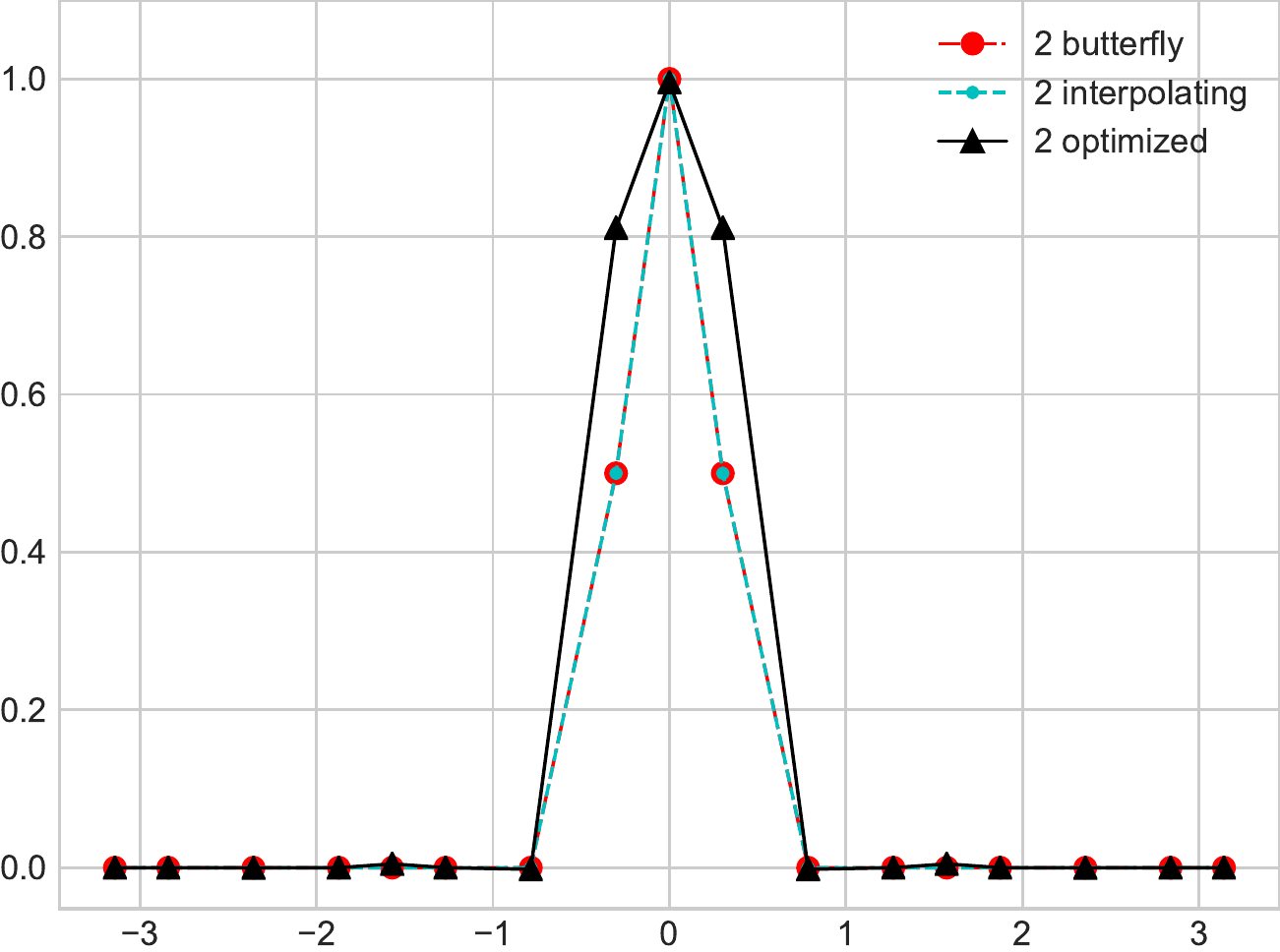}
        \caption{Horizontal representation of one filter of $\mathbf A^2$, for the butterfly, interpolating and optimized filters.}
        \label{fig:A2-opt}
    \end{minipage}

    \begin{minipage}[t]{0.7\textwidth}
        \includegraphics[width=\textwidth,valign=t]{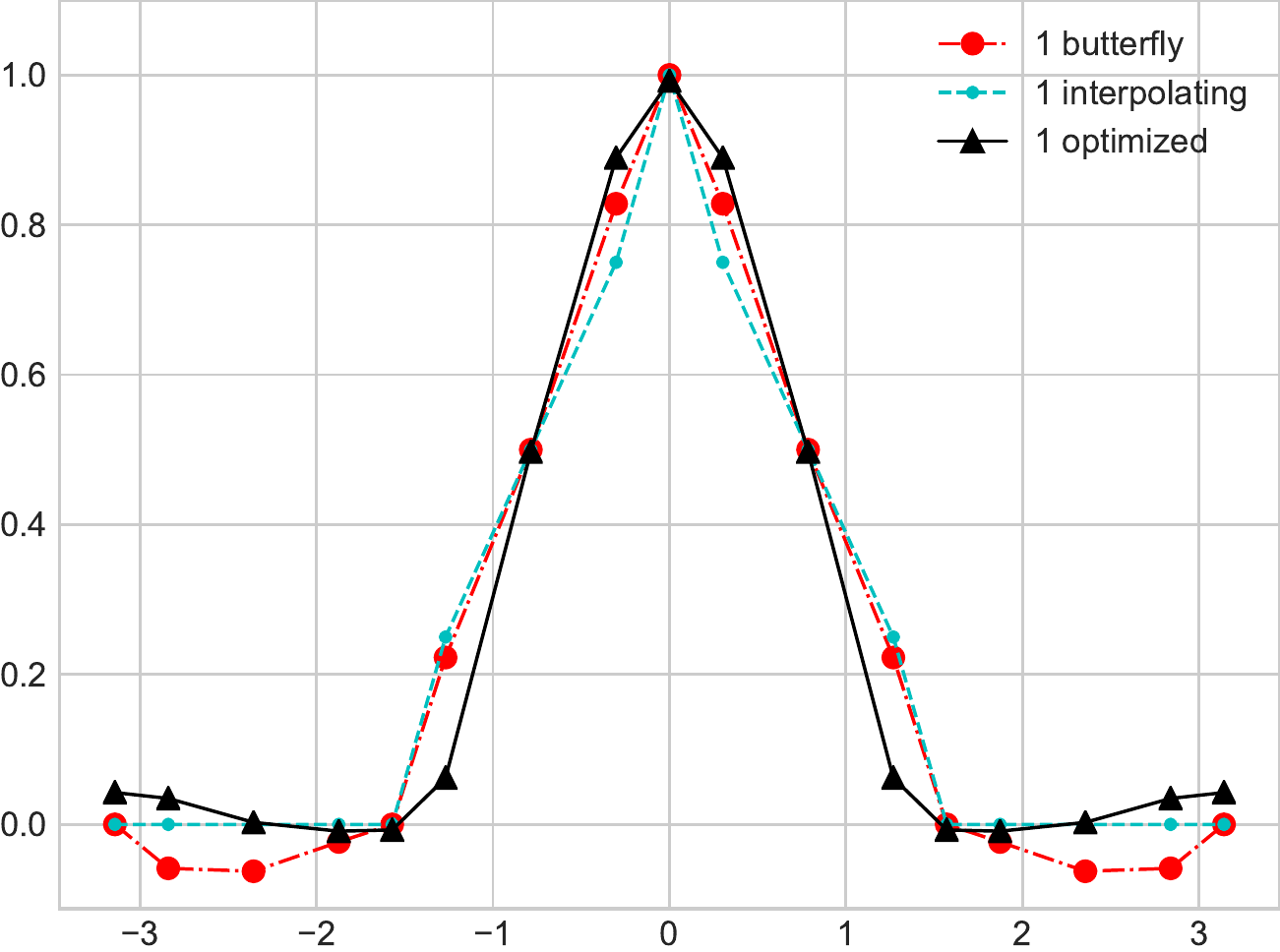}
        \caption{Horizontal representation of one filter of $\mathbf A^1 \mathbf A^2$, for the butterfly, interpolating and optimized filters.}
        \label{fig:A1-upscaled}
    \end{minipage}

\end{figure}

\paragraph{Resulting filters}
In Figures~\ref{fig:A1-opt} and \ref{fig:A2-opt}
we report an example of dual scaling filters ($\mathbf A^1$, $\mathbf A^2$) obtained with this method for the spherical mesh
described in Section~\ref{sec:meshinterpolation}.
These figures show a comparison between the interpolating and butterfly filters with the optimized one.
In these pictures the wavelets and scaling coming from the lifting scheme (interpolating and butterfly) are swapped for their duals
(the reason will be explained in Section~\ref{sec:interpolating-swf-gains}).
The optimization is robust, meaning that we get the same filters by starting from very different initial conditions.
A more interesting result is the Figure~\ref{fig:A1-upscaled} which shows the combined action of cascading
two filters together $\mathbf A^1$ $\mathbf A^2$, that is the operation of downsampling performed during the SWF encoding,
which is a representation of the dual scaling function.
The resulting filters are still in number of 6, that is the number of filters at level 0, but they have a length of 66 taps.

\begin{figure}
    \centering
    \begin{minipage}[t]{0.46\textwidth}
        \includegraphics[width=\textwidth,valign=t]{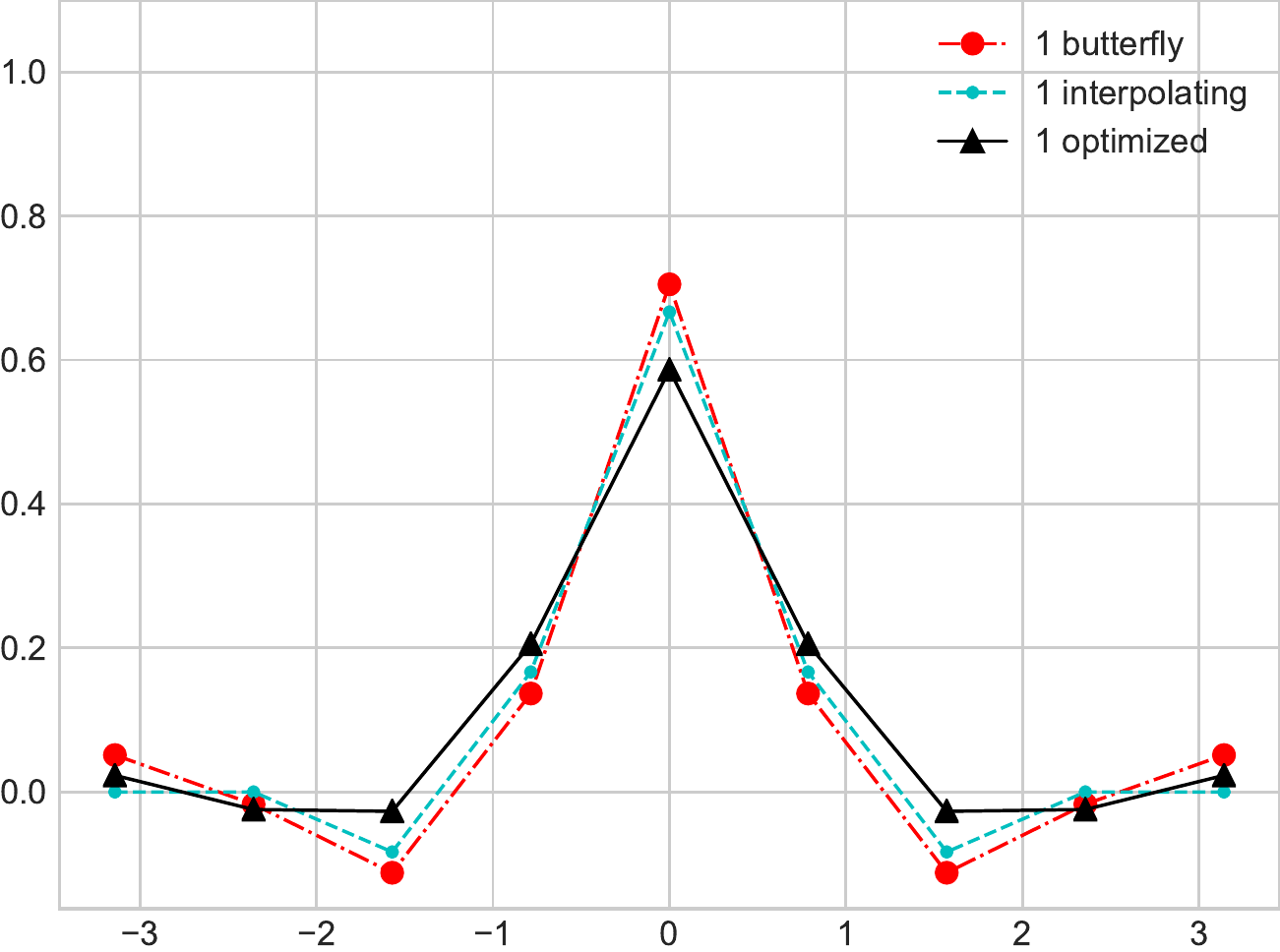}
        \caption{Horizontal representation of one filter of $\mathbf P^1$, for the butterfly, interpolating and optimized filters.}
        \label{fig:P1-opt}
    \end{minipage}
    \hfill
    \begin{minipage}[t]{0.46\textwidth}
        \includegraphics[width=\textwidth,valign=t]{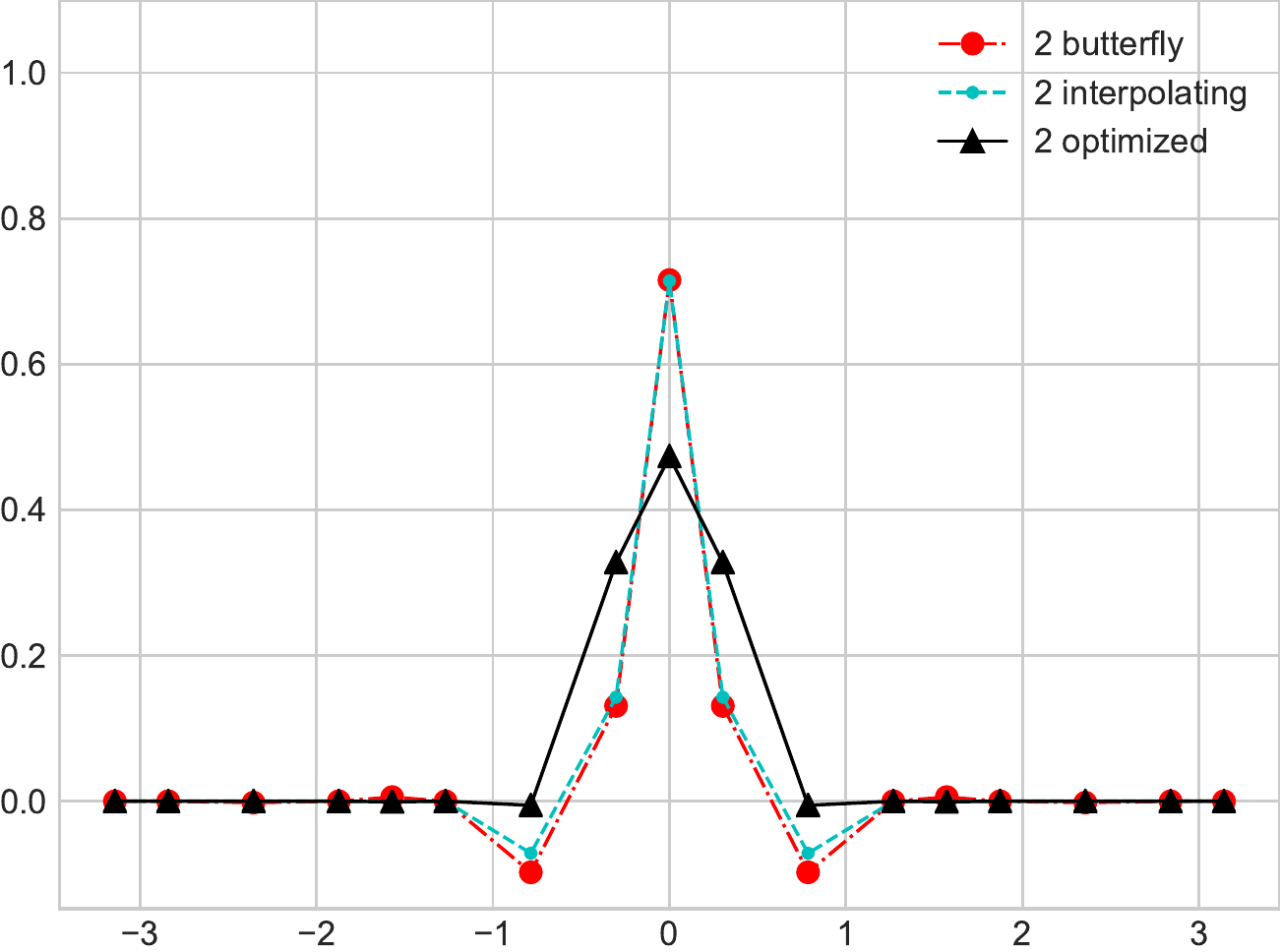}
        \caption{Horizontal representation of one filter of $\mathbf P^2$, for the butterfly, interpolating and optimized filters.}
        \label{fig:P2-opt}
    \end{minipage}

    \begin{minipage}[t]{0.7\textwidth}
        \includegraphics[width=\textwidth,valign=t]{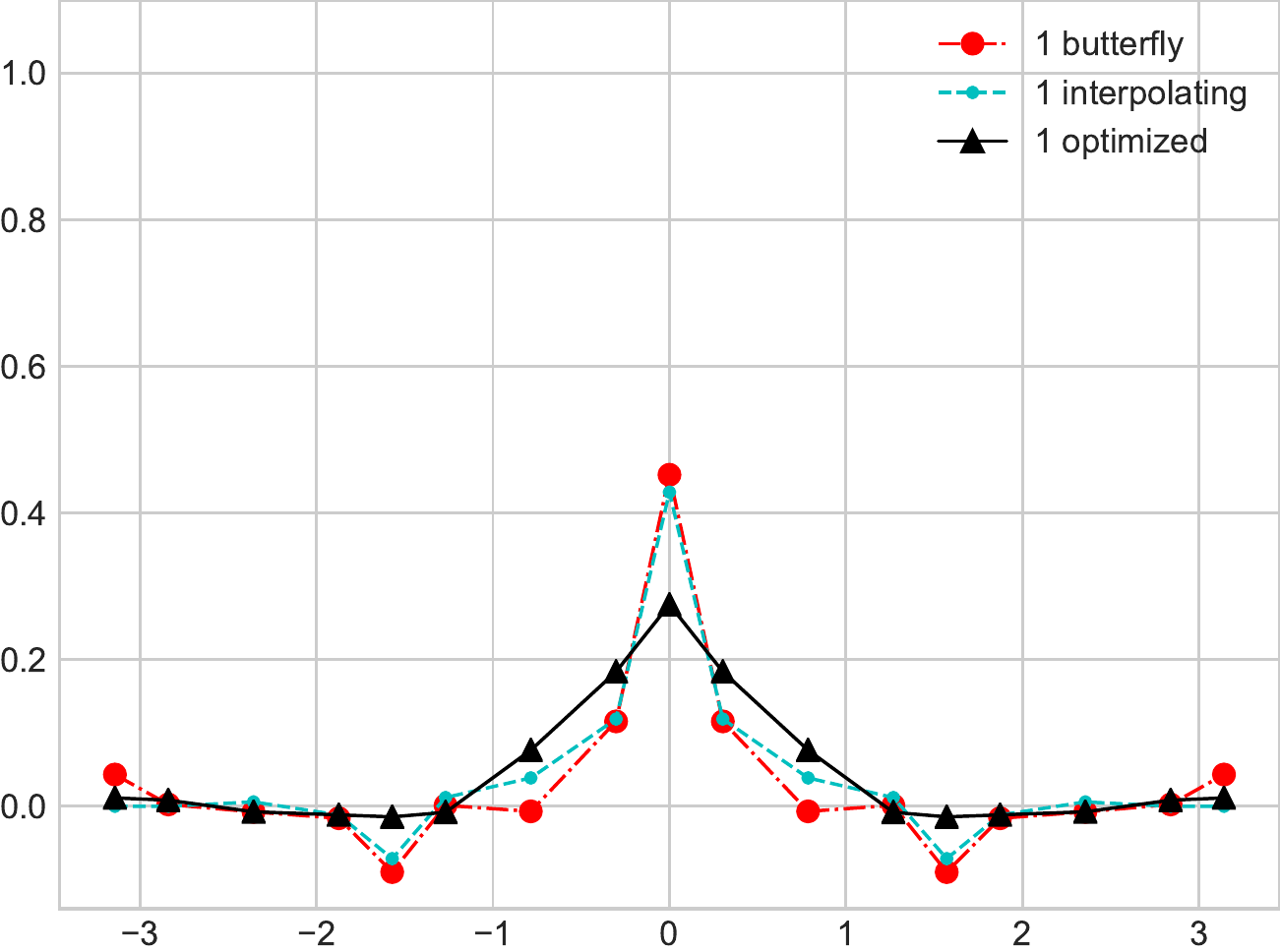}
        \caption{Horizontal representation of one filter of $\mathbf P^1 \mathbf P^2$, for the butterfly, interpolating and optimized filters.}
        \label{fig:P1-upscaled}
    \end{minipage}

\end{figure}

In Figures~\ref{fig:P1-opt}, \ref{fig:P2-opt} and \ref{fig:P1-upscaled}
we report the filters generated by the $\mathbf P$ matrices.
Similar considerations to the $\mathbf A$ filters apply.

In general, the filters generated via the optimization tend to have less negative values and have higher values in the
vicinity of the main peak, we can say that are slightly rounder, but not wider.

\begin{figure}
    \centering
    \begin{minipage}[t]{0.46\textwidth}
        \includegraphics[width=\textwidth,valign=t]{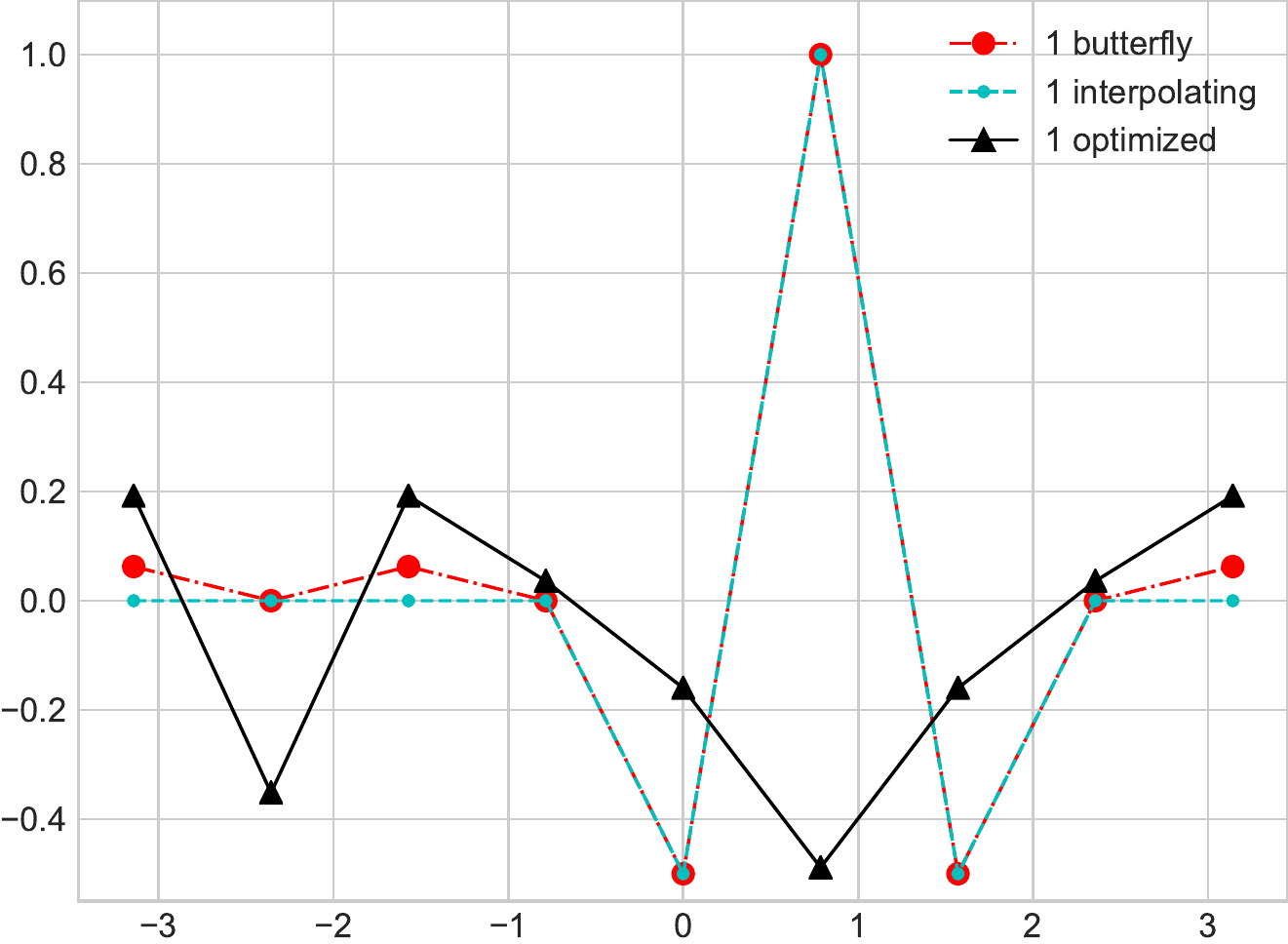}
        \caption{Horizontal representation of one filter of $\mathbf Q^1$, for the butterfly, interpolating and optimized filters.}
        \label{fig:Q1-opt}
    \end{minipage}
    \hfill
    \begin{minipage}[t]{0.46\textwidth}
        \includegraphics[width=\textwidth,valign=t]{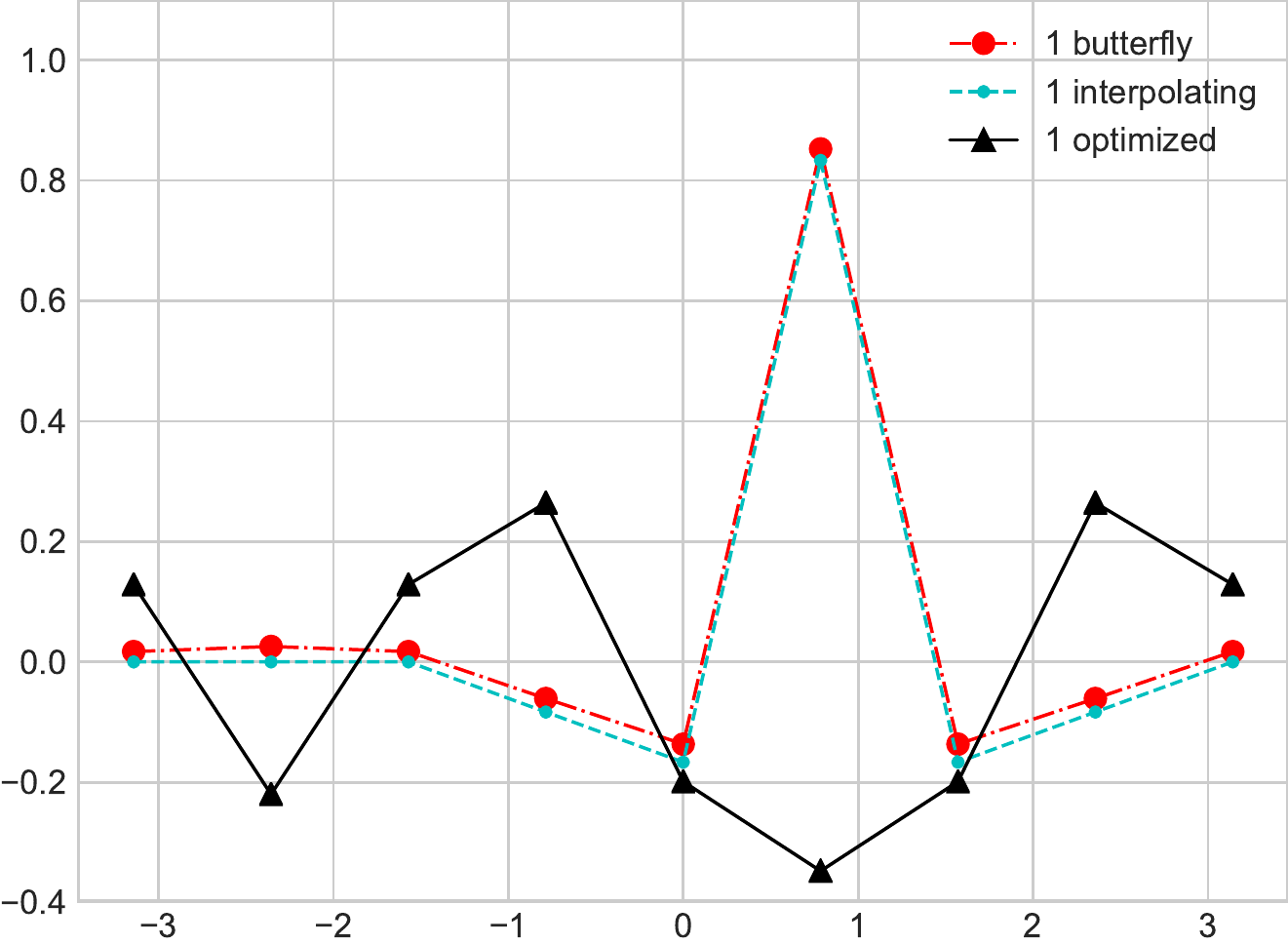}
        \caption{Horizontal representation of one filter of $\mathbf B^1$, for the butterfly, interpolating and optimized filters.}
        \label{fig:B1-opt}
    \end{minipage}

\end{figure}

In Figures~\ref{fig:Q1-opt} and \ref{fig:B1-opt} we report the optimized wavelet filters $\mathbf B$ and $\mathbf Q$,
against the swapped interpolating and butterfly ones.
It is possible to notice that while the $\mathbf A$ and $\mathbf P$ filters are very similar among the
different families, while the optimized $\mathbf B$ and $\mathbf Q$ filters are very different from the ones
coming from the lifting scheme.
Nevertheless, they all satisfy the equations \eqref{eq:biorthogonality-relations} and \eqref{eq:inv2}.
It is apparent that the imposed constraints are not sufficient to determine uniquely the $\mathbf B$ and $\mathbf Q$.
We did not investigate further on their additional properties, since they reach their purpose, and these filters are not central in our construction.
The specific detail of the filters $\mathbf B$ and $\mathbf Q$  affect the efficiency of the
storage representation of the wavelet coefficients, but do not alter the essential
properties of the wavelet framework, which are determined uniquely by the filters $\mathbf A$ and $\mathbf P$.

In Chapter~\ref{ch:comparison-swf-internals} we will analyze in more detail the effects of the differences between the filters
on the SWF audio chain.

\section{Summary}
In this Chapter we have described how to generate numerical wavelet families in the Second Generation framework leveraging the Lifting Scheme.
The definition of the lifting scheme is completely general and can be applied to any multiresolution mesh.
We detailed the generic construction of the interpolating wavelet and then particularized it for a spherical mesh.
Lastly, we report our method for wavelet filters optimization and we illustrate some of its outcomes.

    \part{Evaluation} \label{part:evaluation}

\chapter{Implementation and Evaluation of Specific Spherical Wavelet Formats} \label{ch:comparison-swf-internals}

In this Chapter we will describe three specific SWF implementations: VBAP-SWF based on VBAP interpolation,
SINT-SFW based on the interpolating wavelet, and OPT-SWF is built with the help of a numerical optimized wavelet.
We then proceed to compare their properties in terms of pressure, energy, velocity and intensity preservation.

\section{Specific SWF Implementations} \label{sec:swf-implementation}

\begin{figure}[t]
    \centering
    \includegraphics[width=\columnwidth]{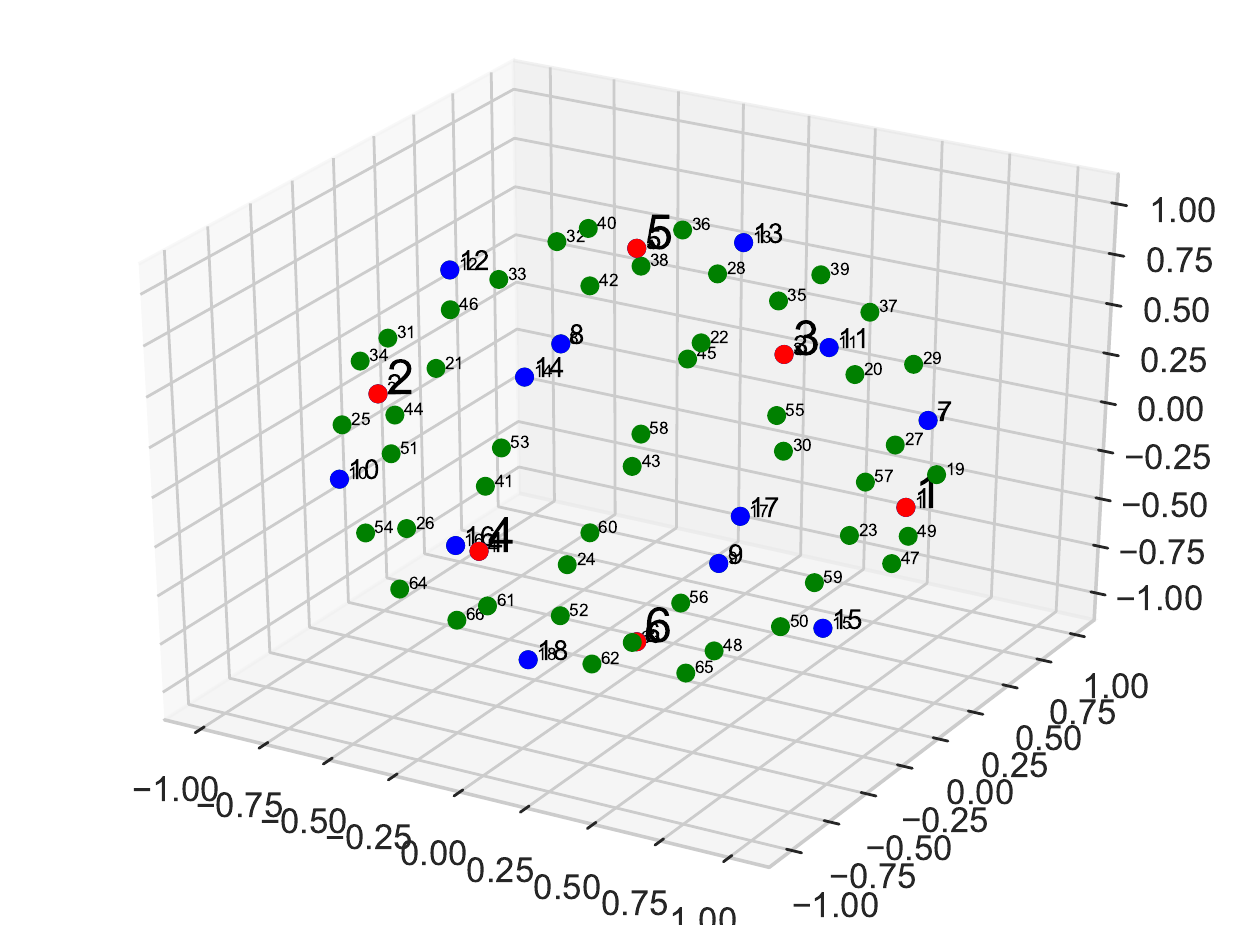}
    \caption{Mesh points at different levels: red level, red+blue level 1, red+blue+green level 2.}
    \label{fig:mesh}
\end{figure}

In Section~\ref{sec:swf} we defined SWF as an audio encoding with three characteristics:
a subdivision mesh defined over the sphere, a set of filters defining a wavelet space, a truncation level.
In this Section we build three different implementations of SFW.
All the three implementations have the first and third point in common: the subdivision mesh starts
from an octahedron and gets refined with the Loop subdivision method (see Figure~\ref{fig:mesh}).
The number of channels is 6 for {level 0} and 18 for {level 1}.
The difference between the three SWF implementations is the set of filters defining the wavelet space.

To compare the three SWF implementations,
we take the finest level to be 2 (66 points),
and we encode a point source rotating over the horizontal plane to a 66 points mesh
using VBAP to interpolate from the continuous space to the discrete mesh, $f\left( \theta, \phi \right) \rightarrow f_{66}$.
The 66 channels are later downsampled to 18 or 6 using the decomposition matrices of each format.
At level zero the 6 channels correspond to the six components of the coarser dual scaling function $\mathbf{c}^0$,
and at level 1 the 18 channels are the 18 components of the dual scaling function at level 1, $\mathbf{c}^1$.
As discussed in Section~\ref{sec:swf},
the signals in SWF can be represented in two alternative equivalent ways: 
either as the scaling function coefficients at the coarser level plus a set of successively finer wavelet details,
$ \{ \mathbf c^0, \mathbf{d}^0, \dots, \mathbf{d}^{\ell-1} \}$,
or as the scaling functions at the truncation level $\ell$: $\{ \mathbf c^\ell \}$.
In the first representation we need the matrices $\mathbf{A}^j$ and $\mathbf{Q}^j$, 
and in this Section $\ell$ will be $\ell=2$.
In the second representation only the downsampling $\mathbf{A}^j$ and upsampling $\mathbf{P}^j$ filters are strictly needed.
By looking at $\mathbf{c}^0$ and $\mathbf{c}^1$ we investigate the properties of $\mathbf A^1$ and $\mathbf A^2$,
if we limit ourselves to the downscaling operation.
If we reconstruct $\mathbf{c}^1$ with the help of $\mathbf{d}^0$,
$\mathbf{c}^1 = \mathbf P^1 \mathbf{c}^0 + \mathbf Q^1 \mathbf{d}^0$ (Eq.~\eqref{eq:iwt}),
we can inspect separately the effect of $\mathbf P^1$ and especially $\mathbf Q^1$.
Upsampling to $\tilde{\mathbf{c}}^0$ from $\mathbf{c}^0$, gives us the possibility to analyze the action of $\mathbf P^1$ alone.
In Section~\ref{sec:multiresolution} we presented $\mathbf A$  and $\mathbf P$ as the dual scaling filters and scaling filters respectively,
and $\mathbf{c}$ as the coarse coefficients resulting from the wavelet transform.
Now, we can interpret $\mathbf{c}$, $\mathbf A$  and $\mathbf P$ in a different way.
If we place a set of virtual speakers on the points defining the mesh,
then the signals carried by $\mathbf{c}$ would be the feeds for the speakers.
If the encoded signal is a point source in a certain position, then the $\mathbf{c}$ represent the gains
for each virtual speaker needed to represent a point source in that position.
For this reason, and in this specific case, we could call the $\mathbf{c}$ the `panning functions' of the SWF.
With this interpretation, the matrices $\mathbf A$  and $\mathbf P$ connect layouts with different numbers of virtual speakers;
in spatial audio this kind of matrices are called downmixing and upmixing matrices, respectively.

\afterpage{%
\clearpage%
\begin{landscape}%
    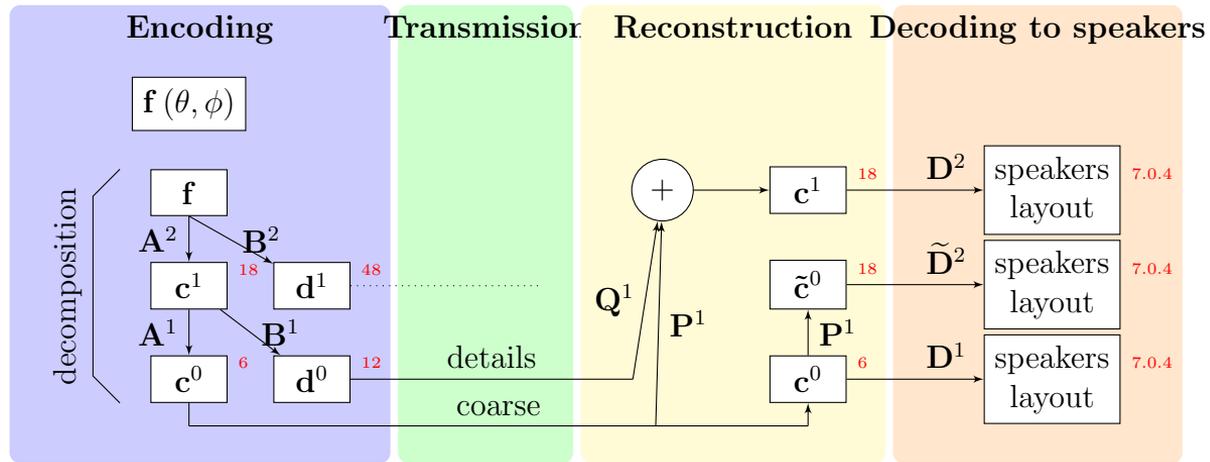
\begin{figure*}
    \centering
    \begin{tikzpicture}[>=latex']
    \tikzset{block/.style= {draw, rectangle, align=center,minimum width=1cm,minimum height=0.6cm},
    }
        \node [draw=none,fill=none] (start) {};
        \node (rect-encoding) [fill=blue!20, shape=rectangle, rounded corners, minimum width=5cm, minimum height=6cm,
                font=\sffamily, right=0. of start, anchor=north]{};
        \node[anchor=north] at (rect-encoding.north) {\textbf{Encoding}};
        \node (rect-transmission) [fill=green!20, shape=rectangle, rounded corners, minimum width=2.3cm, minimum height=6cm,
                font=\sffamily, right=3.75cm of start, anchor=north]{};
        \node[anchor=north] at (rect-transmission.north) {\textbf{Transmission}};
        \node (rect-reco) [fill=yellow!20, shape=rectangle, rounded corners, minimum width=4cm, minimum height=6cm,
                font=\sffamily, right=7cm of start, anchor=north]{};
        \node[anchor=north] at (rect-reco.north) {\textbf{Reconstruction}};
        \node (rect-deco) [fill=orange!20, shape=rectangle, rounded corners, minimum width=3.8cm, minimum height=6cm,
                font=\sffamily, right=11cm of start, anchor=north]{};
        \node[anchor=north] at (rect-deco.north) {\textbf{Decoding to speakers}};

        \node [block,fill=white, below=0.8cm of start] (function) {$ \mathbf{f} \left( \theta, \phi \right) $};
        \node [block,fill=white, below=0.5cm of function] (fn) {$ \mathbf{f} $};

        \node [coordinate, below=0 of fn](newbegin){};
        \node [block,fill=white, below=0.6cm of newbegin] (cn1) {$ \mathbf{c}^1 $};
        \node [block,fill=white, below=0.6cm of newbegin, right=0.6cm of cn1] (dn1) {$ \mathbf{d}^1 $};
        \node [block,fill=white, below=0.6cm of cn1] (cn0) {$ \mathbf{c}^0 $};
        \node [block,fill=white, below=0.6cm of cn1, right=0.6cm of cn0] (dn0) {$ \mathbf{d}^0 $};

        \node [draw=none,fill=none, above right=0cm of cn0.east, text=red] (c6){\tiny6};
        \node [draw=none,fill=none, above right=0cm of cn1.east, text=red] (c18){\tiny18};
        \node [draw=none,fill=none, above right=0cm of dn0.east, text=red] (d12){\tiny12};
        \node [draw=none,fill=none, above right=0cm of dn1.east, text=red] (d48){\tiny48};

        \node [coordinate, left=0.4cm of fn.north west](lineorigin){};
        \node [coordinate, below left=0.5cm of lineorigin](lineleft){};
        \node [coordinate, left=0.4cm of cn0.south west](linebott){};
        \node [coordinate, above left=0.5cm of linebott](lineleftbott){};
        \path[draw] (lineorigin) -- (lineleft) -- node[midway, rotate=90, above]{decomposition}
                    (lineleftbott) -- (linebott);

        \path[->]
            (newbegin) edge node [midway, left]{$\mathbf{A}^2$} (cn1)
            (newbegin) edge node [midway, right]{$\mathbf{B}^2$} (dn1)
            (cn1) edge node [midway, left]{$\mathbf{A}^1$} (cn0)
            (cn1) edge node [midway, right]{$\mathbf{B}^1$} (dn0);

        \node[coordinate, below=0.3 of cn0](belowcn1){};

        \path[draw]
            (cn0) -- (belowcn1);

        \node[block,fill=white, right=5.5cm of dn0](reco1){$\mathbf{c}^0$};
        \node[coordinate, below=0.3 of reco1](belowreco1){};
        \node[block,fill=white, above=0.6cm of reco1](upscaled){$\mathbf{\tilde{c}}^0$};
        \node[block,fill=white, above=0.6cm of upscaled](reco2){$\mathbf{c}^1$};

        \node [draw=none,fill=none, above right=0cm of reco1.east, text=red] (c6reco){\tiny6};
        \node [draw=none,fill=none, above right=0cm of upscaled.east, text=red] (c6upreco){\tiny18};
        \node [draw=none,fill=none, above right=0cm of reco2.east, text=red] (c18reco){\tiny18};

        \node[coordinate, left=2cm of belowreco1](toreco2){};
        \node[coordinate, left=2.3cm of reco1.center](dtoreco2){};
        \node[coordinate, left=2cm of reco2.center](toreco2left){};
        \node[coordinate, left=2.3cm of reco2.center](dtoreco2left){};

        \path[draw]
            (belowcn1) -- node [midway, above]{coarse} (belowreco1);
        \path[->]
            (belowreco1) edge (reco1)
            (reco1) edge node [midway, right]{$\mathbf{P}^1$} (upscaled)
            (toreco2left) edge (reco2);

        \path[draw]
            (dn0) -- node [midway, above]{details} (dtoreco2);

        \node[block,fill=white, right=1.8cm of reco1]    (spk){speakers\\layout};
        \node[block,fill=white, right=1.8cm of upscaled] (spkup){speakers\\layout};
        \node[block,fill=white, right=1.8cm of reco2]    (spk2){speakers\\layout};
        \node [draw=none,fill=none, above right=0cm of spk.east, text=red] (spklabel){\tiny7.0.4};
        \node [draw=none,fill=none, above right=0cm of spkup.east, text=red] (spklabel){\tiny7.0.4};
        \node [draw=none,fill=none, above right=0cm of spk2.east, text=red] (spklabel){\tiny7.0.4};

        \path[->] (reco1) edge node    [above right]{$\mathbf{D}^1$} (spk);
        \path[->] (upscaled) edge node [above right]{$\widetilde{\mathbf{D}}^2$} (spkup);
        \path[->] (reco2) edge node    [above right]{$\mathbf{D}^2$} (spk2);

        \node [circle, left=1.5cm of reco2.center, fill=white, draw, minimum height=0.4cm] (circle-plus){$+$};
        \path[->] (toreco2) edge node [midway,right]{$\mathbf{P}^1$}(circle-plus);
        \path[->] (dtoreco2) edge node [midway,left]{$\mathbf{Q}^1$}(circle-plus);

        \node[coordinate, right=2.5cm of dn1](dn1-line){};
        \path[draw,dotted] (dn1) -- (dn1-line);

    \end{tikzpicture}
    \caption{Example of a possible SWF workflow, corresponding to the particular implementation described in the text.
    The number of channels is noted in red.}
    \label{scheme:encode-decode}
    \end{figure*}

    \end{landscape}
    \clearpage%
}

The objective of this Section is to see how different types of filters behave, in terms of pressure preservation
and encoding gains for a panning around the horizontal axis.
These quantities would be the ones reconstructed by a set of speakers (6 or 18)
placed exactly on the location of the mesh vertices.
Referring to Figure~\ref{scheme:encode-decode}, we will compare the different flavours of SWF after the reconstruction
and before the decoding to speakers.

In the following we will look at the coefficients of the encoding, or the gains of the virtual speakers,
$\mathbf{c}^0$, $\mathbf{c}^1$ and $\tilde{\mathbf{c}}^0$
for different versions of SWF:
in Section~\ref{sec:vbap-swf-gains} the VBAP-SWF, where the interpolation between meshes is the trilinear interpolation used in VBAP;
in Section~\ref{sec:interpolating-swf-gains} the SINT-SWF, where the interpolation between meshes is given by the swapped-interpolating lifting wavelet,
which is a modification of the interpolating wavelet family illustrated in Section~\ref{sec:spherical-lifting-scheme};
and finally in Section~\ref{sec:opt-swf-gains} the OPT-SWF, which is the result of the optimization procedure described in Chapter~\ref{ch:our-wavelets}.
All the gains plots are reported in linear scale to make visible the eventual negative gains, even if the logarithmic scale is probably more common.
\subsection{VBAP-SWF} \label{sec:vbap-swf-gains}
In Section~\ref{sec:swf} we defined the matrices $\mathbf A_{\text{VBAP}}$ and $\mathbf P_{\text{VBAP}}$ for the VBAP-SWF.
These matrices are the only ones we can design with the procedure described there.
To get the full set
$\left\{  \mathbf A_{\text{VBAP}}, \mathbf B_{\text{VBAP}}, \mathbf P_{\text{VBAP}}, \mathbf Q_{\text{VBAP}}, \right\} $,
we should apply the othonormality relations \eqref{eq:biorthogonality-relations} and \eqref{eq:inv2}.
For several reasons, the filters $\mathbf B$ and $\mathbf Q$ typically
are not obtained directly from Eqs.~\eqref{eq:biorthogonality-relations} and \eqref{eq:inv2},
but with procedures (e.g.\ Lifting Scheme) that guarantee to satisfy those equations.
Since the VBAP-SWF is not obtained via the Lifting Scheme, but is somewhat built procedurally, we skip the construction
of $\mathbf B_{\text{VBAP}}$ and $\mathbf Q_{\text{VBAP}}$
and we analyze only $\mathbf{c}^0_{\text{VBAP}}$, $\mathbf{c}^1_{\text{VBAP}}$ and $\tilde{\mathbf{c}}^0_{\text{VBAP}}$
coming from $\mathbf A_{\text{VBAP}}$ and $\mathbf P_{\text{VBAP}}$.

\begin{figure}
    \vspace*{-1cm}
    \begin{minipage}[t]{\textwidth}
        \centering
        \includegraphics[width=0.6\textwidth]{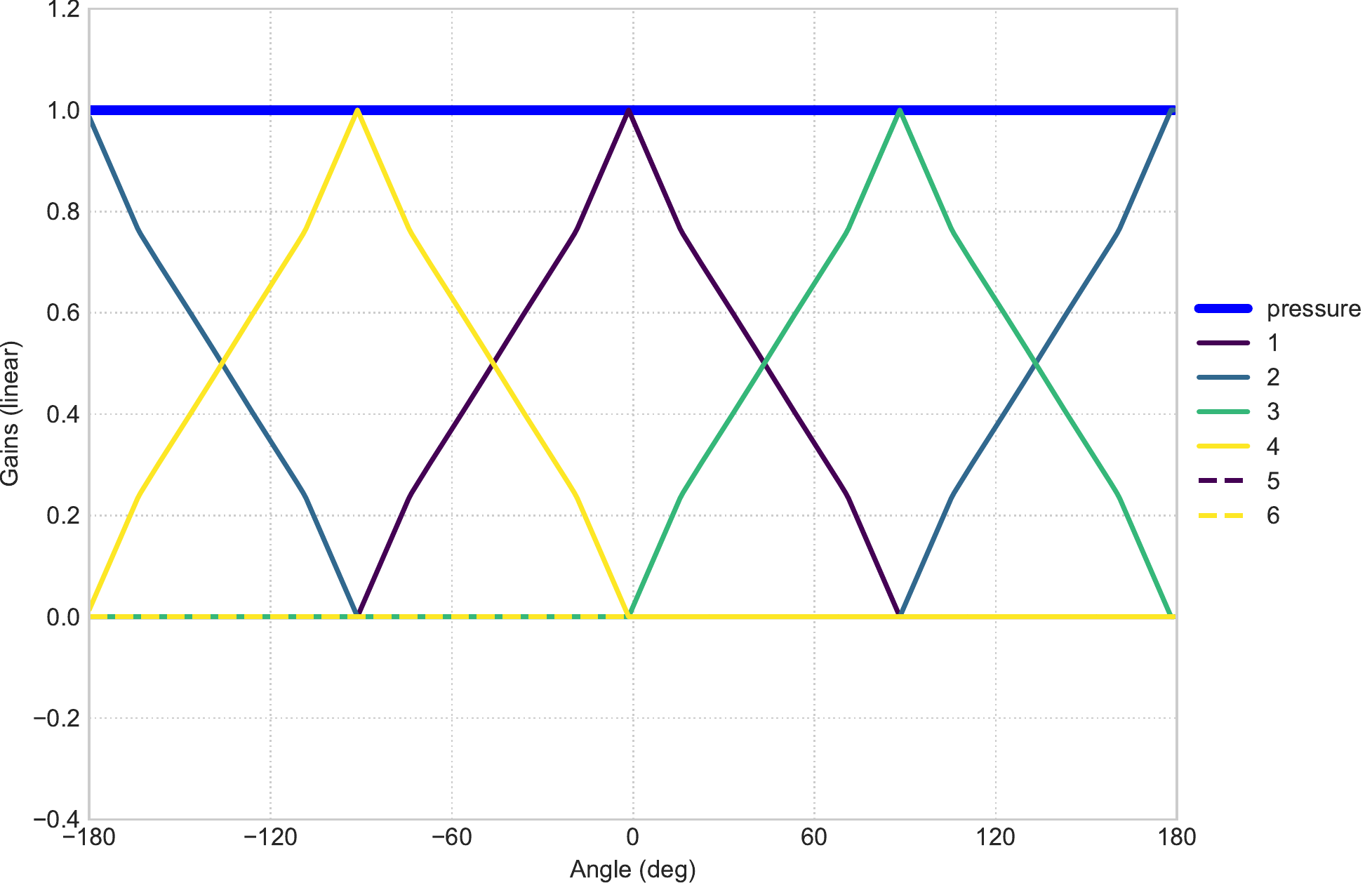}
        \caption{Scaling coefficients $\mathbf c^0_{6}$ for an horizontal panning of VBAP-SWF at level 0.
                These can be interpreted as the gains of 6 virtual speakers located on the mesh.
                ($f_{66} \rightarrow  \mathbf c^0_{6}$).}
        \label{fig:66-6_vbap-swf}
    \end{minipage}
    \vspace*{0.5cm}

    \begin{minipage}[t]{\textwidth}
        \centering
        \includegraphics[width=0.6\textwidth]{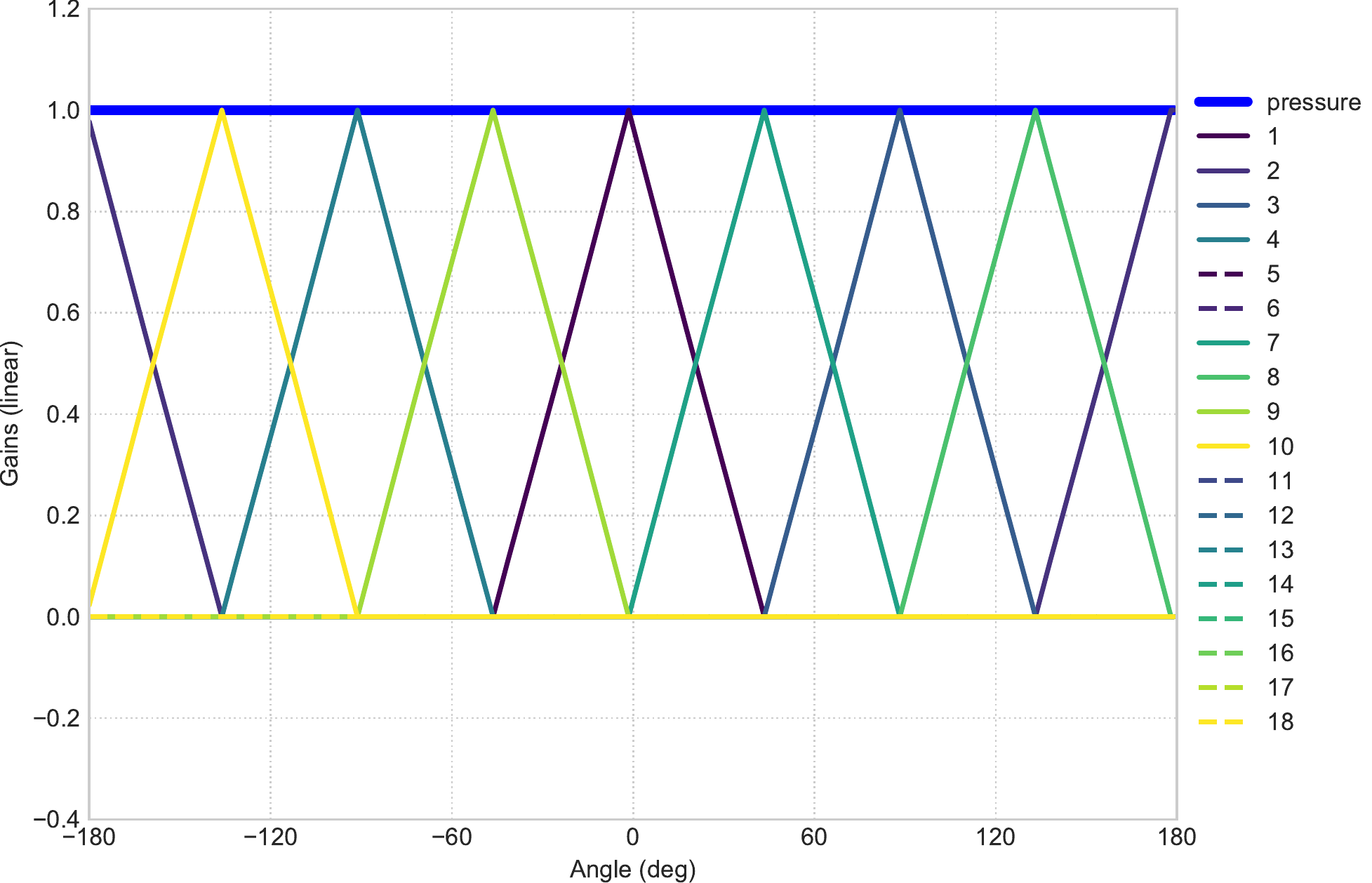}
        \caption{Scaling coefficients $\mathbf c^1_{18}$ for an horizontal panning of VBAP-SWF at level 1.
                These can be interpreted as the gains of 18 virtual speakers located on the mesh.
                ($f_{66} \rightarrow \mathbf c^1_{18}$).}
        \label{fig:66-18_vbap-swf}
    \end{minipage}
    \vspace*{0.5cm}

    \begin{minipage}[t]{\textwidth}
        \centering
        \includegraphics[width=0.6\textwidth]{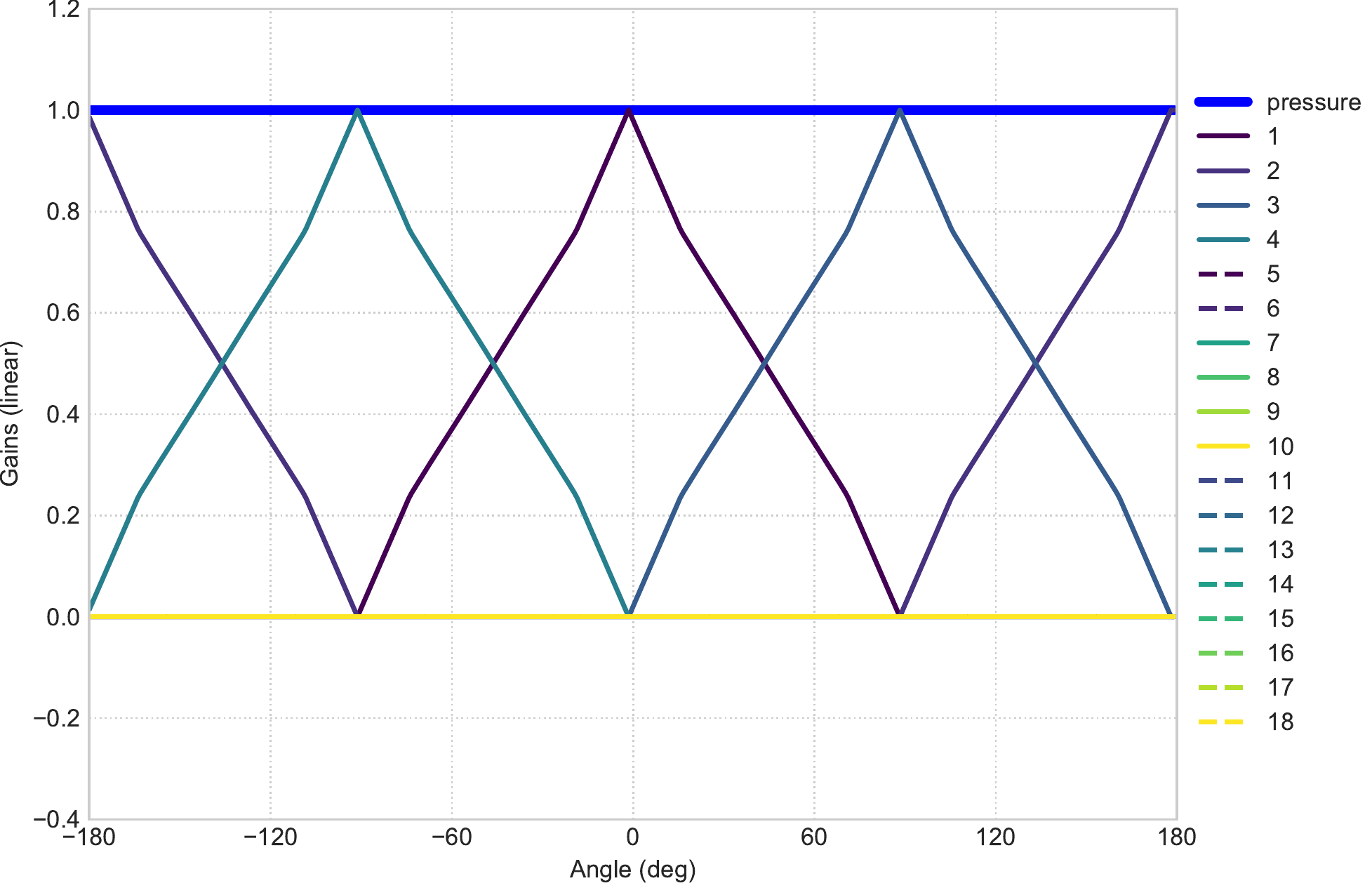}
        \caption{Upsampled scaling coefficients $ \mathbf c^0_{6}$ for an horizontal panning of VBAP-SWF at level $\tilde{0}$.
                These can be interpreted as the gains of 18 virtual speakers located on the mesh.
                ($f_{66} \rightarrow \mathbf c^0_{6} \rightarrow \tilde{\mathbf c}^0_{18}$).}
        \label{fig:66-6-18_vbap-swf}
    \end{minipage}
\end{figure}

The gains produced by the format VBAP-SWF defined in~\ref{sec:swf} for an horizontal panning at level 0 and 1 are reported in
Figures~\ref{fig:66-6_vbap-swf}, \ref{fig:66-18_vbap-swf}, \ref{fig:66-6-18_vbap-swf}.
In the plots representing the gains, we represent with solid lines the (virtual) speakers located in the horizontal plane
(zero elevation) and with dashed lines the (virtual) speakers with non-zero elevation.

In particular, Figure~\ref{fig:66-18_vbap-swf} displays the effect of $\mathbf A^2_{\text{VBAP}}$ alone, downsampling from 66 to 18 points.
Schematically: $f\left( \theta, \phi \right) \rightarrow f_{66} \rightarrow c^1_{18}$.
It is possible to notice that the downsampling is in fact a linear interpolation.

Figure~\ref{fig:66-6_vbap-swf} shows the joint action of $\mathbf A^1_{\text{VBAP}} \mathbf A^2_{\text{VBAP}}$ downsampling the panning from 66 to 6 points.
Schematically: $f\left( \theta, \phi \right) \rightarrow f_{66} \rightarrow c^0_{6}$.
Pressure is preserved at each application of $\mathbf A_{\text{VBAP}}$.

Finally, in Figure~\ref{fig:66-6-18_vbap-swf} we present the effect of upsampling the $\mathbf c^0$ (6 points) to $\tilde{\mathbf c}^0$ (18 points),
which is the result of applying $\mathbf P^1_{\text{VBAP}} \mathbf A^1_{\text{VBAP}} \mathbf A^2_{\text{VBAP}}$ to the horizontal rotating delta,
which is our signal.
Schematically: $f\left( \theta, \phi \right) \rightarrow f_{66} \rightarrow c^0_{6} \rightarrow \tilde{c}^0_{18}$.
It is possible to note how the pressure is preserved also after this stage.
The interesting fact to notice is that only the four horizontal points (or virtual speakers) at level 0 are activated, even if we effectively upsampled to level 1.
This is due to the fact that the $\mathbf P_{\text{VBAP}}$ matrices are trivial (an identity for the points at the lower resolution and zero elsewhere).

In this particular version of SWF all the gains are strictly positive.

\subsection{SINT-SWF} \label{sec:interpolating-swf-gains}
We mentioned in~\ref{sec:swf} that the
VBAP-inspired wavelets are close in spirit to the interpolating wavelets,
with the difference of having a different set of neighbours and with the dual and direct spaces swapped.
Here we show the result of swapping the wavelets and the scaling with their duals in the case of the
lifted spherical interpolating wavelets presented in Section~\ref{sec:spherical-lifting-scheme}.
In the literature several types of lifted wavelets families are available, other than the interpolating one,
but in general they don't preserve pressure even when swapping direct and dual spaces.
In this dissertation we chose the swapped interpolating wavelets for the sake of simplicity and for the analogy with the VBAP case.

\begin{figure}
    \centering
    \vspace*{-1cm}
    \begin{minipage}[t]{\textwidth}
        \centering
        \includegraphics[width=0.6\textwidth]{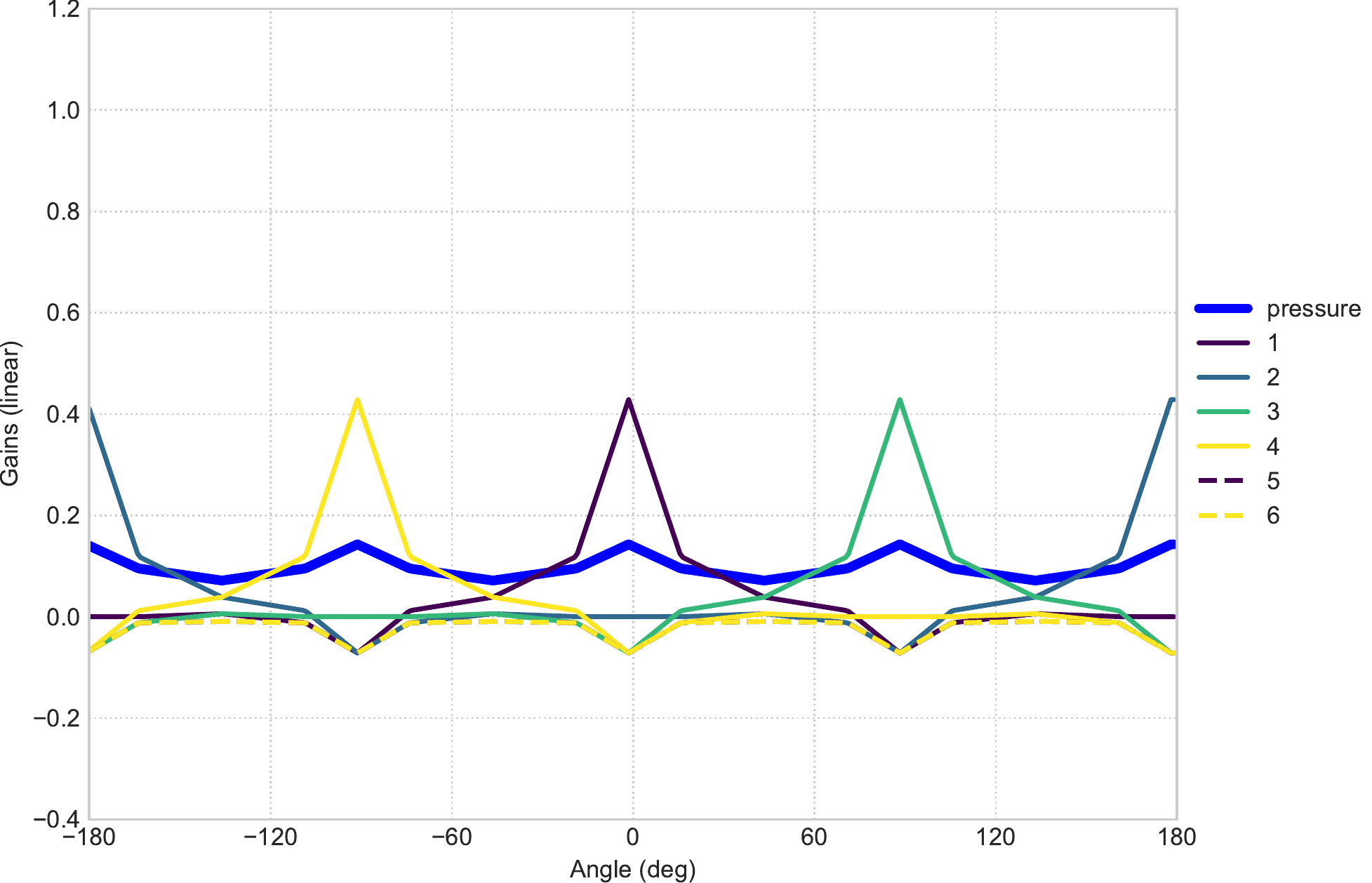}
        \caption{Scaling coefficients $\mathbf c^0_{6}$ for an horizontal panning of interpolating SWF at level 0.
                These can be interpreted as the gains of 6 virtual speakers located on the mesh.
                ($f_{66} \rightarrow  \mathbf c^0_{6}$).}
        \label{fig:66-6_lifting-swf}
    \end{minipage}
    \vspace*{0.5cm}

    \begin{minipage}[t]{\textwidth}
        \centering
        \includegraphics[width=0.6\textwidth]{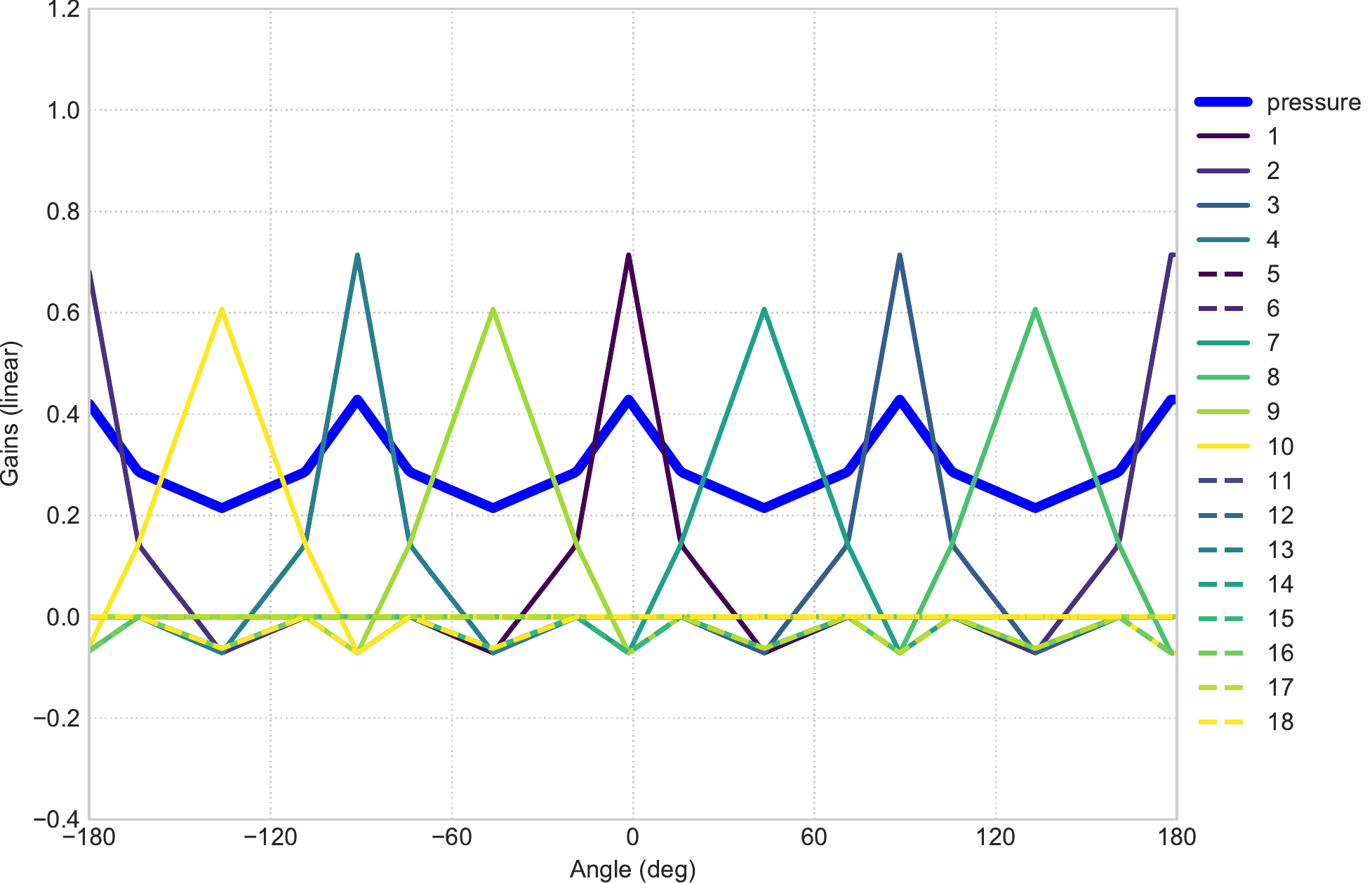}
        \caption{Scaling coefficients $\mathbf c^1_{18}$ for an horizontal panning of interpolating SWF at level 1.
                These can be interpreted as the gains of 18 virtual speakers located on the mesh.
                ($f_{66} \rightarrow \mathbf c^1_{18}$).}
        \label{fig:66-18_lifting-swf}
    \end{minipage}
    \vspace*{0.5cm}

    \begin{minipage}[t]{\textwidth}
        \centering
        \includegraphics[width=0.6\textwidth]{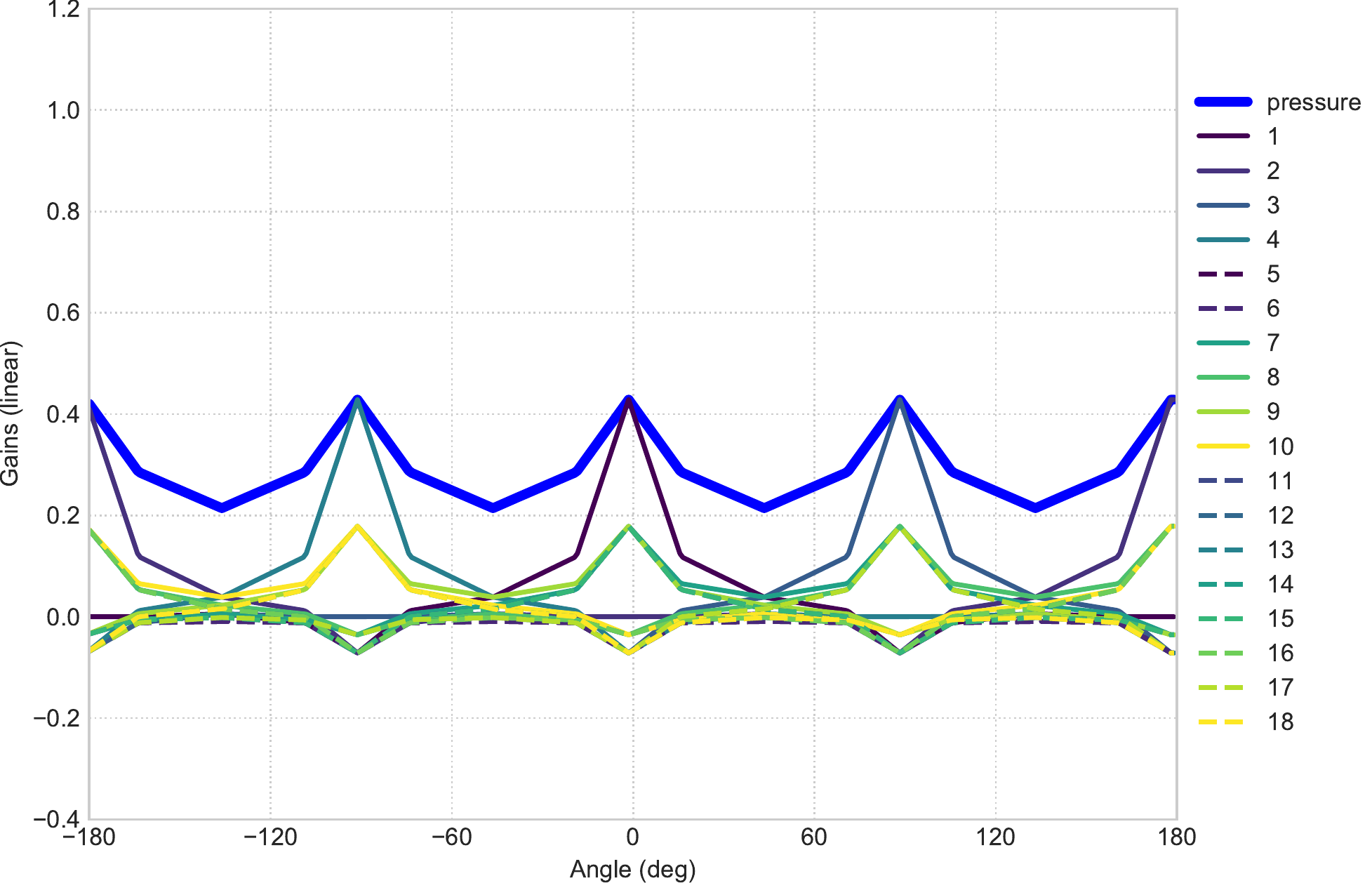}
        \caption{Upsampled scaling coefficients $ \mathbf c^0_{6}$ for an horizontal panning of interpolating SWF at level $\tilde{0}$.
                These can be interpreted as the gains of 18 virtual speakers located on the mesh.
                ($f_{66} \rightarrow \mathbf c^0_{6} \rightarrow \tilde{\mathbf c}^0_{18}$).}
        \label{fig:66-6-18_lifting-swf}
    \end{minipage}
\end{figure}

In Figures~\ref{fig:66-6_lifting-swf}, \ref{fig:66-18_lifting-swf} and \ref{fig:66-6-18_lifting-swf}
we report the $\mathbf{c}^0$, $\mathbf{c}^1$, $\tilde{\mathbf{c}}^0$ for the lifted spherical interpolating wavelets
(as defined in Section~\ref{sec:spherical-lifting-scheme}).
It is evident that the pressure is not preserved by these $\mathbf A_{\text{INT}}$ (Eq.~\eqref{eq:sph-dual-lift-analysis}) and this type of scaling functions
is not interesting for our application.

\begin{figure}
    \centering
    \vspace*{-1cm}
    \begin{minipage}[t]{\textwidth}
        \centering
        \includegraphics[width=0.6\textwidth]{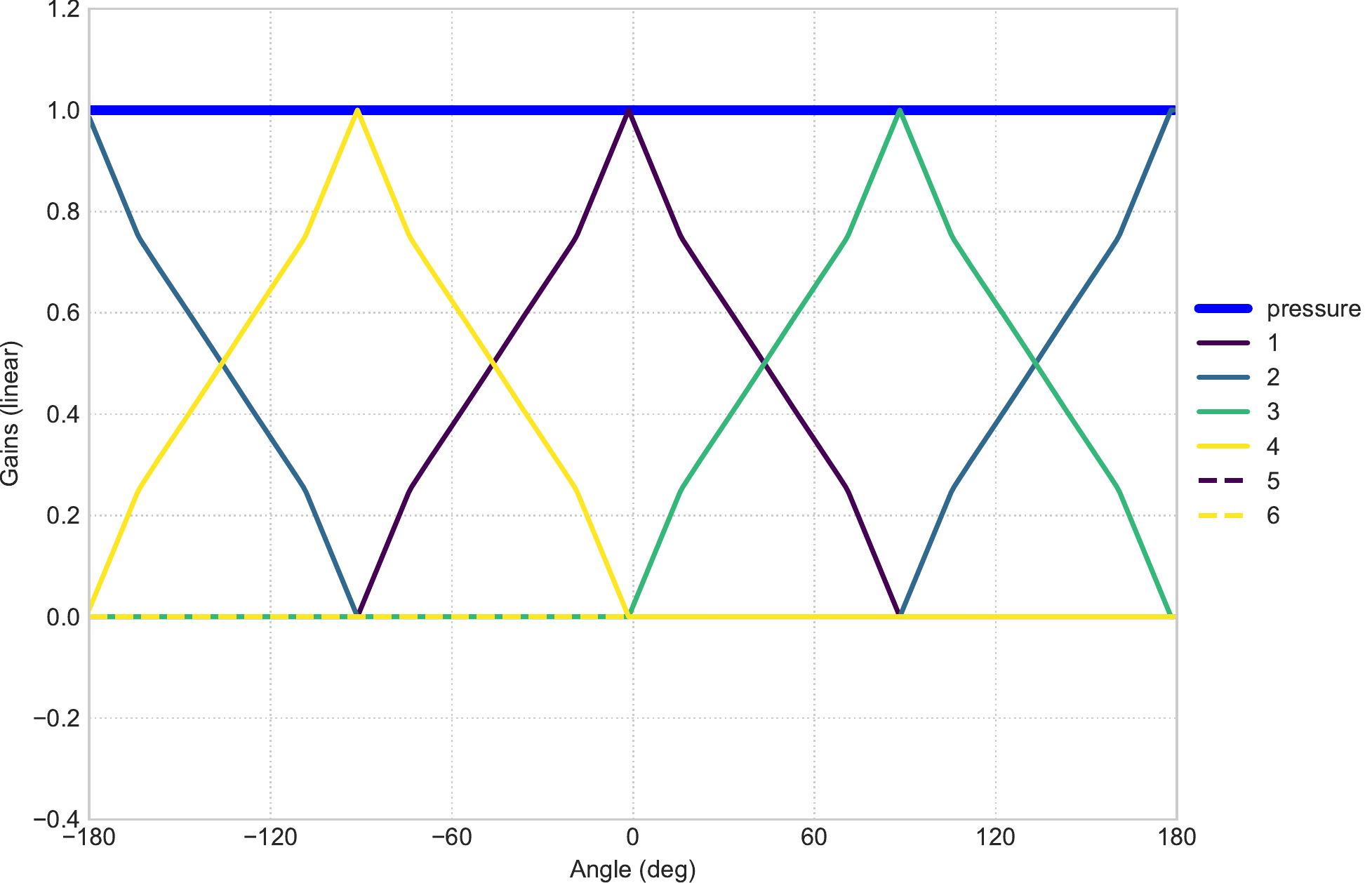}
        \caption{Scaling coefficients $\mathbf c^0_{6}$ for an horizontal panning of SINT-SWF at level 0.
                These can be interpreted as the gains of 6 virtual speakers located on the mesh.
                ($f_{66} \rightarrow  \mathbf c^0_{6}$).}
        \label{fig:66-6_swapped-lifting-swf}
    \end{minipage}
    \vspace*{0.5cm}

    \begin{minipage}[t]{\textwidth}
        \centering
        \includegraphics[width=0.6\textwidth]{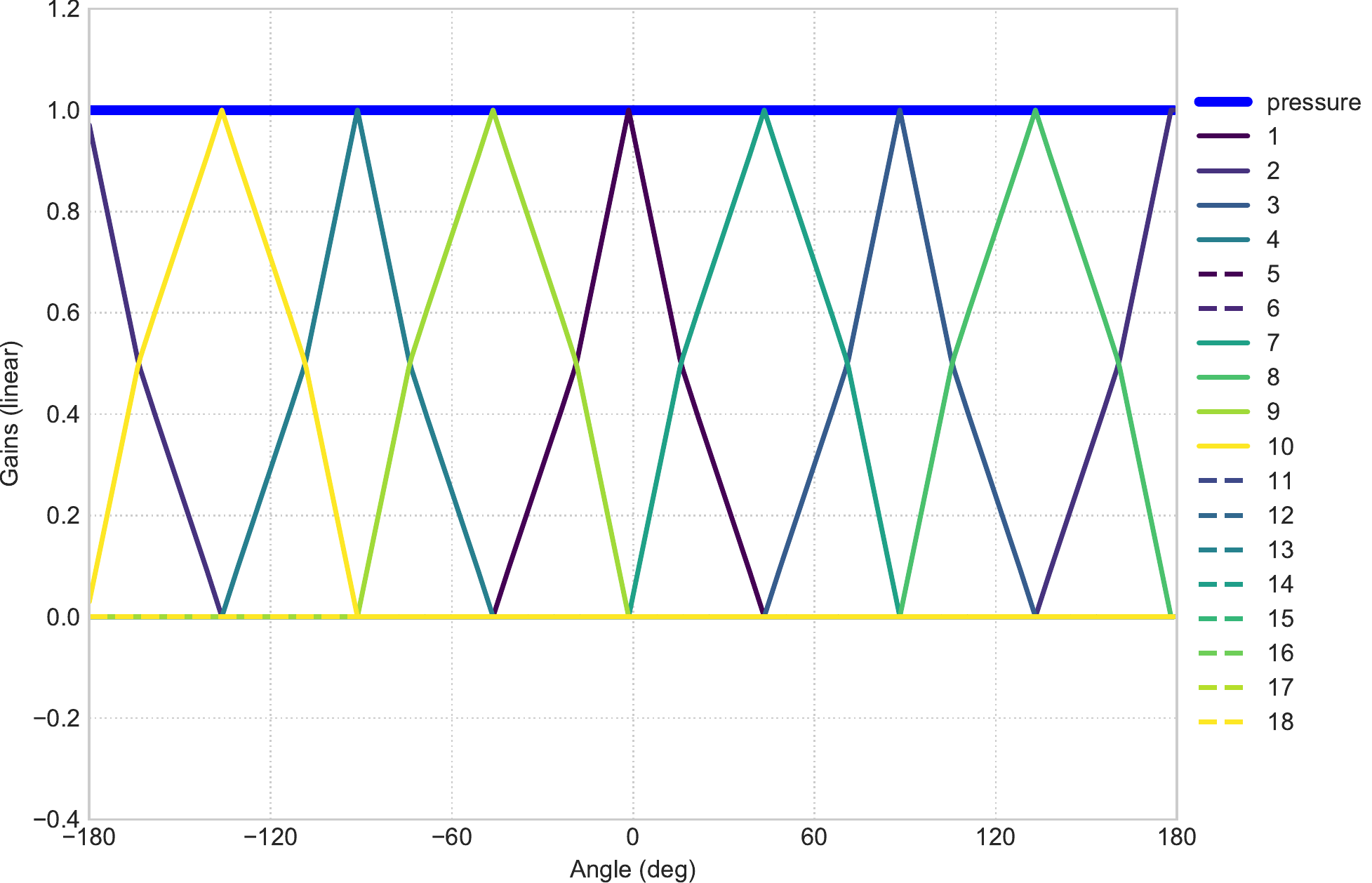}
        \caption{Scaling coefficients $\mathbf c^1_{18}$ for an horizontal panning of SINT-SWF at level 1.
                These can be interpreted as the gains of 18 virtual speakers located on the mesh.
                ($f_{66} \rightarrow \mathbf c^1_{18}$).}
        \label{fig:66-18_swapped-lifting-swf}
    \end{minipage}
    \vspace*{0.5cm}

    \begin{minipage}[t]{\textwidth}
        \centering
        \includegraphics[width=0.6\textwidth]{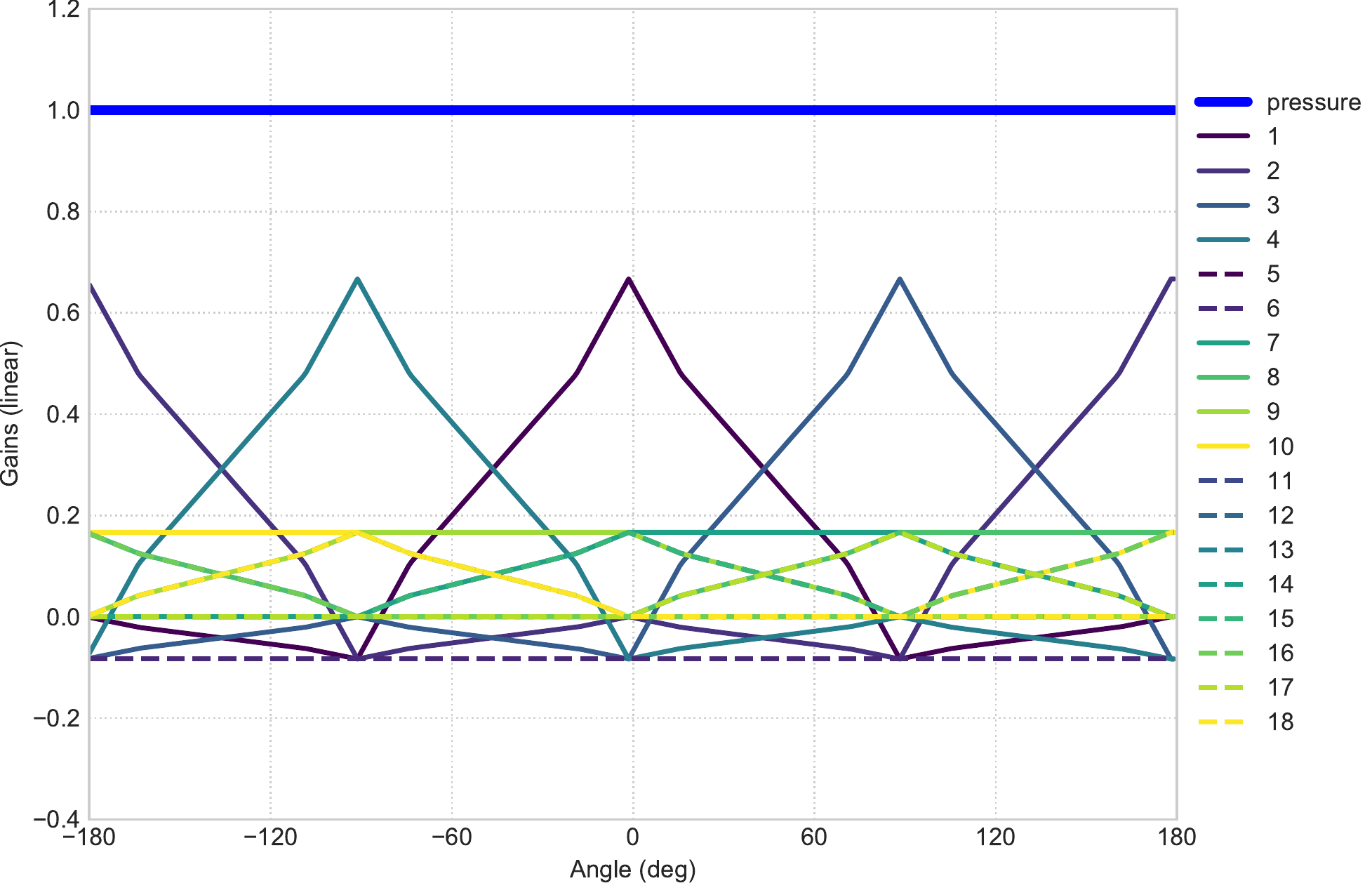}
        \caption{Upsampled scaling coefficients $ \mathbf c^0_{6}$ for an horizontal panning of SINT-SWF at level $\tilde{0}$.
                These can be interpreted as the gains of 18 virtual speakers located on the mesh.
                ($f_{66} \rightarrow \mathbf c^0_{6} \rightarrow \tilde{\mathbf c}^0_{18}$).}
        \label{fig:66-6-18_swapped-lifting-swf}
    \end{minipage}
\end{figure}

In Figures~\ref{fig:66-6_swapped-lifting-swf}, \ref{fig:66-18_swapped-lifting-swf} and \ref{fig:66-6-18_swapped-lifting-swf}
we show what happens to the $\mathbf{c}^0$, $\mathbf{c}^1$, $\tilde{\mathbf{c}}^0$
when we swap the dual for the direct, essentially $\mathbf A_{\text{SINT}} = \mathbf P_{\text{INT}}^\intercal $
and $\mathbf P_{\text{SINT}} = \mathbf A_{\text{INT}}^\intercal $ (Eqs.~\eqref{eq:sph-dual-lift-analysis}, \eqref{eq:sph-dual-lift-synthesis}).
We will call this format SINT-SWF.
Similarly to the VBAP-SWF, pressure is preserved at all moments. Figure~\ref{fig:66-18_swapped-lifting-swf} shows
the result of applicating $\mathbf A^2_{\text{SINT}}$ interpolation alone.
Unlike $\mathbf P_{\text{VBAP}}$, $\mathbf P_{\text{SINT}}$ is not trivial, and we get a meaningful upsampling, see Figure~\ref{fig:66-6-18_swapped-lifting-swf}.
Small negative gains are introduced in the upsampling procedure.

\begin{figure}[t!]
    \centering
    \vspace*{-1cm}
    \begin{minipage}[t]{\textwidth}
        \centering
        \includegraphics[width=0.6\textwidth]{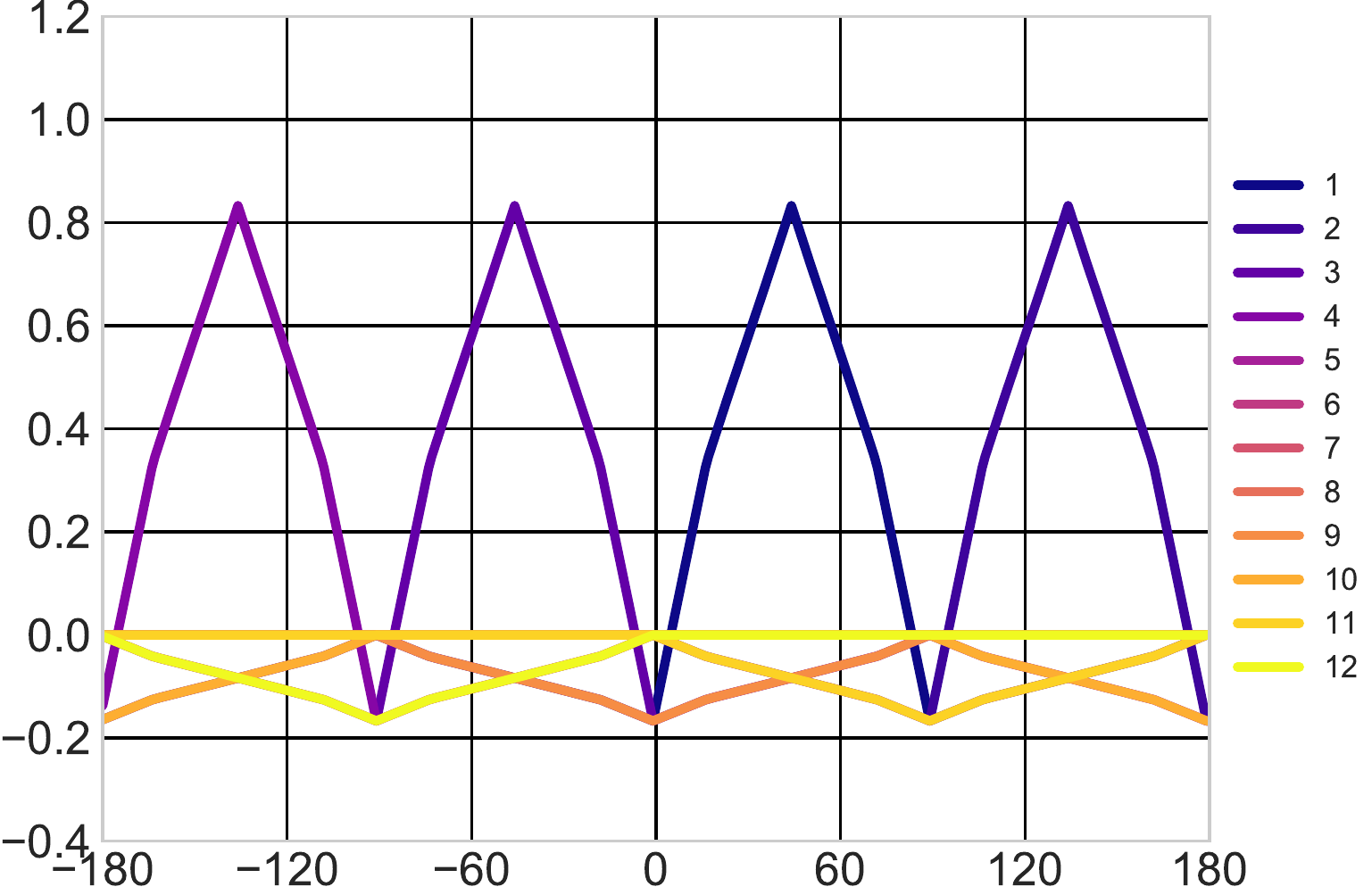}
        \caption{Wavelet coefficients, $\mathbf d^0$, for an horizontal panning of SINT-SWF at level 0.}
        \label{fig:d0_swapped-lifting-swf}
    \end{minipage}
    \vspace*{0.5cm}

    \begin{minipage}[t]{\textwidth}
        \centering
        \includegraphics[width=0.6\textwidth]{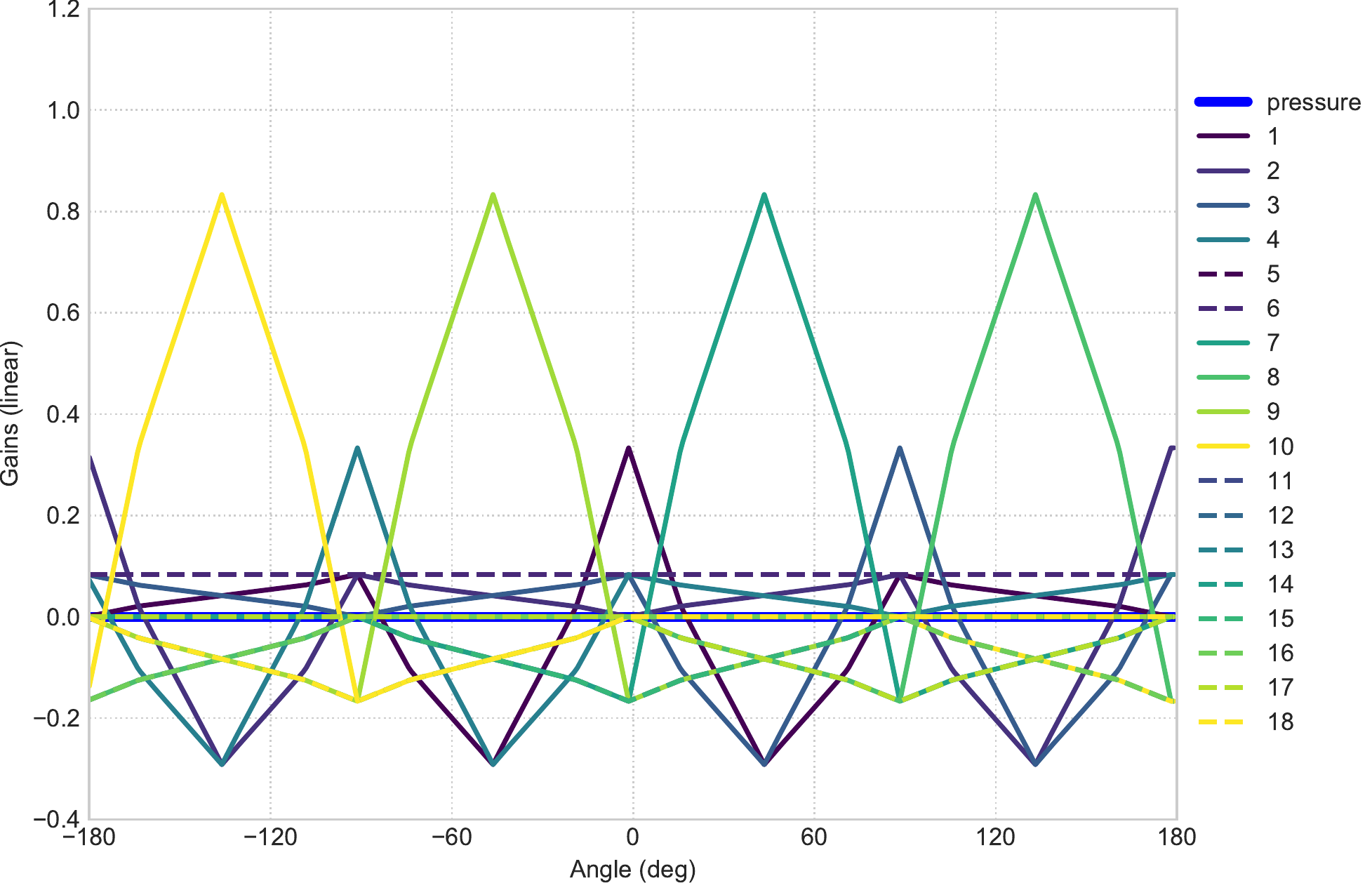}
        \caption{Upscaled wavelet coefficients, $ \mathbf Q^1 \mathbf d^0$, for an horizontal panning of SINT-SWF at level 0..}
        \label{fig:d0up_swapped-lifting-swf}
    \end{minipage}
\end{figure}

It is interesting to see which is the contribution of the details $\mathbf d^0$ when the $\mathbf c^1$ is obtained
via the mentioned reconstruction equation $\mathbf{c}^1 = \mathbf P^1 \mathbf{c}^0 + \mathbf Q^1 \mathbf{d}^0$.
In Figure~\ref{fig:d0_swapped-lifting-swf} we show the bare wavelet coefficients $\mathbf d^0$.
These $\mathbf d^0$ do not have an interpretation per-se, but they have to be uplifted via $\mathbf Q^1$ to make sense in the virtual speaker interpretation.
In Figure~\ref{fig:d0up_swapped-lifting-swf} we illustrate the isolated effect of the upsampled $\mathbf Q^1 \mathbf d^0$
over the 18 points of the mesh at level 1.
If we look at these upsampled details in the interpretation where the mesh is a set of virtual speakers,
then the gains carried by $\mathbf Q^1 \mathbf{d^0}$ are summed linearly to the ones coming from $\mathbf P^1 \mathbf{c}^0$.
It is important to notice that these details do not carry pressure, which is already preserved by the $\mathbf A$ and $\mathbf P$ matrices.

\subsection[Optimized-SWF Gains]{Optimized-SWF Gains for Horizontal Panning} \label{sec:opt-swf-gains}
\begin{figure}
    \centering
    \vspace*{-1cm}
    \begin{minipage}[t]{\textwidth}
        \centering
        \includegraphics[width=0.6\textwidth]{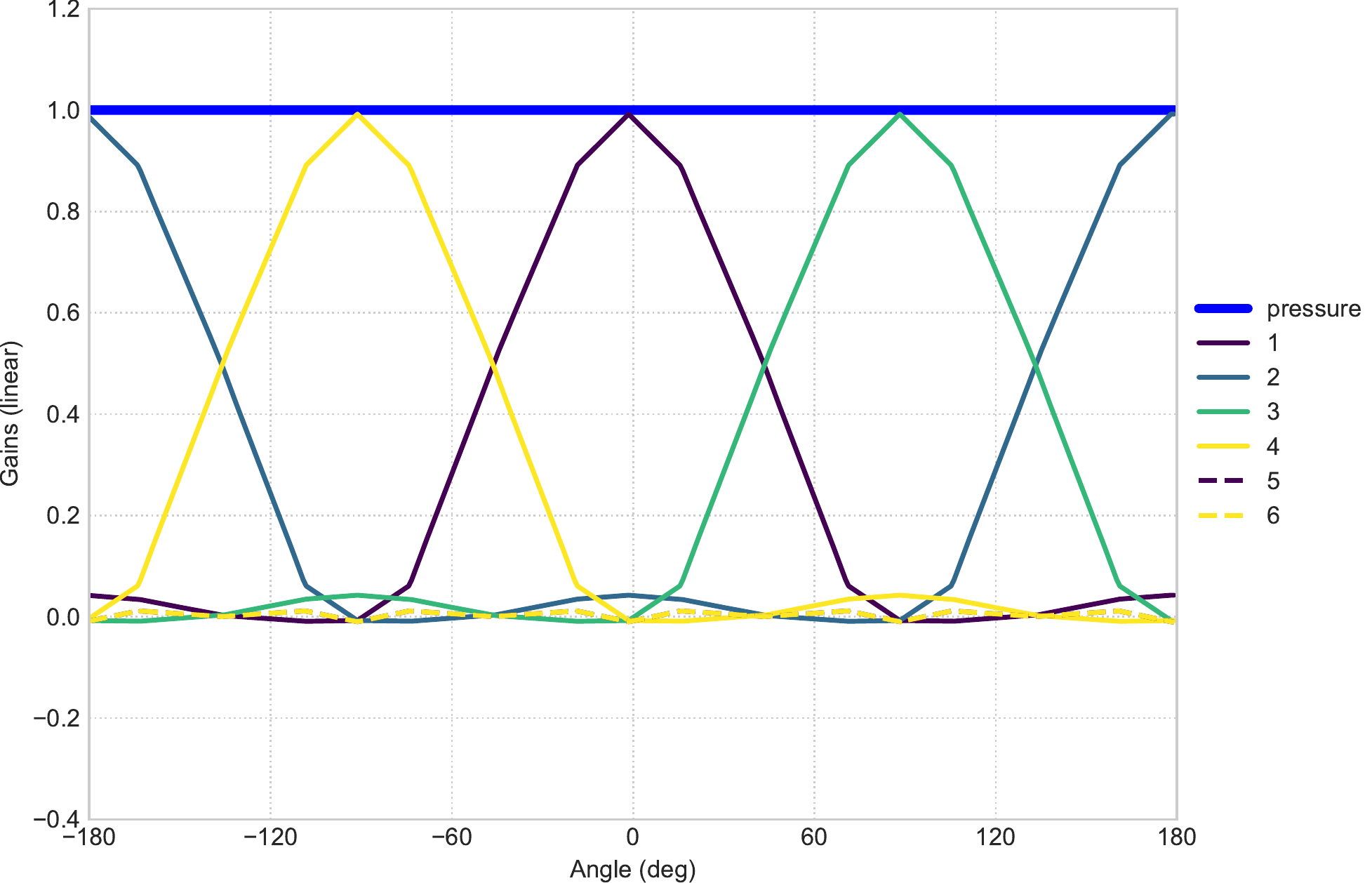}
        \caption{Scaling coefficients $\mathbf c^0_{6}$ for an horizontal panning of OPT-SWF at level 0.
                These can be interpreted as the gains of 6 virtual speakers located on the mesh.
                ($f_{66} \rightarrow  \mathbf c^0_{6}$).}
        \label{fig:66-6_optimized-swf}
    \end{minipage}
    \vspace*{0.5cm}

    \begin{minipage}[t]{\textwidth}
        \centering
        \includegraphics[width=0.6\textwidth]{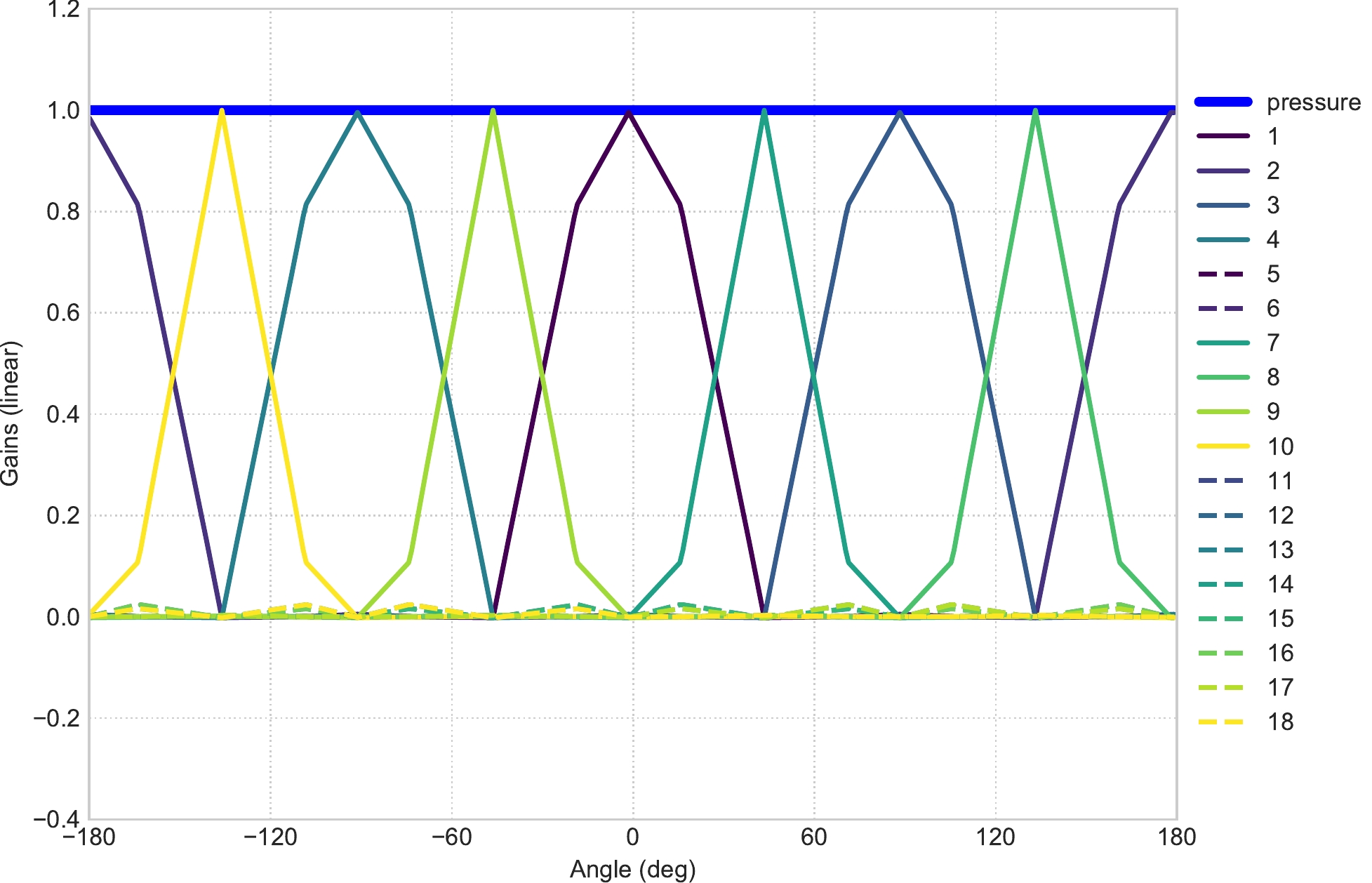}
        \caption{Scaling coefficients $\mathbf c^1_{18}$ for an horizontal panning of OPT-SWF at level 1.
                These can be interpreted as the gains of 18 virtual speakers located on the mesh.
                ($f_{66} \rightarrow \mathbf c^1_{18}$).}
        \label{fig:66-18_optimized-swf}
    \end{minipage}
    \vspace*{0.5cm}

    \begin{minipage}[t]{\textwidth}
        \centering
        \includegraphics[width=0.6\textwidth]{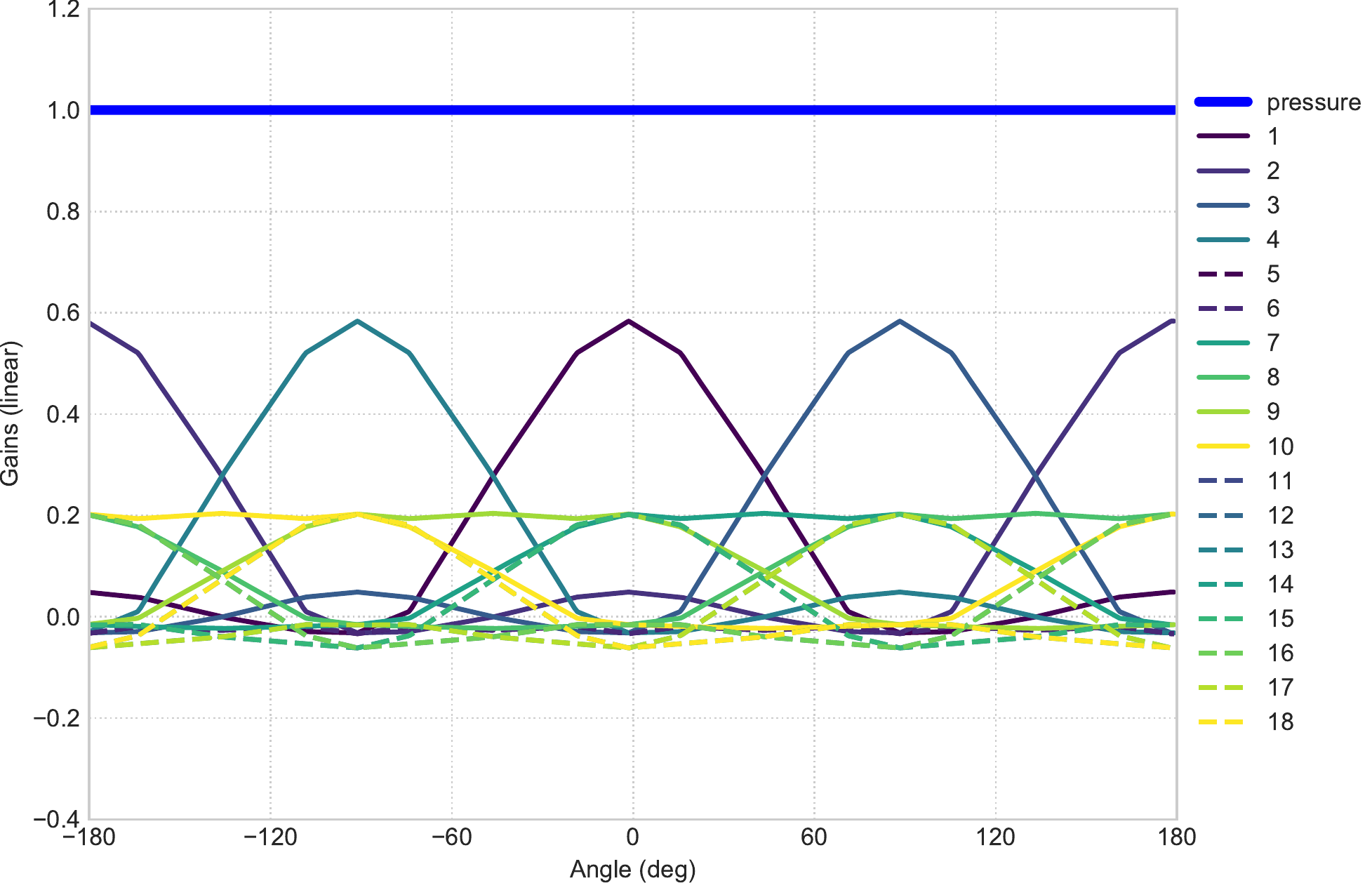}
        \caption{Upsampled scaling coefficients $ \mathbf c^0_{6}$ for an horizontal panning of OPT-SWF at level $\tilde{0}$.
                These can be interpreted as the gains of 18 virtual speakers located on the mesh.
                ($f_{66} \rightarrow \mathbf c^0_{6} \rightarrow \tilde{\mathbf c}^0_{18}$).}
        \label{fig:66-6-18_optimized-swf}
    \end{minipage}
\end{figure}

In Figures~\ref{fig:66-6_optimized-swf}, \ref{fig:66-18_optimized-swf} and \ref{fig:66-6-18_optimized-swf}
we report the signals $\mathbf{c}^0$, $\mathbf{c}^1$, $\tilde{\mathbf{c}}^0$ for the optimized set of filters of the OPT-SWF.
It is possible to notice that pressure is properly preserved at all stages, similarly to what happens for VBAP-SWF and SINT-SWF.
Noticeable negative gains are introduced only in the upsampling stage, Figure~\ref{fig:66-6-18_optimized-swf},
and are smaller than in the equivalent processing for SINT-SWF.
The panning functions have a more `round' shape than in VBAP-SWF and INT-SWF.

\begin{figure}[t!]
    \centering
    \vspace*{-1cm}
    \begin{minipage}[t]{\textwidth}
        \centering
        \includegraphics[width=0.6\textwidth]{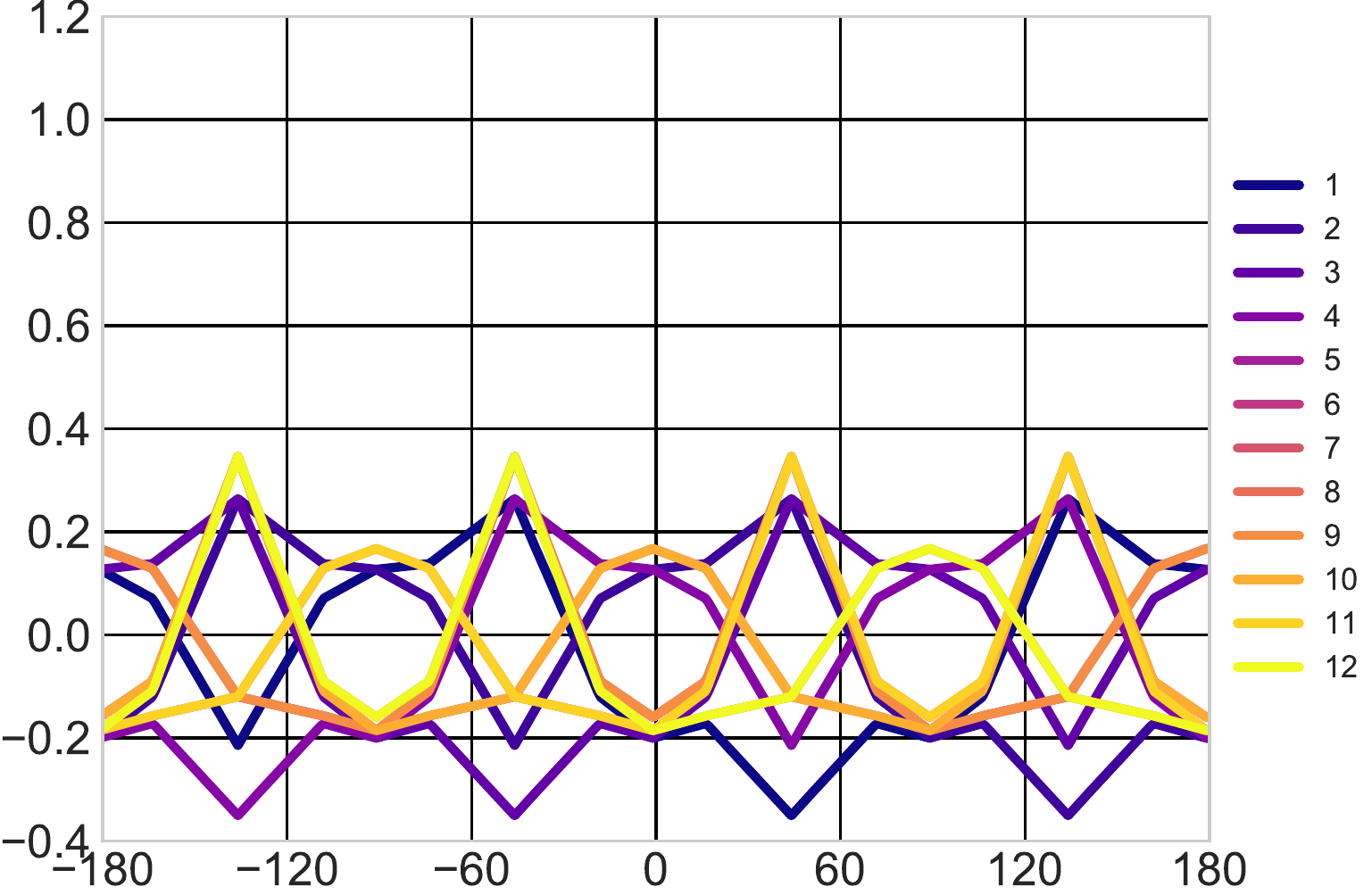}
        \caption{Wavelet coefficients, $\mathbf d^0$, for an horizontal panning of OPT-SWF at level 0.}
        \label{fig:d0_opt-swf}
    \end{minipage}
    \vspace*{0.5cm}

    \begin{minipage}[t]{\textwidth}
        \centering
        \includegraphics[width=0.6\textwidth]{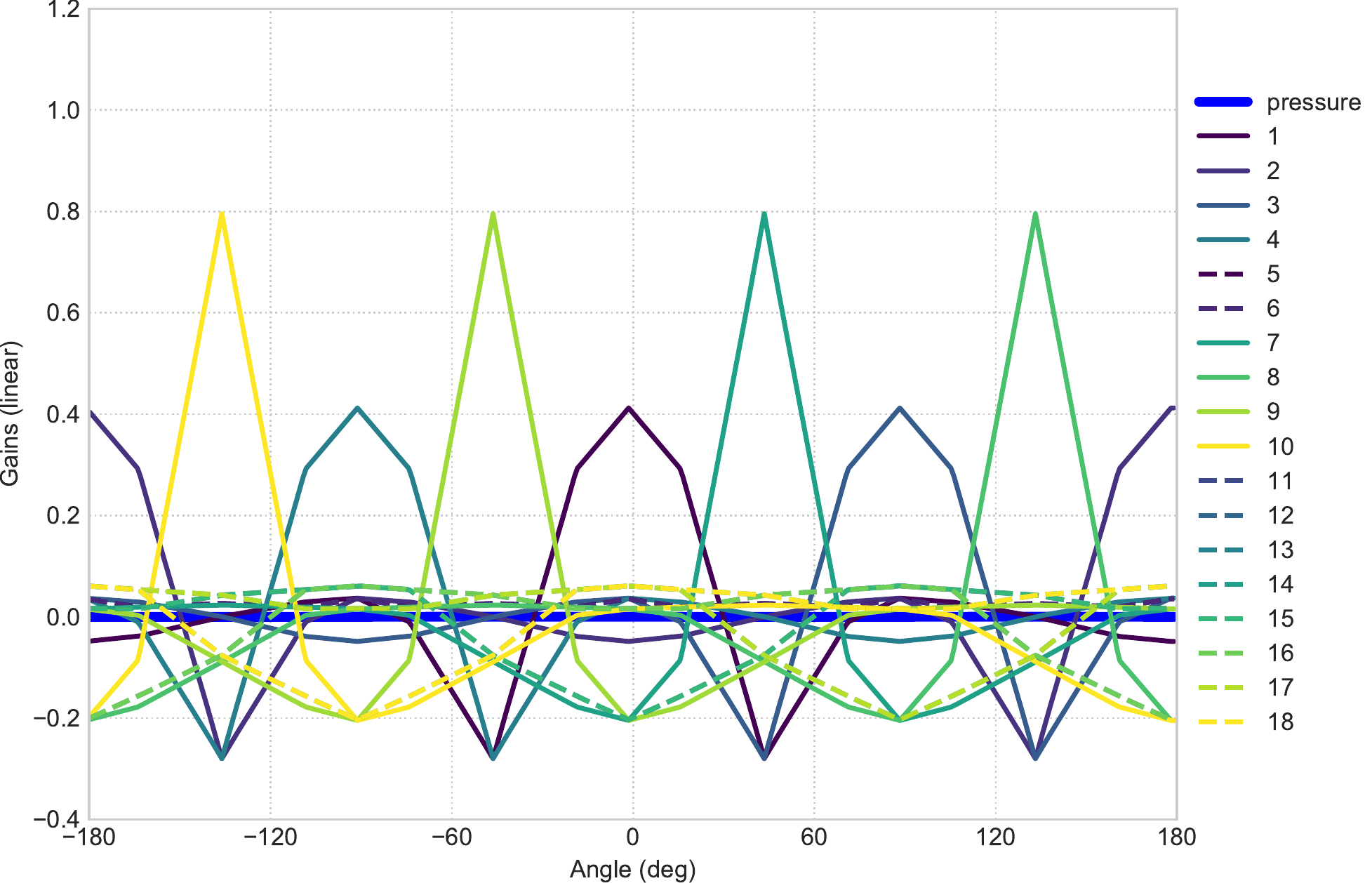}
        \caption{Upscaled wavelet coefficients, $ \mathbf Q^1 \mathbf d^0$, for an horizontal panning of OPT-SWF at level 0.}
        \label{fig:d0up_opt-swf}
    \end{minipage}
\end{figure}

As before, it is interesting to see which is the contribution of the details $\mathbf d^0$ when the $\mathbf c^1$ is obtained
via the mentioned reconstruction equation $\mathbf{c}^1 = \mathbf P^1 \mathbf{c}^0 + \mathbf Q^1 \mathbf{d}^0$.
In Figure~\ref{fig:d0_opt-swf} we report the bare $\mathbf d^0$ wavelet coefficients for OPT-SWF.
It is interesting to note that the $\mathbf d^0$ from OPT-SWF are less sparse than the one coming from SINT-SWF.
This effect is due to the fact that we did not ask for special properties of $\mathbf B^1$ and $\mathbf Q^1$,
a part from the orthonormality with $\mathbf A^1$ and $\mathbf P^1$.
Notice that the properties of $\mathbf d^0$ do not influence the localization properties of the wavelet-based audio format,
because those are given by $\mathbf Q \mathbf d$.
The $\mathbf d^0$ influence only the spatial compression qualities of the audio format.
In Figure~\ref{fig:d0up_opt-swf} we illustrate the isolated effect of the upsampled $\mathbf Q^1 \mathbf d^0$
over the 18 points of the mesh at level 1.
It is important to notice that these details do not carry pressure, that is already preserved by the $\mathbf A$ and $\mathbf P$ matrices.

The main difference with respect to the upsampled details of SINT-SWF in Figure~\ref{fig:d0up_swapped-lifting-swf}
is the shape of the contribution of the details.
In the next Section we will investigate more on this subject.

\section{Reconstructed Velocity, Energy and Intensity for Different SWF Implementations}
In the previous Section we have seen that the analyzed SWF flavours preserve pressure along the whole audio chain,
and present some differences in the panning functions' shape.
In this Section we will investigate the differences between the three SWF flavours, VBAP/SINT/OPT,
in terms of reconstruction of some relevant observables: velocity, energy and intensity.
These quantities would be the ones reconstructed by a set of speakers (6 or 18)
placed exactly on the location of the mesh vertices.

We want to stress the fact that we are looking at these values as reconstructed on the points of the original mesh,
as if they were virtual speakers, and no actual decoding is involved.
Making a parallel with Ambisonics, it is possible to interpret the reconstructed physical quantities we are investigating
as the quantities reproduced by a basic decoding with a regular layout.
With this exploration we are effectively looking at the internals of the format, inspecting which quantities
are preserved and which are not along the encoding chain.

\begin{figure}
    \centering
    \vspace*{-1cm}
    \begin{minipage}[t]{0.46\textwidth}
        \centering
        \includegraphics[width=\textwidth]{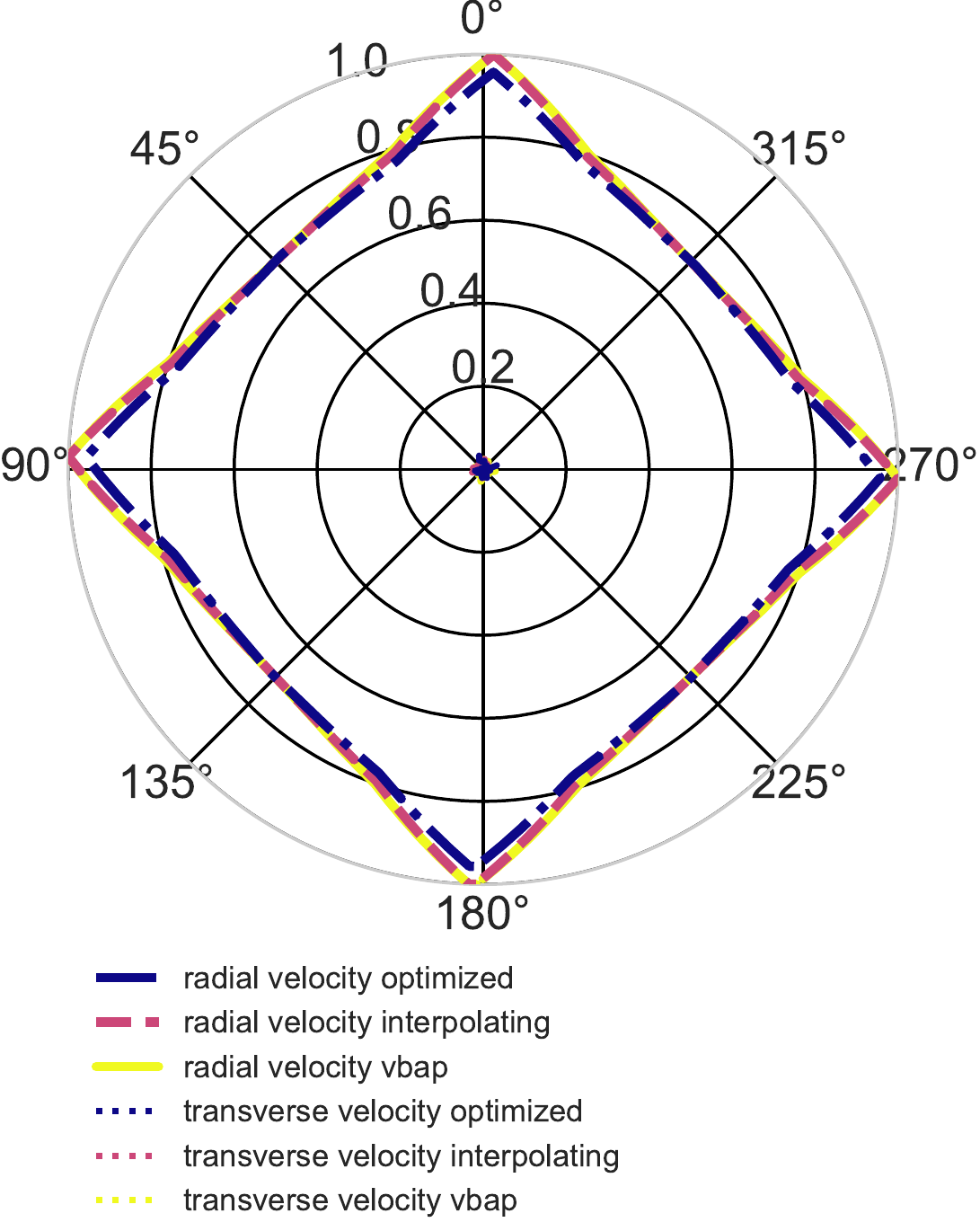}
        \caption{Velocity at level 0; comparison between OPT/SINT/VBAP-SWF flavours.}
        \label{fig:mesh-velocity-level0}
    \end{minipage}
    \hfill
    \begin{minipage}[t]{0.46\textwidth}
        \centering
        \includegraphics[width=\textwidth]{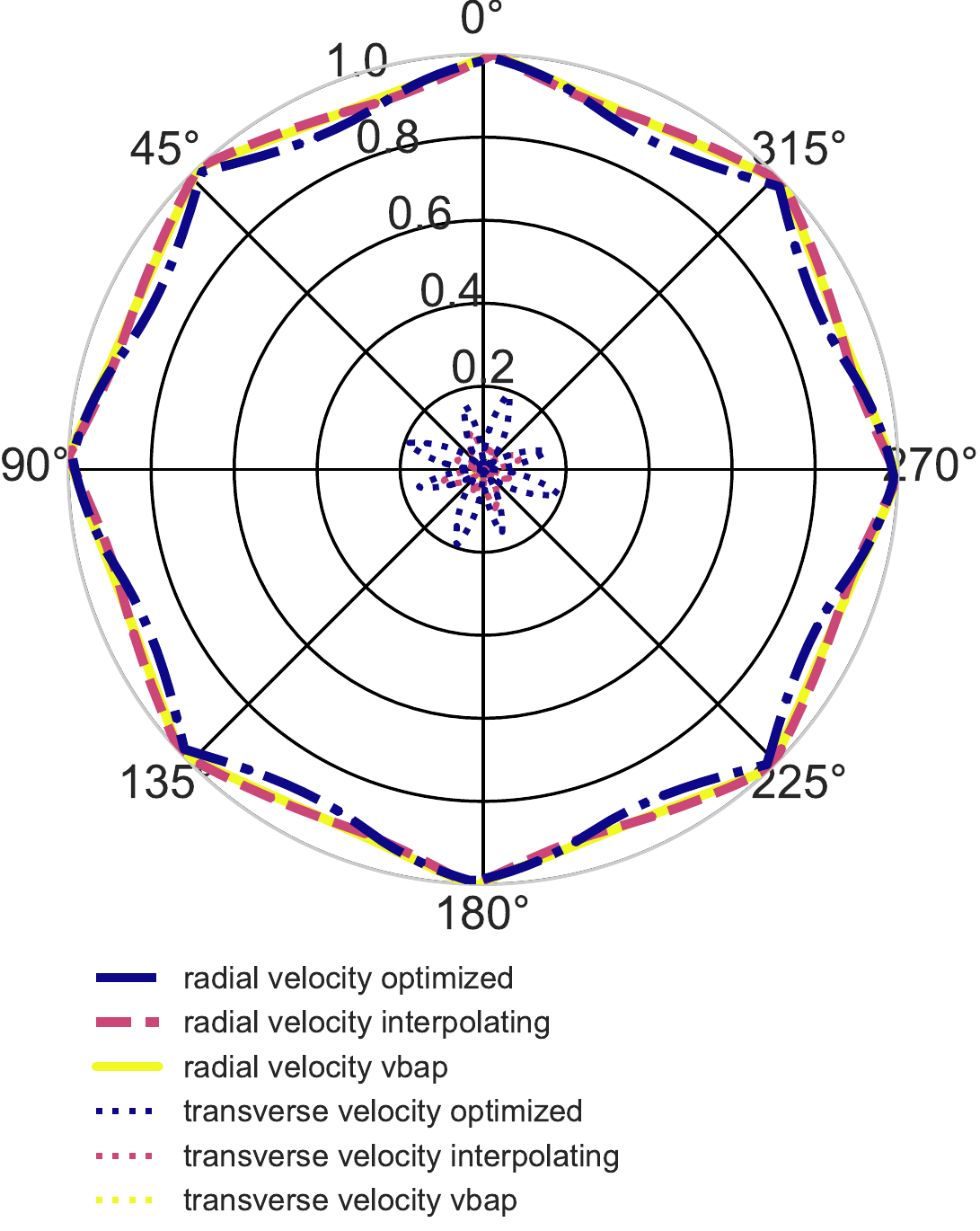}
        \caption{Velocity at level 1; comparison between OPT/SINT/VBAP-SWF flavours.}
        \label{fig:mesh-velocity-level1}
    \end{minipage}
    \begin{minipage}[t]{0.46\textwidth}
        \centering
        \includegraphics[width=\textwidth]{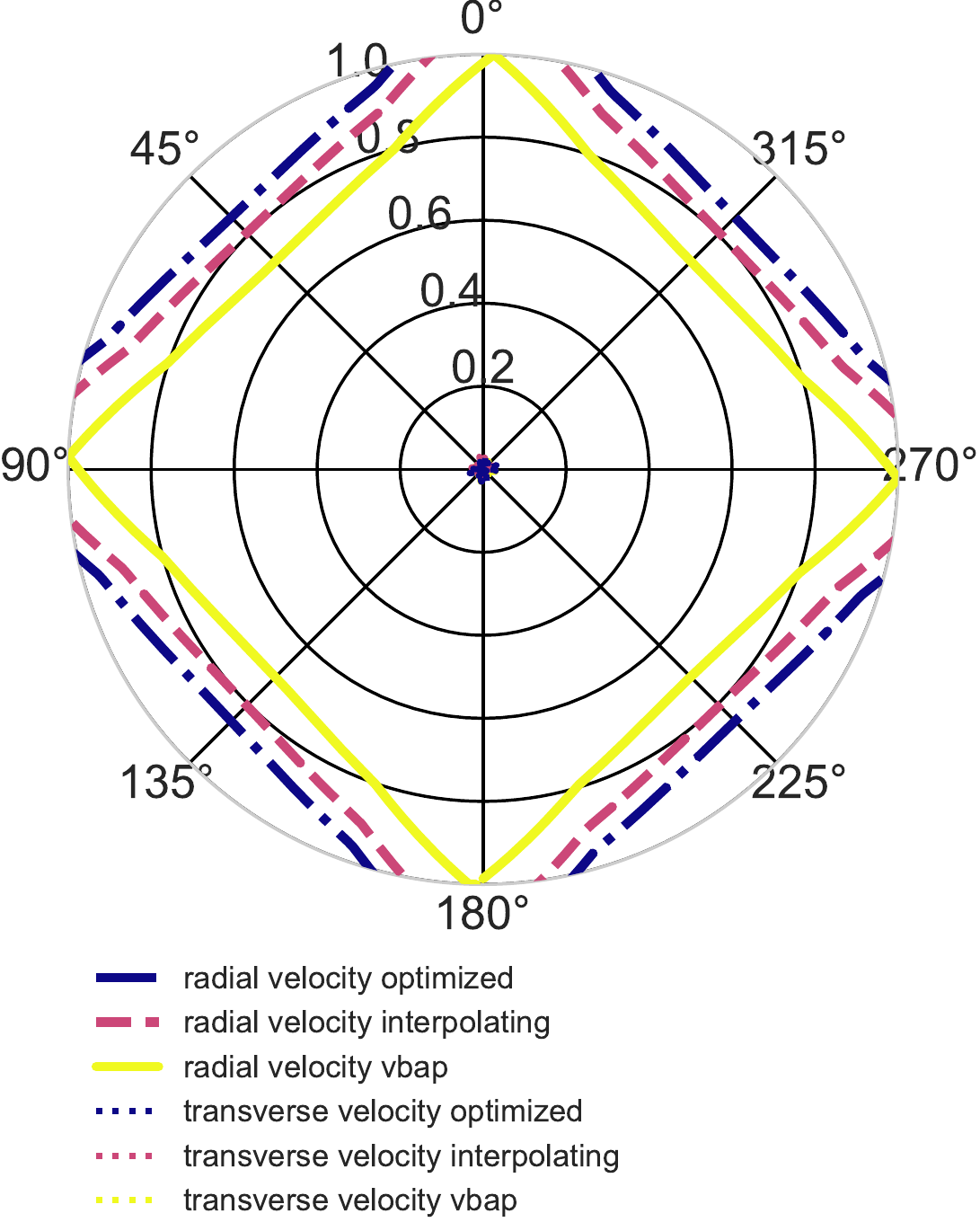}
        \caption{Velocity at level $\tilde{0}$; comparison between OPT/SINT/VBAP-SWF flavours.}
        \label{fig:mesh-velocity-level0tilde}
    \end{minipage}
\end{figure}

In Figures~\ref{fig:mesh-velocity-level0}, \ref{fig:mesh-velocity-level1} and \ref{fig:mesh-velocity-level0tilde}
we report the velocity reconstruction, separated in its radial and transverse components, for the three SWF flavours that are object of this analysis.
In Figures~\ref{fig:mesh-velocity-level0} and \ref{fig:mesh-velocity-level1} (where only $\mathbf{A}$ is involved)
it is possible to see that the velocity (in particular the radial component) is properly reconstructed only at the points of the mesh;
in between the mesh points, the type of interpolation is responsible for the preserved properties.
In Figure~\ref{fig:mesh-velocity-level0tilde} the action of $\mathbf{P}$ adds up,
modifying the reconstruction of the radial component of velocity for OPT-SWF and SINT-SWF. VBAP-SWF component remains unchanged,
with respect to Figure~\ref{fig:mesh-velocity-level0}, since $\mathbf{P}_{\text{VBAP}}$ is trivial.

\begin{figure}
    \centering
    \vspace*{-1cm}
    \begin{minipage}[t]{0.46\textwidth}
        \includegraphics[width=\textwidth]{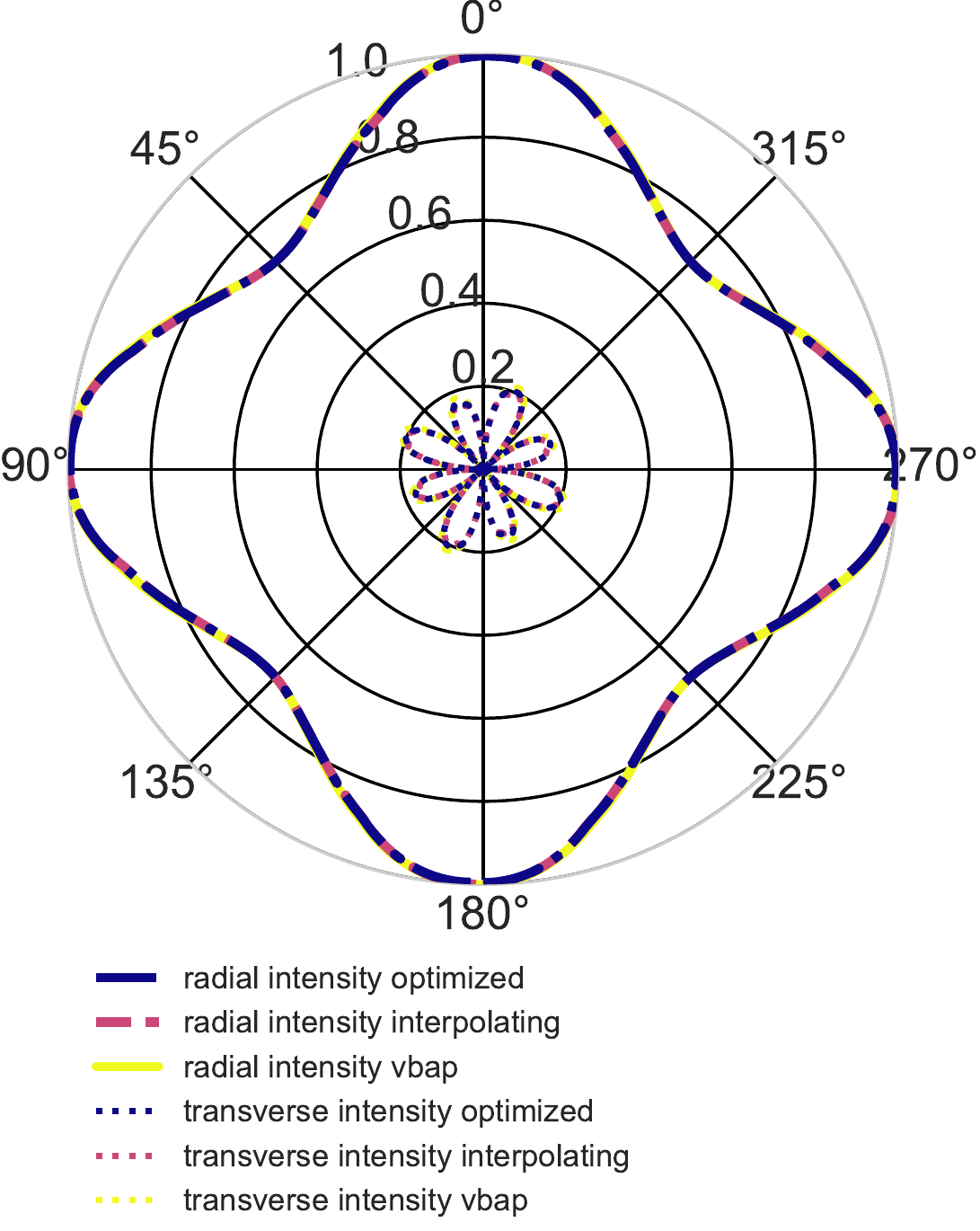}
        \caption{Intensity at level 0; comparison between OPT/SINT/VBAP-SWF flavours.}
        \label{fig:mesh-intensity-level0}
    \end{minipage}
    \hfill
    \begin{minipage}[t]{0.46\textwidth}
        \includegraphics[width=\textwidth]{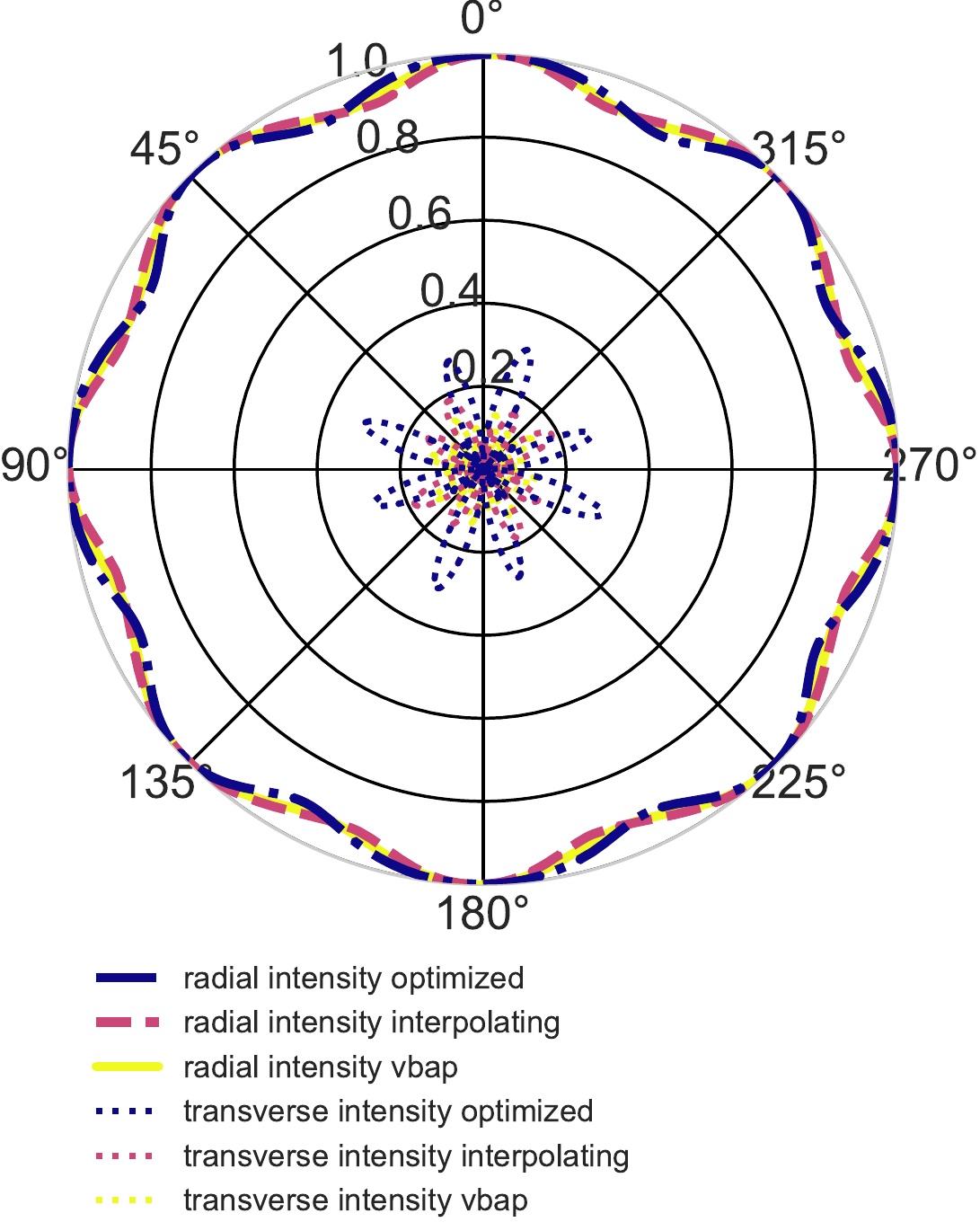}
        \caption{Intensity at level 1; comparison between OPT/SINT/VBAP-SWF flavours.}
        \label{fig:mesh-intensity-level1}
    \end{minipage}
    \begin{minipage}[t]{0.46\textwidth}
        \includegraphics[width=\textwidth]{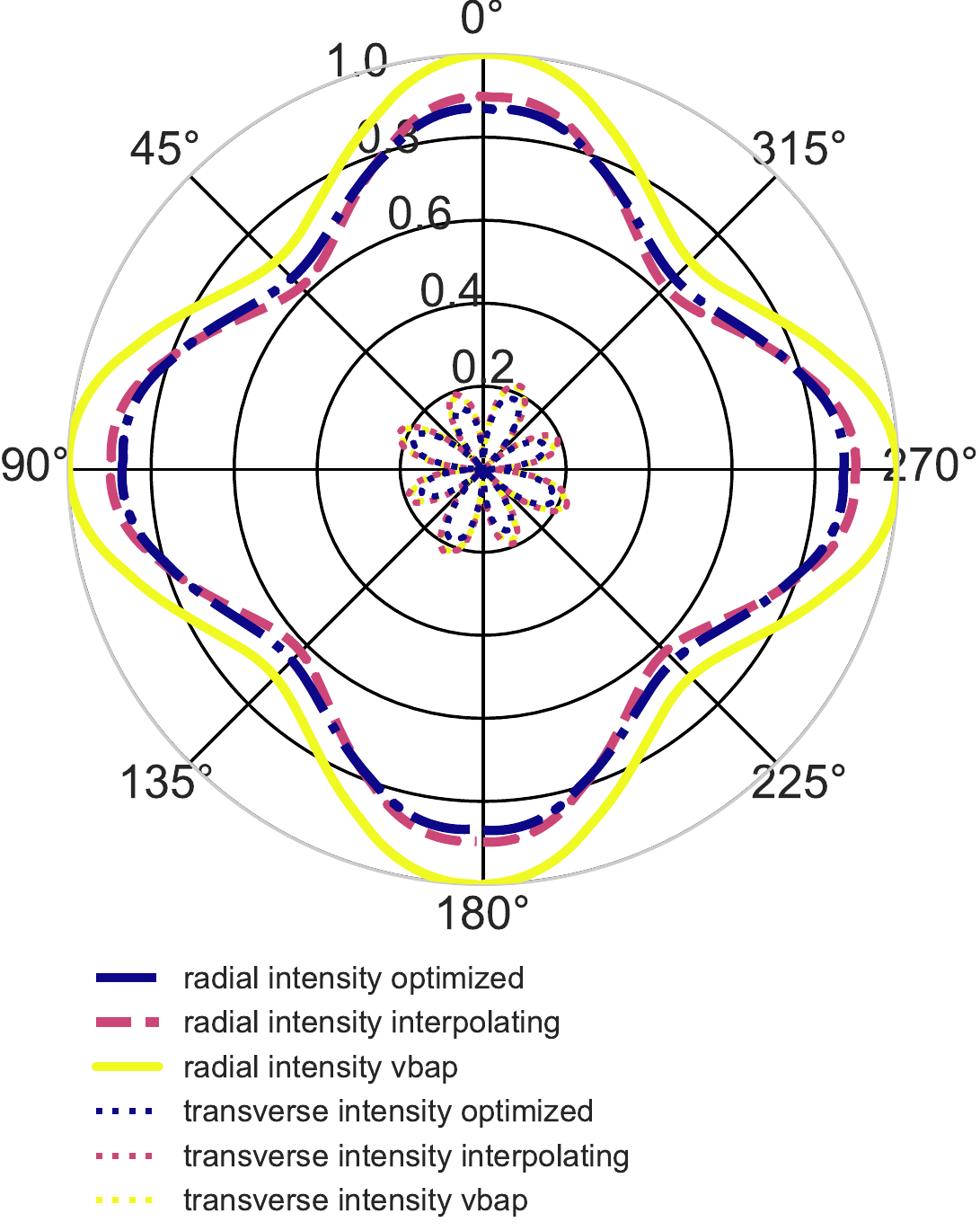}
        \caption{Intensity at level $\tilde{0}$; comparison between OPT/SINT/VBAP-SWF flavours.}
        \label{fig:mesh-intensity-level0tilde}
    \end{minipage}
\end{figure}

Very similar considerations to the velocity reconstruction can be done for the intensity, reported in
Figures~\ref{fig:mesh-intensity-level0}, \ref{fig:mesh-intensity-level1} and \ref{fig:mesh-intensity-level0tilde},
with the only difference of a more relevant transverse component starting to appear in between the mesh points.
The value of the radial intensity is very good already at level 0.

\begin{figure}
    \centering
    \vspace*{-1cm}
    \begin{minipage}[t]{0.46\textwidth}
        \includegraphics[width=\textwidth]{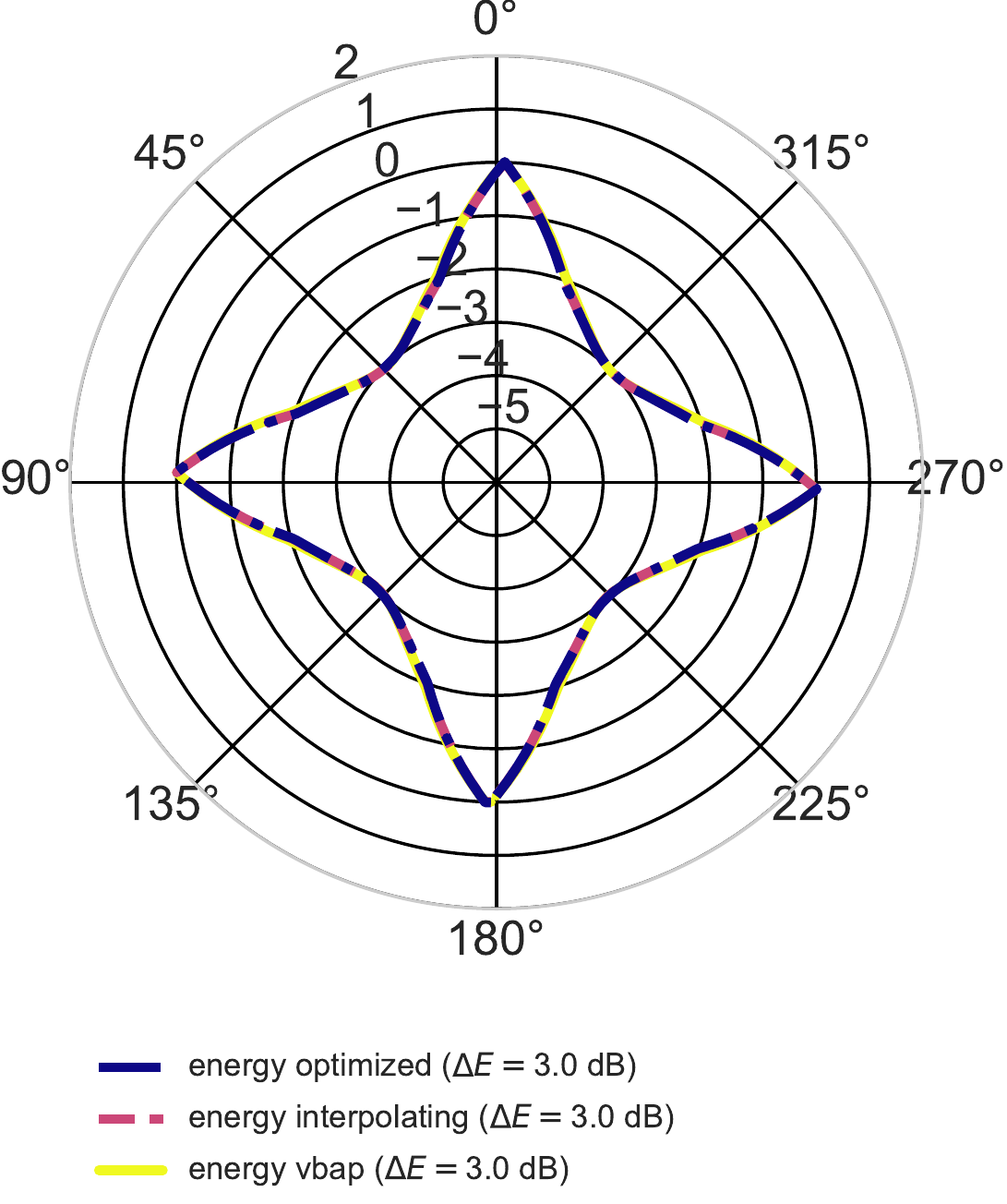}
        \caption{Energy at level 0; comparison between OPT/SINT/VBAP-SWF flavours. Scale is in dB.}
        \label{fig:mesh-energy-level0}
    \end{minipage}
    \hfill
    \begin{minipage}[t]{0.46\textwidth}
        \includegraphics[width=\textwidth]{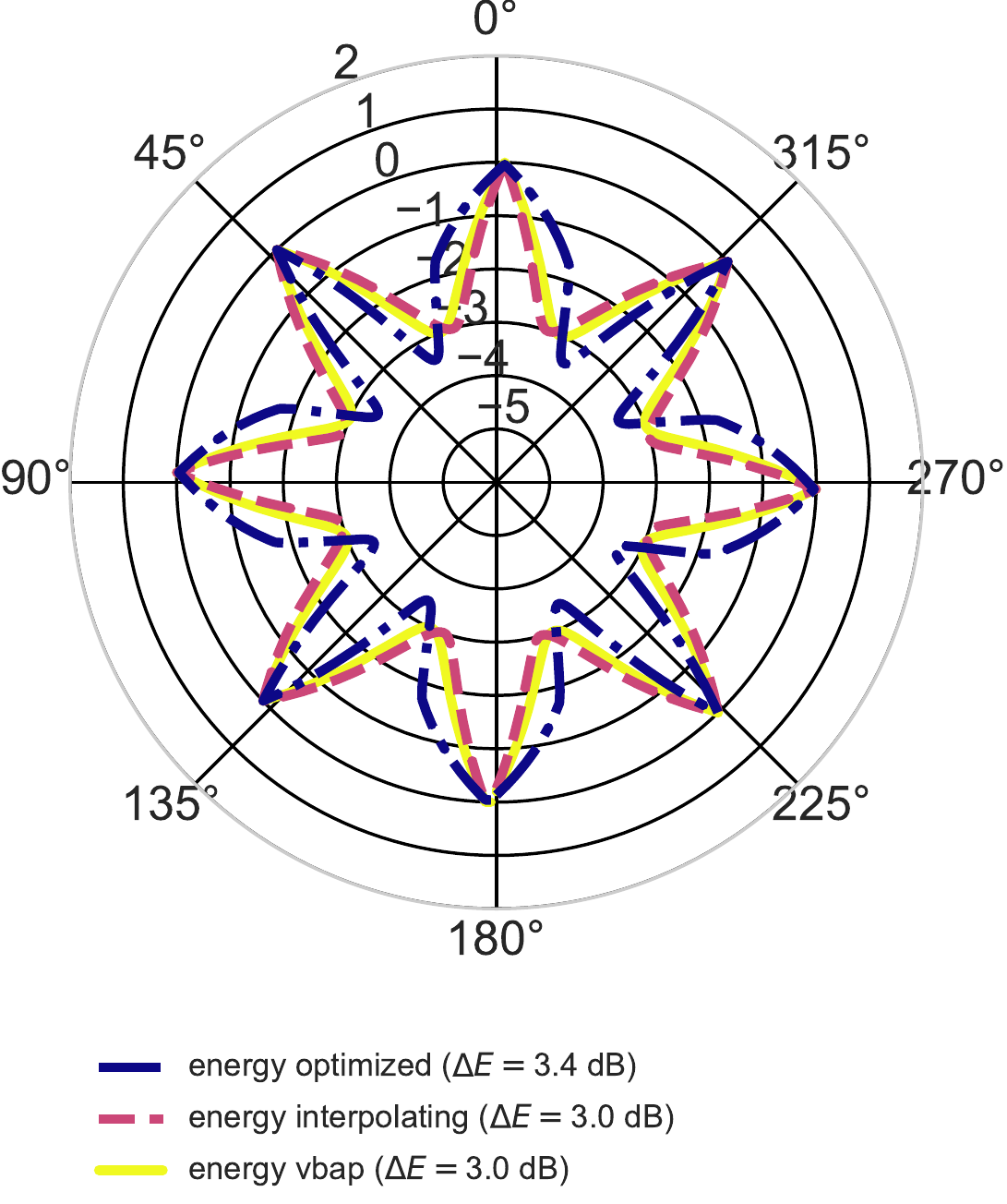}
        \caption{Energy at level 1; comparison between OPT/SINT/VBAP-SWF flavours. Scale is in dB.}
        \label{fig:mesh-energy-level1}
    \end{minipage}
    \begin{minipage}[t]{0.46\textwidth}
        \includegraphics[width=\textwidth]{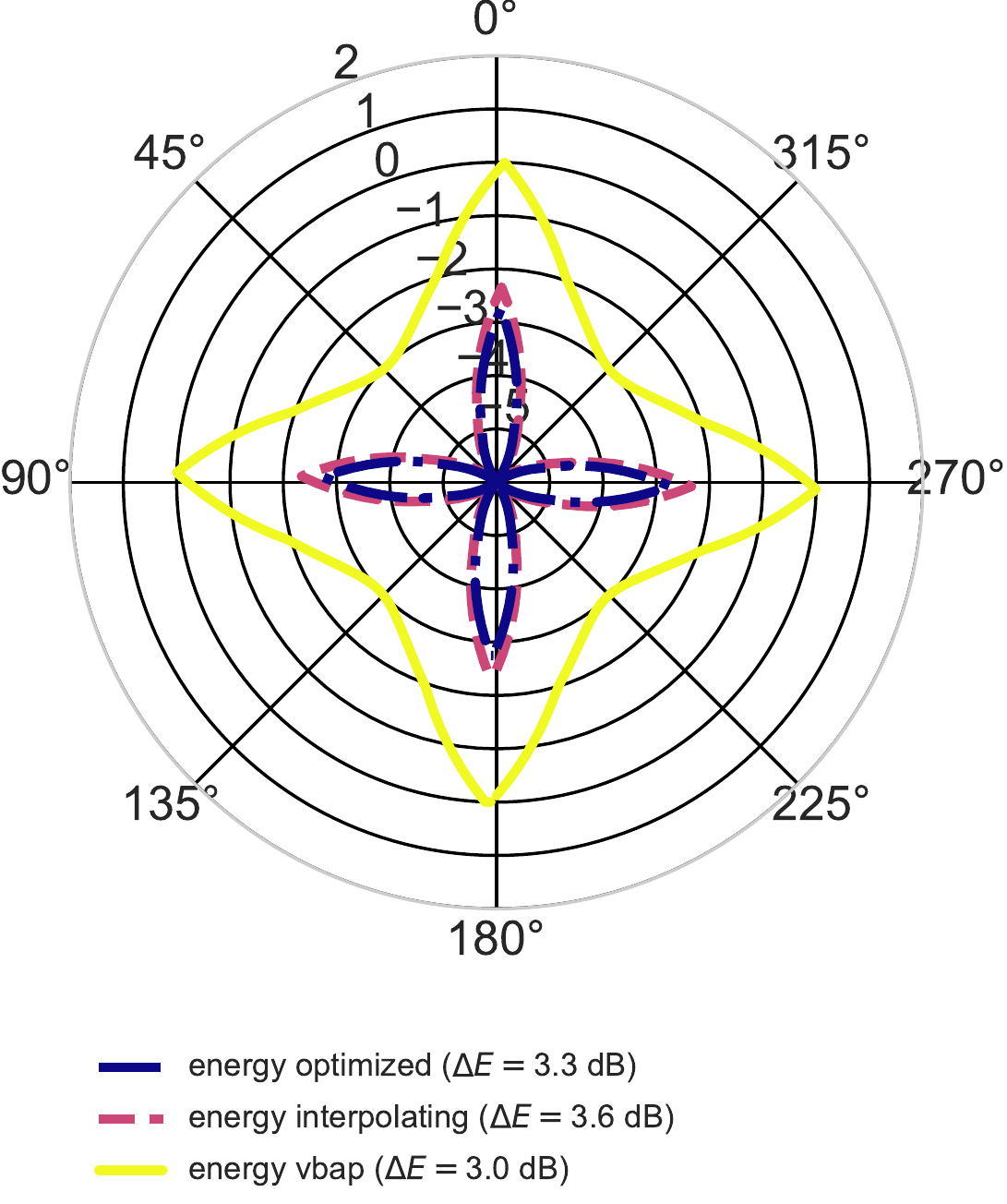}
        \caption{Energy at level $\tilde{0}$; comparison between OPT/SINT/VBAP-SWF flavours. Scale is in dB.}
        \label{fig:mesh-energy-level0tilde}
    \end{minipage}
\end{figure}

\begin{figure}
    \centering
    \begin{minipage}[t]{0.46\textwidth}
        \includegraphics[width=\textwidth,valign=t]{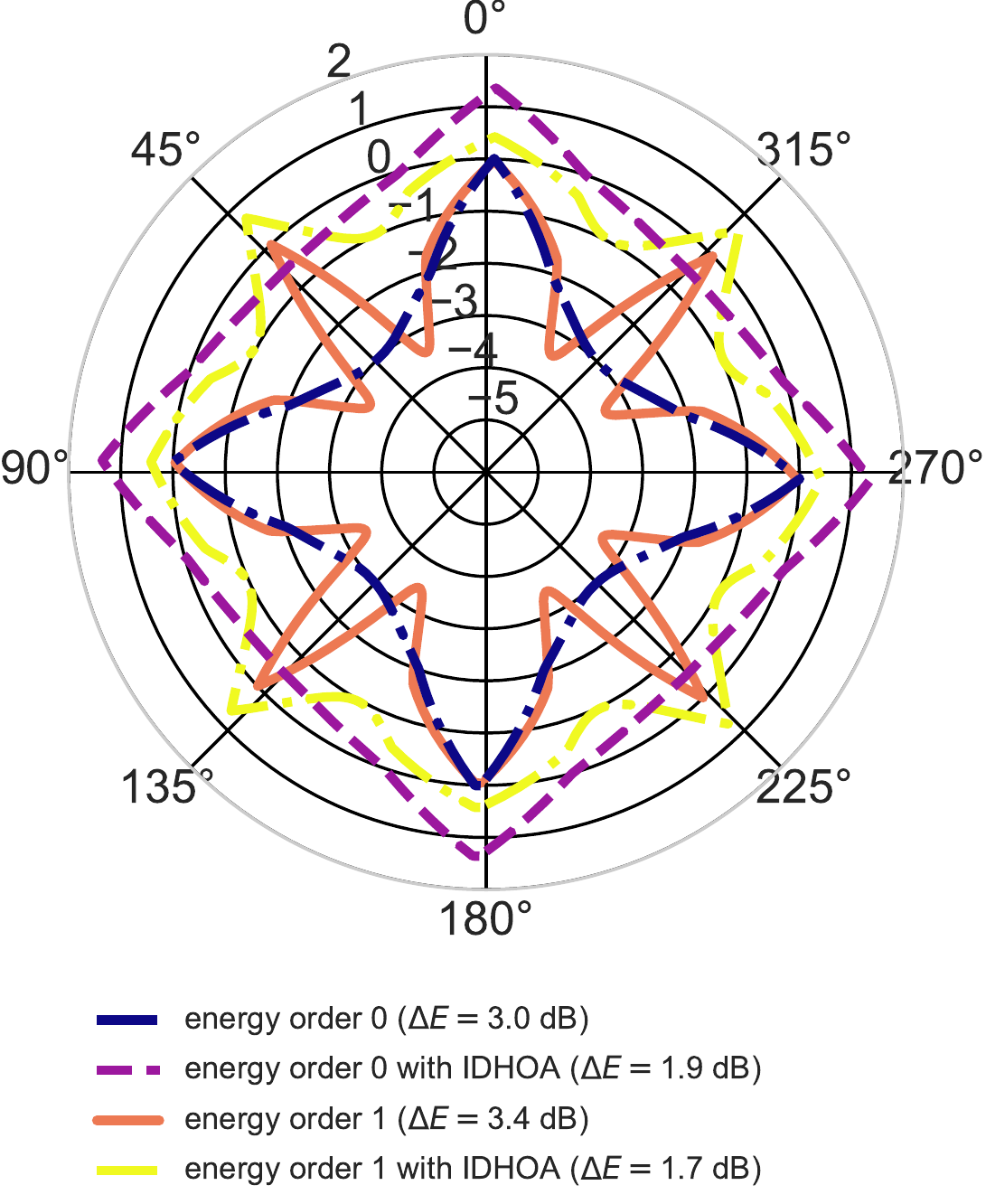}
        \caption{Energy as reconstructed by OPT-SWF alone and with the addition of the IDHOA decoding at levels 0 and 1. Scale is in dB.}
        \label{fig:mesh-energy-idhoa}
    \end{minipage}
    \hfill
    \begin{minipage}[t]{0.46\textwidth}
        \includegraphics[width=\textwidth,valign=t]{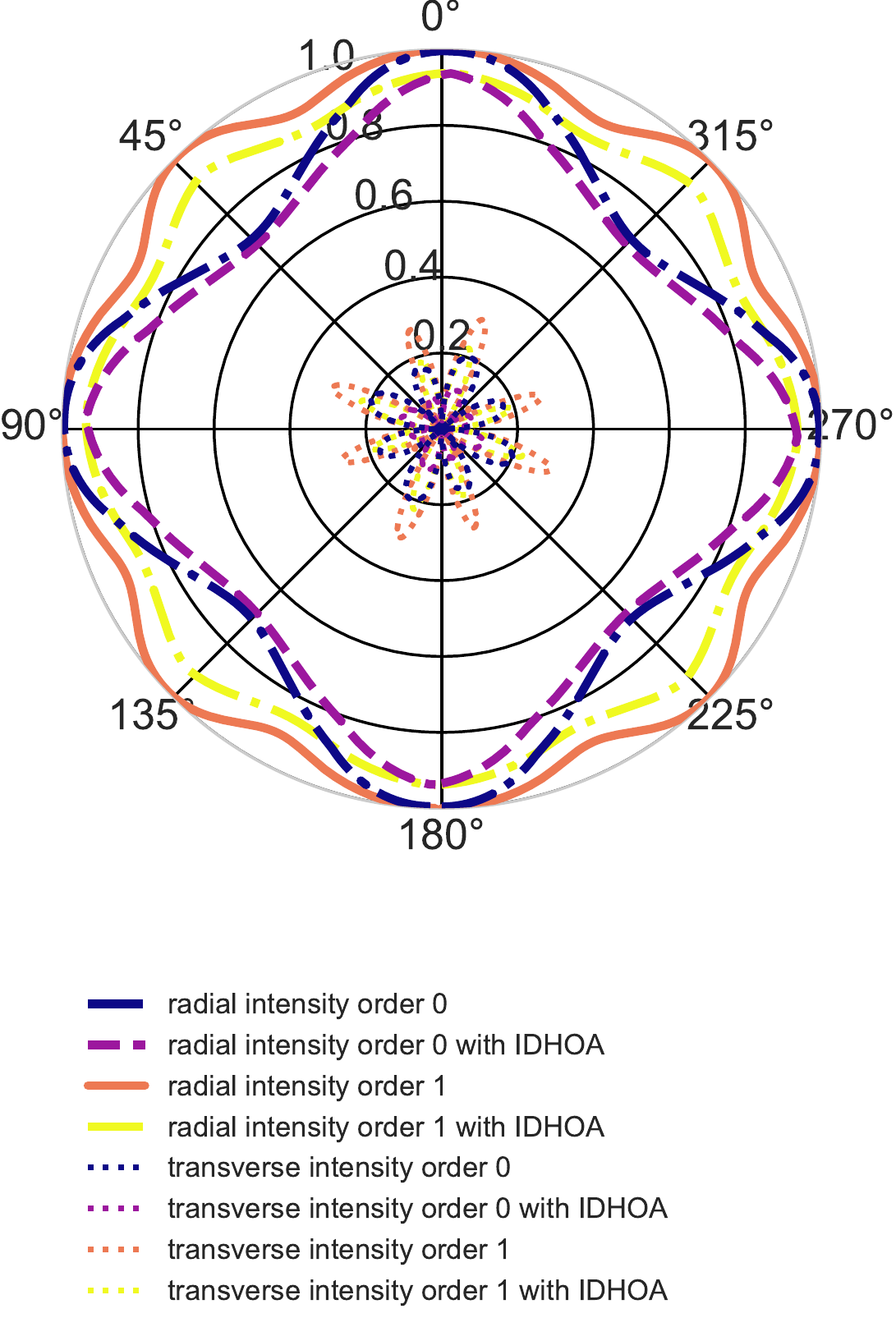}
        \caption{Intensity as reconstructed by OPT-SWF alone and with the addition of the IDHOA decoding at levels 0 and 1. Scale is in dB.}
        \label{fig:mesh-intensity-idhoa}
    \end{minipage}
\end{figure}

The downside of a good intensity reconstruction, is a quite non-uniform energy reconstruction, see
Figures~\ref{fig:mesh-energy-level0}, \ref{fig:mesh-energy-level1} and \ref{fig:mesh-energy-level0tilde}.
The variation in energy across the horizontal plane is greater or equal than 3~dB,
which is expected for a pressure-preserving panning technique but not ideal for reproduction.
In a typical workflow (e.g.\ Ambisonics) when decoding to speakers, two decoders are produced:
one that aims at pressure and velocity  reconstruction for low frequencies,
and one that aims at energy and intensity reconstruction for high frequencies,
as explained in Section~\ref{sec:obj-function}.
For this task we built, and later modified to cope with wavelets decoding, IDHOA.
We can, for example generate a decoding that attempts to properly recover the energy, while maximizing the radial intensity (max-$r_E$).
As en example, in Figures~\ref{fig:mesh-energy-idhoa} and \ref{fig:mesh-intensity-idhoa}
we demonstrate the action of IDHOA when generating a decoder for the virtual speakers on the mesh.
The result is a reduced reconstructed energy variation, less than 2~dB, at the cost of a slight reduction in the radial intensity.

These considerations are quantified numerically in Tables~\ref{tab:swf-summary-0} and \ref{tab:swf-summary-1},
where we report the 
the minimum, maximum and average values
for each observable and SWF version.
In the last column of each Table we show the action of applying the max-$r_E$ IDHOA decoding shown in
Figures~\ref{fig:mesh-energy-idhoa} for the energy and \ref{fig:mesh-intensity-idhoa} for the intensity.
It is possible to generate a different IDHOA decoder for low frequencies that preserves pressure and optimizes the velocity,
and we report the values for the improved velocity in italics in the last column of the mentioned tables.
The reconstructed pressure in this decoding scheme is exactly 1, so we preferred to omit it from the Table.

\begin{sidewaystable}

        \centering
    \begin{tabular}{lcccc}
        \toprule
        observable              & VBAP-SWF & SINT-SWF & OPT-SWF & OPT-SWF + IDHOA \\
        \midrule
        $E$ energy (dB)                  & $-1.91^{0.00}_{-3.01}$ & $-1.95^{0.00}_{-3.01}$ & $-1.95^{-0.01}_{-3.00}$ & $0.06^{1.36}_{-0.54}$ \\ \addlinespace[0.2em] \hline \addlinespace[0.2em]
        $I_{\text R}$ intensity          & $0.85^{1.00}_{0.71}  $ & $0.85^{1.00}_{0.71}  $ & $0.85^{1.00}_{0.71}   $ & $0.79^{0.94}_{0.69} $ \\ \addlinespace[0.20em]
        $I_{\text T}$ intensity          & $0.13^{0.21}_{0.00}  $ & $0.12^{0.20}_{0.00}  $ & $0.13^{0.20}_{0.00}   $ & $0.06^{0.11}_{0.00} $ \\ \addlinespace[0.20em]
        $I_{\text T}$ intensity ($\deg$) & $7.5^{12.1}_{0.0}    $ & $6.9^{11.5}_{0.0}    $ & $7.5^{11.5}_{0.0}     $ & $3.4^{6.3}_{0.0}    $ \\ \addlinespace[0.2em] \hline \addlinespace[0.2em]
        $v_{\text R}$ velocity           & $0.80^{1.00}_{0.71}  $ & $0.79^{1.00}_{0.71}  $ & $0.78^{0.96}_{0.71}   $ & $\textit{0.95}^{1.18}_{0.87}$ \\ \addlinespace[0.20em]
        $v_{\text T}$ velocity           & $0.01^{0.03}_{0.0}   $ & $0.01^{0.03}_{0.0}   $ & $0.01^{0.03}_{0.0}    $ & $\textit{0.02}^{0.04}_{0.00}$ \\ \addlinespace[0.20em]
        $v_{\text T}$ velocity ($\deg$)  & $0.6^{1.7}_{0.0}     $ & $0.6^{1.7}_{0.0}     $ & $0.6^{1.7}_{0.0}      $ & $\textit{1.1}^{2.3}_{0.0}$ \\
        \bottomrule
    \end{tabular}
    \caption{Level 0 - comparison between different SWF formats.
    Each entry reports the average and maximum and minimum values of the specified observable, $\text{avg}^{\text{max}}_{\text{min}}$.
    In italics the results for a decoder optimizing pressure and velocity.}
    \label{tab:swf-summary-0}
\end{sidewaystable}

\begin{sidewaystable}
        \centering
    \begin{tabular}{lcccc}
        \toprule
        observable              & VBAP-SWF & SINT-SWF & OPT-SWF & OPT-SWF + IDHOA \\
        \midrule
        $E$ energy (dB)                  & $-1.76^{0.00}_{-3.01}$ & $-1.76^{0.00}_{-3.01}$ & $-1.67^{0.00}_{-3.44}$ & $-0.30^{0.70}_{-0.97}$ \\ \addlinespace[0.2em] \hline \addlinespace[0.2em]
        $I_{\text R}$ intensity          & $0.96^{1.00}_{0.92}  $ & $0.96^{1.00}_{0.92}  $ & $0.96^{1.00}_{0.91}  $ & $0.90^{0.94}_{0.86}  $ \\ \addlinespace[0.20em]
        $I_{\text T}$ intensity          & $0.08^{0.14}_{0.00}  $ & $0.09^{0.18}_{0.00}  $ & $0.13^{0.30}_{0.00}  $ & $0.09^{0.22}_{0.00}  $ \\ \addlinespace[0.20em]
        $I_{\text T}$ intensity ($\deg$) & $4.6^{8.0}_{0.0}     $ & $5.2^{10.4}_{0.0}    $ & $7.5^{17.5}_{0.0}    $ & $5.2^{12.7}_{0.00}   $ \\  \addlinespace[0.2em]\hline \addlinespace[0.2em]
        $v_{\text R}$ velocity           & $0.95^{1.00}_{0.92}  $ & $0.95^{1.00}_{0.92}  $ & $0.93^{0.99}_{0.89}  $ & $\textit{0.99}^{1.07}_{0.95}$ \\ \addlinespace[0.20em]
        $v_{\text T}$ velocity           & $0.02^{0.02}_{0.00}  $ & $0.04^{0.10}_{0.00}  $ & $0.09^{0.19}_{0.0}   $ & $\textit{0.09}^{0.20}_{0.00}$ \\ \addlinespace[0.20em]
        $v_{\text T}$ velocity ($\deg$)  & $1.1^{1.1}_{0.00}    $ & $2.3^{5.7}_{0.00}    $ & $5.2^{10.9}_{0.0}    $ & $\textit{5.2}^{11.5}_{0.0}$ \\
        \bottomrule
    \end{tabular}
    \caption{Level 1 - comparison between different SWF formats.
    Each entry reports the average and maximum and minimum values of the specified observable, $\text{avg}^{\text{max}}_{\text{min}}$.
    In italics the results for a decoder optimizing pressure and velocity.}
    \label{tab:swf-summary-1}
\end{sidewaystable}

\begin{figure}
        \centering
    \begin{minipage}[t]{0.46\textwidth}
        \includegraphics[width=\textwidth,valign=t]{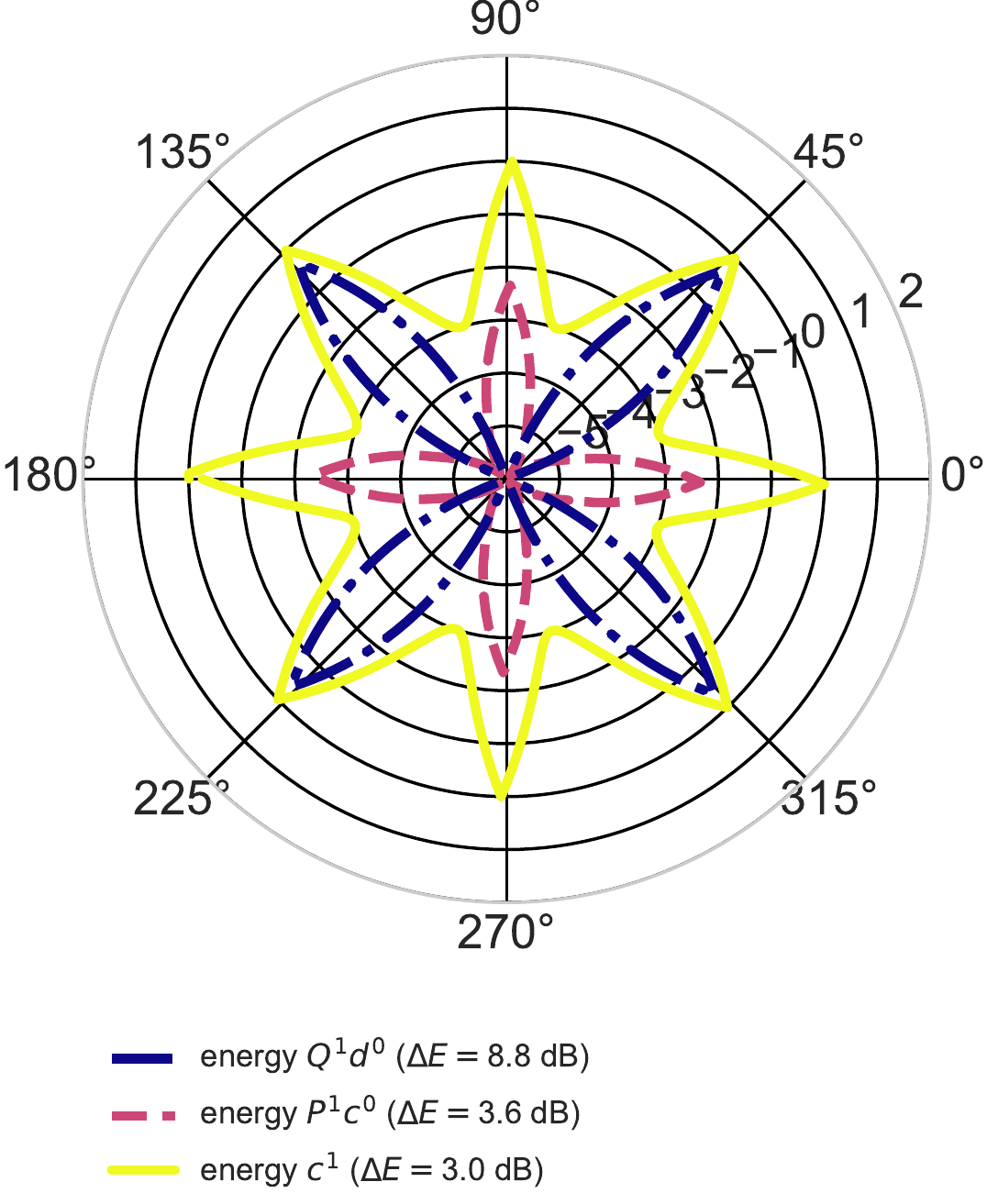}
        \caption{Energy contributions from $\tilde{\mathbf{c}}^0 = \mathbf P^1 \mathbf c^0 $ and
        $\tilde{\mathbf{d}}^0 = \mathbf Q^1 \mathbf d^0 $, which constitute  $\mathbf{c}^1$, for SINT-SWF. Scale is in dB.}
        \label{fig:mesh-energy-d0interp}
    \end{minipage}
    \hfill
    \begin{minipage}[t]{0.46\textwidth}
        \includegraphics[width=\textwidth,valign=t]{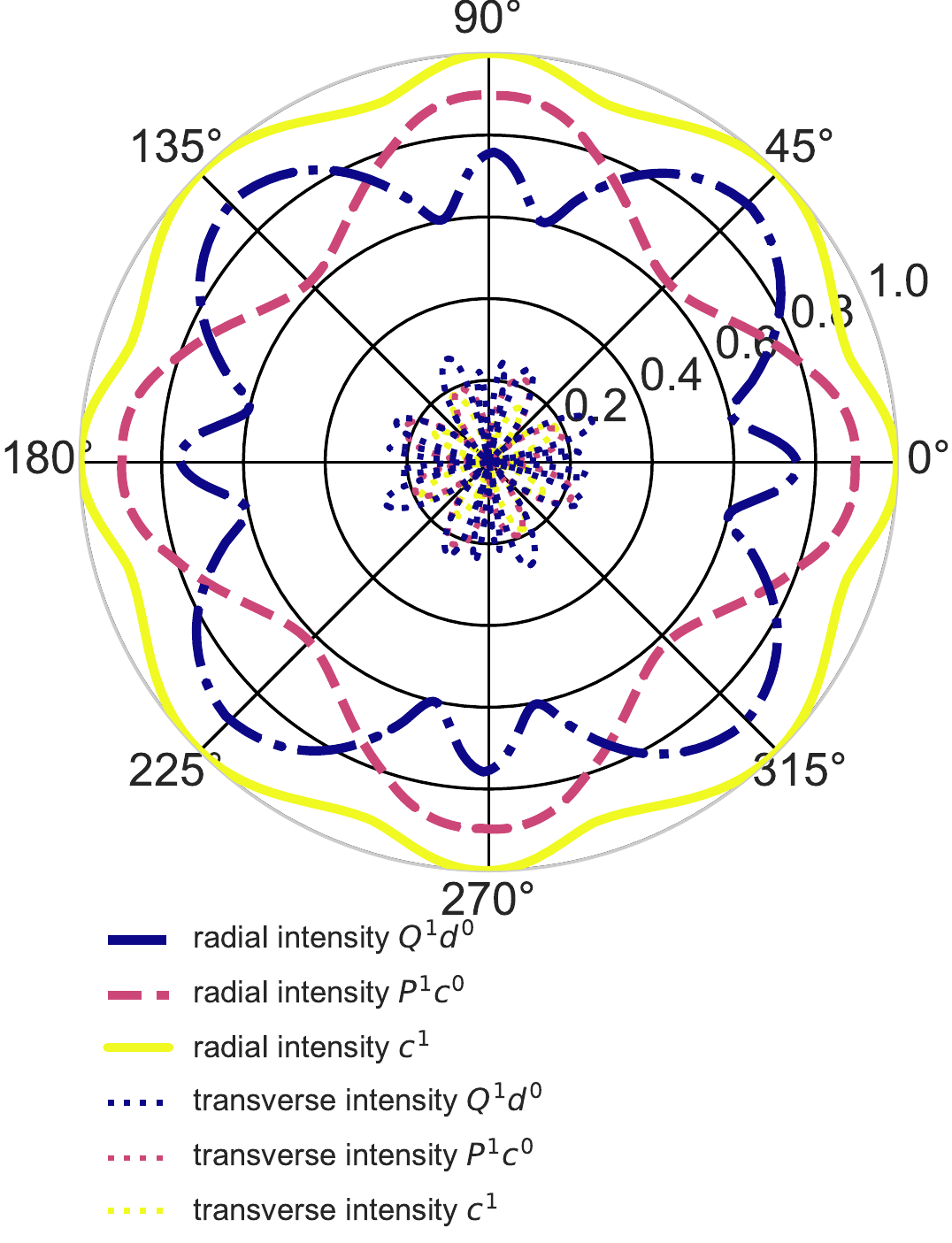}
        \caption{Intensity contributions from $\tilde{\mathbf{c}}^0 = \mathbf P^1 \mathbf c^0 $ and
        $\tilde{\mathbf{d}}^0 = \mathbf Q^1 \mathbf d^0 $, which constitute  $\mathbf{c}^1$, for SINT-SWF.}
        \label{fig:mesh-intensity-d0interp}
    \end{minipage}
\end{figure}

\begin{figure}
        \centering
    \begin{minipage}[t]{0.46\textwidth}
        \includegraphics[width=\textwidth,valign=t]{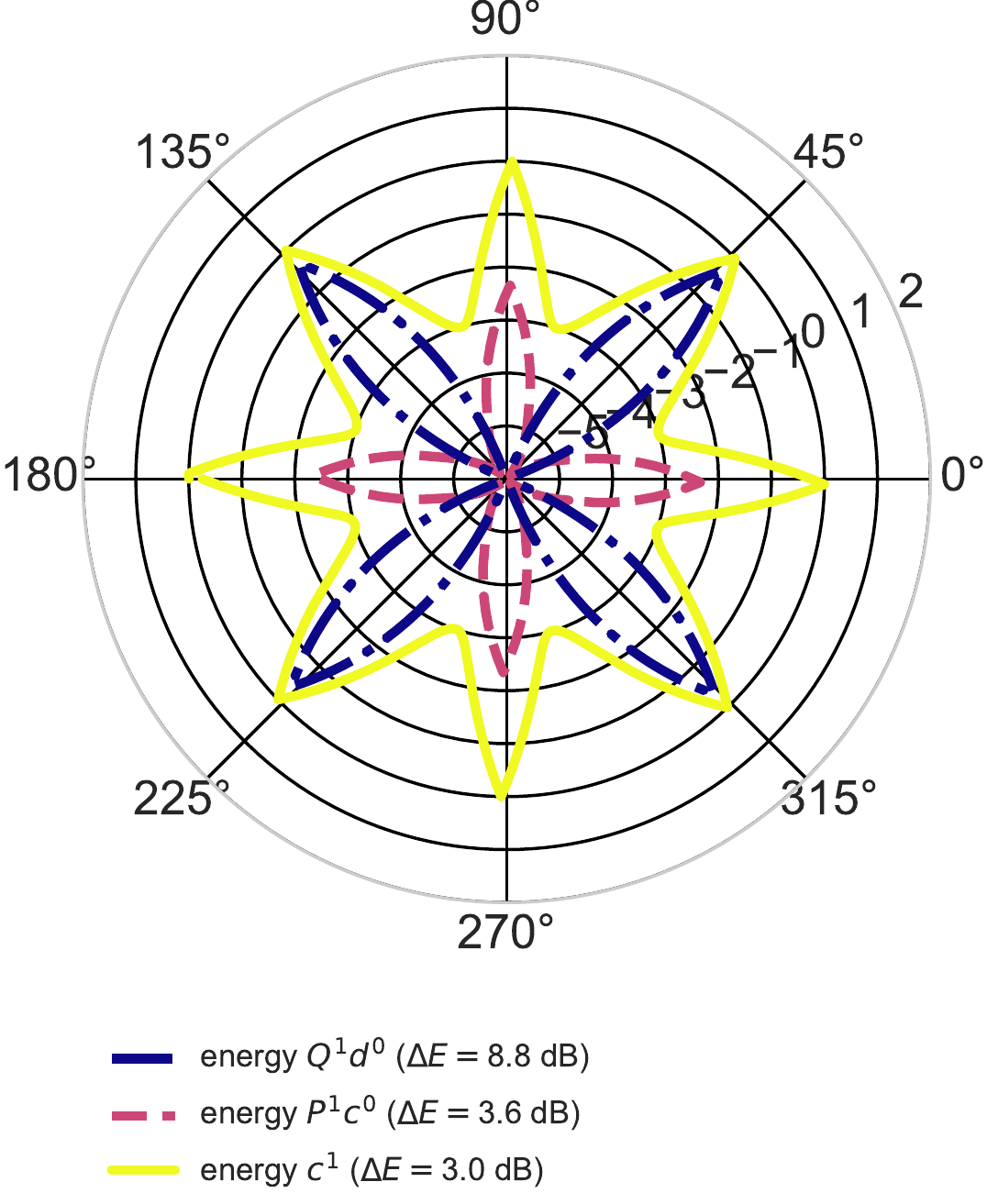}
        \caption{Energy contributions from $\tilde{\mathbf{c}}^0 = \mathbf P^1 \mathbf c^0 $ and
        $\tilde{\mathbf{d}}^0 = \mathbf Q^1 \mathbf d^0 $, which constitute  $\mathbf{c}^1$, for OPT-SWF. Scale is in dB.}
        \label{fig:mesh-energy-d0opt}
    \end{minipage}
    \hfill
    \begin{minipage}[t]{0.46\textwidth}
        \includegraphics[width=\textwidth,valign=t]{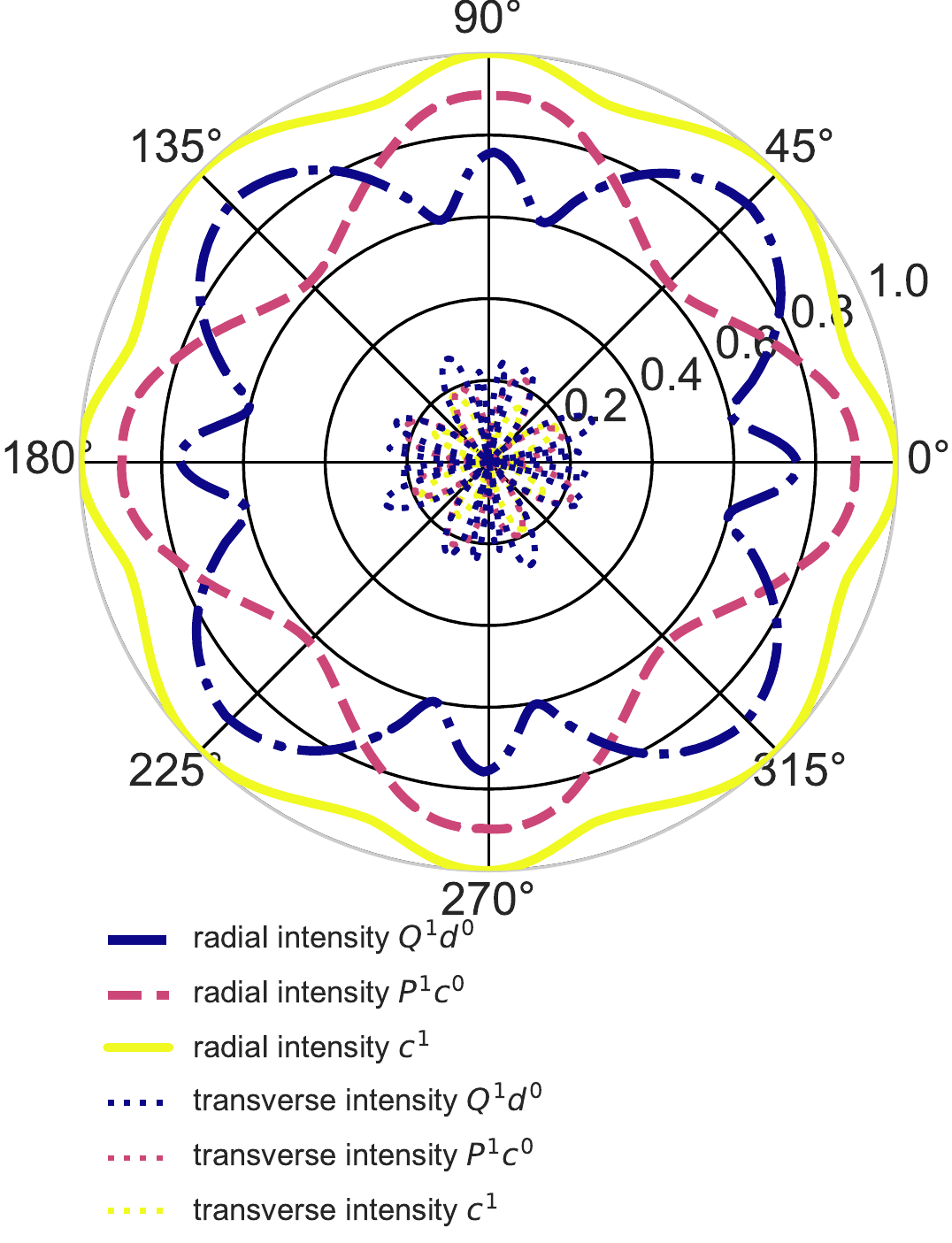}
        \caption{Intensity contributions from $\tilde{\mathbf{c}}^0 = \mathbf P^1 \mathbf c^0 $ and
        $\tilde{\mathbf{d}}^0 = \mathbf Q^1 \mathbf d^0 $, which constitute  $\mathbf{c}^1$, for OPT-SWF.}
        \label{fig:mesh-intensity-d0opt}
    \end{minipage}
\end{figure}

As a final task, it is interesting to inspect what are the physical quantities carried by the $\mathbf{d}^0$, for SINT-SWF and OPT-SWF.
In Figures~\ref{fig:mesh-energy-d0interp} and Figures~\ref{fig:mesh-intensity-d0interp} we report the reconstructed energy and intensity carried by
$\tilde{\mathbf{c}}^0 = \mathbf P^1 \mathbf c^0 $, $\tilde{\mathbf{d}}^0 = \mathbf Q^1 \mathbf d^0 $ and $\mathbf{c}^1$
for SINT-SWF. The same quantities are shown in Figures~\ref{fig:mesh-energy-d0opt} and \ref{fig:mesh-intensity-d0opt} for OPT-SWF.
We can conclude that the physical quantities carried by the $\mathbf d^0$ are mainly energy and intensity, while their contribution
to the pressure is zero by design.

\section{Summary}
In this Chapter we constructed three versions of SWF that share the same subdivision mesh, but differ in the wavelet families that constitute the SWF filters.
Even if the three wavelet families are built in very different ways (VBAP interpolation, lifting scheme, numerical optimization),
when looking at physical reconstructed quantities, they perform in a remarkably similar way.

\chapter{Objective Evaluation of the Decoding of SWF and Ambisonics} \label{ch:idhoa-evaluation-swf-ambi}
In this Chapter we compare a VBAP-SWF and OPT-SWF implementations with Higher Order Ambisonics, both decoded with IDHOA to a standard layout.

\section{Objective Evaluation of VBAP-SWF, OPT-SWF and Ambisonics for the 7.0.4 Layout}\label{sec:eval-swf-ambi}

\begin{figure}[h]
    \centering
    \includegraphics[width=0.7\textwidth]{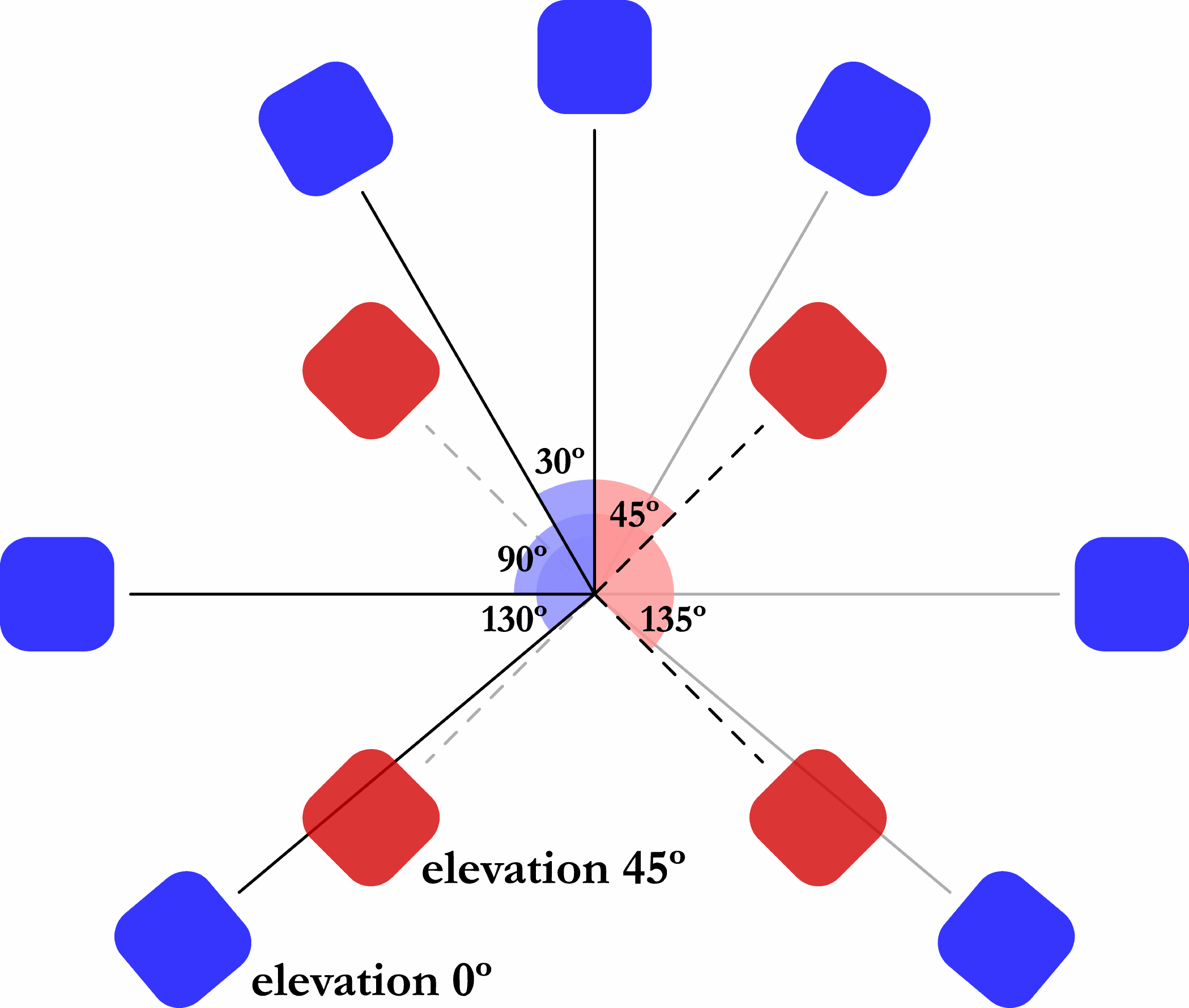}
    \caption{Detailed view of 7.0.4 speakers' layout.}
    \label{fig:704layout}
\end{figure}

As a first comparison,
the chosen destination speakers' layout is a standard 7.0.4, meaning: 7 speakers on the horizontal plane, 0 LFE, 4 ceiling speakers,
located as shown in Figure~\ref{fig:704layout}.
The benefits of this layout are that it is an industry-standard layout with loudspeakers above the horizontal plane,
that can be easily replicated and is meaningful.
With meaningful we mean than has a sufficient number of elevated speakers to distinguish front, back, left and right
in the upper part of the layout (which is not the case for the 5.0.2).
Moreover, the number of loudspeakers is still moderate with respect to other industry standards, e.g.\ 9.0.6, 22.2,
and it is increasingly common in mixing facilities.
\enlargethispage{\baselineskip} %

With reference to the signal processing scenarios described in the introduction of this Chapter,
for the sake of clarity we can represent them schematically as follows.
The first scenario $f \rightarrow f_{66} \rightarrow c_{6} \rightarrow \mathbf s^0_{7.0.4}$:
\begin{align*}
    f \left( \theta, \phi \right) &\rightarrow \mathbf{f}_{66} \\
    \mathbf{c}^0_{6} &= \mathbf{A}^1_{6\times18} \mathbf{A}^2_{18\times66} \mathbf{f}^2_{66} \\
    \mathbf s^0_{7.0.4} &= \mathbf D^{0}_{7.0.4\times6} \mathbf{c}^0_{6}
\end{align*}
The second scenario $f \rightarrow f_{66} \rightarrow c_{18} \rightarrow \mathbf s^1_{7.0.4}$:
\begin{align*}
    f \left( \theta, \phi \right) &\rightarrow \mathbf{f}_{66} \\
    \mathbf{c}^1_{18} &= \mathbf{A}^2_{18\times66} \mathbf{f}^2_{66} \\
    \mathbf s^1_{7.0.4} &= \mathbf D^{1}_{7.0.4\times18} \mathbf{c}^1_{18}
\end{align*}
Finally, the third scenario $f \rightarrow f_{66} \rightarrow c_{6} \rightarrow \tilde{c}_{18} \rightarrow \tilde{\mathbf s}^0$:
\begin{align*}
    f \left( \theta, \phi \right) &\rightarrow \mathbf{f}_{66} \\
    \mathbf{c}^0_{6} &= \mathbf{A}^1_{6\times18} \mathbf{A}^2_{18\times66} \mathbf{f}^2_{66} \\
    \mathbf{\tilde{c}}^0_{18} &= \mathbf{P}^1_{18\times6} \mathbf{c}^0_{6} \\
    \tilde{\mathbf s}^0_{7.0.4} &= \mathbf D^{1}_{7.0.4\times18} \mathbf{\tilde{c}}^0_{18}
\end{align*}

The stages outlined here, encoding (consisting of interpolation and decomposition), transmission,
upsampling (optional) and decoding to speakers, are graphically illustrated in Figure~\ref{scheme:encode-decode}
with the number of channels annotated in red close to each box.
Note that the downsampling, upsampling and decoding steps are kept separate for clarity, but they can be performed as one single matrix product.

\begin{table}[t]
    \centering
    \begin{tabular}{lcc}
        \toprule
        type & order & \# channels \\
        \midrule
        Ambisonics & 1 & 4  \\
        Ambisonics & 2 & 9  \\
        Ambisonics & 3 & 16 \\
        \bottomrule
    \end{tabular}
    \hspace{2ex}
    \begin{tabular}{lcc}
        \toprule
        type & level & \# channels \\
        \midrule
        Wavelet & 0 & 6 \\
        Wavelet & 1 & 18 \\
        Wavelet & $\tilde{0}$ & 6 ($\tilde{18}$) \\
        \bottomrule
    \end{tabular}
    \caption{Comparison of number of channels per order/level for Ambisonics, on the left, and Wavelets, on the right.}
    \label{tab:numchans-swf-ambi}
\end{table}

Given the difference in number of channels between Ambisonics orders and SWF levels, see Table~\ref{tab:numchans-swf-ambi} it is difficult to fairly compare the two.
Moreover, the destination layout has 11 speakers and theoretically, for a regular Ambisonics layout, only up to second order 3D Ambisonics could be decoded to it.
Given these facts,
we show the activation gains, reconstructed energy and intensity in the horizontal plane only for the following formats decoded to 7.0.4:
$1^{st}$, $2^{nd}$, $3^{rd}$ order Ambisonics, 0, 1, $\tilde{0}$ levels in OPT-SWF and VBAP-SWF.
For the SWF (both flavours) we report the resulting gains obtained with two fairly extreme configurations of IDHOA,
and hence two very different decoders.
The first decoder is designed to mimic the smoothness of Ambisonics (at $1^{st}$, $2^{nd}$ order),
we call it \emph{smooth} in the following.
The second one is built with the intent of activating less (neighbouring) speakers possible, mimicking a VBAP-like behaviour,
and we call it \emph{focus}.
The decoders are obtained by balancing the terms of energy reconstruction,
and the request for focused sources represented by the radial intensity, both described in Section~\ref{sec:obj-function}.

%
%
\subsection{Comparison between SWF Levels \texorpdfstring{$\mathbf{0}$}{0} and \texorpdfstring{$\mathbf{\tilde{0}}$}{0-tilde} with Ambisonics Orders 1 and 2}
\label{sec:comparison0}

\begin{figure}[t!]
    \centering
    \begin{minipage}[t]{0.46\textwidth}
    \includegraphics[width=\textwidth,valign=t]{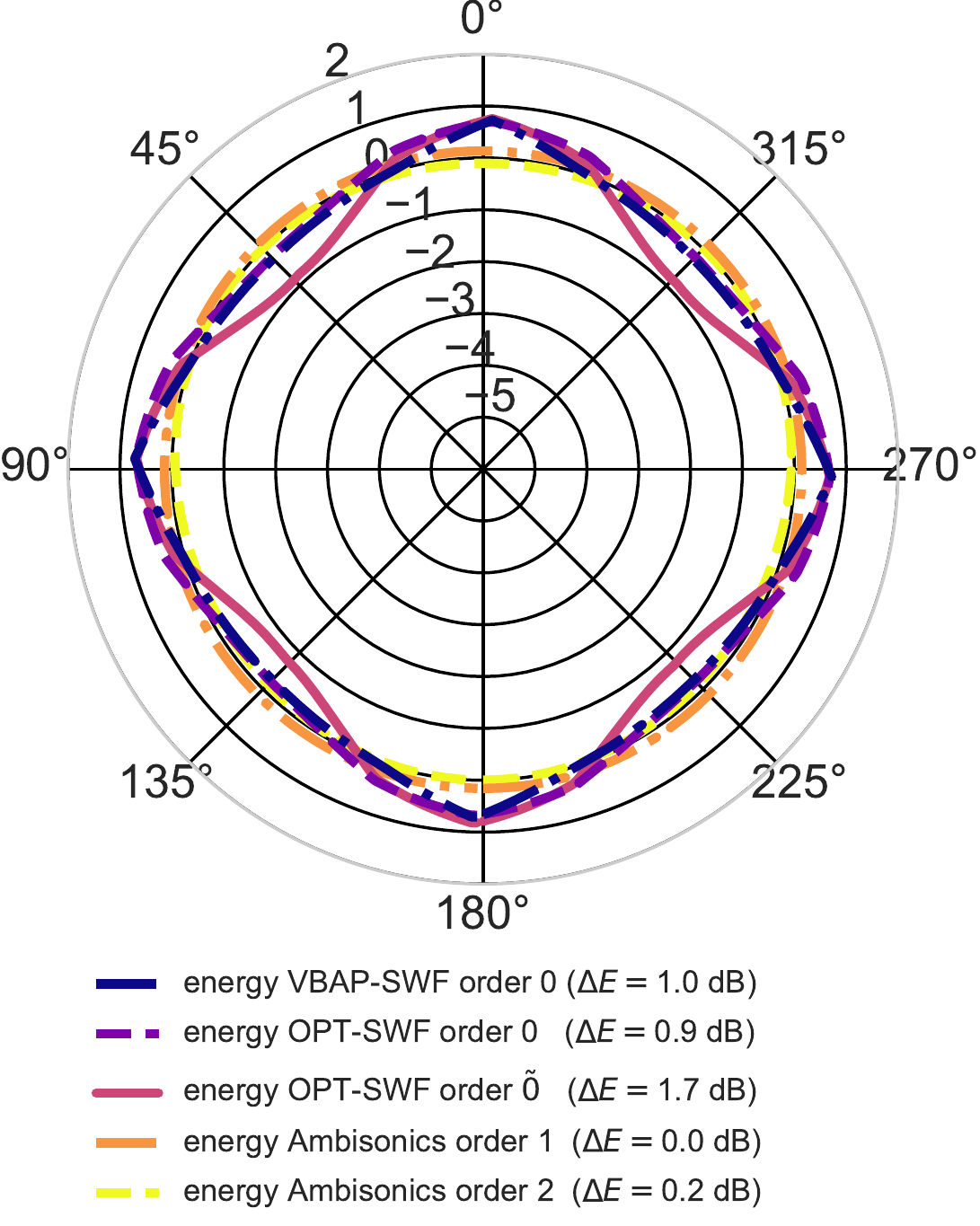}
    \caption{Energy comparison for levels and orders with similar channel count:
             SWF at level 0 with the \emph{smooth} decoding preset and Ambisonics at {order 1} and {2}, decoded to a 7.0.4 layout. Scale is in dB.}
    \label{fig:energy-level0}
    \end{minipage}
    \hfill
    \begin{minipage}[t]{0.46\textwidth}
    \includegraphics[width=\textwidth,valign=t]{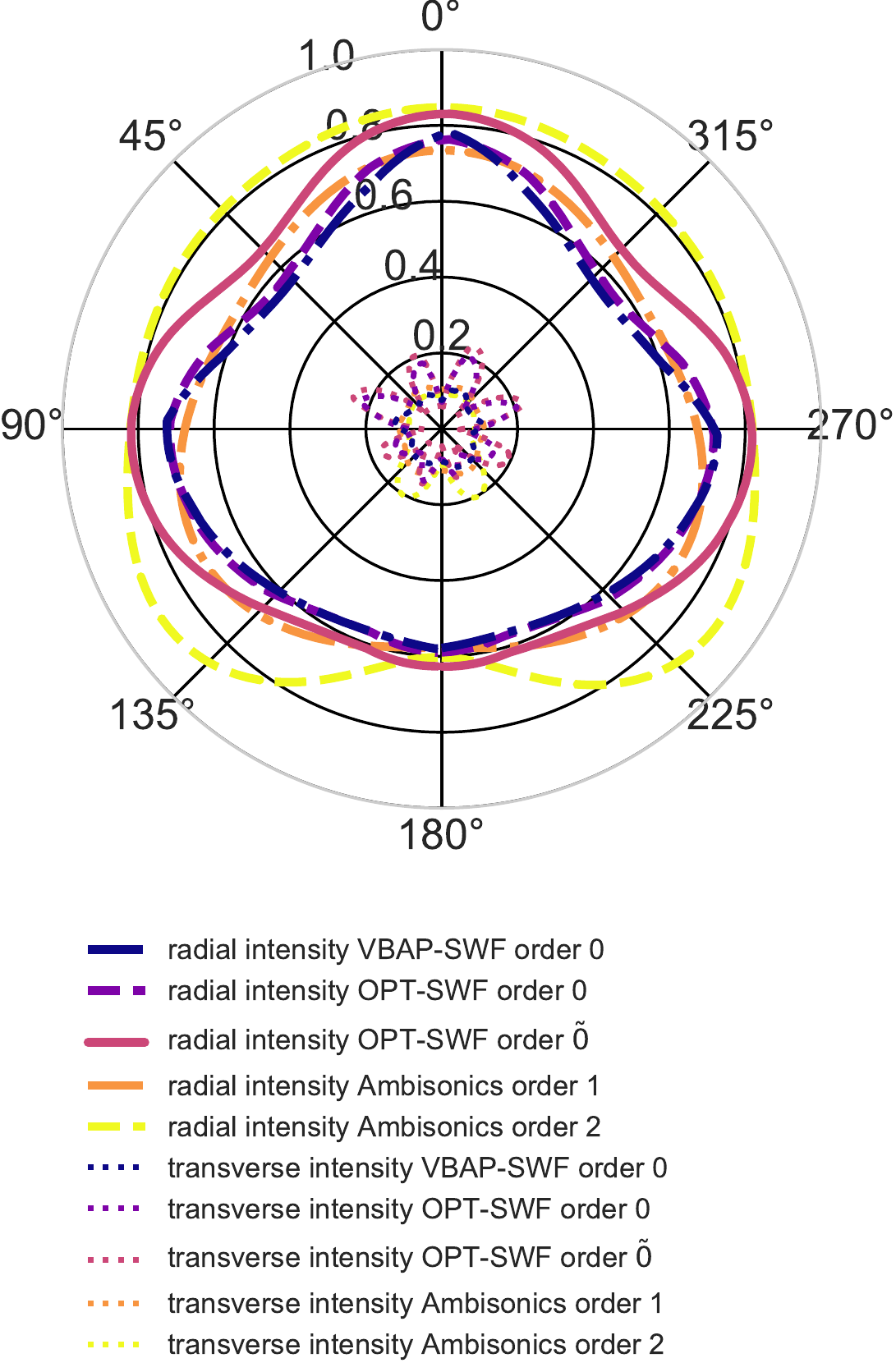}
    \caption{Intensity comparison for levels (with the \emph{smooth} decoding preset) and orders with similar channel count.
            The point and dash lines represent radial intensity and the dashed ones the transverse intensity component.}
    \label{fig:intensity-level0}
    \end{minipage}
\end{figure}

\begin{figure}
    \centering
    \includegraphics[width=\textwidth,valign=t]{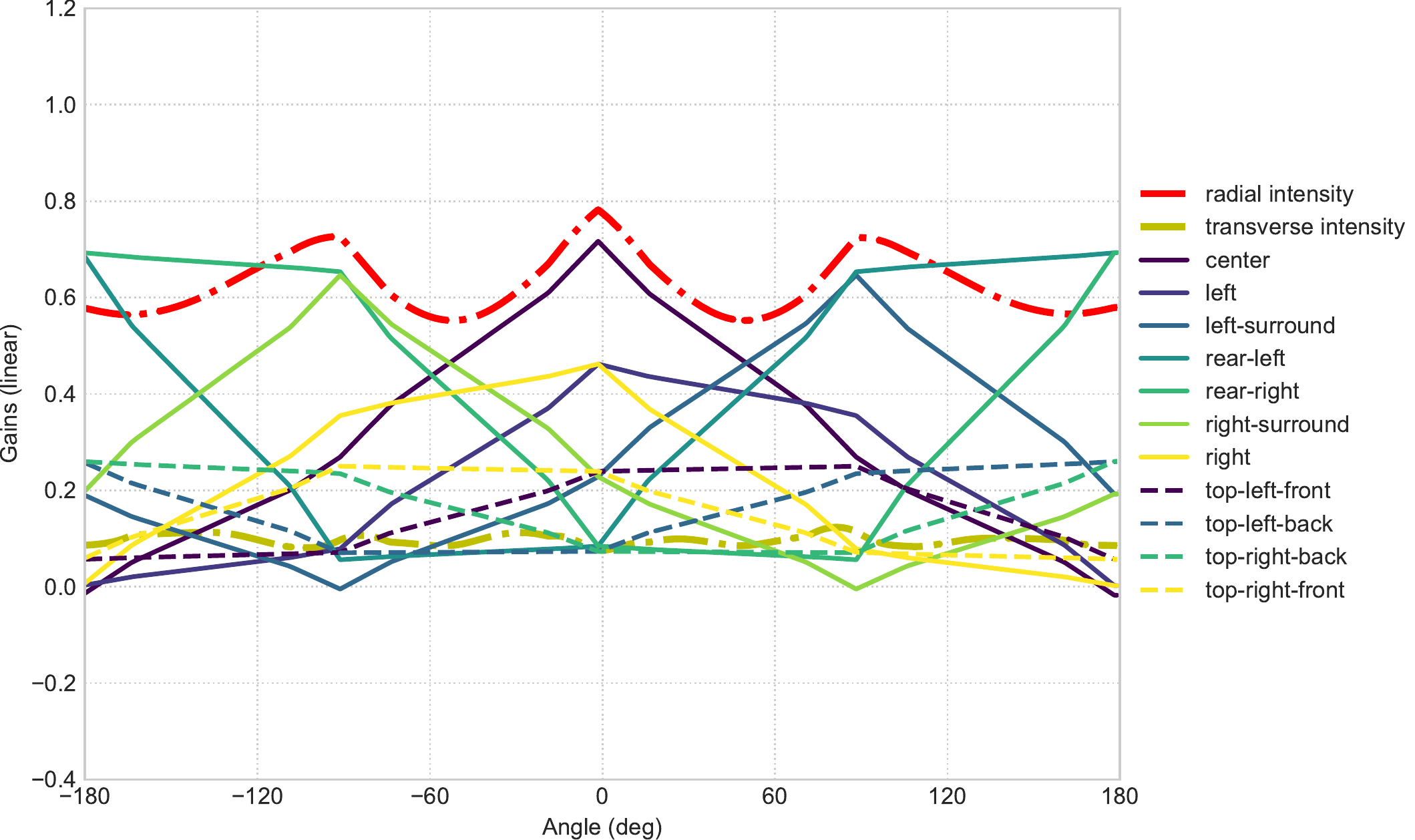}
    \caption{Horizontal panning for VBAP-SWF at level 0 decoded to 7.0.4 layout, using the \emph{smooth} decoding preset.}
    \label{fig:vbap-panfunc-66-6-11-linear}

    \includegraphics[width=\textwidth,valign=t]{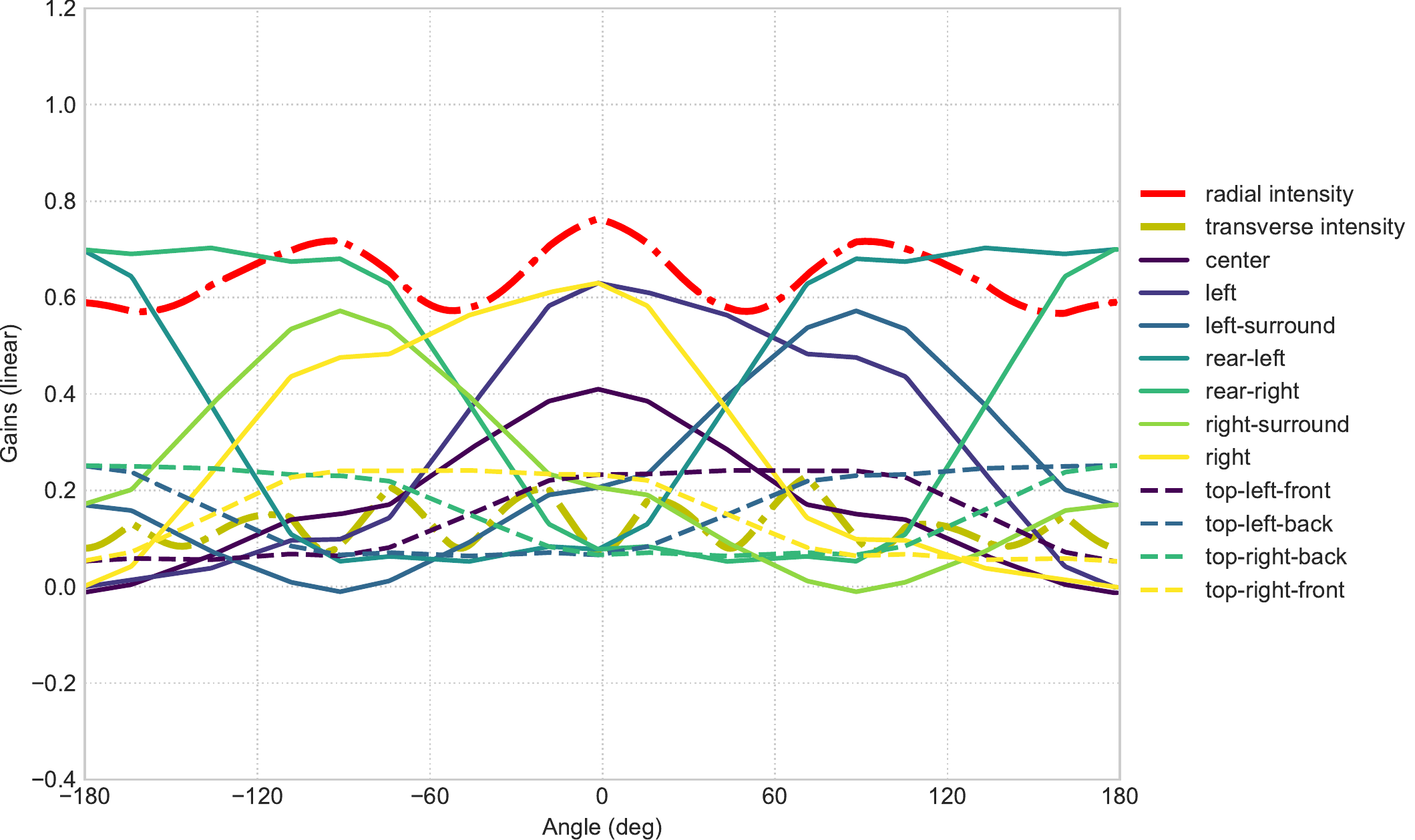}
    \caption{Horizontal panning for OPT-SWF at level 0 decoded to 7.0.4 layout, using the \emph{smooth} decoding preset.}
    \label{fig:wv-panfunc-66-6-11-linear}
\end{figure}

\begin{figure}
    \includegraphics[width=\textwidth,valign=t]{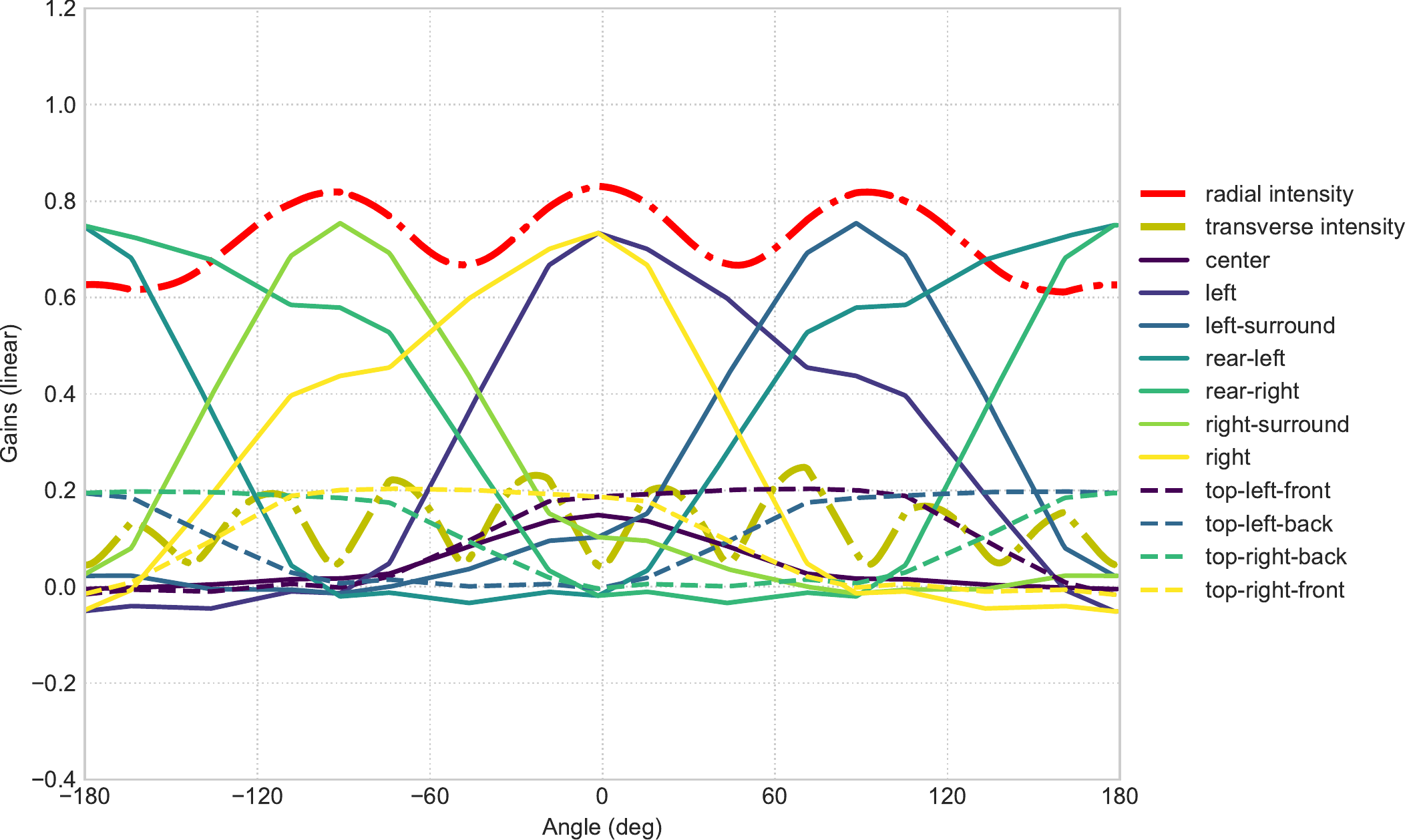}
    \caption{Horizontal panning for OPT-SWF at level $\tilde{0}$ decoded to 7.0.4 layout, using the \emph{smooth} decoding preset.}
    \label{fig:wv-panfunc-66-6-18-11-linear}

    \includegraphics[width=\textwidth,valign=t]{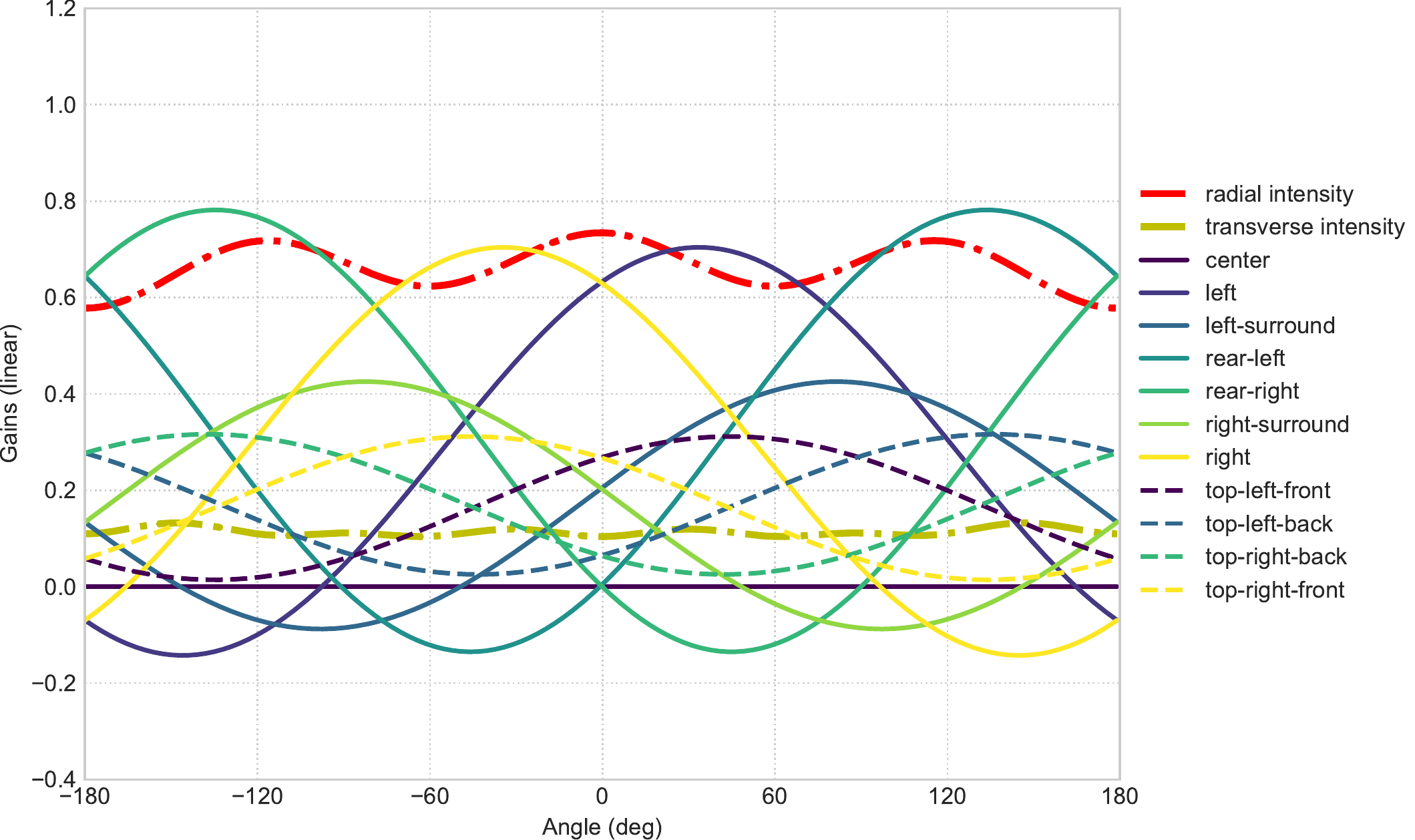}
    \caption{Horizontal panning for Ambisonics at order 1 decoded to 7.0.4 layout.}
    \label{fig:ambi1-panfunc-again-linear}
\end{figure}

\begin{figure}[t!]
    \centering
    \begin{minipage}[t]{0.46\textwidth}
        \includegraphics[width=\textwidth,valign=t]{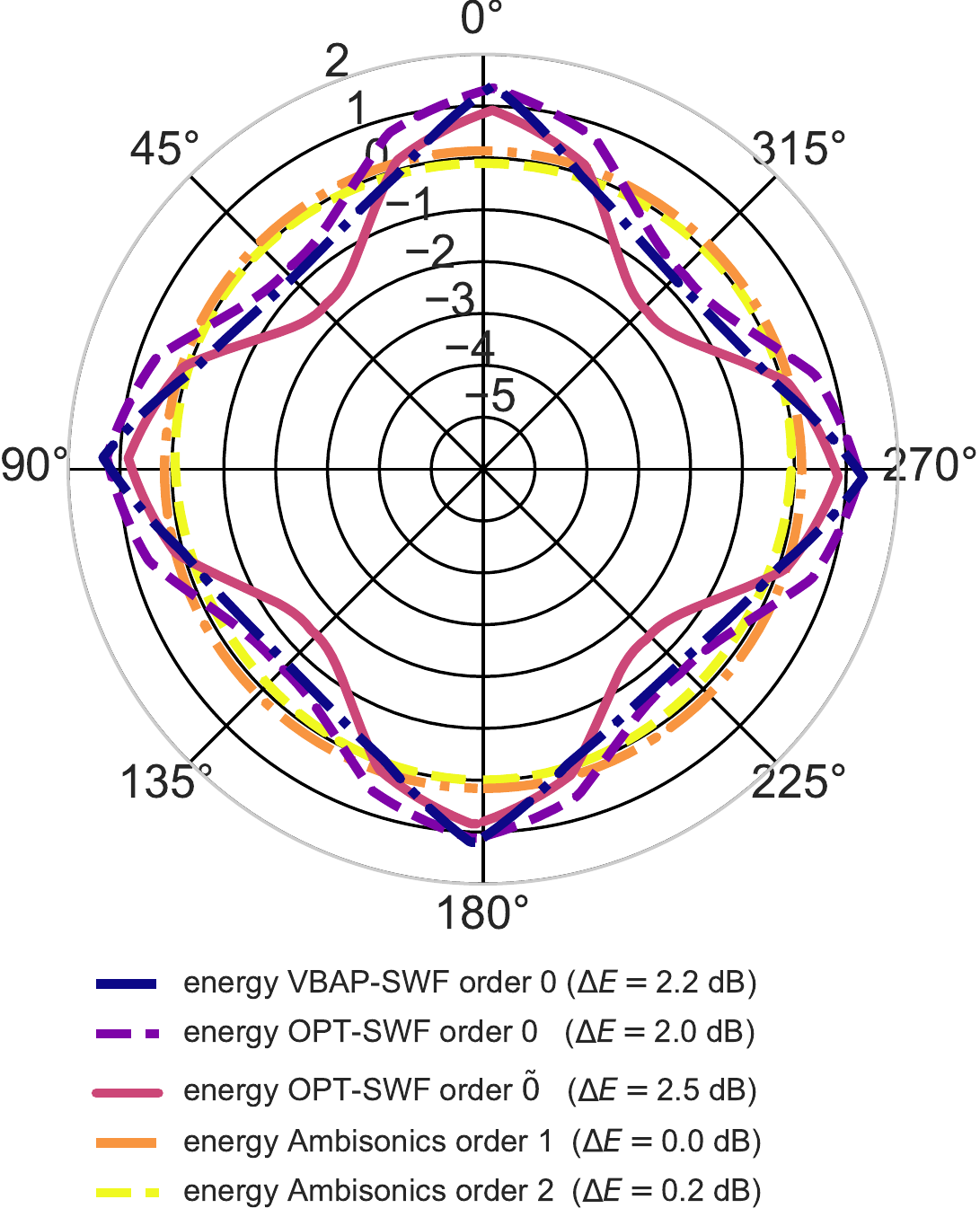}
        \caption{Energy comparison for levels and orders with similar channel count:
                 SWF at level 0 with the \emph{focus} decoding preset and Ambisonics at {order 1} and {2}, decoded to a 7.0.4 layout. Scale is in dB.}
        \label{fig:energy-level0-focus}
    \end{minipage}
    \hfill
    \begin{minipage}[t]{0.46\textwidth}
        \includegraphics[width=\textwidth,valign=t]{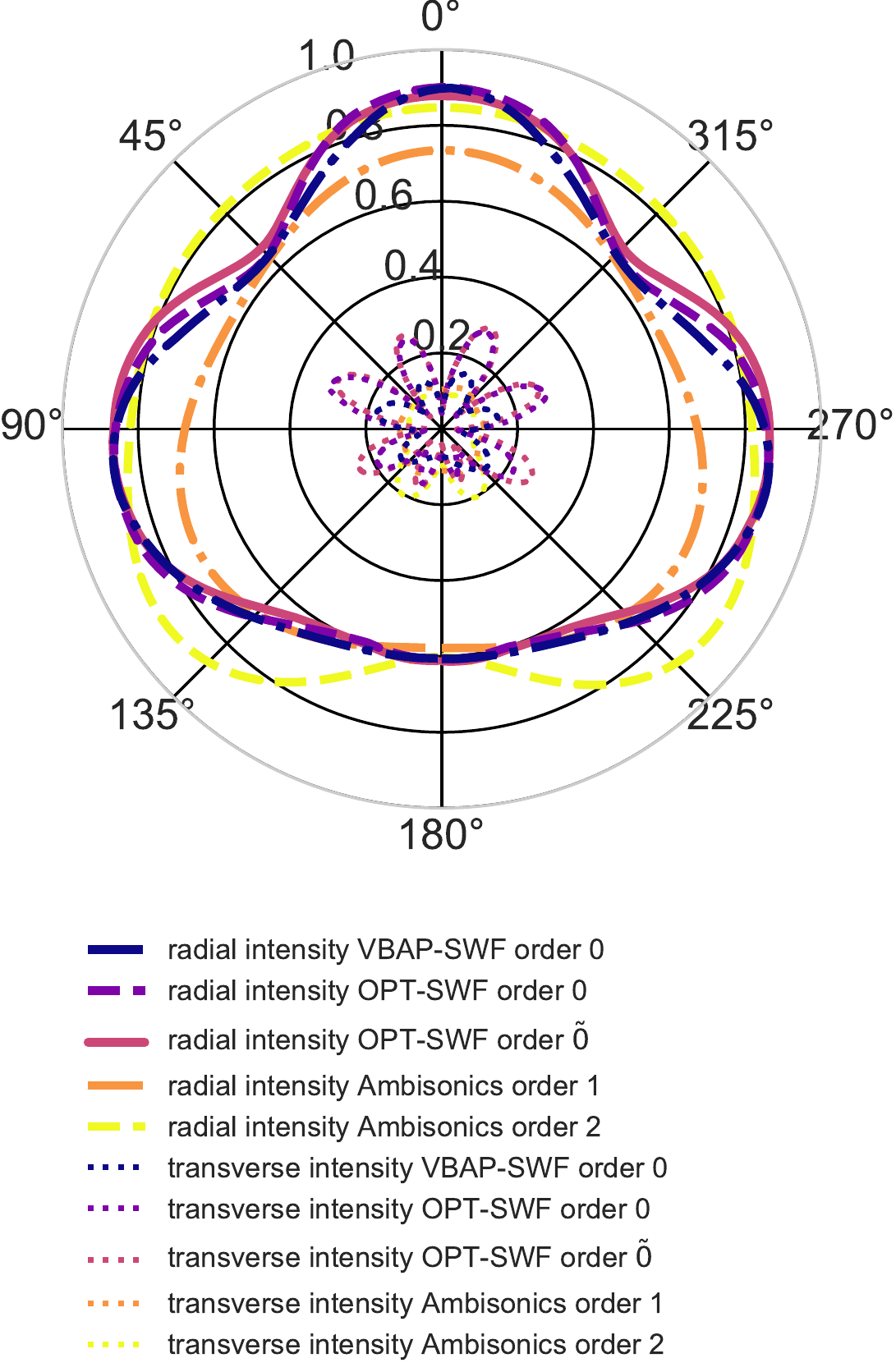}
        \caption{Intensity comparison for levels (with the \emph{focus} decoding preset) and orders with similar channel count.
                The dashed lines represent radial intensity and the dotted ones the transverse intensity component.}
        \label{fig:intensity-level0-focus}
    \end{minipage}
\end{figure}

\begin{figure}
    \centering
    \includegraphics[width=\textwidth]{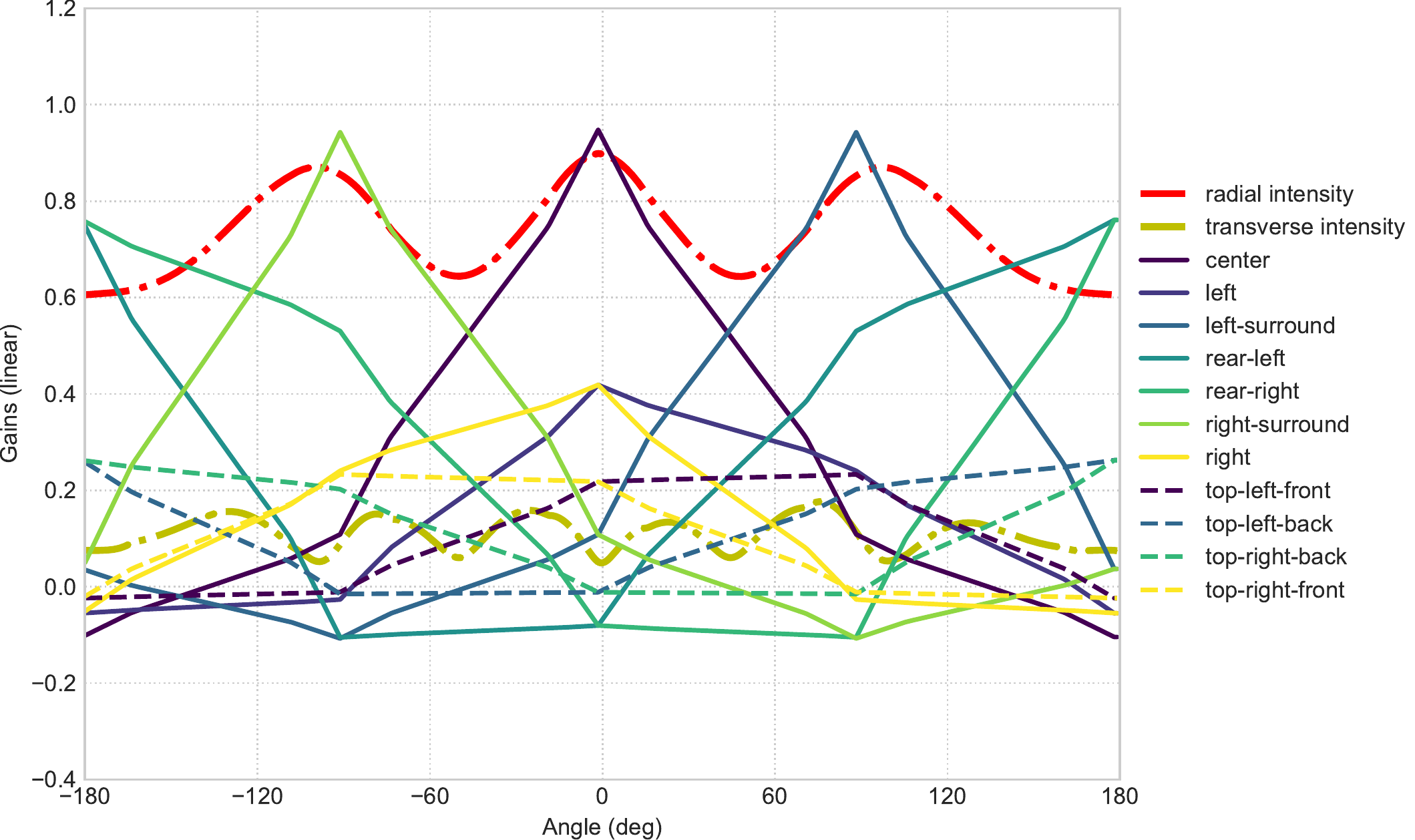}
    \caption{Horizontal panning for VBAP-SWF at level 0 decoded to 7.0.4 layout, using the \emph{focus} decoding preset.}
    \label{fig:vbap-panfunc-66-6-11-linear-focus}

    \includegraphics[width=\textwidth]{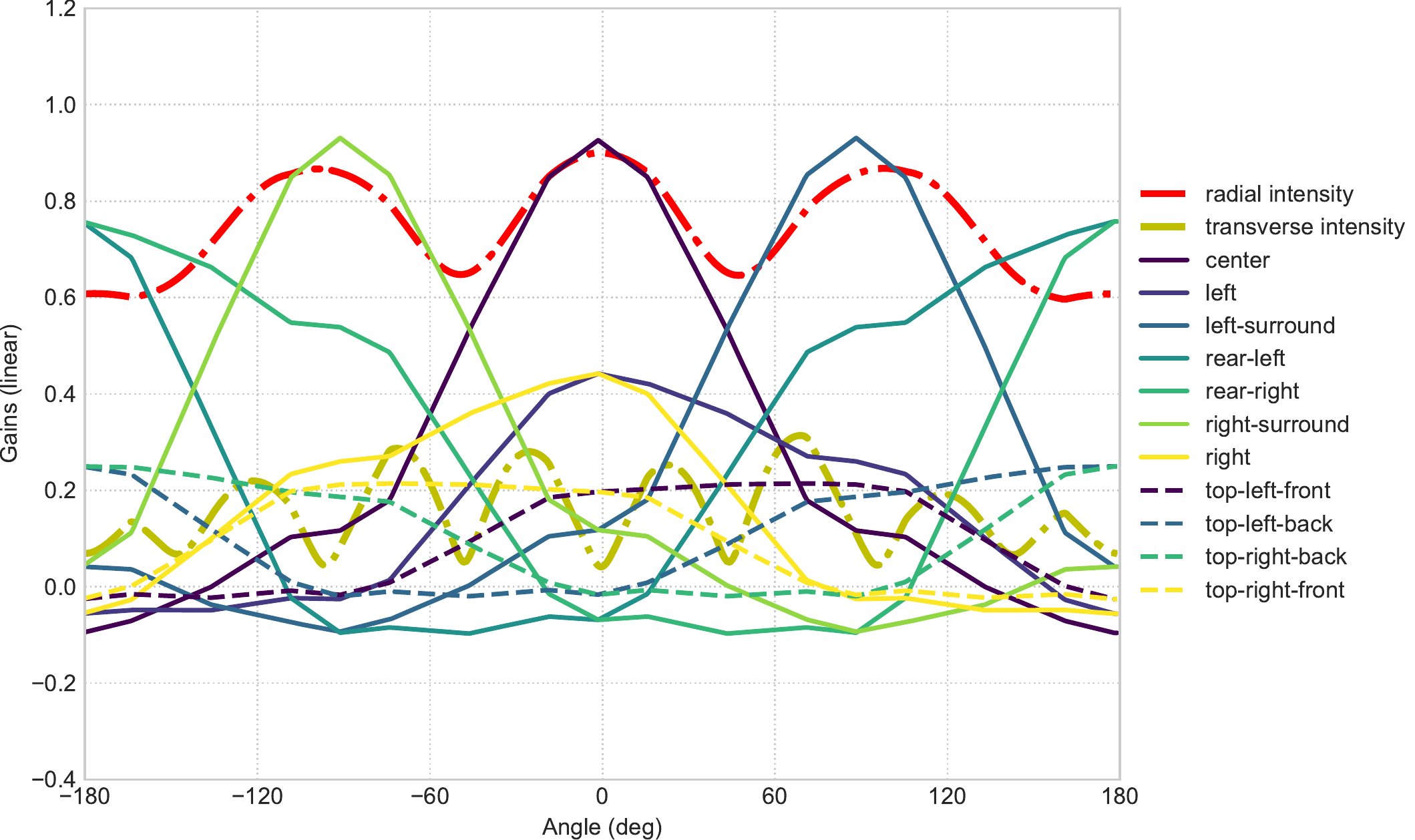}
    \caption{Horizontal panning for OPT-SWF at level 0 decoded to 7.0.4 layout, using the \emph{focus} decoding preset.}
    \label{fig:wv-panfunc-66-6-11-linear-focus}
\end{figure}

\begin{figure}
    \includegraphics[width=\textwidth]{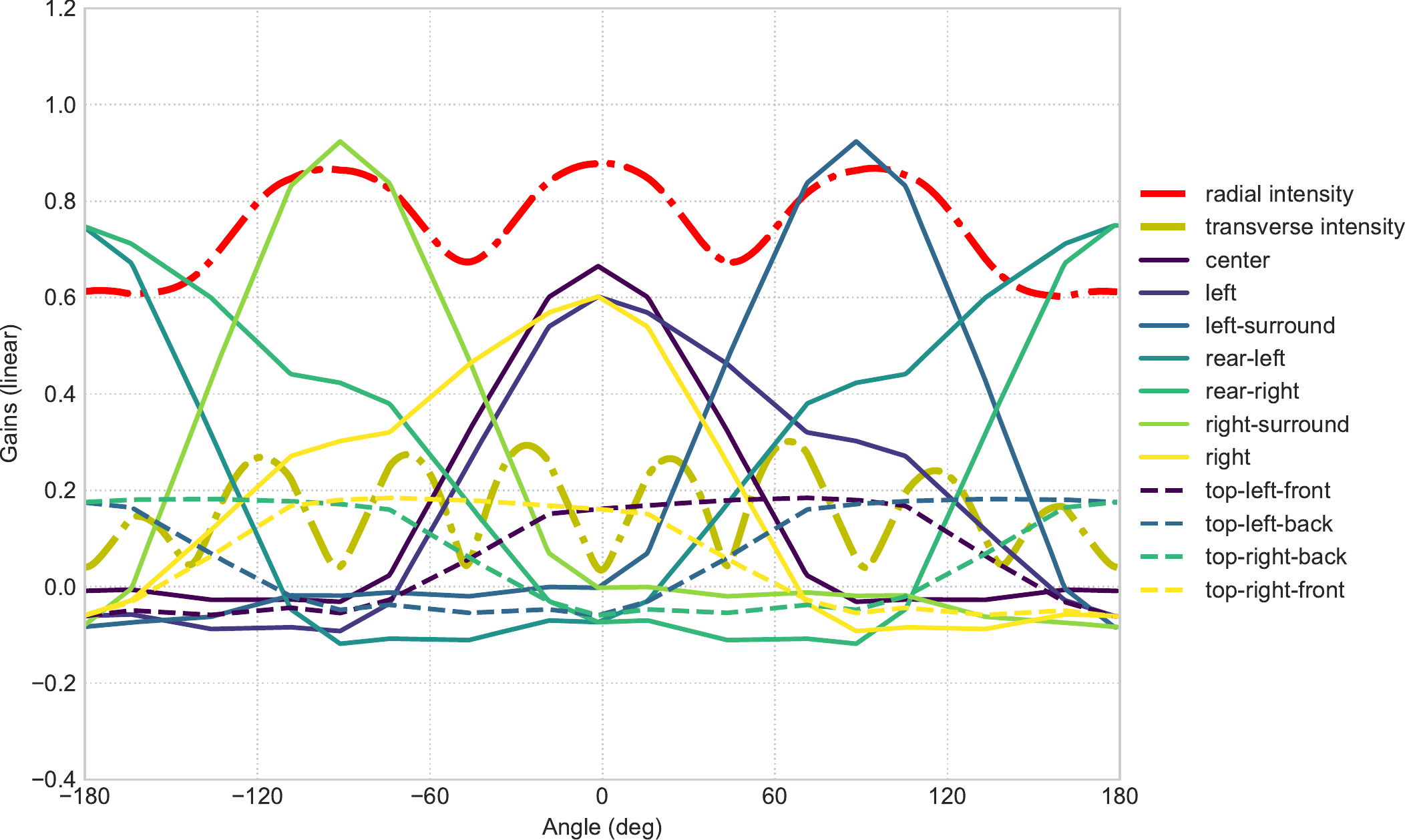}
    \caption{Horizontal panning for OPT-SWF at level $\tilde{0}$ decoded to 7.0.4 layout, using the \emph{focus} decoding preset.}
    \label{fig:wv-panfunc-66-6-18-11-linear-focus}
\end{figure}

For VBAP-SWF
the differences between {level 0} and level ${\tilde{0}}$
are given only by the decoding matrix $\mathbf{D}$, and not by the upsampling,
since the matrix $\mathbf{P}$ generated by VBAP is trivial.
For this reason (and for a clearer presentation), for VBAP-SWF we show only level 0, for example Figure~\ref{fig:vbap-panfunc-66-6-11-linear}.

For OPT-SWF, instead, the differences are given by the combination of $\mathbf{P}$ and $\mathbf{D}$ matrices.
Figures~\ref{fig:wv-panfunc-66-6-11-linear} and~\ref{fig:wv-panfunc-66-6-18-11-linear} show the activation gains for the
\emph{smooth} decoding preset, and Figures~\ref{fig:wv-panfunc-66-6-11-linear-focus} and~\ref{fig:wv-panfunc-66-6-18-11-linear-focus}
show the activation gains for the \emph{focus} decoding preset.

It is interesting to observe how the two SWF formats compare with Ambisonics in terms of apparent source width,
which is related to the radial intensity reported in Figures~\ref{fig:intensity-level0} and~\ref{fig:intensity-level0-focus}
for the two decodings.
Figure~\ref{fig:intensity-level0}, for the \emph{smooth} decoder, shows that VBAP-SWF, OPT-SWF-0 and Ambisonics-1 are very similar.
Interestingly, $\text{OPT-SWF-}\tilde{0}$ performs better than the plain level 0
and performs in some areas (front and sides) as Ambisonics-2.
The improvement comes with a worsened reconstruction error, see Figure~\ref{fig:energy-level0}.
Figure~\ref{fig:intensity-level0-focus}, for the \emph{focus} decoder, shows that
it is possible to improve SWF's focusing of the sources in some areas (front and sides, where the points of the original mesh are located)
at the expense of some worsening in the energy reconstruction.
Given the activation gains for this decoder, Figures~\ref{fig:vbap-panfunc-66-6-11-linear-focus},
\ref{fig:wv-panfunc-66-6-11-linear-focus} and~\ref{fig:wv-panfunc-66-6-18-11-linear-focus},
and the experience during listening, this decoder sounds more ``jumpy'' than the \emph{smooth} one.
This effect is given by the fact that the L and R speakers are less active during the transition to the side-surrounds,
with respect to the \emph{smooth} decoder.

These considerations are summarized in Tables~\ref{tab:level0_704_summary_smooth} and \ref{tab:level0_704_summary_focus}.
These observations are confirmed in informal listening tests described in Section~\ref{sec:informal-listening}.

Note that it is legit to compare level $\tilde{0}$ with Ambisonics order 1 and {2} because {level $\tilde{0}$}
has effectively only 6 channels, since the upsampling operation consists of a matrix product that can
be precalculated and embedded into the decoding matrix, leading to a new decoding matrix.

Ambisonics, at all orders considered here, is the one with largest negative gains.
VBAP-SWF, by construction, generates no negative gains during the encoding, but the decoding to speakers can introduce them.
OPT-SWF has some negative gains embedded in the downsampling and upsampling matrix, and the decoding to speakers can further increase them.
For these reasons, in general, the two SWF flavours are better behaved than Ambisonics ({order 1} and {2})
when listening out of the sweet spot
and are closer to a pure amplitude panner.

\afterpage{%
    \begin{landscape}%
        \centering %
    \begin{tabular}{lccccc}
    \toprule
    observable & VBAP-SWF 0 & OPT-SWF 0 & OPT-SWF $\tilde 0$ & Ambi 1 & Ambi 2 \\
    \midrule
    $E$ energy (dB)                  & $0.05^{0.73}_{-0.27}$ & $0.19^{0.70}_{-0.22}$ & $-0.07^{0.82}_{-0.84}$ & $0.14^{0.16}_{0.14}$ & $-0.06^{0.03}_{-0.16}$  \\ \addlinespace[0.2em] \hline \addlinespace[0.2em]
    $I_{\text R}$ intensity          & $\textit{0.63}^{0.78}_{0.55} $ & $\textit{0.64}^{0.76}_{0.57} $ & $0.72^{0.83}_{0.61}  $ & $\textit{0.66}^{0.73}_{0.58}$ & $\textbf{0.81}^{0.89}_{0.60}  $  \\ \addlinespace[0.2em]
    $I_{\text T}$ intensity          & $0.10^{0.12}_{0.08} $ & $0.13^{0.23}_{0.07} $ & $0.13^{0.25}_{0.04}  $ & $0.11^{0.13}_{0.10}$ & $0.11^{0.20}_{0.09}  $  \\ \addlinespace[0.2em]
    $I_{\text T}$ intensity ($\deg$) & $5.7^{6.9}_{4.6}    $ & $7.5^{13.3}_{4.0}   $ & $7.47^{14.5}_{2.3}   $ & $6.3^{7.5}_{5.5}   $ & $6.3^{11.5}_{5.2}    $  \\
    \bottomrule
    \end{tabular}
    \captionof{table}{Summary Table for the comparison between SWF levels $0$ and $\tilde{0}$ and Ambisonics order 1 and 2,
            decoded to a 7.0.4 layout with the smooth preset.
            Each entry reports the average and maximum and minimum values of the specified observable, $\text{avg}^{\text{max}}_{\text{min}}$.
            Highlighted in italic the values of mean radial intensity that are similar across different formats.
            Highlighted in bold the highest value for the mean radial intensity.}
    \label{tab:level0_704_summary_smooth}
    \vspace{1cm}
    \begin{tabular}{lccccc}
    \toprule
    observable & VBAP-SWF 0 & OPT-SWF 0 & OPT-SWF $\tilde 0$ & Ambi 1 & Ambi 2 \\
    \midrule
    $E$ energy (dB)                  & $-0.04^{1.40}_{-0.80}$ & $0.30^{1.35}_{-0.63}$ & $-0.33^{0.92}_{-1.62}$ & $0.14^{0.16}_{0.14}$ & $-0.06^{0.03}_{-0.16}$ \\ \addlinespace[0.2em] \hline \addlinespace[0.2em]
    $I_{\text R}$ intensity          & $\textit{0.74}^{0.90}_{0.60}  $ & $\textit{0.75}^{0.90}_{0.60} $ & $\textit{0.75}^{0.88}_{0.60}  $ & $0.67^{0.73}_{0.58}$ & $\textbf{0.81}^{0.89}_{0.60}  $ \\ \addlinespace[0.2em]
    $I_{\text T}$ intensity          & $0.11^{0.18}_{0.05}  $ & $0.16^{0.31}_{0.04} $ & $0.16^{0.30}_{0.03}  $ & $0.11^{0.13}_{0.10}$ & $0.11^{0.20}_{0.09}  $ \\ \addlinespace[0.2em]
    $I_{\text T}$ intensity ($\deg$) & $6.3^{10.4}_{2.9}    $ & $9.2^{18.1}_{2.3}   $ & $9.2^{17.5}_{1.7}    $ & $6.3^{7.5}_{5.7}   $ & $6.3^{11.5}_{5.2}    $ \\ \addlinespace[0.2em]
    \bottomrule
    \end{tabular}
    \captionof{table}{Summary Table for the comparison between SWF levels $0$ and $\tilde{0}$ and Ambisonics order 1 and 2,
            decoded to a 7.0.4 layout with the focus preset.
            Each entry reports the average and maximum and minimum values of the specified observable, $\text{avg}^{\text{max}}_{\text{min}}$.
            Highlighted in italic the values of mean radial intensity that are similar across different formats.
            Highlighted in bold the highest value for the mean radial intensity.}
    \label{tab:level0_704_summary_focus}
    \end{landscape}
    \clearpage%
}

\subsection{Comparison between SWF Level 1 with Ambisonics Orders 2 and 3} \label{sec:comparison1}

\begin{figure}[t!]
    \centering
    \begin{minipage}[t]{0.46\textwidth}
        \includegraphics[width=\textwidth,valign=t]{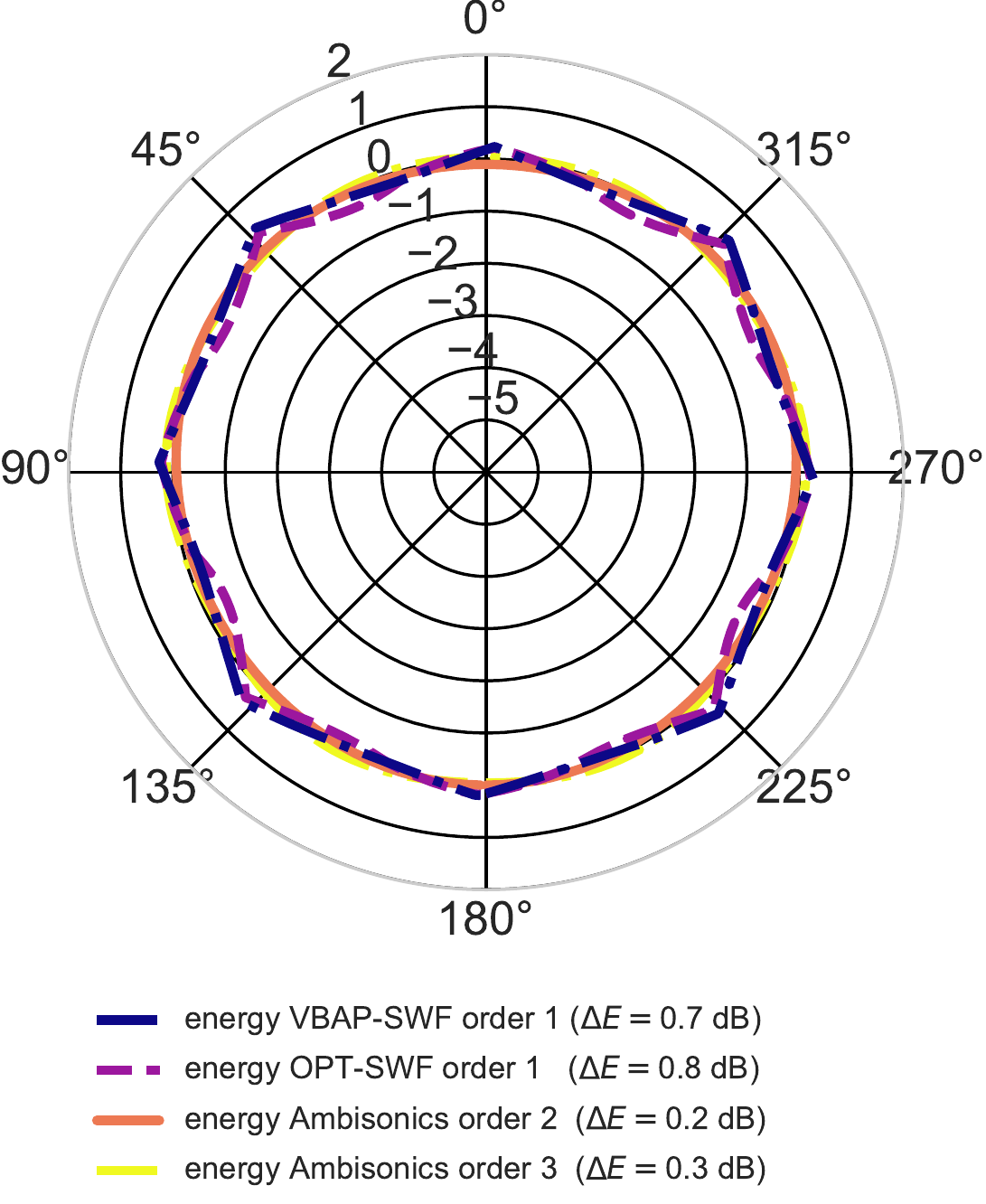}
        \caption{Energy comparison for levels and orders with similar channel count:
                SWF at level 1 with the \emph{smooth} decoding preset and Ambisonics at {order 2} and {3}, decoded to a 7.0.4 layout. Scale is in dB.}
        \label{fig:energy-level1}
    \end{minipage}
    \hfill
    \begin{minipage}[t]{0.46\textwidth}
        \includegraphics[width=\textwidth,valign=t]{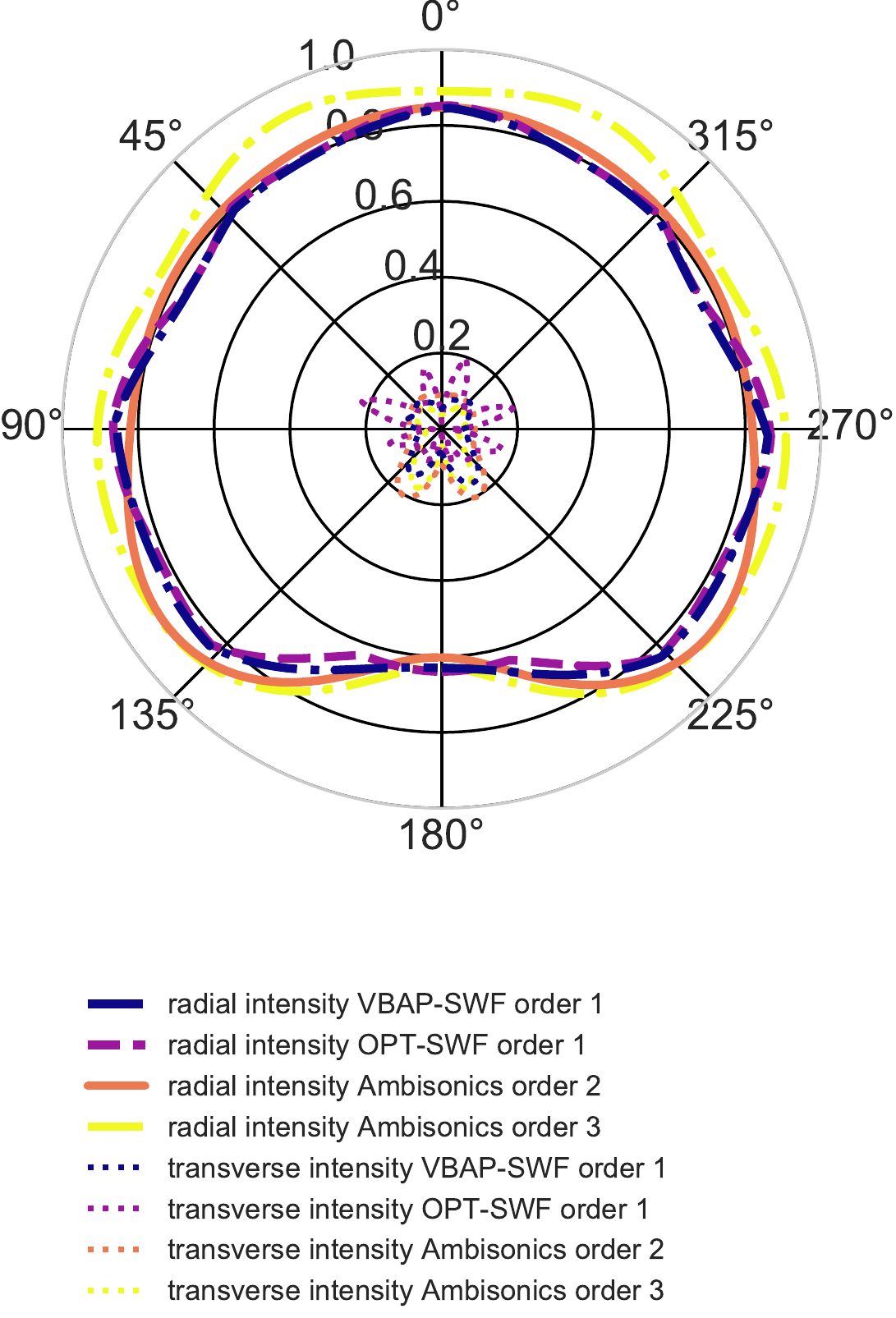}
        \caption{Intensity comparison for levels (with the \emph{smooth} decoding preset) and orders with similar channel count.
                The point and dash lines represent radial intensity and the dashed ones the transverse intensity component.}
        \label{fig:intensity-level1}
    \end{minipage}
\end{figure}

\begin{figure}
    \centering
    \includegraphics[width=\textwidth]{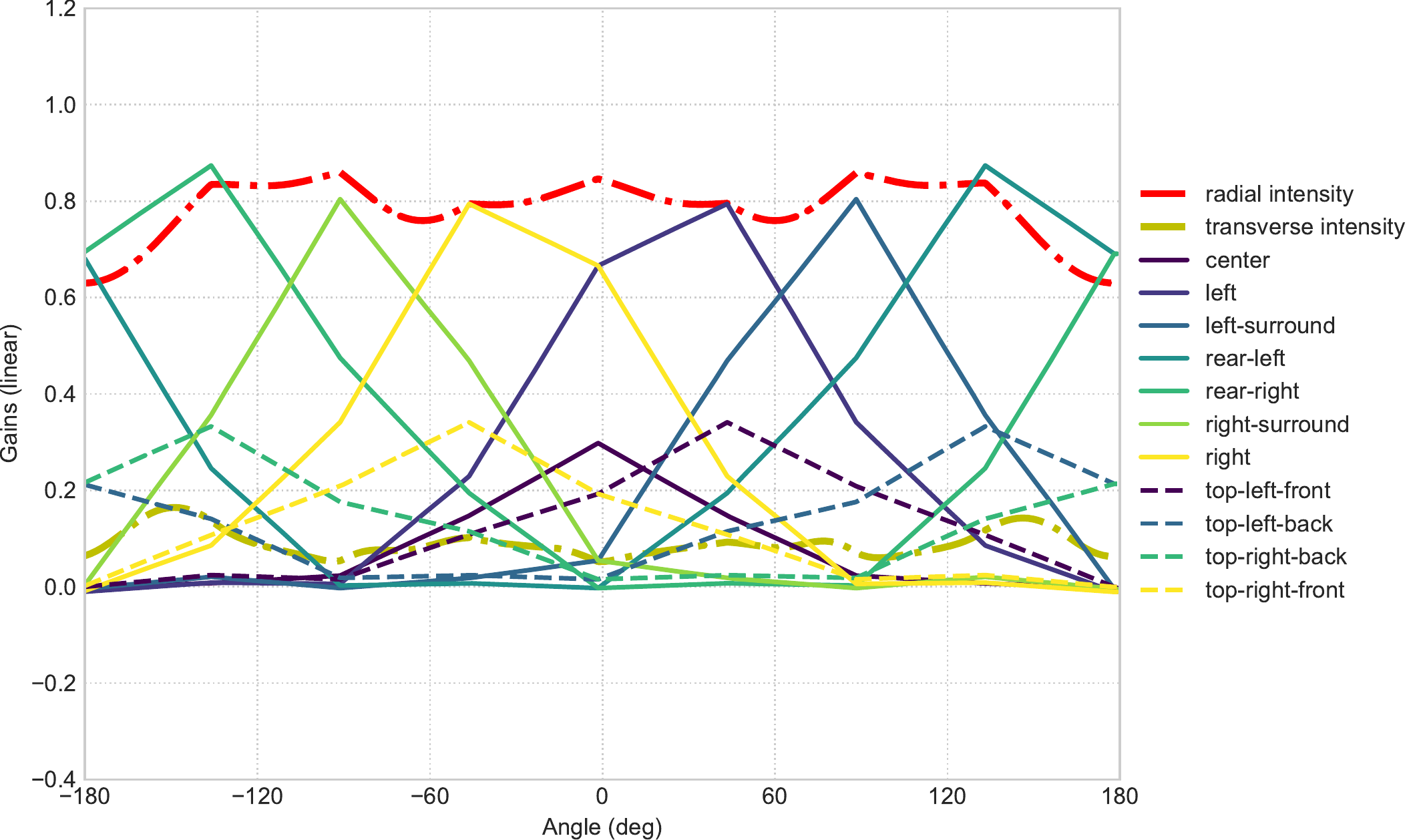}
    \caption{Horizontal panning for VBAP-SWF at level 1 decoded to 7.0.4 layout, using the \emph{smooth} decoding preset.}
    \label{fig:vbap-panfunc-66-18-11-linear}

    \includegraphics[width=\textwidth]{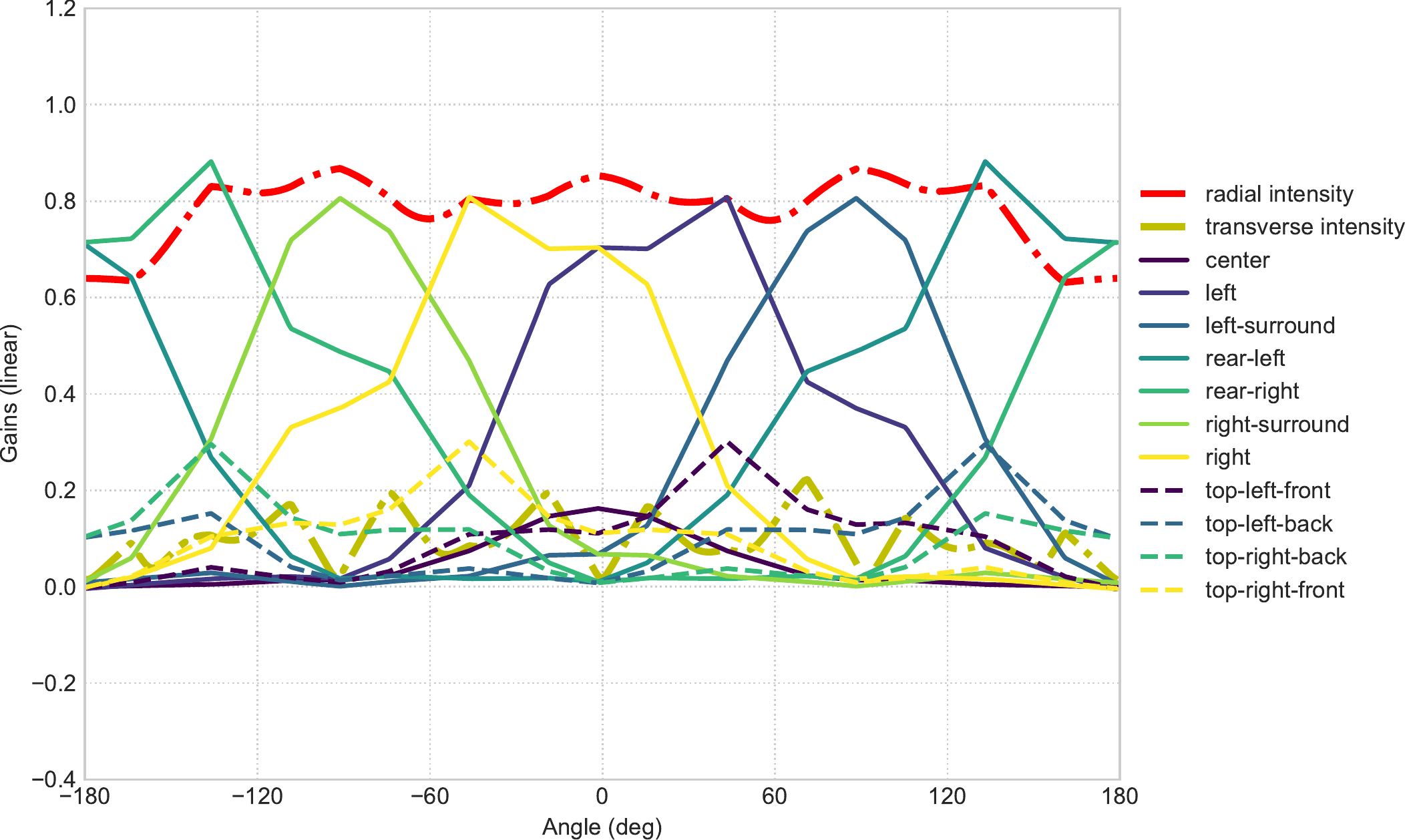}
    \caption{Horizontal panning for OPT-SWF at level 1 decoded to 7.0.4 layout, using the \emph{smooth} decoding preset.}
    \label{fig:wv-panfunc-66-18-11-linear}
\end{figure}

\begin{figure}
    \includegraphics[width=\textwidth]{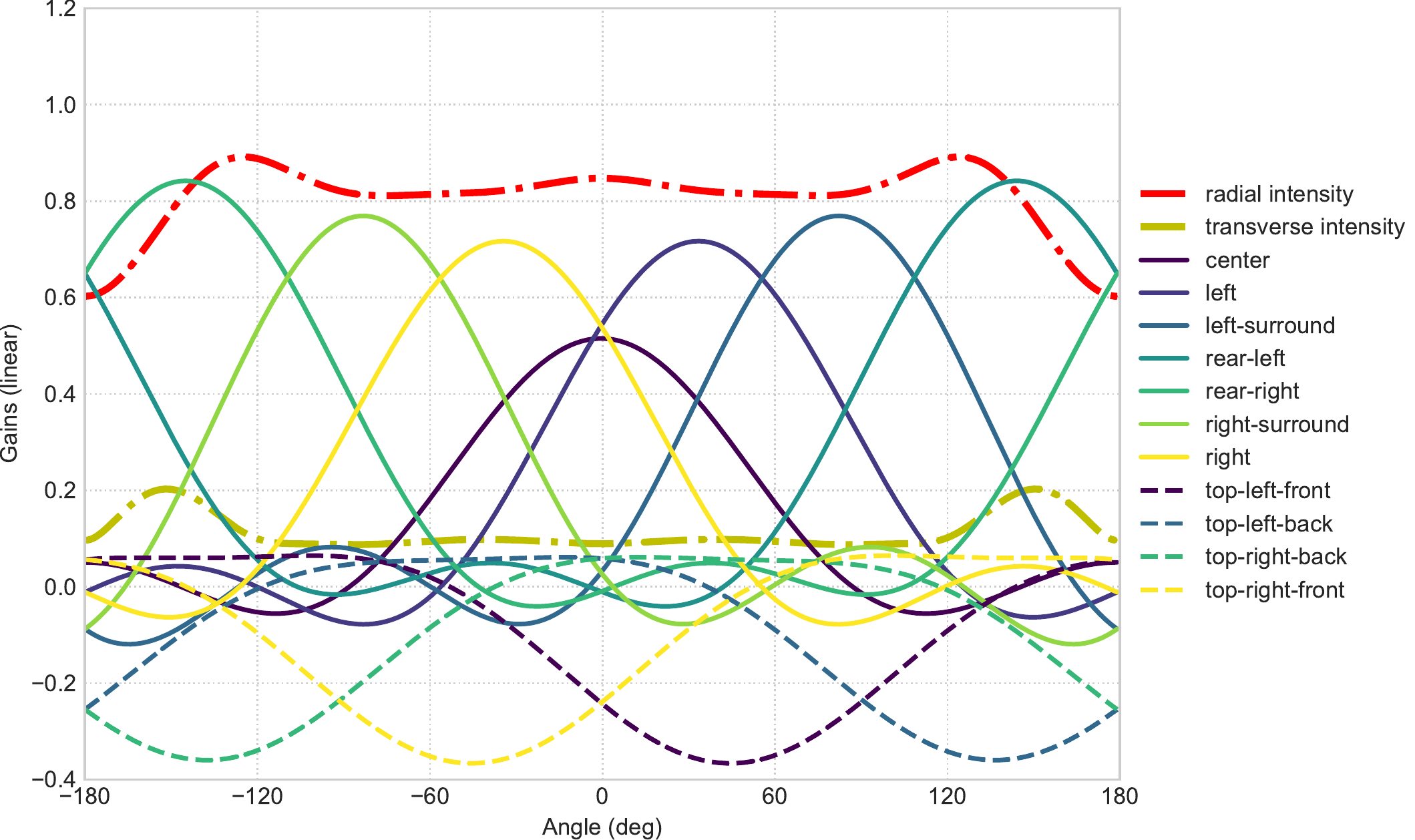}
    \caption{Horizontal panning for Ambisonics at {order 2} decoded to 7.0.4 layout.}
    \label{fig:ambi2-panfunc-linear}

    \includegraphics[width=\textwidth]{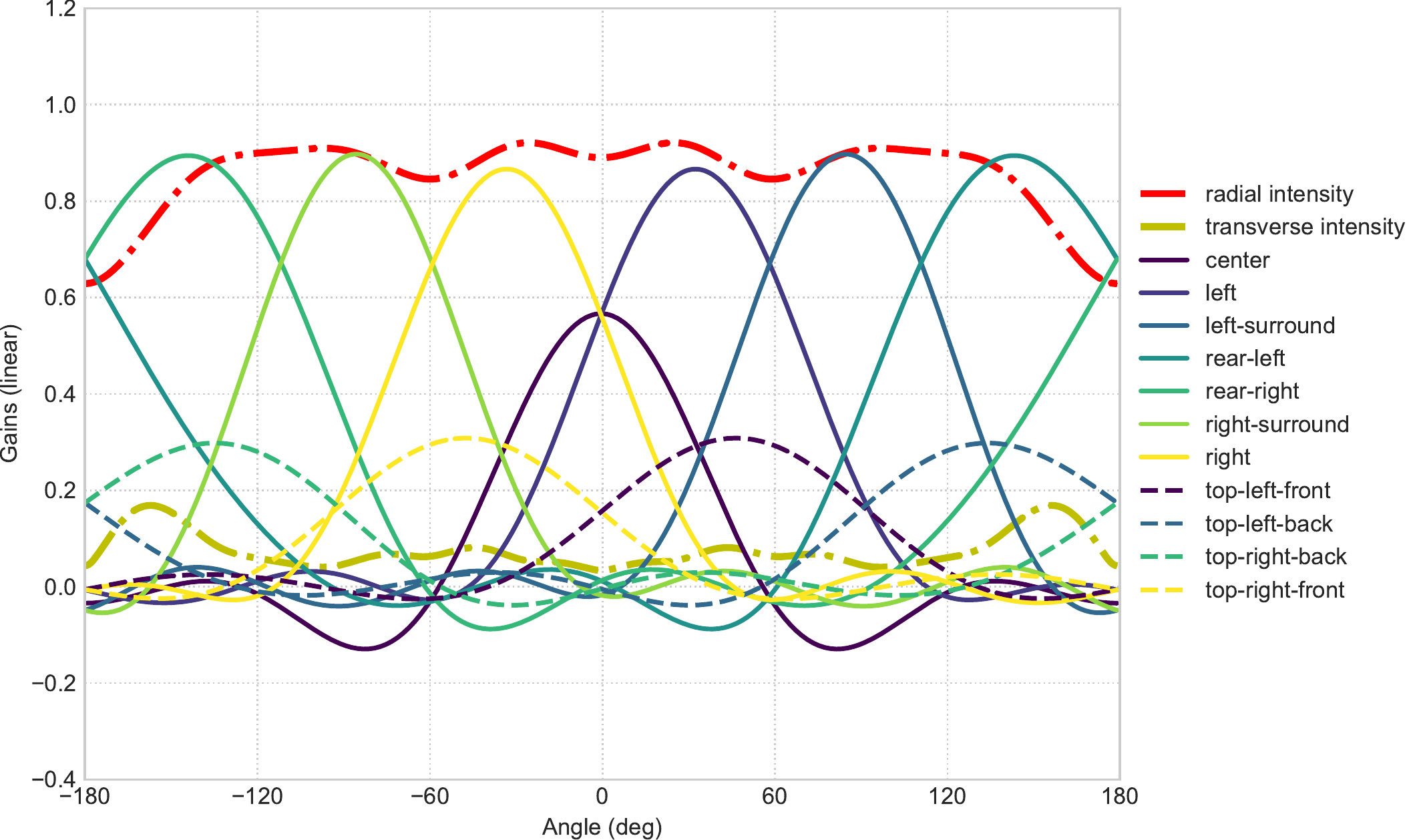}
    \caption{Horizontal panning for Ambisonics at {order 3} decoded to 7.0.4 layout.}
    \label{fig:ambi3-panfunc-linear}
\end{figure}

\begin{figure}[t!]
    \centering
    \begin{minipage}[t]{0.46\textwidth}
        \includegraphics[width=\textwidth,valign=t]{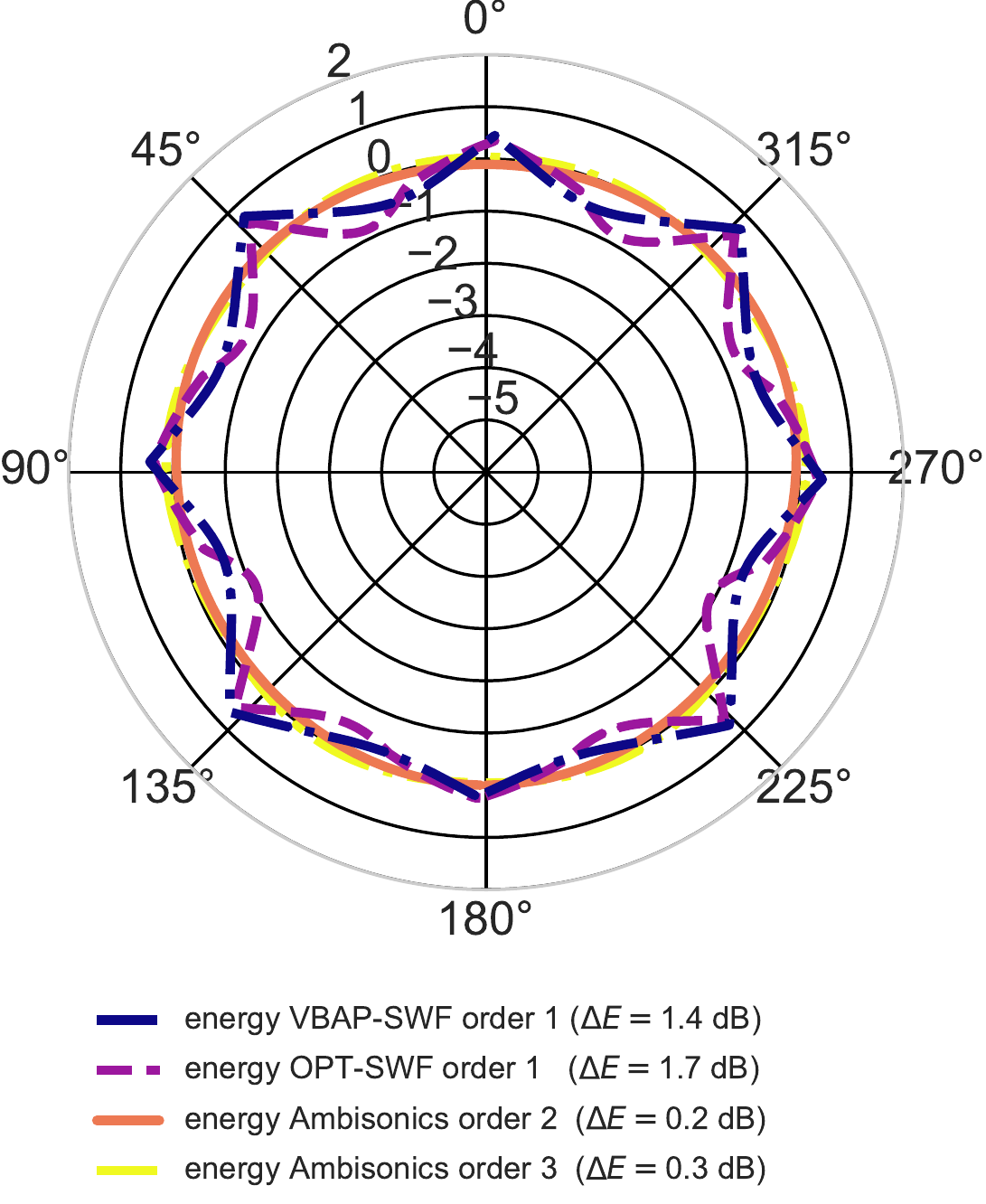}
        \caption{Energy comparison for levels and orders with similar channel count:
                SWF at level 1 with the \emph{focus} decoding preset and Ambisonics at {order 2} and {3}, decoded to a 7.0.4 layout. Scale is in dB.}
        \label{fig:energy-level1-focus}
    \end{minipage}
    \hfill
    \begin{minipage}[t]{0.46\textwidth}
        \includegraphics[width=\textwidth,valign=t]{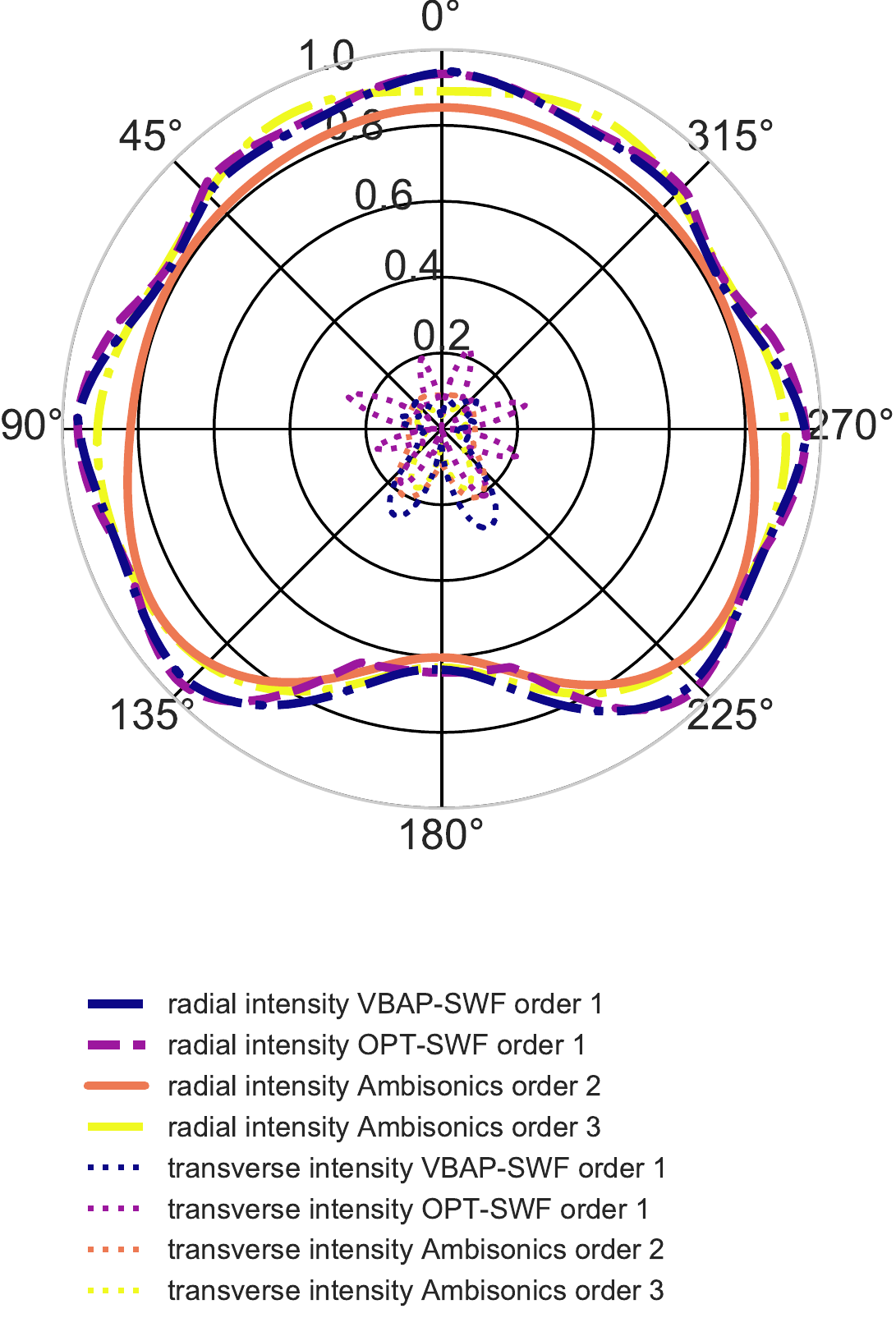}
        \caption{Intensity comparison for levels (with the \emph{focus} decoding preset) and orders with similar channel count.
                The dashed lines represent radial intensity and the dotted ones the transverse intensity component.}
        \label{fig:intensity-level1-focus}
    \end{minipage}
\end{figure}

\begin{figure}[tbh]
    \centering
    \includegraphics[width=\textwidth]{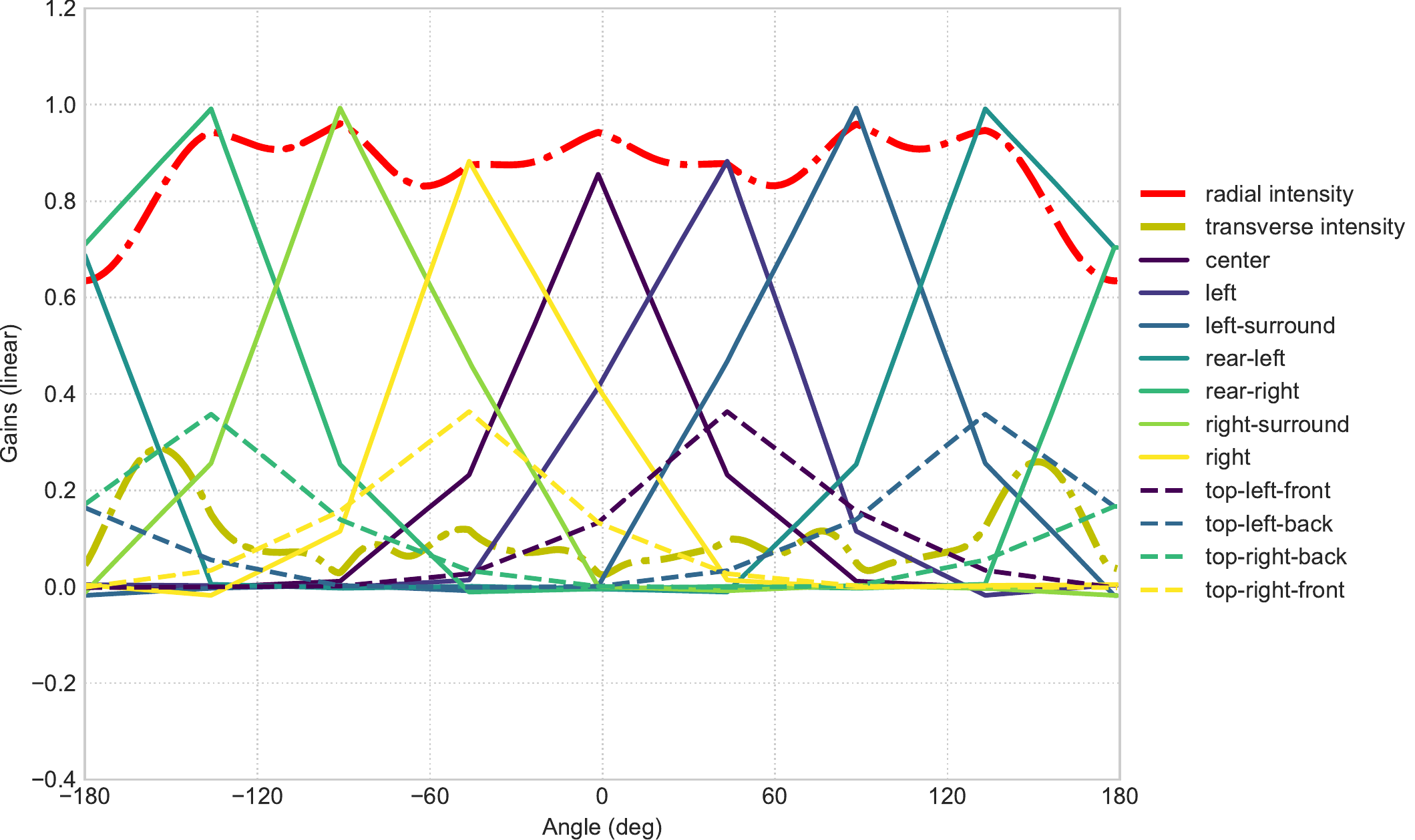}
    \caption{Horizontal panning for VBAP-SWF at level 1 decoded to 7.0.4 layout, using the \emph{focus} decoding preset.}
    \label{fig:vbap-panfunc-66-18-11-linear-focus}

    \includegraphics[width=\textwidth]{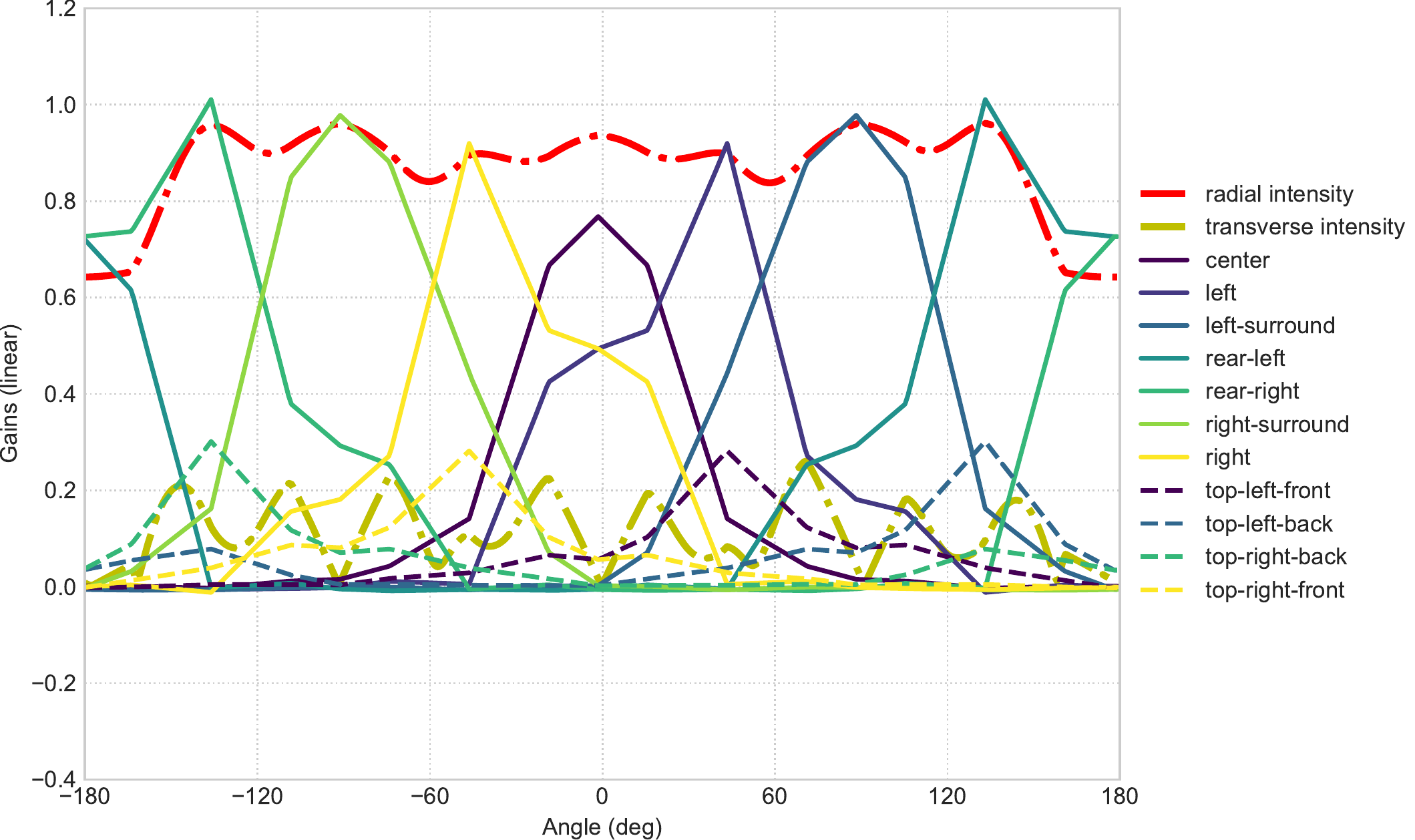}
    \caption{Horizontal panning for OPT-SWF at level 1 decoded to 7.0.4 layout, using the \emph{focus} decoding preset.}
    \label{fig:wv-panfunc-66-18-11-linear-focus}
\end{figure}

In this case VBAP-OPT-1 (Figures~Figures~\ref{fig:intensity-level1},~\ref{fig:vbap-panfunc-66-18-11-linear})
and OPT-SWF-1 (Figures~\ref{fig:intensity-level1},~\ref{fig:wv-panfunc-66-18-11-linear})
performs comparably to Ambisonics-2 (Figures~\ref{fig:intensity-level1},~\ref{fig:ambi2-panfunc-linear}) for the \emph{smooth} decoding.
For the \emph{focus} decoding SWF-1 performs on par (or slightly better) than Ambisonics-3, see Figures~\ref{fig:intensity-level1-focus},
\ref{fig:wv-panfunc-66-18-11-linear-focus} and~\ref{fig:ambi3-panfunc-linear}.
In particular, looking at the linear gains plots, it is possible to notice that SWF-1 \emph{focus} makes use
of the central loudspeaker in a neater way than Ambisonics-3 does.

These reflections are summarized in Tables~\ref{tab:level1_704_summary_smooth} and \ref{tab:level1_704_summary_focus}.

In general, the price to pay for a greater radial intensity is a worsened reconstructed energy (and vice versa).
This effect is highlighted in the polar plots showing the reconstructed energy by reporting the
$E_{max}-E_{min} = \Delta E$ in the legend.
At the levels indicated here, $\Delta E<2$~dB, this effect does not seem to be noticeable, but further investigation is needed.

\afterpage{%
    \begin{landscape}%
        \centering %
    \begin{tabular}{lcccc}
    \toprule
    observable & VBAP-SWF 1 & OPT-SWF 1 & Ambi 2 & Ambi 3 \\
    \midrule
    $E$ energy (dB)                  & $0.00^{0.43}_{-0.23}$ & $-0.10^{0.32}_{-0.46}$ & $-0.06^{0.03}_{-0.15}$ & $0.01^{0.14}_{-0.12}$ \\ \addlinespace[0.2em] \hline \addlinespace[0.2em]
    $I_{\text R}$ intensity          & $\textit{0.79}^{0.86}_{0.63} $ & $\textit{0.79}^{0.87}_{0.63}  $ & $\textit{0.81}^{0.89}_{0.60}  $ & $\textbf{0.86}^{0.92}_{0.63} $ \\ \addlinespace[0.2em]
    $I_{\text T}$ intensity          & $0.09^{0.16}_{0.05} $ & $0.10^{0.22}_{0.01}  $ & $0.11^{0.20}_{0.09}  $ & $0.08^{0.17}_{0.03} $ \\ \addlinespace[0.2em]
    $I_{\text T}$ intensity ($\deg$) & $5.2^{9.2}_{2.9}    $ & $5.7^{12.7}_{0.6}    $ & $6.3^{11.5}_{5.2}    $ & $4.6^{9.8}_{1.7}    $ \\ \addlinespace[0.2em]
    \bottomrule
    \end{tabular}
    \captionof{table}{Summary Table for the comparison between SWF level 1 and Ambisonics order 1 and 2,
            decoded to a 7.0.4 layout with the smooth preset.
            Each entry reports the average and maximum and minimum values of the specified observable, $\text{avg}^{\text{max}}_{\text{min}}$.
            Highlighted in italic the values of mean radial intensity that are similar across different formats.
            Highlighted in bold the highest value for the mean radial intensity.}
    \label{tab:level1_704_summary_smooth}
    \vspace{1cm}
    \begin{tabular}{lcccc}
    \toprule
    observable & VBAP-SWF 1 & OPT-SWF 1 & Ambi 2 & Ambi 3 \\
    \midrule
    $E$ energy (dB)                  & $-0.12^{0.74}_{-0.70}$ & $-0.24^{0.59}_{-1.08}$ & $-0.06^{0.03}_{-0.15}$ & $0.01^{0.14}_{-0.12}$ \\ \addlinespace[0.2em] \hline \addlinespace[0.2em]
    $I_{\text R}$ intensity          & $\textbf{0.87}^{0.96}_{0.63}  $ & $\textbf{0.87}^{0.96}_{0.64}  $ & $0.81^{0.89}_{0.60}  $ & $\textbf{0.86}^{0.92}_{0.63} $ \\ \addlinespace[0.2em]
    $I_{\text T}$ intensity          & $0.10^{0.29}_{0.02}  $ & $0.11^{0.26}_{0.00}  $ & $0.11^{0.20}_{0.09}  $ & $0.08^{0.17}_{0.03} $ \\ \addlinespace[0.2em]
    $I_{\text T}$ intensity ($\deg$) & $5.7^{16.9}_{1.1}    $ & $6.3^{15.1}_{0.0}    $ & $6.3^{11.5}_{5.2}    $ & $4.6^{9.8}_{1.7}    $ \\ \addlinespace[0.2em]
    \bottomrule
    \end{tabular}
    \captionof{table}{Summary Table for the comparison between SWF level 1 and Ambisonics order 1 and 2,
            decoded to a 7.0.4 layout with the focus preset.
            Each entry reports the average and maximum and minimum values of the specified observable, $\text{avg}^{\text{max}}_{\text{min}}$.
            In bold we highlight the highest values of mean radial intensity, which in this case are also similar across different formats.}
    \label{tab:level1_704_summary_focus}
    \end{landscape}
    \clearpage%
}

\subsection{Informal Listening}
\label{sec:informal-listening}
During the evaluation of the formats and decodings, we carried out some informal listening tests.
The tests where performed in a critical listening room with RT60~$< 0.4$~s below 200~Hz and RT60~$< 0.25$~s above 200~Hz.  %
The listening tests assessed the quality of the decoders and the different audio chains.

The differences in the perceived localization properties of the audio chains described in Sections~\ref{sec:comparison0} and \ref{sec:comparison1}
are confirmed also during the subjective listenings.
This supports the fact that the radial intensity models well the perceived source size.
With the tests we performed, it is difficult to quantify how relevant is the small non-zero transverse intensity for the incorrect positioning of the source.
Specific tests should be carried out.
As for the loudness across the horizontal trajectory of the audio source,
we did not notice any variation with these moderate energy differences.

\section{Objective Evaluation of OPT-SWF and Ambisonics for the Hamasaki 22.2 Layout} \label{sec:nhk-eval}
In this Section we would like to briefly illustrate a reduced comparison between OPT-SWF and Ambisonics for a layout of the channel-based format
that has the greatest number of speakers to date, the Hamasaki 22.2~\cite{hamasaki2005the}.
For this comparison we are not in the position to be able to assess the differences also via listening tests, even if limited and informal.
We rely only on the psychoacoustic indicators described in the manuscript.
For this comparison, we created a decoding for OPT-SWF at level 1 and Ambisonics at order 3.
In both cases the number of loudspeakers exceeds the number of format's channels to decode.

The Hamasaki 22.2 layout is composed of 22 speakers disposed in three levels and 2 subwoofers.
The lower level has 3 loudspeakers in the frontal area (between $-15^\circ$ and $-25^\circ$ of elevation).
The middle layer has 10 loudspeakers, and it is possible to imagine it as an enriched 9.0 layout with an added speaker
right in the back of the sweet spot (or opposite to the center channel).
The top layer (between $30^\circ$ and $45^\circ$ of elevation) has 8 loudspeakers at an equal angular distance between each other.
To reach the number of 22 loudspeakers, the last loudspeaker is placed at the zenith of the layout ($90^\circ$ elevation),
typically called ``voice of god''.

\begin{figure}[t!]
    \centering
    \begin{minipage}[t]{0.46\textwidth}
        \includegraphics[width=\textwidth,valign=t]{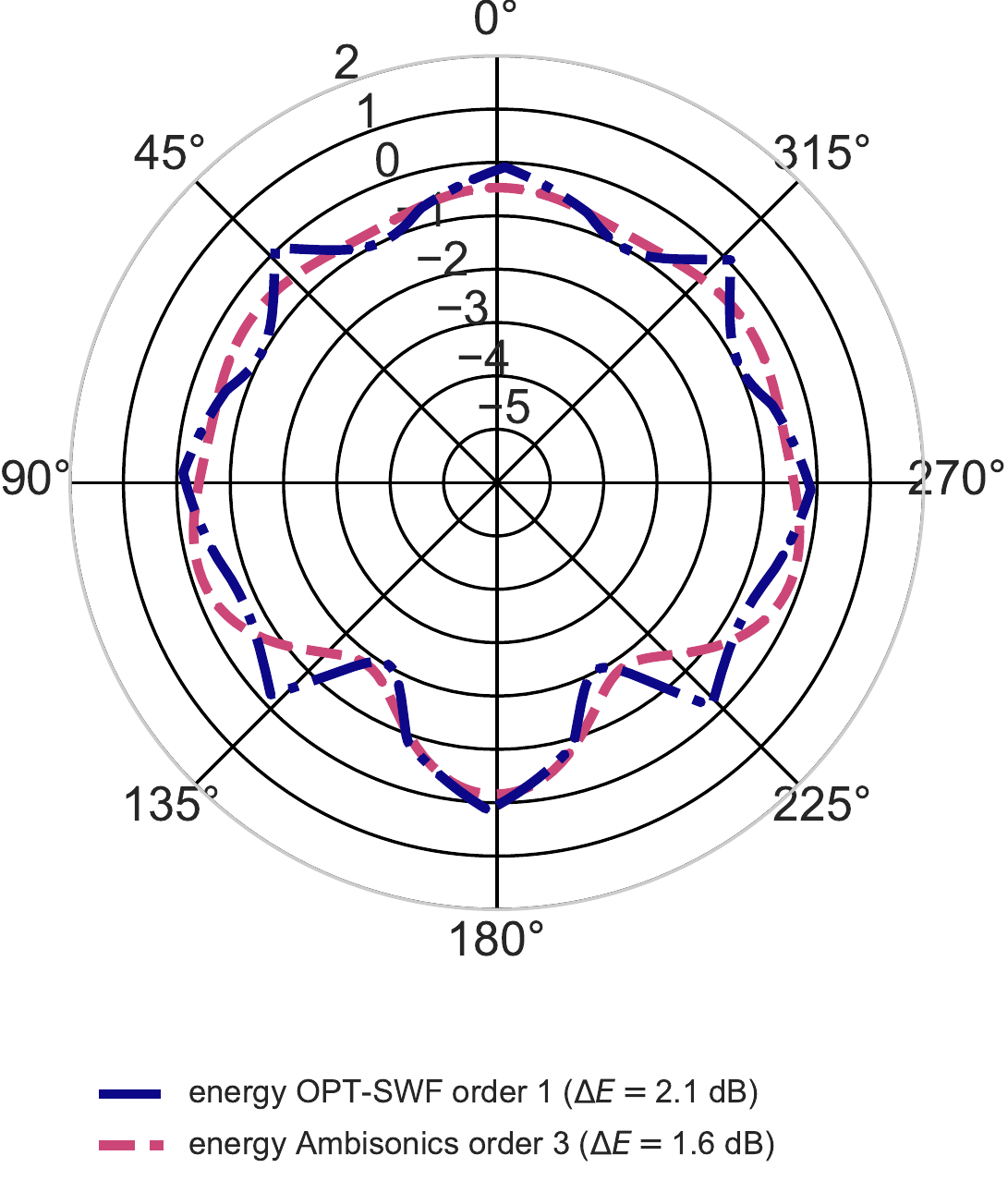}
        \caption{Energy comparison for OPT-SWF at level 1 and Ambisonics at order 3, decoded to an Hamasaki 22.0 layout, on the horizontal plane.}
        \label{fig:energy-22-h}
    \end{minipage}
    \hfill
    \begin{minipage}[t]{0.46\textwidth}
        \includegraphics[width=\textwidth,valign=t]{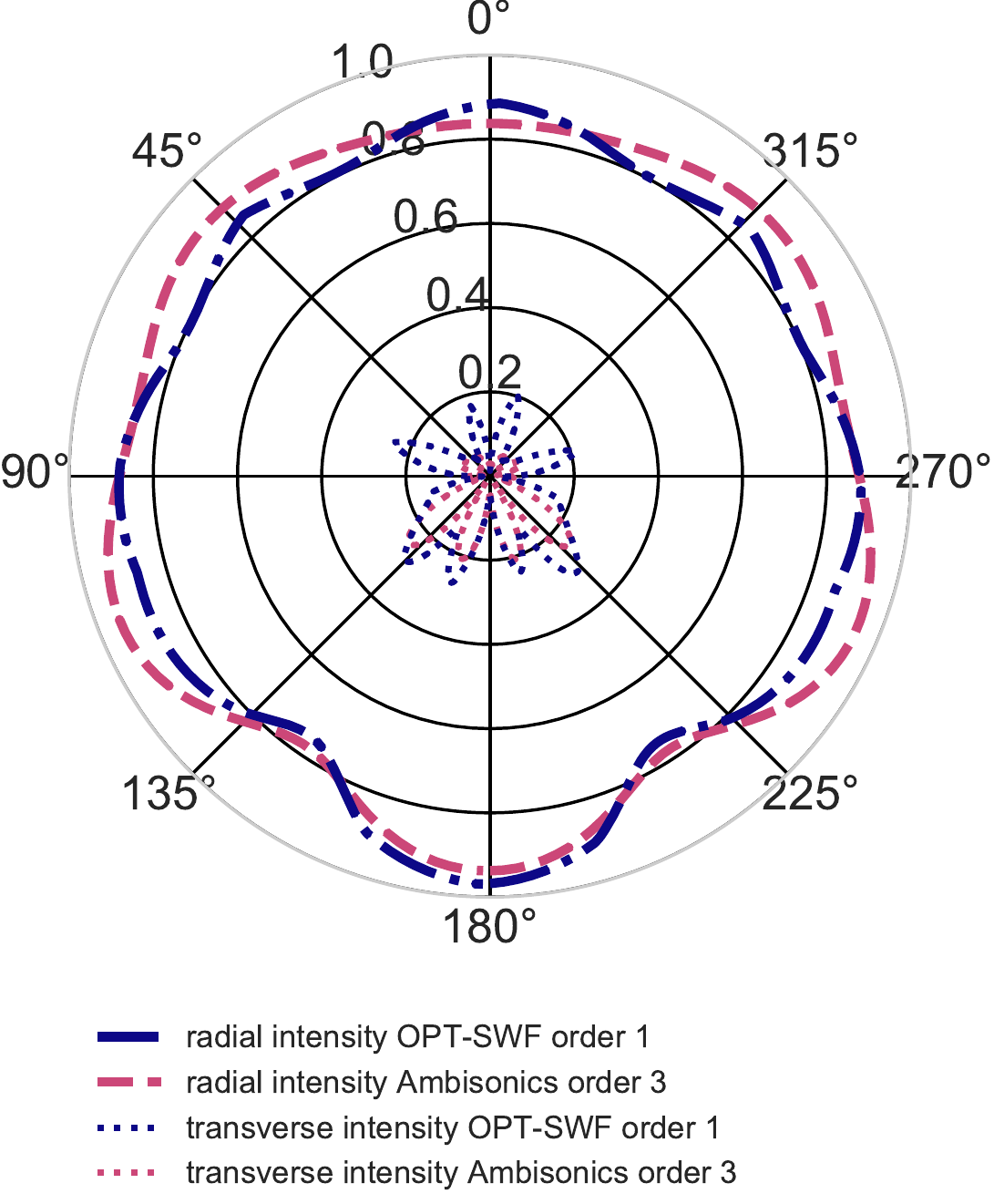}
        \caption{Intensity comparison for OPT-SWF at level 1 and Ambisonics at order 3, decoded to an Hamasaki 22.0 layout, on the horizontal plane.
                The dashed lines represent radial intensity and the dotted ones the transverse intensity component.}
        \label{fig:intensity-22-h}
    \end{minipage}
\end{figure}

\begin{figure}
    \centering
    \begin{minipage}[t]{\textwidth}
        \includegraphics[width=\textwidth]{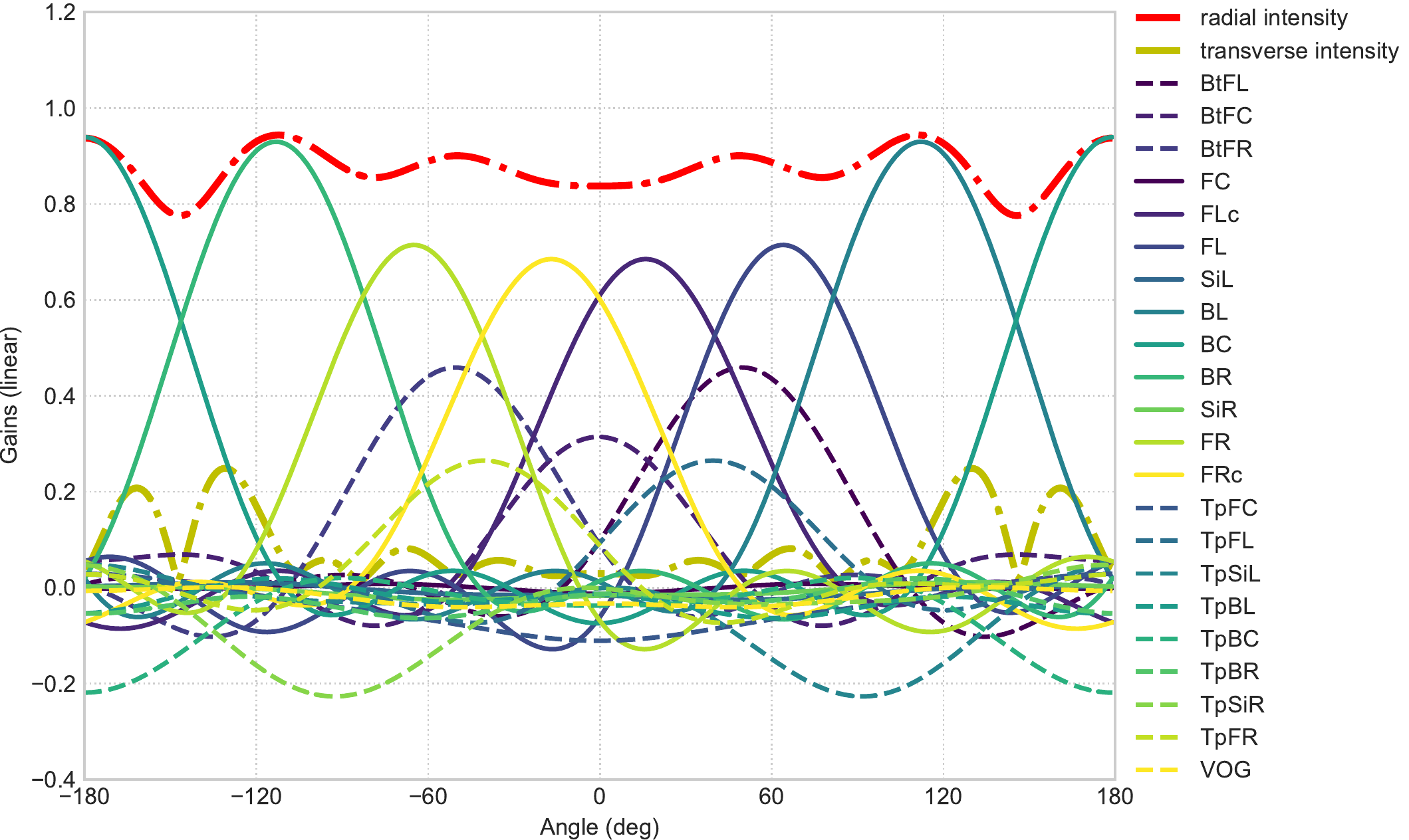}
        \caption{Horizontal panning for Ambisonics at order 3 decoded to an Hamasaki 22.0 layout.}
        \label{fig:22.2-ambi3-h}
    \end{minipage}
    \begin{minipage}[t]{\textwidth}
        \includegraphics[width=\textwidth]{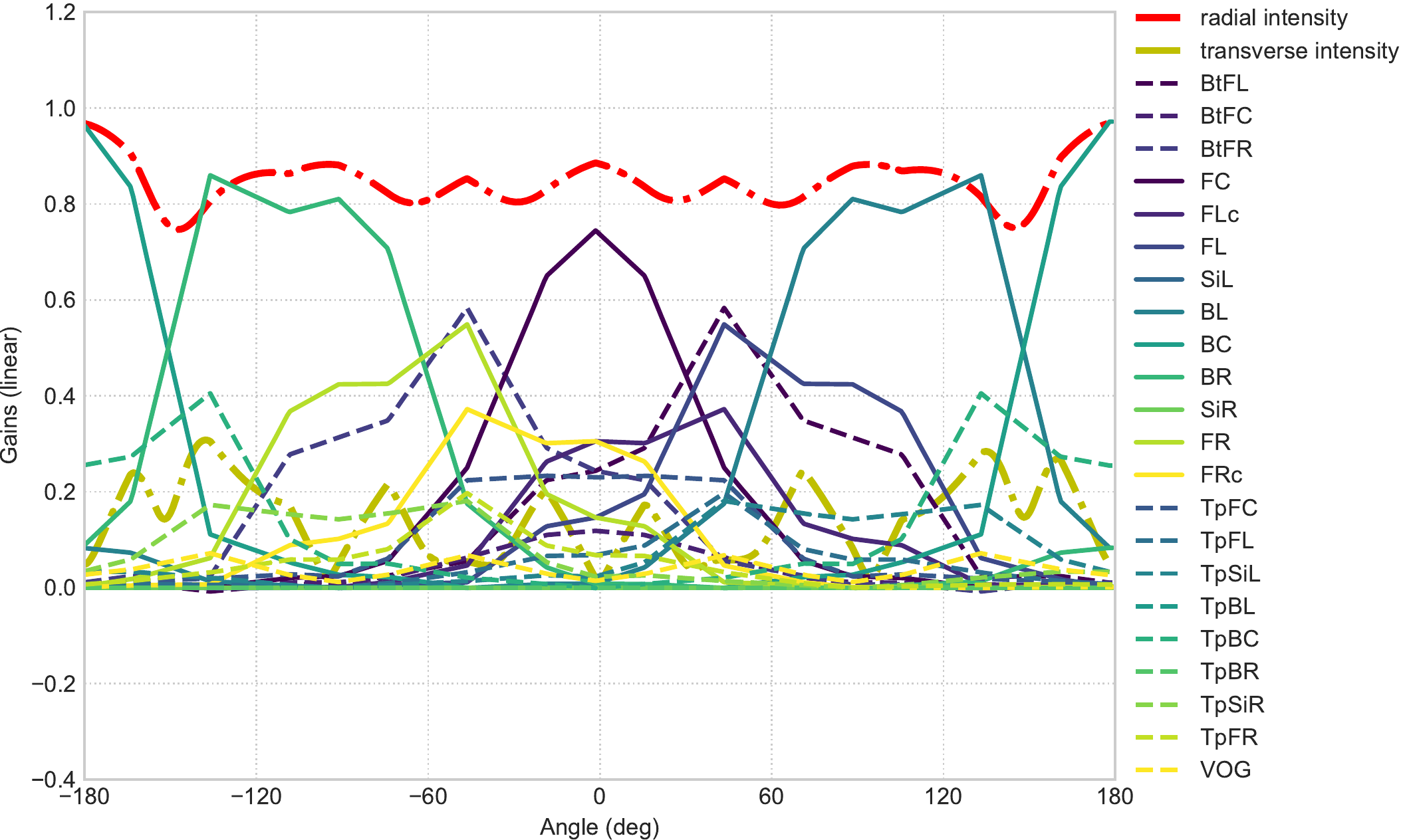}
        \caption{Horizontal panning for OPT-SWF at level 1, decoded to an Hamasaki 22.0 layout.}
        \label{fig:22.2-optswf-h}
    \end{minipage}
\end{figure}

In Figures~\ref{fig:22.2-ambi3-h} and~\ref{fig:22.2-optswf-h} we report the panning functions for an horizontal panning
for Ambisonics at order 3 and OPT-SWF at level 1 decoded to the Hamasaki 22.0 layout previously described.
The solid lines represent the gains of the speakers located in the horizontal plane
(zero elevation) while the dashed lines depict the gains of the speakers with non-zero elevation.
This representation makes apparent that some speakers in the top layer are activated even if the panning is purely
in the horizontal plane, that overlaps with the Hamasaki's middle layer.
The number of speakers activated by Ambisonics and OPT-SWF is very similar.
The only difference, that in this specific case we consider minimal, is the type of speakers activated:
for a source located in the center in OPT-SWF the central speaker is activated
while in Ambisonics the left and right speakers generate a phantom center.

In Figures~\ref{fig:energy-22-h} and~\ref{fig:intensity-22-h} depict
the reconstructed energy and intensity on the horizontal plane, respectively,
for Ambisonics at order 3 and OPT-SWF at level 1 decoded to the Hamasaki 22.0.
It is possible to see that both techniques achieve similar performances.
The $\Delta E$ is limited around 1~dB for both techniques except to the left and right of the back speaker (BC).
This is an effect due to the distance between the speaker at $\pm110^\circ$ and the speaker at $180^\circ$,
and is controlled by a (tunable) parameter of IDHOA.
We will elaborate more in the discussion about the vertical plane plots.

\begin{figure}
    \centering
    \begin{minipage}[t]{0.46\textwidth}
        \includegraphics[width=\textwidth,valign=t]{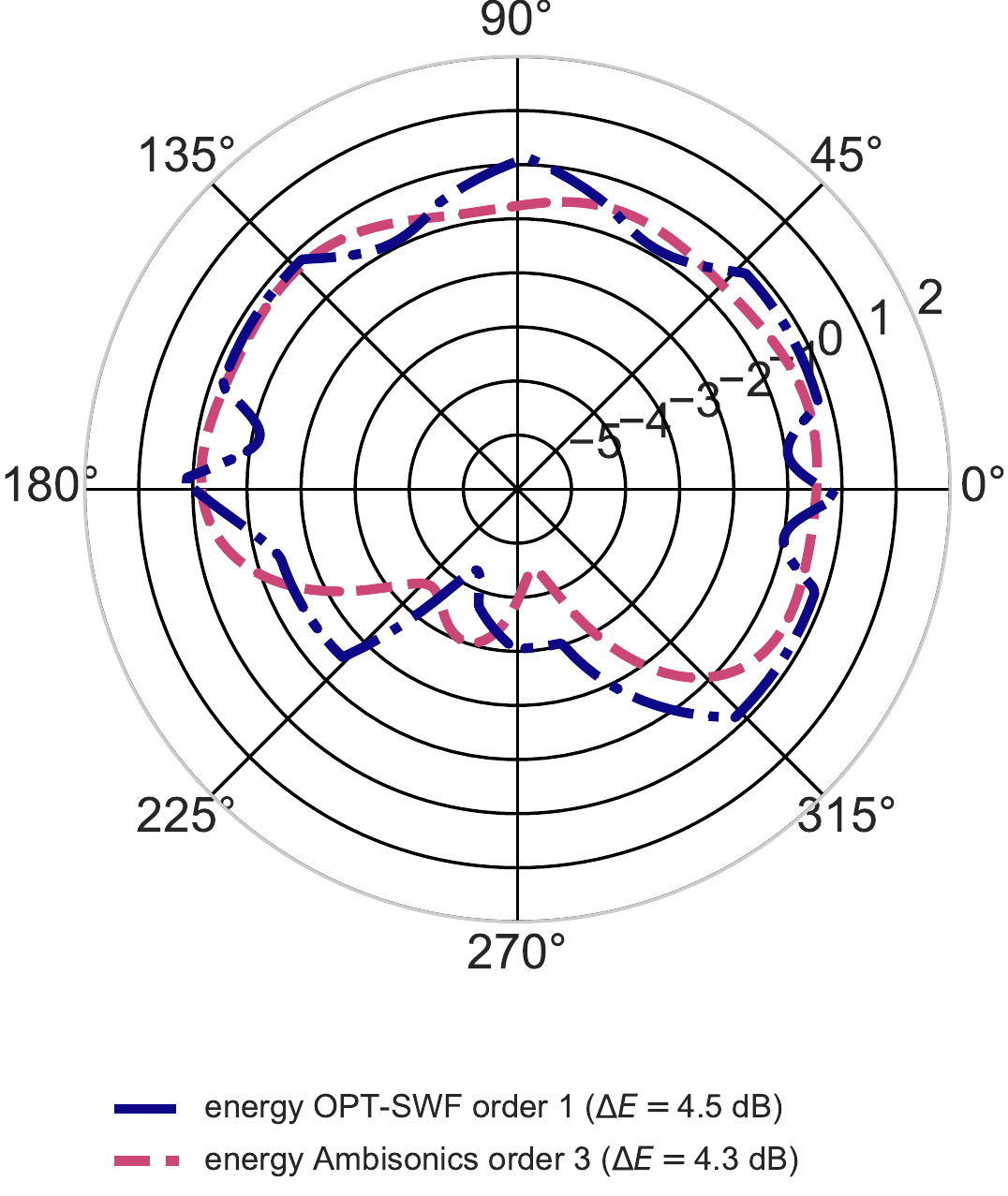}
        \caption{Energy comparison comparison for OPT-SWF at level 1 and Ambisonics at order 3, decoded to an Hamasaki 22.2 layout, on the vertical plane.}
        \label{fig:energy-22-v}
    \end{minipage}
    \hfill
    \begin{minipage}[t]{0.46\textwidth}
        \includegraphics[width=\textwidth,valign=t]{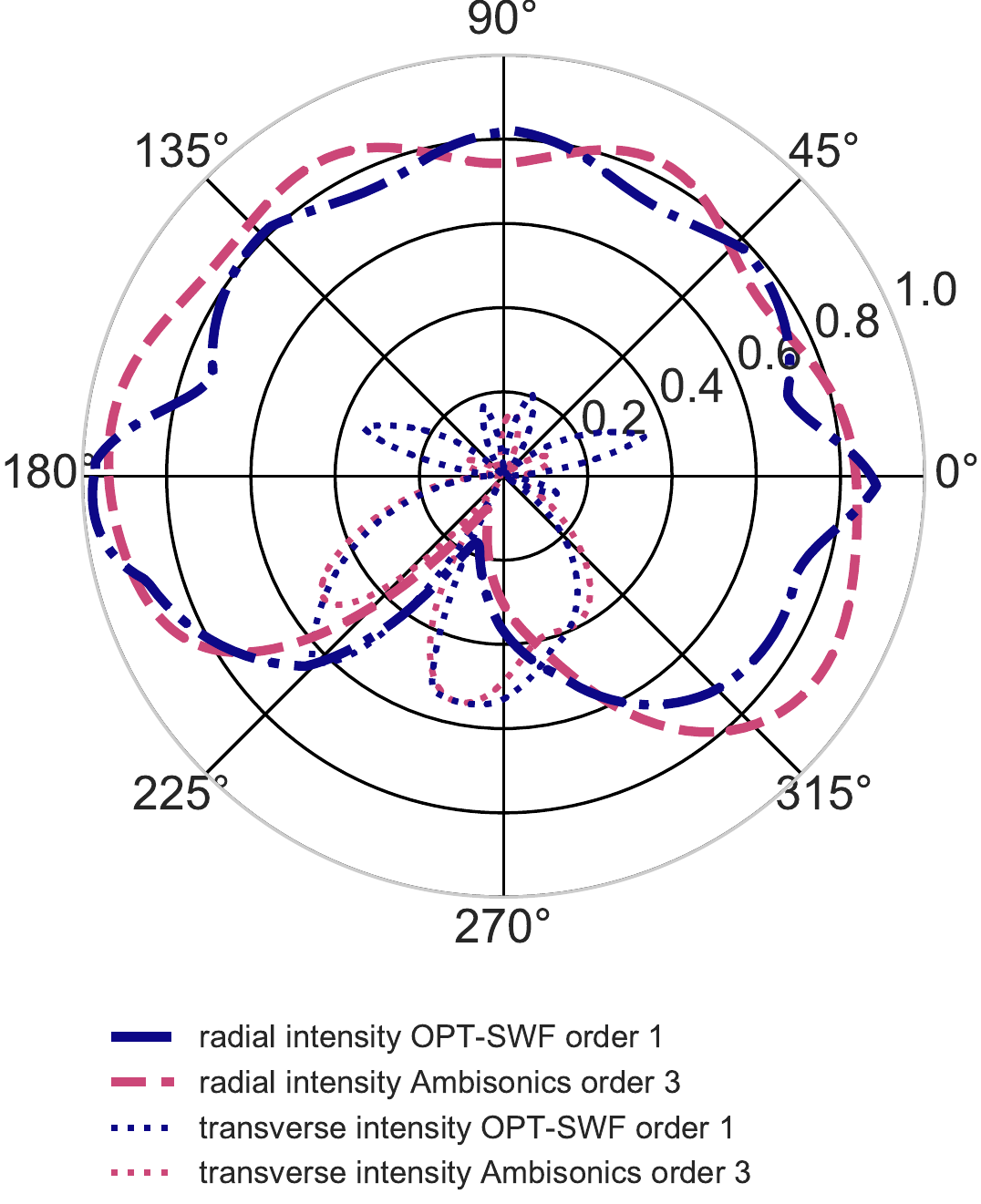}
        \caption{Intensity comparison comparison for OPT-SWF at level 1 and Ambisonics at order 3, decoded to an Hamasaki 22.2 layout, on the vertical plane.
                The dashed lines represent radial intensity and the dotted ones the transverse intensity component.}
        \label{fig:intensity-22-v}
    \end{minipage}
\end{figure}

\begin{figure}
    \centering
    \begin{minipage}[t]{\textwidth}
        \includegraphics[width=\textwidth]{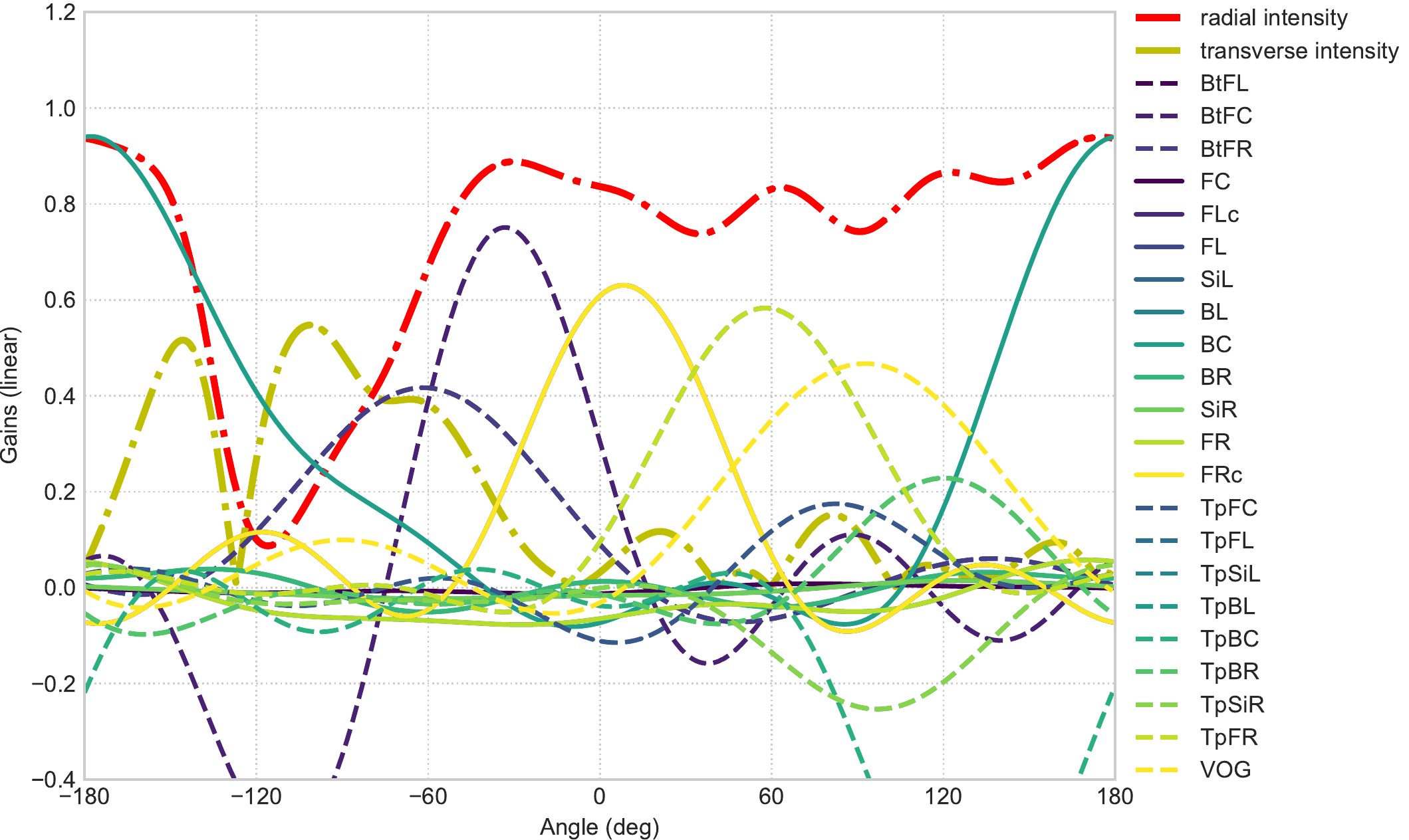}
        \caption{Vertical panning for Ambisonics at order 3 decoded to an Hamasaki 22.2 layout.}
        \label{fig:22.2-ambi3-v}
    \end{minipage}
    \begin{minipage}[t]{\textwidth}
        \includegraphics[width=\textwidth]{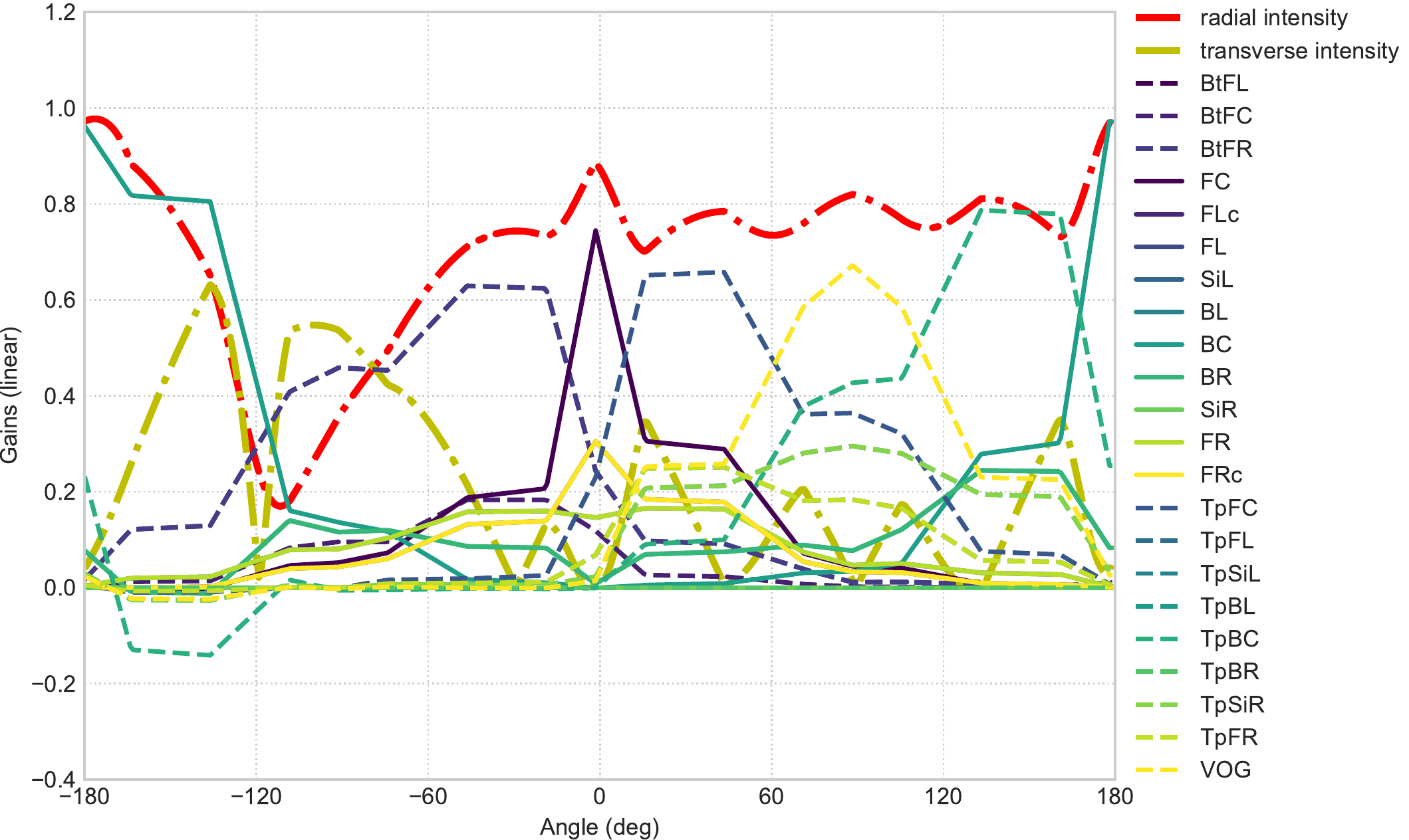}
        \caption{Vertical panning for OPT-SWF at level 1, decoded to an Hamasaki 22.2 layout.}
        \label{fig:22.2-optswf-v}
    \end{minipage}
\end{figure}

A relevant difference that is not shown in the horizontal plots is
the actual difficulty to generate an Ambisonics decoding with very limited negative gains.
The tweaking of parameters for the minimization parameters is non trivial,
especially for layouts that cover only partially the surface of the sphere.
This is probably one of the reasons why the vast majority of commercial tools to decode Ambisonics
typically provide static matrices\footnote{The only commercial tool we managed to find that offers
the possibility of generating custom decoders for irregular layouts is the
Rapture3D ``Advanced'' edition software by Blue Ripple Sound~\cite{rapture3d}.}.  %
To stress this point, we report the same plots, gains, energy and intensity, for the vertical plane.

In Figures~\ref{fig:22.2-ambi3-v} and~\ref{fig:22.2-optswf-v}  we report the panning functions for a panning around the vertical plane
for Ambisonics at order 3 and OPT-SWF at level 1 decoded to the Hamasaki 22.0 layout described.
It is apparent from Figure~\ref{fig:22.2-ambi3-v} that Ambisonics struggles to maintain positive gains.
For OPT-SWF essentially any decoder produced will have very limited negative gains by construction.
To try to not compensate for the areas without loudspeakers, i.e.~the lower half of the sphere, we implemented in IDHOA
a mechanism to reduce the contribution of these areas to the value of the cost function,
and so their contribution to the overall optimization.
This is the motivation for the drop in the reconstructed energy between the $180^\circ$ and $315^\circ$ elevation, see Figure~\ref{fig:energy-22-v}.
The same happens for the radial intensity, Figure~\ref{fig:intensity-22-v}.

Both Ambisonics and the SWF implementation we show here have difficulties dealing with irregularly spaced speakers.
It is quite evident in the 22.2 layout, since the layout is hemispherical and the frontal area is more dense in speakers than the rear zone.
The Hamasaki distribution of speakers is more concentrated where the human perception accuracy is greater: in the horizontal plane
and especially in the frontal area.
Both Ambisonics and this implementation of SWF treat all directions equally: Ambisonics because of the support of the SH
and SWF because our mesh has (almost) equispaced points on the sphere.
The improvement that SWF brings is the easiness of producing a decoding with positive gains.
Another implementation of SWF with an irregular sampling by design (irregular mesh) could deal with irregular layouts natively.

\section{Summary}
The SWF implementation described in Chapter~\ref{sec:swf-implementation} is designed
to be compared directly with Ambisonics: covers the full sphere, has equispaced sampling points that correspond to
one of the platonic solids (octahedron or 3-design) to which Ambisonics has closed form decoding equations.
In this Chapter we compared the decoding of SWF and Ambisonics to two irregular layouts of speakers.
We summarize this comparison with a list of
advantages and disadvantages of SWF decoding over Ambisonics decoding:
\begin{itemize}
    \item[+] More control of negative gains.
             Implies bigger sweet spot and more robust imaging.
    \item[+] More predictable fine tuning.
             Implies that we can push the decoder in the \emph{smooth} or \emph{focus} direction without adding significant out-of-phase contributions.
    \item[+] Possibly more directional for a similar number of channels (for the \emph{focus} decoder).
    \item[-] The energy reconstruction is more irregular.
             It is the price to pay for not having negative gains and preserving pressure during encoding.
    \item[-] SWF at level 1 is mostly equivalent to Ambisonics at order 3.
             Because the construction of this specific SFW is made along the lines of Ambisonics,
             the two techniques are not significantly different.
\end{itemize}

If we look at the performance in terms of radial intensity, the two formats perfom similarly for an irregular array of speakers.
Ambisonics stands out for the smooth energy reconstruction, while SWF for the absence of negative gains and decoding flexibility.

\chapter{Conclusions and Future Work} \label{ch:results-futurework}

\section{Conclusions and Discussion}
In this thesis we have defined a new generic framework for spatial audio encodings based on wavelet filters.
We have described the complete audio workflow that makes use of this new tool.
Then, we have particularized the framework to the spherical case for a specific mesh construction,
resulting in a practical realization: the spherical wavelet format (SWF).
Similarly to Ambisonics, this format is channel-agnostic.
Unlike Ambisonics, in the case of SWF the signals that compose the format have a particular spatial localization.
On the encoding side of the audio chain,
we have devised a numerical method for wavelets optimization (with short filter length),
enabling the creation of a possibly infinite set of core filters.
On the decoding side of the audio chain,
we have built and made publicly available\footnote{The final version of IDHOA used in this work will be made public soon.
Previous versions are already available here \cite{github-idhoa} and here \cite{github-idhoa-new}.}
a universal decoding method, based on the numerical optimization of some psychoacoustical observables.

\paragraph{SWF}
The objective of this thesis was to build a new channel agnostic format, that is homogeneous and coherent,
but also has good localization with few channels, easily handles irregular layouts
and holds well when moving out of the sweet spot.
We have depicted the full audio chain: encoding to mesh and downsampling, transmission, upsampling and decoding to speakers.
The new format is effectively channel agnostic, there is no reference to the destination layout in the definition of the format.
The homogeneity and coherence characteristics need a more detailed discussion.

If we look at homogeneity of a particular format, we should distinguish between the format homogeneity,
and whether this homogeneity is retained during decoding.
SWF is homogeneous in the specific implementation defined in the thesis, since the mesh over which the format is defined is in fact homogeneous.
Nevertheless, it could be perfectly possible to define a wavelet format with a non-homogeneous mesh and obtain a non-homogeneous format.
Ambisonics is by construction homogeneous, and the decoders for regular layouts are also (typically) homogeneous.
However, when decoding to irregular layouts both Ambisonics and SWF do not assure that the resulting decoding will be homogeneous.
Actually, the best decoders for irregular layouts are not homogeneous,
like the ones produced by IDHOA and presented in this thesis.
In this case SWF and Ambisonics are no different.
The coherence follows a similar train of thought, both SWF and Ambisonics can be coherent in special conditions.
SWF has indeed good localization with few channels and, thanks to the extremely limited negative gains,
holds well when moving out of the sweet spot, like an amplitude panner does.
It has been demonstrated that SWF behaves well when decoding to layouts that are irregular (in the SWF sense)
and with the help of IDHOA it is possible to generate meaningful decoders in a matter of minutes.

We have explored three variations of a particular incarnation of this format.
In both cases
wavelets were defined over a spherical mesh,
created from a primitive solid (an octahedron)
using a Loop subdivision scheme.
In the first variation, the wavelet family is implicitly defined by the VBAP panning rule.
In the second variation, we use an off-the-shelf wavelet family, called interpolating wavelet.
In the third variation, the wavelets were optimized numerically by a brute-force method.
The three methods generate audio formats that have very similar characteristics in terms of energy and intensity reconstruction.
The main differences lay in the shape of the panning functions and in the behaviour of the upsampling matrix, $\mathbf{P}$.
It is to be noted that the three examples explored do not necessarily represent the best possible realizations.
One of the virtues of SWF is precisely the ability to adapt to many different situations.
We believe that there is no such thing as the best possible SWF format, but rather it depends on the particular context and goal.
One of the drawbacks of SWF with respect to Ambisonics is that the acoustical and perceptual interpretation of the format
in terms of pressure and velocity is lost in general (we still retain the notion of the global pressure by a careful wavelet design).
In this context, it is key to have an acoustically and perceptually motivated decoder that can reinstate
the missing physical and perceptual observables.

A three-element comparison, OPT-SWF (using an optimized wavelet),  VBAP-SWF (trivial wavelet from VBAP) and Ambisonics,
has been carried out for two different speakers layouts.
Observations from reconstructed signals, reconstructed energy and intensity indicate that SWF is a format
that, depending on the decoding, can fit between an amplitude panner and Ambisonics.
It has localization characteristics similar to (or in some cases better than) Ambisonics,
with greater control on the negative gains.
Informal listening tests confirm these characteristics.

In our experience, the difference between the two variations of the wavelet format explored are relatively minor when evaluated in terms of the decoding results.
We noticed that final results depend only slightly on the wavelet family as long as this family has been designed with reasonable characteristics.
A possible explanation is that the IDHOA decoding minimizes any possible intrinsic differences between the different encodings.
Also, notice that we have only explored meshes of relatively low order.
It is possible that differences become more apparent when going to higher order meshes, since the filtering effects are cumulative.
Additionally, besides the decoding characteristics, other filter design properties, e.g.\ encoding performance,
can be considered when designing and evaluating the wavelet families.
Anyway, we expect that the different characteristics of the wavelet families will be more evident
when using custom subdivision meshes that represent directly standard speaker positions.
When building custom meshes, the requirement for a spherical format could be lifted,
and we could define a format for meshes with a non-spherical topology (e.g.\ half dome).

As a related remark, notice that the only filter which is strictly essential in our framework is the analysis filter $\mathbf A$,
given that the decoding to speakers is computed separately with IDHOA via a numerical optimization.
However, we still believe that it is important to have a complete wavelet representation.
When optimizing the filters it is important to optimize at the same time the analysis and synthesis filters $\mathbf A$ and $\mathbf P$
to ensure that the wavelet transform can have a well behaved reconstruction.
This fact can later on ease the task of decoding to speakers performed by IDHOA.
Besides, on practical grounds, having a complete wavelet construction can be useful to be able to manipulate the spatial signal;
on theoretical grounds, the wavelet construction is key to understand what is left out by the truncation of SWF,
in the spirit of the sampling theorem, something we leave for future work.

Overall, SWF's encoding, transmission and decoding flexibility and rendering performance make it an interesting family of formats
to explore in real-life conditions.

\paragraph{IDHOA}
A fundamental piece for the new format, and also for the comparison with the reference technology Ambisonics,
is the stage of decoding to a real world layout.
One of the main outputs of the work is the formulation and implementation of the IDHOA decoder.
While initially oriented to solve the problem of decoding Ambisonics to irregular layouts (to date still relevant),
we developed an algorithm that, leveraging psychoacoustic criteria, can generate a decoding matrix for
any linear encoding format, as long as this encoding format allows encoding a point source to a set of directions on the sphere.
We have described to applications of IDHOA to Ambisonics and SWF, but it could be applied to any other format,
e.g.~using a specific multichannel layout as intermediate audio format and decoding it to any other layout.
The main novelty factor is the separation of intensity vectors in radial and tangential components,
making it possible to optimize the two components separately.

Often, in the past literature, there has been an excessive stress, in our opinion,
on reducing the tangential component of velocity and intensity.
This limits the possibility of the radial part to reach relevant values, thus effectively broadening the audio sources
and making them difficult to localize in space.
In the same line of thought, forcing the velocity or intensity vector to have the same value along the area covered by loudspeakers,
which is the homogeneity characteristic of Ambisonics, is not a good requirement for irregular layouts.
We think that separating the velocity and intensity into their radial and tangent part, and trying to maximize
the radial component without trying to make it uniform, while minimizing the tangential one without requiring it to be strictly zero,
generates decodings that are in practice much superior to the commonly available ones.
Along with the IDHOA code we will publish all the decoding matrices used in this thesis to decode Ambisonics to the mentioned layouts.

One final note, we believe that adopting this factorization in components for the velocity and intensity vectors
should become a common practice among the researchers in the area of Ambisonics decoding
especially for reporting the results for different decoders and technologies.
Typically the results are reported with obscure sphere projections and reporting only the modulus of the vectors.
We think that it would be useful to adopt a common format for reporting,
that has proven to be very immediate to relate to the actual listening experience.

\section{Future Work}
Reach out to the community disseminating this work is indeed our first priority.
From there, we hope to get some feedback and shape the upcoming goals.
Ideally, it would be interesting to design a more industry-oriented format,
that does not necessarily need to compare directly with Ambisonics.
Formal listening tests would be required to validate the approach.

Along the line of searching for a more industry-oriented format and in the spirit of (compressing) wavelets,
whose philosophy is ``to model a function, use a function similar to the function you want to model'',
it would be interesting to experiment with meshes similar to the destination speakers' layout.
It should be quite straightforward to test this concept on any mesh using a VBAP-SWF,
while generating an optimized wavelet format requires more work and fine-tuning.

Given the flexibility of our construction,
a possibly interesting application would be using SWF as a spatial encoder for other formats.
For example, it is possible to decode Ambisonics to a SWF mesh with a basic decoding (pressure preserving),
then manipulate the signals via the described SWF tools (if needed),
and decode perceptually with IDHOA to the destination layout.
Note that the spatial operations like rotations, zoom or spatial deformation are quite trivial over the $\mathbf c$ signals,
since they have a spatial meaning in the three-dimensional space, and the usual transformations apply.
Another example, SWF could be used as a transport format for object-based formats to reduce the number of transmitted audio files.
The object based format could be encoded to SWF, manipulated and transmitted, and decoded to the destination loudspeakers' layout via IDHOA.

After format and specific wavelet definition it would be interesting to
understand what is left out by the truncation of SWF, i.e.\ the meaning of the `details',
in the spirit of the sampling theorem.

    \cleardoublepage
    \phantomsection
    \addcontentsline{toc}{chapter}{Bibliography}

    \bibliography{main}
    \bibliographystyle{apalike}

\appendix

\chapter{Minimization Problems in Python via IPOPT and PyTorch} \label{app:ipopt+pytorch}
Both the optimization of the decoding cost function and the wavelets one
are the result of several years trying and experimenting with different technologies.
We feel that it is relevant to leave in writing if not the whole trajectory,
at least the sketch of the final implementation.
During the last 7 years we tried NLopt~\cite{nlopt}, IPOPT~\cite{ipopt} as minimization libraries,
and several libraries for auto-differentiation, namely algopy~\cite{Walter2011algopy}, Theano~\cite{2016arXiv160502688short},
TensorFlow~\cite{tensorflow2015-whitepaper} and PyTorch~\cite{paszke2017automatic}.
The final implementation uses IPOPT over NLopt, and PyTorch over the rest of mentioned libraries.
This combination is the one that proved to have the fastest execution times, ease of debugging
and programming flexibility.

\section{IPOPT Minimization Library}
IPOPT (Interior Point OPTimizer) is an open-source software package for large non-linear optimization.
It can be used to solve general nonlinear programming problems, including arbitrary constraints.
The software itself is written in C++ but has several native or contributed APIs for other languages.
In our case, we wanted to interface with Python and we used PyIpopt~\cite{pyipopt} as the Python API for IPOPT.
IPOPT requires the computation of the Jacobian of the cost function and the constraints.
If provided, IPOPT uses also the Hessian of both cost function and constraints, otherwise it will internally calculate it numerically.

The main obstacle is then calculating the first and second derivatives.
IPOPT just requires the numerical value of the derivatives, so the choice of the method is left to the user.

\section{Calculating Derivatives}
There are essentially three methods available, with their benefits and drawbacks.
We will list them schematically in the following:
\begin{enumerate}
    \item Analytic differentiation: derivatives are computed and implemented once, by hand or with the help of some computer algebra software.
    A short list of pros and cons for this technique is:
    \begin{itemize}
        \item[+] Exact derivatives, the numerical evaluation is fast.
        \item[-] The process of (manually) calculating the derivatives is time consuming, the implementation can be very complicated.
        \item[-] Not flexible: the derivatives have to be recalculated at every change in cost function or constraints.
    \end{itemize}

    \item Numerical differentiation: approximate the derivatives by finite differences.
    A short list of advantages and disadvanteges for this technique are:
    \begin{itemize}
        \item[+] Can be always calculated, even when the cost function does not have a closed analytic expression.
        \item[-] Approximation errors arise (round-off and truncation) and they do accumulate.
        \item[-] The evaluation can be really slow.
    \end{itemize}

    \item Automatic differentiation: numerically evaluates the exact derivative of a computer function with the help of a specific library.
    Some pros and cons of automatic differentiation are:
    \begin{itemize}
        \item[+] Exact derivatives, once calculated the numerical evaluation is fast.
        \item[+] Very flexible: since the derivatives are exact and calculated automatically, it is possible to experiment with cost function and constraints.
        \item[-] Can be calculated only if the cost function or constraints can be expressed in terms of operations whose derivative is known by the library.
        \item[-] Needs some adaptation of the algorithms and introduces an external dependency.
        \item[-] Can be slow or fast depending on the specific library.
    \end{itemize}
\end{enumerate}

We chose the analytic differentiation, since we want the flexibility to experiment with different cost functions and constraints.
In the recent years packages for automatic differentiation have evolved dramatically thanks to the rise of deep learning in artificial intelligence.

\subsection{Automatic Differentiation Packages}
Generally, the automatic differentiation packages are built on the fact that the
the derivative of any expression can be computed using the chain rule.
Applying several times the chain rule, the composite expression is broken
in elementary operations and functions whose derivatives are known.
This way, the derivative of the initial function is computed algorithmically in a finite number of steps.
The automatic differentiation software has to build an internal representation of the derivative of the function, which is called computational graph.
This computational graph can be built statically (first build the graph, then compile it) of dynamically (the graph is built at execution time).
Different libraries tend to use different methods, even if lately this difference is gradually being relaxed.

We initially started our research using Theano, but we hit a wall when starting to use sparse matrices (for reducing the problem's dimensionality)
and Theano team announced that they would cease the development (3 October 2017).
We then moved to TensorFlow, that, at the time, built the computational graph only statically.
A part from the inherent difficulty of debugging (cryptic error messages that are related to internal graph and not the actual code),
the code results difficult to reuse for interfacing with IPOPT.
Ideally the same function that outputs numerical values for IPOT should be
the input to the automatic differentiation algorithm.
We found the interaction with IPOPT much more neat to handle using PyTorch.

\subsection{Examples of Automatic Differentiation in TensorFlow and PyTorch}
In the following, we will present a simple example showing how to calculate the derivative of a function
in the static paradigm, with TensorFlow, and in the dynamic paradigm, with PyTorch.
Starting with TensorFlow:
\begin{minted}[mathescape,
               linenos,
               numbersep=5pt,
               frame=lines,
               framesep=2mm,
                breaklines, fontsize=\footnotesize]{Python}
# derivatives of a function in tensorflow

import tensorflow as tf

# get a number from terminal
print("type a number and press enter")
point = input()
point = float(point)
data = tf.placeholder(dtype=tf.float32, shape=())

# the function you want to calculate the gradient
def function(indata):
    square = tf.pow(indata, 2)
    return square

# start a TensorFlow session
# (you need it to evaluate numerically the symbolic expressions)
sess = tf.Session()

# TF symbolic version of function
tf_function = function(data)
# value of function in your point
value_function = tf_function.eval(feed_dict={data: point}, session=sess)
print("Function value (the value you entered squared): %

# calculate the symbolic gradient of function
grad_function = tf.gradients(tf_function, [data])[0]
# evaluate the gradient in your point
value_grad_funct = grad_function.eval(feed_dict={data: point}, session=sess)
print("Derivative of the function's value (twice the value you entered): %
\end{minted}
The output of the code is:
\begin{minted}[linenos, breaklines, fontsize=\footnotesize]{text}
type a number and press enter
3
Function value (the value you entered squared): 9.000
Derivative of the function's value (twice the value you entered): 6.000
\end{minted}

While with PyTorch:
\begin{minted}[mathescape,
               linenos,
               numbersep=5pt,
               frame=lines,
               framesep=2mm,
                breaklines, fontsize=\footnotesize]{Python}
# derivatives of a function in pytorch

import torch

# get a number from terminal
print("type a number and press enter")
point = input()
point = float(point)
data = torch.tensor(point, requires_grad=True, dtype=torch.float32)

# the function you want to calculate the gradient
def function(indata):
    square = torch.pow(indata, 2)
    return square

# value of function in your point (it's a torch object!)
value_function = function(data)
print("Function value (the value you entered squared): %

# calculate the symbolic gradient of function in your point
value_function.backward()
# get the value of the gradient in your point
value_grad_funct = data.grad.numpy()
print("Derivative of the function's value (twice the value you entered): %
\end{minted}
The output is obviously the same as in the previous formulation.
Even if we tried to maintain the same steps, it is apparent than in the PyTorch case the torch objects and the numerical values
of those objects are carried together.
This makes much easier the (numerical) debugging.

To scale this simple example to matrices and complex operations is trivial.
Nevertheless there is a small detail missing that is very relevant in our context.
When scaling the problem of calculating the derivatives of a function that accepts as input a matrix,
all the libraries for automatic differentiation used in deep learning return the sum of the derivatives,
and not the full Jacobian.
For this reason we have to calculate a derivative for each component of the input matrix, which is called \emph{stride}.
Fortunately, there is a small library that does exactly this striding for us~\cite{hessian-pytorch}.
Calculating Jacobians and Hessians becomes extremely easy, as an example:
\begin{minted}[mathescape,
               linenos,
               numbersep=5pt,
               frame=lines,
               framesep=2mm,
                breaklines, fontsize=\footnotesize]{Python}
# jacobian and hessian of a function in tensorflow

import torch
import hessian as h

# define two variables
x = torch.tensor([1.5, 2.5], requires_grad=True)
y = torch.tensor([5.5, -4.], requires_grad=True)

# define the function
function = x.pow(y)
# calculate its jacobian
jac = h.jacobian(function, [x, y])
# print result
print("Jacobian")
print(jac)

# define the function
function = x.pow(2).prod().sum()
hes = h.hessian(function, x)
# print result
print("Hessian")
print(hes)
\end{minted}
The output is:
\begin{minted}[linenos, breaklines, fontsize=\footnotesize]{text}
Jacobian
tensor([[ 3.4101e+01, -0.0000e+00,  3.7710e+00,  0.0000e+00],
        [ 0.0000e+00, -4.0960e-02,  0.0000e+00,  2.3457e-02]])
Hessian
tensor([[12.5000, 15.0000],
        [15.0000,  4.5000]])
\end{minted}

Having the autodifferentiation machinery sorted out, we can use PyIpopt~\cite{pyipopt} examples to start building
a minimization using IPOPT as a minimization library.

\chapter{Strategies for Problem Dimensionality Reduction} \label{app:dof-reduction}

We have seen in Chapter~\ref{ch:our-wavelets} how the minimization problem for the wavelet filters is built, the cost functions we defined, and their constraints.
In this Appendix we will analyze more in detail the critical implementation aspects of the wavelet optimization.
Without loss of generality, we will focus on the first step of the minimization, where $\mathbf A$ and $\mathbf P$ are optimized together.

In the first step the unknowns are the whole matrices $\mathbf A$ and $\mathbf P$.
The matrices connect two levels of the subdivision, so the number of elements in each of these two matrices is
\emph{(number of mesh points at level $n$)} $\times$ \emph{(number of mesh points at level $n+1$)}.
To give some numbers, and referring to the mesh used for SWF (e.g.~see Chapter~\ref{ch:comparison-swf-internals}),
the $\mathbf A^1$ and $\mathbf P^1$ have $ 6 \times 18 = 108 $ each, and the number of unknowns would be then $108 \times 2$.
$\mathbf A^2$ has $18 \times 66 = 1188$ elements, and $\mathbf A^3$ has $66 \times 258 = 17028$.
The dimensionality of the Jacobian and Hessian of the cost function grows with the growing number of variables:
linearly for the Jacobian and quadratically for the Hessian.
To this count, we have to add the constraints, in the case of the first step we have only $\mathbf{A}^j \mathbf{P}^j = \mathbf{1}$ (see Chapter~\ref{ch:our-wavelets}),
their Jacobian and Hessian.
In the case of the already mentioned SWF, we have $6^2=36$ constraints at $j=1$, $18^2=324$ at $j=2$ and so on.
Even at small dimension this brute-force approach, where all the elements of $\mathbf A$ and $\mathbf P$ are considered
(to some extent) independent unknowns, tends to blow up quickly.

\section{Loop Spherical Subdivision Symmetries and Reduction of Degrees of Freedom}
A part from the constraints we impose, $\mathbf{A}^j \mathbf{P}^j = \mathbf{1}$,
we do not impose any particular symmetry and we treat the points of the mesh as if they were independent.
Nevertheless, the mesh has its inherent symmetry which is given by the symmetry of the solid that we chose as initial mesh.
The idea is then to explicitly impose the (original) mesh symmetry and get an effective reduction of the number of unknowns of the problem,
resulting in a global reduction in dimensionality: less unknowns, less constraints, less derivatives.

As already discussed, there are approaches where the optimization affects the prediction and update operators alone and not the full refinement matrices.
In this case we wanted complete flexibility and possibly avoid the fractal shapes introduced by the recursion in the
lifting together with the dual-lifting (see Subsection~\ref{subs:dual-lifting}, and especially Eq.~\eqref{eq:lifting-from-lazy}.

\begin{figure}
    \centering
    \begin{minipage}[t]{0.46\textwidth}
        \centering
        \includegraphics[width=0.8\textwidth]{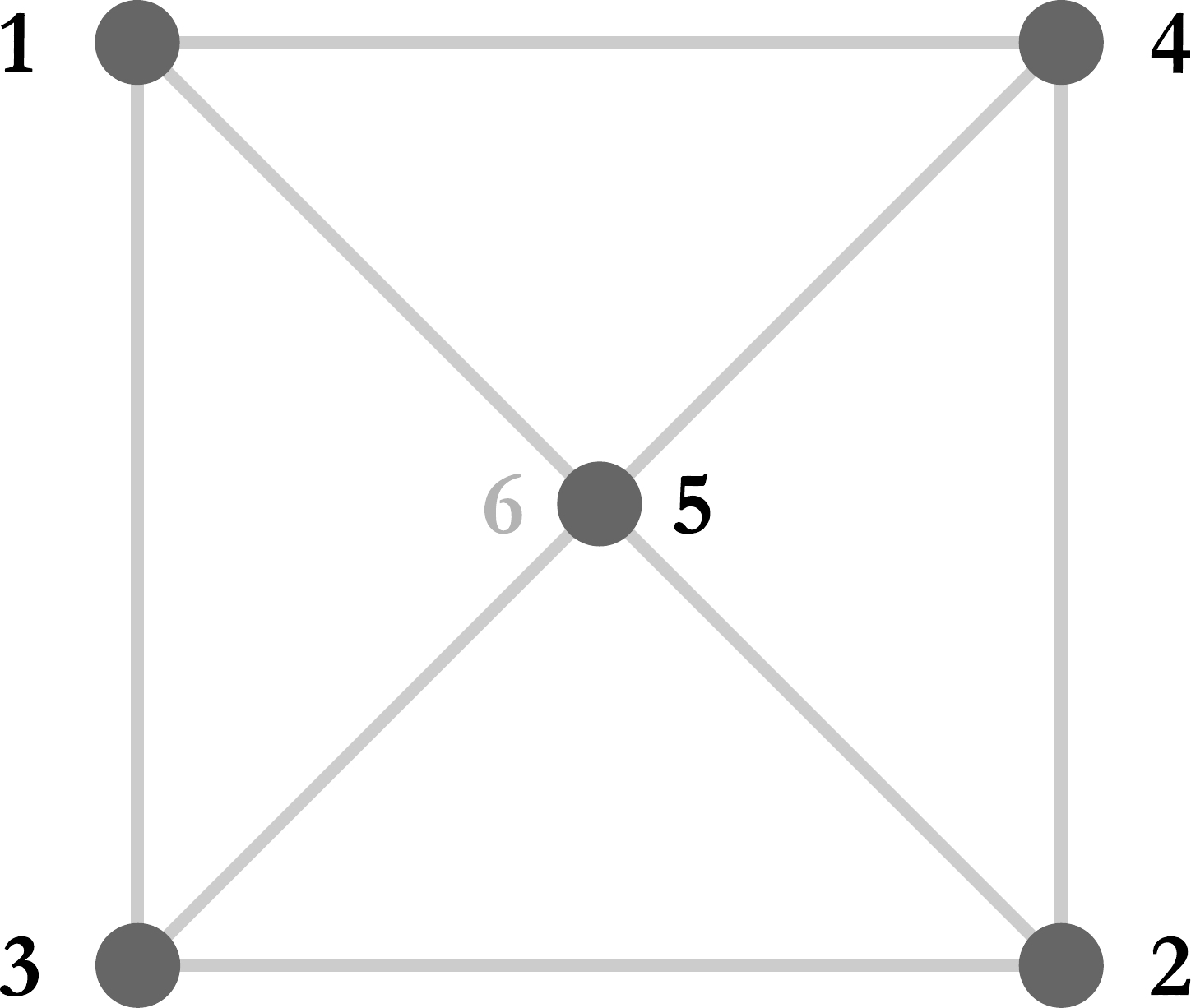}
        \caption{View of the original octahedron from above.
            The visible vertices (dots) are numbered in solid black, the hidden vertex (6) is numbered in gray.
            The original mesh has 6 vertices.}\label{fig:octa-l0}
    \end{minipage}
    \hfill
    \begin{minipage}[t]{0.46\textwidth}
        \centering
        \includegraphics[width=0.8\textwidth]{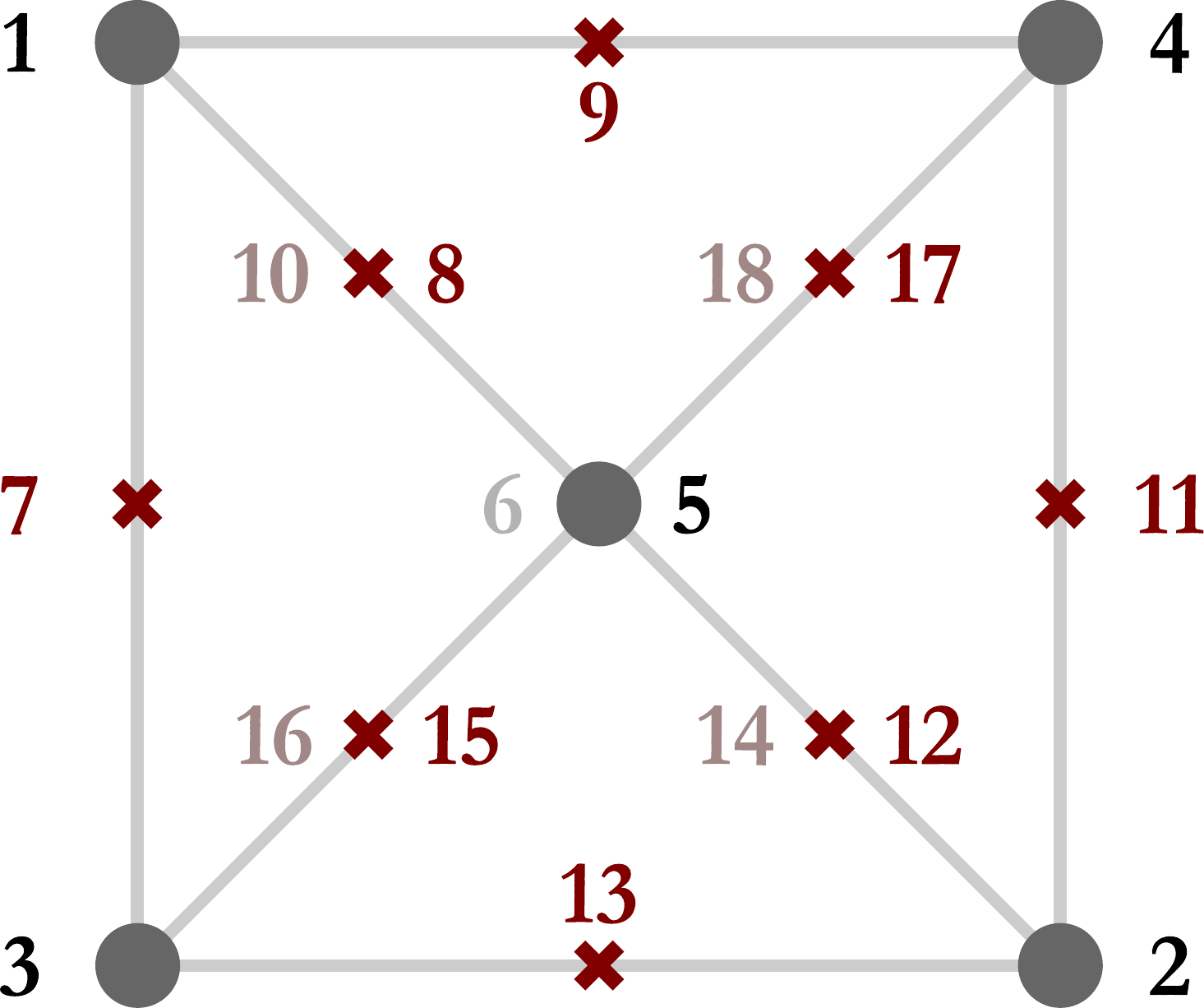}
        \caption{View of the original octahedron with the first Loop subdivision.
            The original vertices are the dots, while the crosses represent the new verices produced by the Loop subdivision.
            The new visible vertices are numbered in solid dark red, while the remaining hidden ones are numbered in light red.
            The new mesh has in total 18 vertices.}\label{fig:octa-l1}
    \end{minipage}
\end{figure}

With the help of a couple of figures we will illustrate the concept behind the reduction of the unknowns.
In Figure~\ref{fig:octa-l0} we report the octahedral mesh (level 0) as seen from above.
The visible vertices are the gray circles numbered in solid black $\set{ 1, 2, 3, 4, 5 }$, the first four are positioned in the horizontal plane,
while the number 5 is the vertex at the top.
The vertex number 6 is the vertex at the bottom and is hidden in this figure, for this reason is indicated with a gray number.
If we apply the Loop subdivision to this initial mesh, we obtain the mesh (at level 1) reported in Figure~\ref{fig:octa-l1}
(the exact numbering of the vertices might not be the one reported in the Figure, but it is not relevant in this context).
We report with a dark red cross the new visible vertices, accompanied with their numbering.
The vertices hidden by this type of projection are indicated in light red.
The vertices added at this level are $\splitatcommas{ \set{ 7, 8, 9, 10, 11, 12, 13, 14, 15, 16, 17, 18 } }$.

Now, if we take the vertex 1 at level 0 and search for its immediate neighbours at level 1, we will find the four vertices
$\left\{ 7, 8, 9, 10 \right\}$, so this vertex has valence\footnote{The vertices at level 1 will have valence 6,
when searching for their immediate neighbours at level 2.
The Loop subdivision is a subdivision with valence 6.} 4.
We can schematically indicate this concept as:
\begin{center}
\begin{tabular}{rcl}
    level 0 & has neighbours & level 1 \\
    1 & $\rightarrow$ & $\left\{ 7, 8, 9, 10 \right\}$
\end{tabular}
\end{center}

And we can write the same for the remaining vertices:
\begin{center}
\begin{tabular}{rcl}
    2 & $\rightarrow$ & $\left\{ 11, 12, 13, 14 \right\}$ \\
    3 & $\rightarrow$ & $\left\{ 13, 15, 7, 16  \right\}$ \\
    4 & $\rightarrow$ & $\left\{ 9, 17, 11, 18  \right\}$ \\
    5 & $\rightarrow$ & $\left\{ 8, 12, 15, 17  \right\}$ \\
    6 & $\rightarrow$ & $\left\{ 10, 14, 16, 18 \right\}$ \\
\end{tabular}
\end{center}

Since the original mesh has rotational symmetry, we can build a function that maps each of the $\set{ 2, 3, 4, 5, 6 }$
to the vertex number 1, and that maps their neighbours to 1's neighbours.
Graphically, for the vertex 2:
\begin{center}
\begin{tabular}{rcl}
    2 & $\rightarrow$ & $\left\{ 11, 12, 13, 14 \right\}$ \\
    $\downarrow$ &   & \multicolumn{1}{c}{$\downarrow$} \\
    1 & $\rightarrow$ & $\left\{ 7, 8, 9, 10 \right\}$
\end{tabular}
\end{center}
With this symmetry we reduce by 6 the number of unknowns (at the level of this example).
Essentially the six rows of $\mathbf{A}$ (and the columns of $\mathbf{P}$) are shifted copies of each other.

Moreover, we can further reduce the dimensionality imposing left/right and up/down symmetries, this way the free parameters represented by the neighbours effectively reduce
from 4 (or 6 in the next levels) to 1.
The same concept extends to further neighbours.
The neighbours are grouped by their distance from the original vertices.
Typically we require a new parameter per each group of neighbours.
With this symmetry we reduce the number of parameters along the columns of $\mathbf{A}$ (and rows of $\mathbf{P}$).
The reduction factor in this case depends on the definition of the neighbour groups.
With this method we obtain a neighbour structure for each matrix, we will call them $\mathbf{A}_\text{stru}$ and $\mathbf{P}_\text{stru}$.

This approach can be obviously made recursive along the different levels of the subdivision.

Introducing these symmetries we reduce considerably the number of independent parameters in the problem.
The actual matrices $\mathbf A$ and $\mathbf P$ can be reduced to a list of degrees of freedom to be fed to the minimization algorithm.
We can design two functions: one that reduces the two matrices to a vector (downscale)
and one that recovers the full matrices from the unknowns vector (upscale).

As a consequence of this reduction of degrees of freedom, we face two challenges:
\begin{itemize}
    \item Implementation challenge: the minimization algorithm sees only the independent parameters that now are reduced in number,
        but to calculate the cost function needs the full matrices are needed.
        The jacobian and hessian of the cost function have to be calculated only with respect to the independent parameters.
        How does this fit into the automatic differentiation implementation?
    \item The number of constraints reduces in a non-trivial manner: the constraints involve a matrix product of the full matrices.
        We have to figure out which constraints are effectively independent after the reduction of degrees of freedom.
\end{itemize}
In the Sections~\ref{sec:upscale_downscale_pytorch} and \ref{sec:constraints_reduction} we will illustrate our approach to these problems.

\section{Implementation of DOF Reduction Techniques Inside the Automatic Differentiation} \label{sec:upscale_downscale_pytorch}
In this Section we will show how to calculate the automatic derivatives of a function, with respect to its `true variables'.
In the following code example, we define our vector of unknowns as $\mathbf x$, that has dimension 3.
The \texttt{function} mimics the cost function of our minimization problem, in a much more simple fashion.
The first operation performed inside the \texttt{function} is to grow the vector $\mathbf x$ into two matrices,
\texttt{upscaled\_A} and \texttt{upscaled\_B}.
Then the two matrices are multiplied together (@) and an identity matrix (\texttt{torch.eye}) is subtracted.
The result of these operations is summed up to obtain a scalar, as for any typical cost function.
The \texttt{function} returns the value of the cost function and $\mathbf x$, that somehow, in disguise, is gone through all the operations described.
We then can take the derivative of the function and see if PyTorch is able to correctly calculate the derivatives of this function
with respect to $\mathbf x$.

\begin{minted}[mathescape,
               linenos,
               numbersep=5pt,
               frame=lines,
               framesep=2mm,
                breaklines, fontsize=\footnotesize]{Python}
# upscaling of variables
# and calculate derivatives of a function incorporating up/downscaling

import torch
import hessian as h

# define two variables
x = torch.tensor([1.5, 2.5, -5.5], requires_grad=True, dtype=torch.float32)

def function(x):
    # upscale tensor
    rows = 2
    cols = 2
    upscaled_A = torch.zeros(rows, cols, requires_grad=True, dtype=torch.float32)
    upscaled_B = torch.zeros(rows, cols, requires_grad=True, dtype=torch.float32)

    upscaled_A[0] = x[[0, 1]]
    upscaled_A[1] = x[[1, 0]]
    print("Matrix A:")
    print(upscaled_A)

    upscaled_B[0] = x[[2, 1]]
    upscaled_B[1] = x[[1, 2]]
    print("Matrix B:")
    print(upscaled_B)

    cost = upscaled_A @ upscaled_B - torch.eye(rows)
    cost = torch.sum(cost)
    return cost, x

f_torch, x = function(x)
print("Function value:")
print(f_torch.data)

# calculate its jacobian
jac = h.jacobian(f_torch, [x])
# print result
print("Jacobian")
print(jac)
\end{minted}
Gives this output:
\begin{minted}[linenos, breaklines, fontsize=\footnotesize]{text}
Matrix A:
tensor([[1.5000, 2.5000],
        [2.5000, 1.5000]], grad_fn=<CopySlices>)

Matrix B:
tensor([[-5.5000,  2.5000],
        [ 2.5000, -5.5000]], grad_fn=<CopySlices>)

Function value:
tensor(-26.)

Jacobian
tensor([[-6.,  2.,  8.]])
\end{minted}
It is important to note that the Jacobian has the correct dimension, having $\mathbf x$ dimension 3,
the Jacobian will be of dimension 3 as well:
\begin{equation*}
Jf(\mathbf x) = ( \partial f / \partial x_1, \partial f / \partial x_2, \partial f / \partial x_3 )
\end{equation*}
and it is exactly what we get.

This example is quite simple, and it might look trivial, but it is not.
The non-trivial part is the ``upscaling'', where $\mathbf x$ is grown into two different matrices.
PyTorch is able to propagate the derivatives through this operation, which is not one of the elementary operations we mentioned in Appendix~\ref{app:ipopt+pytorch}.

The same method works for the derivatives of the constraints, that are defined in a very similar way, e.g.\ $ \mathbf{AP} = \mathbf{1} $.
The challenge with the constraints is to identify the independent constraints, now that the number of DOFs has been reduced dramatically,
i.e.\ not all the equations produced by $ \mathbf{AP} = \mathbf{1} $ are linearly independent.

\section{Method for Reduction of Constraints} \label{sec:constraints_reduction}
From the reduction of variables (just described), we obtain a neighbour structure for each matrix.
With this structure we can calculate the constraints and then isolate the independent ones.

As an example, we will take the first stage of the minimization.
The constraints for this stage are the ones defined by $ \mathbf{AP} - \mathbf{1} = \mathbf{0}$, as already mentioned.
Having this neighbour structure, we rewrite the constraints as $ \mathbf{A}_\text{stru} \mathbf{P}_\text{stru} - \mathbf{1} = \mathbf{0}$.
This equation defines a set of non-independent equations.
We want to figure out which are the independent equations, that are our remaining constraints after the reduction of variables.
Operatively, we put this linear system of equations in matrix representation and use the reduced row-echelon form~\cite{rref}
to identify the independent constraints.
This procedure gives us the independent elements of the system of equations that defines the constraints.
This information is also propagated to the calculation of the Jacobian and Hessian of the constraints,
limiting considerably the amount of derivatives to calculate.

    \backmatter

    \printindex

\end{document}